\documentclass[a4paper,12pt,oneside]{book}

\usepackage{latexsym,amsmath,amsfonts,amssymb,amsopn,mathdots,graphicx}
\usepackage{thesis}		
\title{Aspects of Conformal and Superconformal Field Theory}
\author{Christopher J.~Rayson}
\college{Gonville \& Caius College}
\university{University of Cambridge}
\crest{\includegraphics[width=4cm]{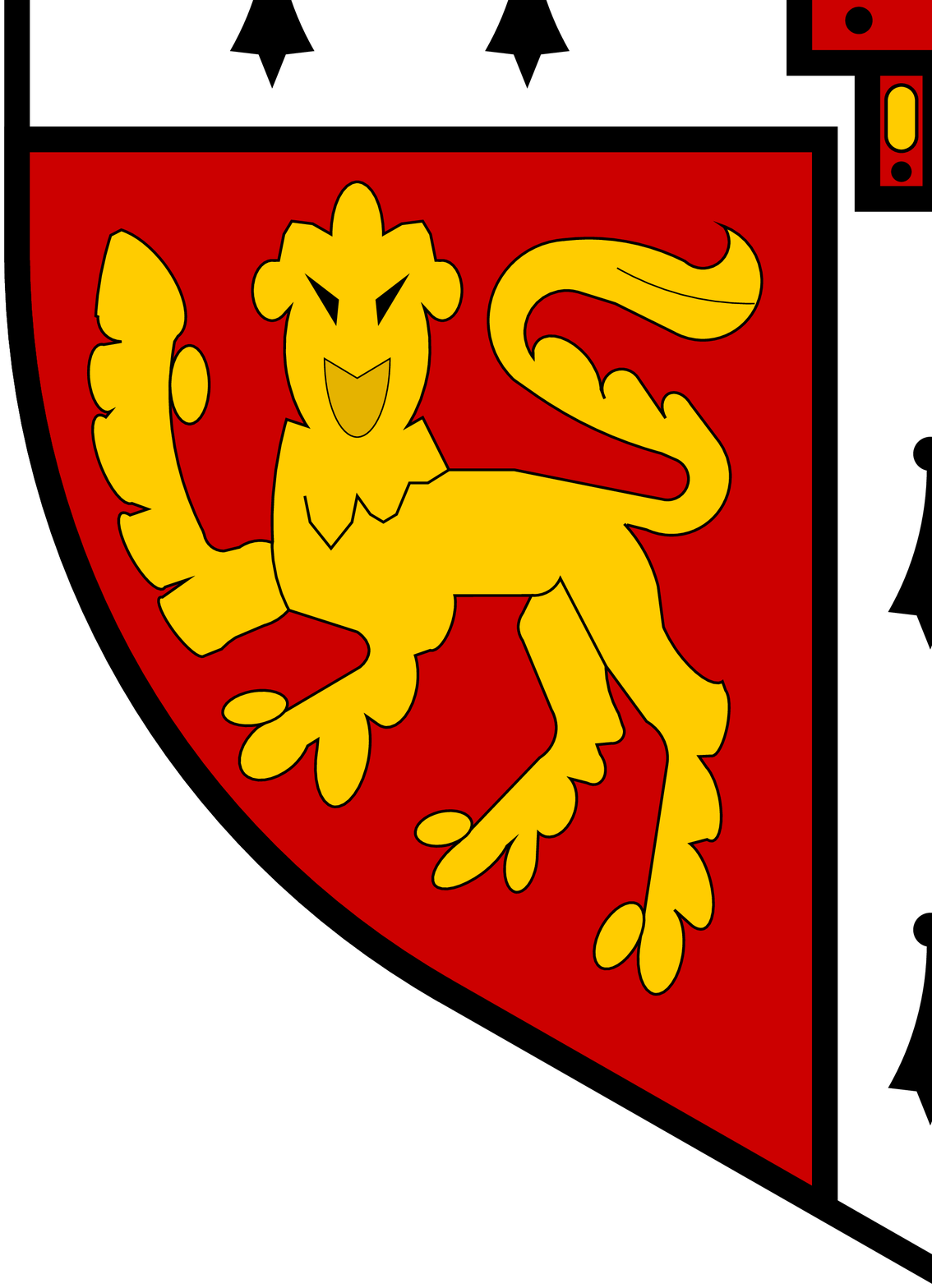}}
\degree{Doctor of Philosophy}
\degreedate{September 2008}
\usepackage{suffix,ifthen,coolstr,stringstrings,labelcas}
\usepackage{braket,array,nicefrac}

\newcommand{\bfrac}[2]{\genfrac(){}{}{#1}{#2}}
\newcommand{\btfrac}[2]{\genfrac(){}{1}{#1}{#2}}
\newcommand{\subtext}[1]{_{\text{#1}}} \newcommand{\supertext}[1]{^{\text{(#1)}}}
\newcommand{\mitx}[1]{\raisebox{0pt}[0pt][0pt]{\ensuremath{#1}}}
\newcommand{\curly}[1]{\mathcal{#1}}
\newcommand{\blg}[1]{\mathchoice{\Bigl{#1}}{#1}{#1}{#1}}
\newcommand{\brg}[1]{\mathchoice{\Bigr{#1}}{#1}{#1}{#1}}
\newcommand{\textbracket}[1]{\left({\textstyle #1}\right)}
\newcommand{\subbox}[2][0.7]{\makebox[#1\width]{$\scriptstyle #2$}}
\newcommand{\LeftLetter}{L} \newcommand{\RightLetter}{R}
\newcommand{\BPS}[2][1]{$\tfrac{#1}{#2}$-BPS}
\newcommand{\nicespace}[2]{%
 \stringlength[q]{#1}%
 \ifthenelse{\theresult=1}{%
  \stringlength[q]{#2}%
  \ifthenelse{\theresult=1}{#1 #2}{#1 \, #2}}%
 {#1 \, #2}}
\newcommand{\ifcomma}[2][]{%
 {#1{#2}}\ifthenelse{\equal{#2}{}}{}{, }}
\newcommand{\secref}[2][Section]{%
 \substring[q]{#2}{1}{3}%
 \ifthenelse{\equal{\thestring}{ch:}}{Chapter}{Section}~\ref{#2}}
\newcommand{\tabref}[2][Table]{#1~\ref{#2}}
\newcommand{\figref}[2][Figure]{\tabref[#1]{#2}}

\newcommand{\corel}[1]{\langle{#1}\rangle}
\newcommand{\expect}[1]{\langle{#1}\rangle}
\newcommand{\comm}[2]{[#1,#2]}
\newcommand{\acomm}[2]{\{#1,#2\}}
\newcommand{\dpr}[2]{#1\cdotp#2}
\newcommand{\repr}[1]{\mathbf{#1}}
\newcommand{\dynk}[3]{[#1,#2,#3]}
\newcommand{\pochhammer}[2]{{(#1)}_{#2}}
\newcommand{\falling}[2]{{#1}^{(#2)}}
\newcommand{\hyperg}[4]{F(#1,#2;#3;#4)}
\newcommand{\HYPERF}[2]{{{}_{#1}\mspace{-1.0mu}F_{#2}\mspace{-3mu}}}
\newcommand{\hyperf}[3]{\HYPERF{#1}{#2}\bigl(#3\bigr)}
\newcommand{\hyperF}[5]{\HYPERF{#1}{#2}\left[\genfrac{}{}{0pt}{}{#3}{#4};#5\right]}
\newcommand{\parp}{p}
\newcommand{\Legendre}[1]{P_{#1}}
\newcommand{\Jacobi}[3]{\Legendre{#3}^{(#1,#2)}}
\newcommand{\JJacobi}[4]{\Jacobi{#1}{#2}{\nicespace{#3}{#4}}}
\newcommand{\productlog}{W} \newcommand{\plog}{\productlog}
\newcommand{\order}{O}
\newcommand{\orderof}[1]{\order\!\left(#1\right)}
\newcommand{\floor}[1]{\left\lfloor\!{#1}\!\right\rfloor}
\newcommand{\polygamma}[1]{\psi^{(#1)}}
\newcommand{\evalat}[2]{\left.#2\right|_{#1}}
\newcommand{\abs}[1]{\left\lvert#1\right\rvert}
\newcommand{\ident}{\mathbf1}
\newcommand{\metric}{\eta} \newcommand{\met}[2]{\metric_{#1#2}}
\newcommand{\jb}{\bar{\jmath}}
\newcommand{\N}{\curly{N}}
\newcommand{\Neq}{\simeq}
\newcommand{\Meq}{\simeq}
\newcommand{\repeq}{\simeq}
\newcommand{\contrib}{\sim}
\newcommand{\goeslike}{\sim}
\newcommand{\grouplike}{\sim}
\newcommand{\anomalousdimension}[1]{\eta_{#1}}
\newcommand{\ad}[1][\ell]{\anomalousdimension{#1}}
\newcommand{\ado}[2][\ell]{\anomalousdimension{#1,#2}}
\DeclareMathOperator{\tr}{tr} \newcommand{\Tr}[1]{\tr\left(#1\right)}
\DeclareMathOperator{\Polylogarithm}{Li} \newcommand{\Li}[1]{\Polylogarithm_{#1}}
\newcommand{\dd}[2][]{\mathrm{d}^{#1}#2}

\newcommand{\thereals}{\mathbb{R}}
\newcommand{\diff}[2][]{\frac{d^{#1}}{d{#1}^{#2}}}
\newcommand{\pdiff}[2][]{\frac{\partial^{#1}}{\partial{#2}^{#1}}}
\WithSuffix\newcommand\pdiff2[2]{\frac{\partial^2}{\partial{#1}\,\partial{#2}}}
\newcommand{\ddt}[1][]{\pdiff[#1]t} \newcommand{\tddt}[1][]{{\textstyle\ddt[#1]}}

\newcommand{\AdS}[1]{\text{AdS}_{#1}}
\newcommand{\Sphere}[1]{\text{S}^{#1}}
\newcommand{\cross}{\times}

\newcommand{\goesto}{\rightarrow}
\newcommand{\tendsto}{\rightarrow}
\newcommand{\swap}[2]{#1\leftrightarrow#2}
\newcommand{\switch}[2]{\ensuremath{\swap{x_#1}{x_#2},\swap{t_#1}{t_#2}}}
\newcommand{\ch}[4]{\ensuremath{#1#2 \goesto #3#4}}

\newcommand{\sq}{\tfrac1{\sqrt2}}
\newcommand{\fps}{(4\pi^2)}
\newcommand{\lemn}[1]{\ensuremath{0\le m\le n\le#1}}
\newcommand{\Ns}[1][1]{N^2-#1}
\newcommand{\Nsb}[1]{(\Ns[#1])}
\newcommand{\D}[1]{\Delta_{#1}}
\newcommand{\DD}[2]{\D{#1#2}}
\newcommand{\DDD}[4][]{\D{#2}+\D{#3}-\D{#4}#1}
\newcommand{\DDl}[4][\ell]{\Delta #2 \DD{#3}{#4} - #1}
\newcommand{\hDDD}[4][]{\frac12(\DDD[#1]{#2}{#3}{#4})}
\newcommand{\hDDl}[4][\ell]{\frac12(\DDl[#1]{#2}{#3}{#4})}
\newcommand{\jj}{j\jb} \newcommand{\jjb}{(j,\jb)}
\newcommand{\nN}[1][p]{\frac{2#1}{N^2}}
\newcommand{\dpt}[2]{\dpr{t_{#1}}{t_{#2}}}

\newcommand{\Op}{\curly{O}}

\newcommand{\OpI}[1][I]{\Op^{#1}}
\newcommand{\muBRACKET}[4][\mu]{_{{#3}{#1}_1\dots{#1}_{#2}{#4}}}
\newcommand{\muSERIES}[5][\mu]{{#3}_{{#4}{#1}_1}\dots{#3}_{{#1}_{#2}{#5}}}
\newcommand{\muel}[1][\ell]{\muBRACKET{#1}{}{}}
\newcommand{\nuel}[1][\ell]{\muBRACKET[\nu]{#1}{}{}}

\newcommand{\muelst}[1][\ell]{\muBRACKET{#1}{\{}{\}}}
\newcommand{\xmuelst}[2][\ell]{\muSERIES{#1}{#2}{\{}{\}}}
\newcommand{\xmunuelst}[5][\ell]{{#4}_{\{{#2}_1|{#3}_1}{#5}\dotsm{#4}_{{#2}_{#1}\}{#3}_{#1}}{#5}}
\newcommand{\rpee}[1][p]{\muBRACKET[r]{#1}{}{}}
\newcommand{\trp}{\muSERIES[r]pt{}{}}
\newcommand{\Tap}[3][p]{\muSERIES[a]{#1}T{#2}{#3}}
\newcommand{\apb}[2][p]{_{a_1\dots a_{#1}{#2}}}
\newcommand{\ap}[1][p]{\apb[#1]{}}
\newcommand{\moo}[2][\ell]{\mu_{#2}\dots\mu_{#1}}
\newcommand{\scalar}{\phi}
\newcommand{\chiral}{\varphi}
\newcommand{\complexscalar}{\phi} \newcommand{\cscal}{\complexscalar}
\newcommand{\adscalar}{\chiral} \newcommand{\adgscalar}{\chi}
\newcommand{\chiralfermion}{\lambda}
\newcommand{\gaugino}[2][\alpha]{\chiralfermion_{#2}{}_{#1}}
\WithSuffix\newcommand\gaugino*[2][\alphd]{\bar{\chiralfermion}^{#2}\!_{#1}}
\newcommand{\multiplet}[1]{\curly{#1}}
\newcommand{\genericmultiplet}{\multiplet{M}}
\newcommand{\M}{\genericmultiplet}
\newcommand{\rep}[5]{\multiplet{R}^{(#4,#5)}_{\dynk{#1}{#2}{#3}}}
\newcommand{\hrep}[5]{\rep{#1}{#2}{#3}{\frac{#4}2}{\frac{#5}2}}
\newcommand{\nrep}[4][n]{\hrep{#2}{#3}{#4}{#1}{#1}}
\newcommand{\shorten}[2]{\ensuremath{(\tfrac1{#1},\tfrac1{#2})}}
\newcommand{\Nfourmult}[7]{\multiplet{#1}^{#2}_{\dynk{#3}{#4}{#5}(#6,#7)}}
\newcommand{\Longmult}[6]{\Nfourmult{A}{#6}{#1}{#2}{#3}{#4}{#5}}
\newcommand{\shortmult}[3]{\Nfourmult{B}{\ifthenelse{\equal{#1}{}}{}{#1,#1}}{#2}{#3}{#2}00}
\newcommand{\semimult}[7][\DefaultArg]{%
  \def\DefaultArg{#2}%
  \Nfourmult{C}{\ifthenelse{\equal{#2}{}}{}{#1,#2}}{#3}{#4}{#5}{#6}{#7}}
\newcommand{\Long}[4]{\multiplet{A}^{#4}_{#1#2,#3}}
\newcommand{\short}[2]{\multiplet{B}_{#1#2}}
\newcommand{\semi}[3]{\multiplet{C}_{#1#2,#3}}
\newcommand{\idmult}{\multiplet{I}}
\newcommand{\plusbeta}[2][]{\ifthenelse{\equal{#2}{0}}{\cpwParamA}{\substr{#2}{1}{1}+\cpwParamA \substr{#2}{2}{end} #1}}
\newcommand{\pp}[3]{#1{\plusbeta[\,]{#2}}{\plusbeta{#3}}}
\newcommand{\xt}[1]{(x_{#1},t_{#1})}

\newcommand{\hspt}[3]{#2\dimn{#3}\xt{#1}}
\newcommand{\spt}[3]{#2\dimn{#3}}

\newcommand{\sscpt}[2][\DefaultOpt]{%
  \def\DefaultOpt{#2}%
  \scalar_{#1}(x_{#2})}

\newcommand{\fourpt}[9]{\corel{#11{#2}{#3} \, #12{#4}{#5} \, #13{#6}{#7} \, #14{#8}{#9}}}

\newcommand{\fourpointF}{\curly{F}}
\newcommand{\fourpointG}{\curly{G}}
\newcommand{\fourpointH}{\curly{H}}
\newcommand{\plainF}{F}
\newcommand{\plainG}{G}
\newcommand{\makecommand}[3][]{%
  \expandafter\newcommand\csname tptp#2\endcsname{#1{#3}}%
  \expandafter\newcommand\csname ttpp#2\endcsname{#1{\tilde{#3}}}%
  \expandafter\providecommand\csname #2\endcsname{\csname tptp#2\endcsname}%
  \WithSuffix\expandafter\newcommand\csname #2\endcsname*{\csname ttpp#2\endcsname}%
  \expandafter\newcommand\csname #2gen\endcsname[4]{#1{#3}^{(##1,##2,##3,##4)}}%
  \expandafter\newcommand\csname #2genp\endcsname{\ottf[\csname #2gen\endcsname]p}%
}
\makecommand{G}{\fourpointG}
\makecommand{GH}{\fourpointH}
\makecommand[\hat]{GHhat}{\fourpointH}
\makecommand{f}{f}
\makecommand[\hat]{fhat}{f}
\makecommand{fg}{g}
\makecommand{fh}{h}
\makecommand{fk}{k}
\makecommand{HF}{\fourpointF}
\makecommand{AB}{B}

\newcommand{\ord}[1]{_{#1}}

\newcommand{\free}{\supertext{Free}}
\newcommand{\dyn}{_I}
\newcommand{\pert}{\subtext{pert.}}
\newcommand{\dimn}[1]{^{(#1)}}
\newcommand{\spin}[1]{\dimn{#1}}
\newcommand{\Gx}[1]{\G_{#1}}
\WithSuffix\newcommand\Gx*[1]{\G*_{#1}}

\newcommand{\sing}{\subtext{log-free}}
\newcommand{\reg}{\subtext{reg.}}

\newcommand{\x}[2]{{x_{#1#2}}}
\newcommand{\xx}[4][2]{\ifthenelse{\equal{#4}{^}}{(\xx[#1]{#2}{#3}{})^}{x^{#1}_{#2#3}#4}}
\newcommand{\xxxx}{(x_1,x_2,x_3,x_4)}
\newcommand{\uv}{(u,v)}
\newcommand{\uvst}{(u,v;\sigma,\tau)}
\WithSuffix\newcommand\uvst*{u,v,\sigma,\tau}
\newcommand{\uvoo}{(u,v;0,0)}
\newcommand{\uvmod}{(\tfrac1u,\tfrac{v}u)}
\WithSuffix\newcommand\uvmod*{(\tfrac{u}v,\tfrac1v)}
\newcommand{\uvstmod}{(\tfrac1u,\tfrac{v}u;\tfrac1\sigma,\tfrac\tau\sigma)}

\newcommand{\st}{(s,t)}
\WithSuffix\newcommand\st*{(\sigma,\tau)}
\newcommand{\usqsxt}{\frac{u^2}{16}}
\newcommand{\xst}[1][]{(#1 x;\sigma,\tau)}
\newcommand{\xb}{\bar{x}}
\newcommand{\alphb}{\bar\alpha}
\newcommand{\xxb}{x-\xb}
\newcommand{\aab}{\alpha-\alphb}
\newcommand{\ax}{(\alpha x-1)}
\newcommand{\axb}{(\alpha\xb-1)}
\newcommand{\abx}{(\alphb x-1)}
\newcommand{\abxb}{(\alphb\xb-1)}
\newcommand{\amx}{\blg(\alpha-\frac1x\brg)}

\newcommand{\abmx}{\blg(\alphb-\frac1x\brg)}

\newcommand{\xa}{(x,\alpha)}
\newcommand{\xab}{(x,\alphb)}
\newcommand{\xba}{(\xb,\alpha)}
\newcommand{\xbab}{(\xb,\alphb)}
\newcommand{\zb}{\bar{z}}
\newcommand{\yb}{\bar{y}}
\newcommand{\zzb}{z-\zb}
\newcommand{\yyb}{y-\yb}
\newcommand{\yz}{(y-z)}
\newcommand{\yzb}{(y-\zb)}
\newcommand{\ybz}{(\yb-z)}
\newcommand{\ybzb}{(\yb-\zb)}

\newcommand{\XL}{X^\LeftLetter}
\newcommand{\XR}{X^\RightLetter}
\newcommand{\XLXR}{\dpr\XL\XR}
\newcommand{\Xbb}[2]{#1_{#2#2\rvert}}
\newcommand{\XLb}[1][b]{\Xbb\XL{#1}}
\newcommand{\XRd}[1][d]{\Xbb\XR{#1}}
\newcommand{\Xp}[1][p]{X\dimn{#1}}
\newcommand{\Xmp}[1][p]{\bar{X}\dimn{#1}}
\newcommand{\COLcontractX}[7]{\XL_{a_1\dots a_{#1}#2}\,\XR_{a_1\dots a_{#1}#3}\;\tfrac12\delta_{#4#5}\;\tfrac12\delta_{#6#7}}
\newcommand{\COLwibbleX}{\notag\\[-16pt]&\hphantom{\:\propto\: \frac1{2!}\cdot\frac1{p!}\; \COLcontractX{p}{}{}cdcd}} 
\newcommand{\COLfusion}[2]{\tfrac12\bigl(\tr(#1#2) - \tfrac1N\tr(#1)\tr(#2)\bigr)}
\newcommand{\COLfission}[2]{\tfrac12\bigl(\tr(#1)\tr(#2) - \tfrac1N\tr(#1#2)\bigr)}
\newcommand{\COLtrtrace}[3]{\tr(T_{(#1}T_{#2}\,T_{#3)})}
\newcommand{\COLtriple}[6]{\COLtrtrace{#1}{#2}{#3}\COLtrtrace{#4}{#5}{#6}}

\newcommand{\loopint}[1]{\Phi^{(#1)}}
\WithSuffix\newcommand\loopint^[1]{\hat{\Phi}^{(#1)}}
\newcommand{\fB}{\loopint1}
\newcommand{\fP}{\loopint2}
\newcommand{\phipolylog}[1]{\phi_{#1}} \newcommand{\phlog}{\phipolylog}
\WithSuffix\newcommand\xxb'{(x',\xb')}
\WithSuffix\newcommand\lnxxb'[1][]{\ln^{#1} x'\xb'}
\newcommand{\adl}[3][a]{{#1}^{#2}_{#3}}
\newcommand{\adlI}[4][a]{\adl[#1]{#2}{#3,#4}}
\newcommand{\cpwParamA}{\beta}
\newcommand{\cpwParamB}{\gamma}
\newcommand{\bt}{\cpwParamA}
\WithSuffix\newcommand\bt*{\cpwParamB}
\newcommand{\Gpw}[6]{\Gpw.{#1}{#2}(#3,#4;#5,#6)}
\WithSuffix\newcommand\Gpw*[4]{\Gpw.{#1}{#2}(#3,#4)}
\WithSuffix\newcommand\Gpw.[2]{G\spin{#1}_{#2}}
\newcommand{\gpw}[6][]{g#1^{(#2,#3)}_{#4,#5}(#6)}
\WithSuffix\newcommand\gpw*[4][]{g#1_{#2,#3}(#4)}
\newcommand{\gbb}[4][\cpwParamA]{\gpw{#1}{-#1}{#2}{#3}{#4}}
\WithSuffix\newcommand\gbb*[3]{\gpw\cpwParamA\cpwParamB{#1}{#2}{#3}}
\WithSuffix\newcommand\gbb'[4][']{\gpw[#1]{\cpwParamA}{-\cpwParamA}{#2}{#3}{#4}}
\WithSuffix\newcommand\gbb.[2]{g^{(\cpwParamA,-\cpwParamA)}_{#1,#2}}
\renewcommand{\makecommand}[3]{%
  \expandafter\newcommand\csname J#1\endcsname[1]{\Jacobi{#2}{#3}{##1}}%
  \expandafter\newcommand\csname JJ#1\endcsname[2]{\JJacobi{#2}{#3}{##1}{##2}}%
}%
\makecommand{ab}{a}{b}
\makecommand{bb}{0}{2\cpwParamA}
\makecommand{bg}{-\cpwParamA-\cpwParamB}{\cpwParamA-\cpwParamB}
\newcommand{\LL}[2]{\Legendre{\nicespace{#1}{#2}}}
\newcommand{\Jlim}{\Jlimbelow}
\newcommand{\Jlimbelow}[1][r]{_{\subbox{-1\le{#1}\le1\hphantom-}}}

\newcommand{\g}[1]{_{#1}}
\newcommand{\p}[1]{^{(#1)}}

\newcommand{\twist}{\p}
\newcommand{\Twist}[1][]{\Delta_{#1}-\ell}
\newcommand{\Gt}[1][]{\G\g{#1t}}
\newcommand{\Gxt}[1]{\Gt[#1,]}
\newcommand{\Gxtj}[2]{\Gxt{#1}\p{#2}}
\WithSuffix\newcommand\Gxtj*[2]{\hat\G\g{#1,t}\p{#2}}
\newcommand{\tbt}[1]{(2\bt+#1)}
\newcommand{\GHj}[1][j]{\GH\g{#1}}
\newcommand{\fQ}[2]{\curly{Q}^{#1}_{#2}}

\WithSuffix\newcommand\fQ^[2]{\hat{\curly{Q}}^{#1}_{#2}}
\newcommand{\JRecurrenceParam}{\gamma}
\newcommand{\gam}[2]{\JRecurrenceParam_{#1,#2}}
\newcommand{\gpl}[1]{\gam{#1}1}
\newcommand{\gze}[1]{\gam{#1}0}
\newcommand{\gmi}[1]{\gam{#1}{-1}}
\newcommand{\gprod}[1]{\bigl(\gpl0\dotsm\gpl{#1}\bigr)}
\WithSuffix\newcommand\gprod*[1][j]{\frac{(-1)^{#1}#1!}{\pochhammer{2\cpwParamA+#1+1}{#1}}}
\newcommand{\something}{\curly{F}}
\newcommand{\F}{\something}
\renewcommand{\makecommand}[2]{%
  \expandafter\newcommand\csname #1\endcsname[2][\ell]{#2^{(##2)}_{##1}}%
  \expandafter\newcommand\csname #1q\endcsname[1][\ell]{\csname #1\endcsname[##1]q}%
}
\newcommand{\makecommandwithp}[2]{\makecommand{#1}{#2} \makecommand{#1p}{#2'}}
\makecommandwithp{lna}{a}
\makecommandwithp{lib}{b}
\makecommandwithp{lnsa}{\bar{a}}
\makecommandwithp{lnca}{\hat{a}}
\makecommandwithp{lic}{c} \makecommandwithp{licb}{\bar{c}}
\makecommandwithp{lnlie}{e} \makecommandwithp{lnlieb}{\smash{\overset{*}{e}}}
\newcommand{\Ap}[3]{A^{(#1)}_{#2,#3}}

\newcommand{\h}{h}
\WithSuffix\newcommand\h*[1]{\h^{(#1)}}
\newcommand{\hb}{\bar{h}}
\WithSuffix\newcommand\hb*[1]{\hb^{(#1)}}
\newcommand{\pexp}[1]{e_{#1}}
\makecommand{Simga}{\Sigma}
\newcommand{\I}{\curly{I}} \newcommand{\J}{\curly{J}}
\newcommand{\DInt}[1]{D_{#1}}
\newcommand{\Db}[4]{\overline{D}_{#1#2#3#4}}
\newcommand{\DbS}[4]{\Db{#1\,}{#2\,}{#3\,}{#4}}
\newcommand{\DbD}[4]{\Db{\D{#1}}{\D{#2}}{\D{#3}}{\D{#4}}}

\newcommand{\Fr}[3]{r^{(#2)}_{#1,#3}} \WithSuffix\newcommand\Fr*[3]{\tilde{r}^{(#2)}_{#1,#3}}
\newcommand{\Fc}[4]{c^{(#2)}_{#1;#3,#4}}

\newcommand{\ottf}[2][\ottflist]{#1{{#2}_1}{{#2}_2}{{#2}_3}{{#2}_4}}
\newcommand{\OTTF}[2][]{%
 \def\TheCommand ##1{#2}%
 \TheCommand1#1\TheCommand2#1\TheCommand3#1\TheCommand4}
\newcounter{OTTcount}
\newcommand{\OTT}[3][]{%
 \setcounter{OTTcount}{#3}%
 \addtocounter{OTTcount}{-1}%
 \ifthenelse{\value{OTTcount}>0}{\OTT[#1]{#2}{\value{OTTcount}} #1}{}%
 \addtocounter{OTTcount}{1}%
 \def\TheCommand ##1{#2}%
 \TheCommand{\arabic{OTTcount}}}
\newcounter{xprimecount}
\newcommand{\fraff}[3][long]{\ifthenelse{\equal{#1}{long}}{{#2}\frac1{#3}}{\frac{#2}{#3}}}
\newcommand{\Num}[2][]{\##1\!\!\left[{#2}\right]}
\newcommand{\ntr}[1][single]{^{\mbox{\makebox[0pt][l]{\raisebox{\height}{\tiny #1}}\tiny trace\:}}}
\newcommand{\desc}{\subtext{desc.}} \newcommand{\prim}{\subtext{prim.}}

\newcommand{\ep}[1][ij]{\varepsilon_{#1}} \WithSuffix\newcommand\ep*[1][ij]{\varepsilon^{#1}}
\newcommand{\Z}[1]{\curly{Z}^{#1}}
\newcommand{\Zi}{\Z{i}} 
\newcommand{\lm}{\chiralfermion} \WithSuffix\newcommand\lm*{\bar{\lm}}
\newcommand{\alphd}{\dot{\alpha}} \newcommand{\aad}{\alpha\alphd}
\newcommand{\Q}[2][\alpha]{Q^{#2}\!_{#1}}
\WithSuffix\newcommand\Q*[2][\alphd]{\bar{Q}_{#2}{}_{#1}}
\newcommand{\q}[2][\alpha]{S_{#2}{}^{#1}}
\WithSuffix\newcommand\q*[2][\alphd]{\bar{S}^{#2}{}^{#1}}
\newcommand{\R}[2]{R^{#1}\!_{#2}}
\newcommand{\So}[1][1]{S_{#1}} \WithSuffix\newcommand\St*[1][2]{\bar{S}^{#1}}
\newcommand{\dZ}[1][q]{\partial^{#1}Z}
\newcommand{\dY}[1][s]{\partial^{#1}Y}
\newcommand{\dl}[1][s]{\partial^{#1}\lm}
\WithSuffix\newcommand\dl*[1][t]{\partial^{#1}\lm*}
\newcommand{\DZ}[2][]{\partial^{#1}\Z{#2}}
\newcommand{\at}{\tilde{a}}
\newcommand{\ah}{\hat{a}}
\newcommand{\bh}{\hat{b}}
\newenvironment{tablecell}[1][small]{\def\EnvCommand{#1matrix}$\begin{\EnvCommand}{}\\}{\\{}\end{\EnvCommand}$}
\newenvironment{formcell}{\begin{tablecell}\textstyle}{\end{tablecell}}
\newenvironment{cases*}{\begin{customspacearray}{4pt}{ll}}{\end{customspacearray}}
\newenvironment{doublecases}[1][4pt]{\left\{\begin{customspacearray}{#1}{l@{\;\;:\;}l@{\qquad}l@{\;\;:\;}l}}{\end{customspacearray}\right.}
\newlength{\OLDextrarowheight}
\newenvironment{customspacearray}[2]{%
 \setlength{\OLDextrarowheight}{\extrarowheight}%
 \setlength{\extrarowheight}{#1}%
 \begin{array}{#2}}{%
 \end{array}%
 \setlength{\extrarowheight}{\OLDextrarowheight}}
\newcommand{\uberstrut}{\rule[-1.3ex]{0pt}{4.3ex}}
\newcommand{\generatingfunctionsymbol}{g}
\newcommand{\limitgeneratingfunctionsymbol}{\hat{\generatingfunctionsymbol}}
\newcommand{\generatingfunctionform}[3][\generatingfunctionsymbol]{{#1}^{\text{#2}}_{#3}}
\newcommand{\generatingfunction}[4][]{\generatingfunctionform{#1}{\dynk{#2}{#3}{#4}}}
\newcommand{\limitgeneratingfunction}[4][]{\generatingfunctionform[\limitgeneratingfunctionsymbol]{#1}{\dynk{#2}{#3}{#4}}}
\newcommand{\gen}{\generatingfunction} \newcommand{\lgen}{\limitgeneratingfunction}
\newcommand{\tgen}[1][]{%
 \def\ArgI{{#1}}%
 \TgenRelay}
\newcommand{\TgenRelay}[1][\tau]{\generatingfunctionform[\limitgeneratingfunctionsymbol]{\ArgI}{#1}}
\newcommand{\qgen}[1][]{%
 \def\ArgI{{#1}}%
 \QgenRelay}
\newcommand{\QgenRelay}[1][\tau]{\TgenRelay[#1,\text{sym}]}
\newcommand{\character}{\chi}
\WithSuffix\newcommand\character*{\hat{\character}}
\WithSuffix\newcommand\ch4[2]{\character_{(#1,#2)}}
\WithSuffix\newcommand\ch8[2]{\character*_{#1,#2}}
\WithSuffix\newcommand\ch*[4]{\character*^{(#4)}_{\dynk{#1}{#2}{#3}}}
\newcommand{\partition}{Z}
\newcommand{\zyab}[1][]{(\zyab*[#1])}
\WithSuffix\newcommand\zyab-{(z,y,-a,-b)}
\WithSuffix\newcommand\zyab*[1][]{z^{#1},y^{#1},a^{#1},b^{#1}}
\newcommand{\shortpartition}[3][]{\partition^{\frac1{#2},\frac1{#3}}_{\text{#1}}}
\newcommand{\Zmt}{\shortpartition[m.t.]}
\newcommand{\Zsugra}{\shortpartition[sugra]}
\newcommand{\Zsym}[1][m.t.]{\shortpartition[#1,sym]}
\newcommand{\pip}[2][i]{\prod_{#1=1}^{#2} \frac1{1-\sigma^{#1}}}
\WithSuffix\newcommand\pip*[2][i]{\prod_{#1=1}^{#2} \frac{\sigma^{#1-1}}{1-\sigma^{#1}}}

\newcommand{\opstate}[2]{\smash{\ket{#1}}_{#2}}
\newcommand{\Beta}[2]{B\left(#1,#2\right)}

\newcommand{\kop}{\ket{\Op}}
\newcommand{\kopp}{\ket{\Op'}}
\newcommand{\kvac}{\ket{0}}

\newcommand{\xd}{\dpr{x}\partial} \WithSuffix\newcommand\xd*{\dpr{x}{\pdiff{x}}}
\newcommand{\ixp}{(\ixp*)} \WithSuffix\newcommand\ixp*{-i\dpr{x}P}
\newcommand{\xps}{(\xps*)} \WithSuffix\newcommand\xps*{x^2P^2}
\newcommand{\xdx}[2]{(\x{#1}{#2},\partial_{x_{#2}})}

\newcommand{\Dt}{\smash{\tilde{D}}}
\newcommand{\PdP}{\dpr{P}{\dP{}}} \WithSuffix\newcommand\PdP*{\dpr{P}{\dP*{}}}
\newcommand{\dP}[1]{\partial_{P_{#1}}} \WithSuffix\newcommand\dP*[1]{\pdiff{P_{#1}}}
\newcommand{\Cxp}{C\left(\ixp*,\xps*\right)}
\newcommand{\CC}{\curly{C}}
\newcommand{\Cl}[1][C]{{#1}\spin\ell}
\newcommand{\ClD}[1][C]{\Cl[#1]_\Delta}
\newcommand{\Cmult}{C_{\phi\Op\Op'}}
\newcommand{\K}[2]{\bigl(x^2\partial_{#1} - 2x_{#1}\xd - {#2}x_{#1}\bigr)}
\newcommand{\tensor}{\curly{T}}

\newcommand{\Jack}[3]{P^{(#1)}_{\nicespace{#2}{#3}}}
\newcommand{\JD}[1]{D^J_{#1}}

\newcommand{\da}{\partial_\alpha}
\newcommand{\db}{\partial_\beta}
\newcommand{\daa}{\partial_\alpha^2}
\newcommand{\dab}{\da\db}
\newcommand{\dbb}{\partial_\beta^2}
\WithSuffix\newcommand\da*{\pdiff\alpha}
\WithSuffix\newcommand\db*{\pdiff\beta}
\WithSuffix\newcommand\daa*{\pdiff[2]\alpha}
\WithSuffix\newcommand\dab*{\pdiff2\alpha\beta}
\WithSuffix\newcommand\dbb*{\pdiff[2]\beta}

\newcommand{\Dphi}{\D{\phi}}
\newcommand{\Dop}{\D{\Op}\!}
\newcommand{\Dopp}{\Delta'}

\newcommand{\Dl}{\tilde{\Delta}}

\newcommand{\lambdasub}[1]{\ell_{#1}}
\newcommand{\lop}{\lambdasub{\Op}}
\newcommand{\lopp}{\ell}

\newcommand{\QED}{\quad\Box}
\newcommand{\half}{\tfrac12}

\newcommand{\quarter}{\tfrac14}

\newcommand{\thing}{1+\Dopp-\tfrac{d}{2}}

\newcommand{\gammb}{\Bar{\gamma}}
\newcommand{\chip}{\chi_+}
\newcommand{\chim}{\chi_-}

\newcommand{\dg}{\partial_\gamma}
\newcommand{\dgb}{\Bar{\partial}_{\gammb}} 
\newcommand{\dgg}{\partial_\gamma^2}
\newcommand{\dggb}{\dg\dgb}
\newcommand{\dgbgb}{\Bar{\partial}_{\gammb}^2} 
\newcommand{\dgbg}{\partial_{\gorgb}}
\WithSuffix\newcommand\dg*{\pdiff\gamma}
\WithSuffix\newcommand\dgb*{\pdiff\gammb}
\WithSuffix\newcommand\dgg*{\pdiff[2]\gamma}
\WithSuffix\newcommand\dggb*{\pdiff2\gamma\gammb}
\WithSuffix\newcommand\dgbgb*{\pdiff[2]\gammb}

\newcommand{\ggb}{\gamma\gammb}
\newcommand{\gpgb}{\gamma + \gammb}
\newcommand{\gmgb}{\gamma - \gammb}
\WithSuffix\newcommand\ggb*{\gamma,\gammb}
\newcommand{\ofggb}{(\ggb*)}
\newcommand{\gorgb}{\gamma[\gammb]}
\newcommand{\gswapgb}{\gamma\leftrightarrow\gammb}

\newcommand{\Chat}{\smash{\widehat{C}}}
\newcommand{\Cgamma}{\Chat\ofggb}

\newcommand{\xpm}{x_{\pm}}
\newcommand{\xmp}{x_{\mp}}
\newcommand{\dpm}{\partial_{\pm}}
\newcommand{\dmp}{\partial_{\mp}}
\newcommand{\dpl}{\partial_+}
\newcommand{\dmi}{\partial_-}

\newcommand{\Ppm}{P_{\pm}}
\newcommand{\Pmp}{P_{\mp}}
\newcommand{\Kpm}{K_{\pm}}

\newcommand{\hpm}{h_{\pm}}
\newcommand{\hmp}{h_{\mp}}
\newcommand{\hpl}{h_+}
\newcommand{\hmi}{h_-}

\newcommand{\qpm}{q_{\pm}}
\newcommand{\qmp}{q_{\mp}}

\newcommand{\qnb}{q_+}
\newcommand{\qba}{q_-}

\newcommand{\xhxh}{x_+^{\hpl}x_-^{\hmi}}

\newcommand{\gqgqb}{\gamma^{\qnb}\gammb^{\qba}}

\newcommand{\Gqqb}{\Gamma_{\qnb\qba}}

\newcommand{\mopp}[1]{\opstate{\Op'}{#1}}

\newcommand{\SO}[1]{SO({#1})}

\newcommand{\Jb}{\Bar{J}}

\newcommand{\eps}{\varepsilon}

\newcommand{\stf}[1]{\left\{#1\right\}}

\newcommand{\RomI}[1]{\curly{U}_{#1}}
\newcommand{\RomII}[1]{\curly{V}^{#1}}
\newcommand{\plusO}{O^+}
\newcommand{\starl}{\curly{W}}
\newcommand{\starbl}{\overline{\starl}}

\newcommand{\E}[2]{\mathcal{E}_{#1#2}}

\newcommand{\elep}{{\ell,\eps}}
\newcommand{\lamz}{{\lambda,0}}

\newcommand{\Ct}{\tilde{C}}

\newcommand{\Fg}[2]{F^{#1}_{#2}}
\newcommand{\Fgb}[2]{\bar{F}^{#1}_{#2}}

\usepackage[all]{xy}

\CompileMatrices

\newcommand{\FFvrtx}{\bullet}
\newcommand{\FFdiagram}[3]{\xymatrix{\FFvrtx\ar@{}[]+UL*{x_3};[r]+UR*{x_4}\ar@{}[d]+DL*{x_1};[dr]+DR*{x_2}{#1}&\FFvrtx{#2}\\\FFvrtx{#3}&\FFvrtx}}
\newcommand{\FFline}[2]{\ar@{-}[]+0;[#1]+0#2}
\newcommand{\FFtwoline}[3]{\ar@/^#3/@{-}[]+0;[#1]+0#2 \ar@/_#3/@{-}[]+0;[#1]+0#2}
\newcommand{\FFpair}[2]{\FFtwoline{#1}{#2}{0.2pc}}

\newcommand{\FFbundle}[2]{\FFtwoline{#1}{}{} \FFline{#1}{|{#2}}}
\newcommand{\FFspace}{\mspace{9mu}}

\newsavebox{\FFDiagI}
\newsavebox{\FFDiaga}
\newsavebox{\FFDiagaX}
\newsavebox{\FFDiagb}
\newsavebox{\FFDiagbX}
\newsavebox{\FFDiagc}

\savebox{\FFDiagI}{\FFdiagram{\FFbundle{r}{p}}{}{\FFpair{r}{}}}
\savebox{\FFDiaga}{\FFdiagram{\FFbundle{r}{p-1} \FFline{d}{}}{\FFline{d}{}}{\FFline{r}{}}}
\savebox{\FFDiagaX}{\FFdiagram{\FFbundle{r}{p-1} \FFline{dr}{}}{\FFline{ld}{|\hole}}{\FFline{r}{}}}
\savebox{\FFDiagb}{\FFdiagram{\FFbundle{r}{p-2} \FFpair{d}{}}{\FFpair{d}{}}{}}
\savebox{\FFDiagbX}{\FFdiagram{\FFbundle{r}{p-2} \FFpair{dr}{}}{\FFpair{dl}{|\hole}}{}}
\savebox{\FFDiagc}{\FFdiagram{\FFbundle{r}{p-2} \FFline{d}{} \FFline{dr}{}}{\FFline{d}{} \FFline{dl}{|\hole}}{}}

\begin{document}
\pagestyle{empty}
\maketitle
\cleardoublepage
\frontmatter
\onehalfspacing
\setdisplayskipstretch{0.8}
\flushbottom
\pagestyle{plain}

\begin{frontsection}{Declaration}

This dissertation is the result of my own work and includes nothing which is the outcome of work done in collaboration, 
except where specifically indicated in the text.
The introduction sets out established results, sources for which are cited.
Subsequent chapters are based on unpublished work with my supervisor, Prof.~Hugh Osborn.

No part of this dissertation submitted has been, or is concurrently being, submitted for any degree, diploma, or other qualification,
nor is substantially the same as any that I have submitted for a degree or diploma or other qualification at any other University.

\end{frontsection}

\begin{frontsection}{Acknowledgements}

This PhD would not have been possible without the guidance, feedback, advice and patience of my supervisor, Prof.~Hugh Osborn.

I am also most grateful to my tutor, Dr Jonathan Evans, for his support and encouragement;
to my colleagues in DAMTP, particularly my office-mates and Linda Uruchurtu, 
for many helpful discussions, for their friendship, and for \OldCite{rf:linda};
and to my many other friends from my years in Cambridge.

My research was funded by a grant from PPARC (now part of the STFC) 
and a scholarship awarded by Gonville \& Caius College, who also provided me with accommodation, dinner tickets, and free washing machines.

Finally, I would like to thank my parents and my wife Kimberley for their constant support and limitless understanding.

\begin{flushright}
\textit{Cambridge, September 2008}
\end{flushright}

\end{frontsection}

\begin{summary}[Abstract]

Chiral primary \BPS2 operators in $\N=4$ superconformal Yang Mills are defined.
Their four point functions are introduced, 
expressed in a form manifestly satisfying the superconformal Ward identities,
simplified by the use of null vectors.
They are subsequently expanded in terms of conformal partial waves.
Correlation functions of two pairs of identical chiral primaries,
one pair having the lowest possible scale dimension, are considered.
Crossing symmetries help determine their free field value up to numeric constants.
The contributions from different supermultiplets to the partial wave expansion is analysed,
and determined in the case of the free field.
This is compared with established results at strong and weak coupling.
In the large $N$, strong coupling limit, non-trivial cancellations are found between the free field values and results from supergravity, 
providing a strong consistency check.
In the perturbative case values are obtained for the anomalous dimensions of lowest twist operators and the correction to the coupling.
To find these results we compute the necessary conformal wave expansions of certain hypergeometric and logarithmic functions.
Next, we attempt to count shortened $\N=4$ SYM operators,
beginning by constructing from fundamental fields the most general operators belonging to certain $SU(4)_R$ representations at low twists. 
The number of independent solutions to the conditions imposed on such operators is found via a combinatoric approach.
We then take partition functions for shortened operators 
and derive generating functions $\generatingfunctionsymbol(\sigma)$ for the number of operators with spin $\ell=0,1,2,\dotsc$.
Explicit values are obtained for specific $R$-symmetry representations at low twist in various sectors of the theory.
In the more general case we find leading order approximations for $\generatingfunctionsymbol(\sigma)$ as $\sigma\tendsto1$,
and consider their asymptotic behaviour at large twist.
Finally we consider the conformal field theory operator product expansion 
and attempt to find solutions in terms of series expansions,
first in restricted cases (for scalar operators only, and in only two dimensions),
then for the more general problem.

\end{summary}

\cleardoublepage
\begin{singlespacing}
 \tableofcontents
 \newpage
 \listoftables
 \listoffigures
\end{singlespacing}

\mainmatter
\pagestyle{plain}

\chapter*{Introduction}
 \addcontentsline{toc}{chapter}{\numberline {0}Introduction}
 \label{ch:Introduction}

 \section{Motivation}
 \label{se:Motivation}


There has been renewed interest in four-dimensional superconformal quantum field theories since the discovery of the AdS/CFT correspondence 
\cite{rf:9711200}, 
relating Type~IIB string theory on $\AdS5\cross\Sphere5$ to 
4-dimensional Super Yang Mills theory with maximal $\N=4$ supersymmetry and gauge group $SU(N)$.
In particular, the supergravity approximation on the AdS side corresponds to the $N\tendsto\infty$ limit of SYM for large $\lambda=g^2N/4\pi^2$.

$\N=4$ SYM has interesting properties, even viewed purely as a quantum field theory. 
It is superconformal for any $g$. 
Its symmetries are given by the supergroup $SU(2,2|4)$, containing 
the conformal symmetry group $SO(2,4) \grouplike SU(2,2)$,
the $R$-symmetry group $SO(6)_R \grouplike SU(4)_R$,
and the Poincar\'e and conformal supersymmetries generated by the supercharges 
$\Q{i}, \q{i}$, $i=1,\dots,4$, and their conjugates 
$\Q*{i}, \q*{i}$.
The operators in the spectrum of the theory are arranged into supermultiplets, each containing a superconformal primary state, 
which has the lowest scale dimension in the multiplet and is annihilated by the superconformal charges $\q{i}, \q*{i}$.
The action of the supercharges $\Q{i}, \Q*{i}$ on this state generates the other operators of the multiplet.
All these states are conformal primaries; acting on them with the momentum generator $P_\mu$ gives conformal descendant operators.

Four point functions in conformal field theories are of particular interest. 
In a general CFT, the correlation functions of primary operators are significantly constrained by the requirements of conformal invariance. 
Two point functions are, indeed, completely fixed up to normalisation, as are the three point functions of scalar operators.
In general, three point functions are determined up to the value of a finite number of constant parameters.
The four point function is thus the first correlator for which conformal symmetry admits non-trivial space-time dependence,
and even so, its freedom is restricted to a single arbitrary function of the two conformally invariant cross-ratios.

The analysis of four point functions may be used to explore the spectrum of operators in the theory, their scale dimensions, spins, and couplings. 
Using the operator product expansion for any pair of operators appearing in the correlation function, 
we obtain an expansion in terms of conformal partial waves, 
analogous to the partial wave expansion of scattering amplitudes which reveals details of resonant states.

Of particular interest to consider are correlators of chiral primary \BPS2 operators, belonging to the $\dynk0p0$ representation of $SU(4)_R$. 
These are represented by rank $p$ symmetric, traceless $SO(6)_R$ tensors, formed by gauge-invariant traces of elementary scalar fields, 
belonging to the adjoint representation of the gauge group $SU(N)$ and the 6-dimensional representation of the $R$-symmetry group.
These operators satisfy BPS-like constraints, and so are protected from renormalisation effects, and have scale dimension $\Delta=p$.

In \cite{rf:0412335} correlation functions of four identical \BPS2 operators, all belonging to the $\dynk0p0$ representation, are considered,
with explicit calculations made in the cases $p=2,3,4$. 
Similar calculations are made in \cite{rf:0601148}.
We will look at the case of the four point function of two $\dynk0p0$ operators with two operators belonging to the $\dynk020$ representation --- 
the $p=2$ supermultiplet descended from this superconformal primary is the current multiplet, 
containing the energy-momentum tensor, the supersymmetry currents, and the $R$-symmetry currents.
Correlation functions of this form are examined in \cite{rf:0504061}, where calculations are made in the perturbation expansion up to order $g^4$, 
using particular flavour representatives in the $\N=1$ formulation.

Unlike the $\corel{pppp}$ correlator, there are two distinct channels present in these functions, namely $2p\goesto 2p$ and $22\goesto pp$.
Different $SU(4)$ representations contribute in either case, as may be seen by decomposing the relevant tensor products.
The structure of the operator product expansion is reflected in the contributions to the conformal partial wave expansion 
at different $\Delta$, $\ell$
.
It is generally non-trivial to separate contributions of superconformal primary operators from those of descendant operators.
The problem is resolved using the solution to the superconformal Ward identities, given in \cite{rf:0407060}, 
which enables us to isolate the free field results for the four point function from the dynamical part, 
which leads to anomalous dimensions.

 \section{Conformal Field Theory}
 \label{se:Conformal}

The conformal group in $d$ dimensions is well known, extending the Poincar\'e group --- containing 
Lorentz transformations $SO(1,d-1)$, generated by $M_{\mu\nu}$, and translations, generated by $P_\mu$ --- 
by the addition of dilations, generated by $D$, and special conformal transformations, generated by $K_\mu$.
The conformal algebra, which corresponds to $SO(2,d)$, is then
 \begin{equation} \label{eq:ConformalAlgebra}
  \begin{gathered}
   \comm{M_{\mu\nu}}{P_\rho} = i(\met\mu\rho P_\nu - \met\nu\rho P_\mu), \qquad
   \comm{M_{\mu\nu}}{K_\rho} = i(\met\mu\rho K_\nu - \met\nu\rho K_\mu),
\\ \comm{M_{\mu\nu}}{M_{\rho\sigma}} = i(\met\mu\rho M_{\nu\sigma} - \met\nu\rho M_{\mu\sigma} - \met\mu\sigma M_{\nu\rho} + \met\nu\sigma M_{\mu\rho})
\\ \comm{D}{P_\mu} = i P_\mu, \quad
   \comm{D}{K_\mu} = -i K_\mu, \quad
   \comm{K_\mu}{P_\nu} = -2i M_{\mu\nu} - 2i\met\mu\nu D.
  \end{gathered}
 \end{equation}
The theory may also possess gauge symmetries generated by $T_a$.

The space of states on which the conformal representation is defined is spanned by vectors of the form
 \begin{equation} \label{eq:ConformalBasis}
  \prod_n P_{\mu_n} \kop,
 \end{equation}
where $\Op(x)$ is a quasi-primary field and $\kop\equiv\Op(0)\kvac$ the corresponding conformal primary state, satisfying
 \begin{equation} \label{eq:ConformalPrimary}
  K_\mu\kop = 0,
\quad
  D\kop = i\Delta\kop,
 \end{equation}
where $\Delta$ is the scale dimension of $\Op$.
The quasi-primary $\Op$ may also belong to a non-zero spin representation, with the $\kop$ then transforming appropriately under $M_{\mu\nu}$.

\subsection{Conformal two, three and four point functions in $d=4$} \label{se:234pt}

As mentioned above, 
under the restrictions of conformal symmetry two and three point functions 
of scalar operators
are determined completely (up to normalisation) by scaling behaviour.
This may be seen as a consequence of the fact that no conformal invariants may be constructed from fewer than four points;
or, alternatively, that any set of three points may be mapped to any other three points by the action of the conformal group.

For scalar fields $\scalar_i(x)$ with scale dimension $\D{i}$ the two point function may be written
 \begin{equation} \label{eq:2pt}
  \corel{\sscpt[i]1\,\sscpt[j]2} = \delta_{ij} \frac{N}{\xx12^{\D{i}}} \,,
 \end{equation}
where $N$ is an arbitrary normalisation constant, and for convenience we have defined
 \begin{equation} \label{eq:xijDefn}
  \x{i}j = x_i-x_j.
 \end{equation}
%
Similarly, the three point function for scalar fields is given by
 \begin{equation} \label{eq:3pt}
  \corel{\sscpt[i]1\,\sscpt[j]2\,\sscpt[k]3} = \fraff{C_{\scalar_i\scalar_j\scalar_k}}{\xx12^{\hDDD{i}jk} \xx13^{\hDDD{k}ij} \xx23^{\hDDD{j}ki}} \,,
 \end{equation}
where $C_{\scalar_i\scalar_j\scalar_k}$ is a constant, depending only on the fields 
appearing in the correlation function.
The interpretation of these constants in terms of the operator product expansion will be explored below.
The four point function is the first to display non-trivial spatial dependence, and may be written
 \begin{multline} \label{eq:4pt}
  \corel{\OTTF[\,]{\sscpt{#1}}}
\\ = \frac1{\xx12^{\frac12(\D1+\D2)}\xx34^{\frac12(\D3+\D4)}}
     \bfrac{\xx24}{\xx14{}}^{\!\frac12\DD12}\! \bfrac{\xx14}{\xx13{}}^{\!\frac12\DD34}\! \plainF\uv,
 \end{multline}
where $\DD{i}{j}\equiv\D{i}-\D{j}$ and
$u,v$ are the two independent conformal invariants formed from four points,
 \begin{displaystretch}{1.0}
 \begin{equation} \label{eq:uvDefn}
  u = \frac{\xx12\xx34}{\xx13\xx24{}} \,,
 \quad
  v = \frac{\xx14\xx23}{\xx13\xx24{}} \,.
 \end{equation}
 \end{displaystretch}
$\plainF\uv$ is an arbitrary function, unrestricted by conformal invariance; 
its value is determined only by the dynamics of the theory.

\subsection{The Operator Product Expansion} \label{se:OPEs}

For scalar operators $\scalar_i$, such as appear in \secref{se:234pt}, 
we may write the contribution of a conformal primary operator $\OpI$ with spin $\ell$ and scale dimension $\Delta$ to the operator product expansion as
 \begin{equation} \label{eq:OPE}
  \sscpt1\sscpt2 \contrib \fraff{C_{\scalar_1\scalar_2\OpI}}{\xx12^{\hDDD[+\ell]12{}}} \ClD\xdx12\muel \OpI\muel(x_2) .
 \end{equation}
This gives the contribution of the ``conformal block'' of the quasi-primary operator,
including $\OpI$ itself and all its descendants, 
which are formed by the action of derivatives.
The index $I$ labels different operators with the same $\Delta$ and $\ell$.
The form of the differential operator $\ClD(x,\partial)$ is determined by the requirements of compatibility with the three and two point functions 
$\corel{\scalar_i\scalar_j\OpI}, \corel{\OpI\OpI[J]}$.
The first of these, extending \eqref{eq:3pt}, can be written
 \begin{multline} \label{eq:3pt-OI}
  \corel{\sscpt1\,\sscpt2\,\OpI\muel(x_3)}
\\ = \fraff{C_{\scalar_1\scalar_2\OpI}}{\xx12^{\hDDD[+\ell]12{}} \xx13^{\hDDl+12} \xx23^{\hDDl-12}} \xmuelst{Z}
 \end{multline}
where
 \begin{displaystretch}{1.0}
 \begin{equation} \label{eq:OPE-ZDefn}
  Z_\mu = \frac{\x13_\mu}{\xx13{}} - \frac{\x23_\mu}{\xx23{}},
\quad
  Z^2 = \frac{\xx12}{\xx13\xx23{}}
 \end{equation}
 \end{displaystretch}
transforms as a conformal vector at $x_3$, 
and $\tensor\muelst$ denotes the symmetric, trace-free part of rank $\ell$ tensor $\tensor\muel$
, e.g. for a rank 2, $d$-dimensional tensor $\tensor_{\mu\nu}$,
 \begin{equation} \label{eq:Traceless2}
  \tensor_{\{\mu\nu\}} = \tfrac12\bigl(\tensor_{\mu\nu} + \tensor_{\nu\mu}\bigr) - \tfrac1d \, \delta_{\mu\nu} \tensor_{\rho\rho}.
 \end{equation} 
%
%
The two point function is non-zero only for identical operators, given by
 \begin{equation} \label{eq:2pt-OIOJ}
  \corel{\OpI\muel(x_1)\,\OpI[J]\nuel(x_2)}
   = \delta_{IJ} \frac1{\xx12^\Delta} \xmunuelst\mu\nu{I}{(\x12)}
 \end{equation}
where the inversion tensor $I_{\mu\nu}$ is given by
 \begin{equation} \label{eq:OPE-IDefn}
  I_{\mu\nu}(x) = \delta_{\mu\nu} - 2\frac{x_\mu x_\nu}{x^2} \,,
 \end{equation}
and we have chosen the normalisation of scalar fields in \eqref{eq:2pt} to be given by $N=1$.
For \eqref{eq:OPE} to be consistent with \eqref{eq:3pt-OI}, \eqref{eq:2pt-OIOJ}, we require that the differential operator $\ClD(x,\partial)$ satisfies
 \begin{multline} \label{eq:OPE-CCond}
  \ClD\xdx12\nuel \frac1{\xx12^\Delta} \xmunuelst\nu\mu{I}{(\x23)}
\\ = \frac1{\xx13^{\hDDl+12}\xx23^{\hDDl-12}} \xmuelst{Z}.
 \end{multline}
In particular, $\ClD(x,0)\muel = \xmuelst{x}$.

If \eqref{eq:OPE} is applied to the mutual pairs of scalar operators appearing in the four point function \eqref{eq:4pt},
we deduce that the contribution from $\OpI$ is of the form
 \begin{multline} \label{eq:4pt-CPW}
  \corel{\OTTF[\,]{\sscpt{#1}}}
\\ \contrib
   \fraff{C_{\scalar_1\scalar_2\OpI} C_{\scalar_3\scalar_4\OpI}}{\xx12^{\frac12(\D1+\D2)}\xx34^{\frac12(\D3+\D4)}}
   \bfrac{\xx24}{\xx14{}}^{\!\frac12\DD12}\! \bfrac{\xx14}{\xx13{}}^{\!\frac12\DD34}\!
\\ \times u^{\frac12(\Twist)} \Gpw\ell\Delta{u}v{\DD21}{\DD43},
 \end{multline}
where 
$u,v$ are the conformal invariants defined in \eqref{eq:uvDefn},
and 
the functions $\Gpw.\ell\Delta$ are partial wave amplitudes, determined by
 \begin{multline} \label{eq:CPW-GCond}
  \ClD\xdx12\muel \frac1{\xx23^{\hDDl+34}\xx24^{\hDDl-34}} \xmuelst{Y},
\\ =
  \frac1{\xx14^{\hDDl+12}\xx23^{\hDDl-12}} \bfrac{\xx14}{\xx13{}}^{\!\hDDl+34}\! \Gpw\ell\Delta{u}v{\DD21}{\DD43},
 \end{multline}
 \begin{equation} \label{eq:OPE-YDefn}
  Y_\mu = \frac{\x32_\mu}{\xx23{}} - \frac{\x42_\mu}{\xx24{}}.
 \end{equation}
Assuming that running over $\set{\OpI\muel + \text{descendants}}$ gives a complete set,
we may expand $\plainF\uv$, as defined in \eqref{eq:4pt}, in terms of conformal partial waves.
By comparison with \eqref{eq:4pt-CPW}, we obtain
 \begin{equation} \label{eq:CPW-FDl}
  \plainF\uv = \sum_{\Delta,\ell} \adl\Delta\ell u^{\frac12(\Twist)} \Gpw\ell\Delta{u}v{\DD21}{\DD43},
 \end{equation}
with
 \begin{equation} \label{eq:CPW-aDlCooO}
  \adl\Delta\ell = \sum_I C_{\scalar_1\scalar_2\OpI} C_{\scalar_3\scalar_4\OpI}.
 \end{equation}
The $\Gpw.\ell\Delta$ possess power expansions in terms of $u$ and $1-v$.
Explicit forms are known, at least in certain dimensions, 
e.g. solutions for $d=2,4$ are given in \cite{rf:0011040}, and for $d=4,6$ in \cite{rf:0309180}.
An expression for $d=4$, and further properties of the conformal partial wave expansion in this case are given in \secref{se:CPW}.
An alternative derivation of some of these results is attempted in \secref{ch:OPE}.

 \section{Superconformal Theory in $d=4$}
 \label{se:Superconformal}

In four dimensions, the conformal group \eqref{eq:ConformalAlgebra} is $SO(2,4) \grouplike SU(2,2)$.
It may be extended by the inclusion of fermionic supercharges $\Q{i}, \Q*{i}, \q{i}, \q*{i}$ for $i=1,\dots,\N$. 
$\Q{i}$ and $\Q*{i}$ are the generators of Poincar\'e supersymmetries; 
$\q{i},\q*{i}$ generate conformal supersymmetries, and are required to close the superconformal algebra.

There is then an additional bosonic symmetry corresponding to the automorphisms of the SUSY generators, 
namely the $R$-symmetry $U(\N)$, with generators $\R{i}j$.
The superconformal group in $d=4$ is thus $SU(2,2|\N)$.
Full commutation relations are given in \cite{rf:0209056}.

The imposition of superconformal invariance and the superconformal Ward identities places further restrictions upon correlation functions 
in supersymmetric theories.
The correlators most amenable to analysis are those of the simplest operators, chiral primaries, which satisfy BPS-like constraints.
In \cite{rf:0006098}, 
the chiral four point function in the $\N=1$ superconformal theory is shown to be completely determined up to a single overall constant.
Constraints for $\N=2,4$ were derived in \cite{rf:0407060}; 
relevant results are quoted in \secref{se:Ward}.

\subsection{$\N=4$ SYM} \label{se:N=4-SYM}

From now on, we will concentrate on $\N=4$ supersymmetry in four dimensions.
In this case, we may consistently impose $\R{i}i=0$, restricting the $R$-symmetry to $SU(4)$.
The supergroup then reduces to $PSU(2,2|4)$.
Irreducible representations are labelled by the quantum numbers of the maximal bosonic subgroup, which may be mapped to
$SO(1,1) \cross SO(1,3) \cross SU(4)$.
The scale dimension $\Delta$ is the quantum number for $SO(1,1)$;
half-integer `spins' $\jjb$ label representations of $SO(1,3) \grouplike SU(2) \cross SU(2)$;
and $\dynk{k}pq$ are the Dynkin labels for representations of $SU(4)_R$.

The fundamental fields in $\N=4$ supersymmetric Yang Mills are 
six scalars $\adscalar_r$, four gauginos (chiral fermions) $\gaugino{i}, \gaugino*{i}$, 
and the gauge field $A_\mu$ with field strength $F_{\mu\nu}$.
They belong to the $\dynk010$, $\dynk100$, $\dynk001$ and trivial representations of $SU(4)_R$ respectively.
All transform in the adjoint representation of the gauge group, which has generators $\set{T_a}$ and may be arbitrary. 

\subsection{$\N=4$ Superconformal Multiplets} \label{se:N=4-Multiplets}


Representations of the superconformal group take the form of superconformal multiplets.
Each supermultiplet contains a unique operator $\Op$, the superconformal primary, 
which commutes with the conformal supersymmetry generators $\q{i}, \q*{i}$.
The rest of the multiplet is generated by the action of the supercharges on $\Op$.
Each application of $\Q{i},\Q*{i}$ raises the conformal dimension by $\frac12$; 
thus $\Op$ is operator in the multiplet with lowest dimension.
We also note that the descendant operators are conformal primary operators, 
but only $\Op$ is a \emph{super}conformal primary, a stricter condition.
The full space of states is realised as in \eqref{eq:ConformalBasis} through the action of $P_\mu$ on $\Op$ and its descendants.
A complete classification of $\N=4$ supermultiplets is undertaken in \cite{rf:0209056}.
The main results are summarised here.

If the action of the supercharges is unconstrained, the result is a long multiplet,
 \begin{equation} \label{eq:N=4-Long}
  \Longmult{k}pqj\jb\Delta, \quad \Delta \ge \max\bigl(2+2j+\tfrac12(3k+2p+q),2+2\jb+\tfrac12(k+2p+3q)\bigr).
 \end{equation}
The labels correspond to the scale, spin and $SU(4)_R$ representations of the multiplet's superconformal primary.
Long multiplets have continuous scale dimension $\Delta$, which need only satisfy the unitarity constraints above.

Alternatively, the superconformal primary operator may commute with certain supercharges.
This results in multiplet shortening.
We shall be interested in two classes of shortened multiplet, short,
 \begin{alignat}{3} \label{eq:N=4-Short}
  &\shortmult{}qp, & \Delta &= p+2q, &&
\\\intertext{and semi-short,} \label{eq:N=4-Semishort}
  &\semimult{}kpqj\jb, \quad & \Delta &= 2+j+\jb+k+p+q, \quad & k-q &= 2(j-\jb).
 \end{alignat}
Such multiplets satisfying shortening conditions correspond to BPS operators.
Consistency with the superconformal algebra leads to the conditions on their scale dimensions.
Thus they do not receive perturbative corrections, and are free from anomalous dimensions (however see below).

We shall primarily be concerned with symmetric multiplets, i.e. those with $k=q, j=\jb=\frac12\ell$.
Where the superconformal primary 
belongs to a representation of $SU(4)_R$ with Dynkin labels $\dynk{n-m}{2m}{n-m}, m\le n$, 
we will denote the corresponding multiplets by
 \begin{equation} \label{eq:N=4-Multiplets}
  \begin{alignedat}3
   &\text{Long multiplet:} &\qquad& \Long{n}{m}\ell\Delta &\quad \Delta &\ge 2n+\ell+2,
\\ &\text{Short multiplet:} &\qquad& \short{n}{m} &\quad \Delta &= 2n,
\\ &\text{Semi-short multiplet:} &\qquad& \semi{n}{m}\ell &\quad \Delta &= 2n+\ell+2. \qquad \qquad
  \end{alignedat}
 \end{equation}

At the unitarity threshold there is a potential ambiguity between the operator content of a single long multiplet and that of semi-short multiplets.
We may make the decomposition
 \begin{equation} \label{eq:N=4-ADecomp}
  \Long{n}{m}\ell{2n+\ell+2} \Meq \semi{n}{m}\ell \oplus \semi{n+1\,}{m}{\ell-1} \oplus \dotsb
 \end{equation}
(where we neglect the contribution from two additional semi-short multiplets with $k\ne q$).
We may extend \eqref{eq:N=4-ADecomp} to $\ell=0$ if we make the identification
 \begin{equation} \label{eq:N=4-Cnm1}
  \semi{n}{m}{-1} \Meq \short{n+1\,}{m} \,.
 \end{equation}
Thus short multiplets may acquire anomalous dimensions in an interacting theory 
if there exists a second multiplet with which they can pair to form a long multiplet.
The only multiplets guaranteed to be protected are $\short{n}n$ and $\short{n+1\,}n$, which cannot form part of a long multiplet.
The former denotes a \BPS2 short multiplet in the $\dynk0p0$ representation of $SU(4)_R$, $p=2n$.
Details of such operators, know as chiral primaries, are investigated in \secref[Chapter]{ch:4pt}.

\subsection{Products of $SU(4)$ Representations} \label{se:SU4Reps}

In $\N=4$ super Yang Mills, the content of the operator product expansion is restricted to operators 
belonging to $SU(4)_R$ representations present in the tensor product of those of the original operators.
We will be particularly interested in \BPS2 chiral primary operators belonging to the $\dynk0p0$ representation,
for which the tensor product may be written, for $p_1\le p_2$,
 \begin{align} \label{eq:SU4-0p10x0p20}
  \dynk0{p_1}0 \otimes \dynk0{p_2}0
   &\repeq \,
  \bigoplus_{r=0}^{p_1}\bigoplus_{s=0}^{p_1-r} \: \dynk{r}{p_2-p_1+2s}r,
\\\intertext{or alternatively} \label{eq:SU4-0p10x0p20-nm}
  \dynk0{p_1}0 \otimes \dynk0{p_2}0
   &\repeq
  \bigoplus_{\lemn{p_1}}\dynk{n-m}{p_2-p_1+2m}{n-m}.
 \end{align}
The corresponding result for the $p_1\ge p_2$ case is evident.
We note here that the tensor product of the representation $\dynk020$ with itself and also with $\dynk0p0$ both have size 6, and are given by
 \begin{align} \label{eq:SU4-020x020}
 &\begin{aligned}[b]
  \dynk020 \otimes \dynk020
   &\repeq
  \dynk000 \oplus \dynk101 \oplus \dynk020 \\& \qquad\quad \oplus \dynk202 \oplus \dynk040 \oplus \dynk121,
  \end{aligned}
\displaybreak[0]\\ \label{eq:SU4-020x0p0}
 &\begin{aligned}[b]
  \dynk020 \otimes \dynk0p0
   &\repeq
  \dynk0{p-2}0 \oplus \dynk0p0 \oplus \dynk0{p+2}0 \\& \qquad\quad \oplus \dynk1{p-2}1 \oplus \dynk1p1 \oplus \dynk2{p-2}2.
  \end{aligned}
 \end{align}
The dimension of an $SU(4)$ representation $\dynk{k}pq$ is given by
 \begin{equation}
  \tfrac1{12}(k+p+q+3)(k+p+2)(p+q+2)(k+1)(p+1)(q+1);
 \end{equation}
thus \eqref{eq:SU4-020x020} may be written
 \begin{equation}
  \repr{20'}\otimes\repr{20'}\repeq\repr1\oplus\repr{15}\oplus\repr{20'}\oplus\repr{84}\oplus\repr{105}\oplus\repr{175} ,
 \end{equation}
where conventionally $\dynk020$ is denoted $\repr{20'}$ to distinguish it from the
$\dynk110$, $\dynk300$ representations and their conjugates, which also have dimension 20.

 \section{Strong and Weak Coupling Limits}
 \label{se:Limits}

In interacting theories we may write the function $\plainF\uv$ appearing in \eqref{eq:4pt}, 
containing the non-trivial space-time dependence of a conformal scalar four point function, 
as a perturbative series in a small parameter $\epsilon$.
This could be the coupling, in the perturbative regime, or $1/N$ in the strong coupling, large $N$ limit.
The expansion takes the general form
 \begin{equation} \label{eq:eps-FExpn}
  \plainF\uv = \plainF\ord0\uv + \sum_{r=1}^\infty \epsilon^r \sum_{s=0}^r \ln^s u \, \plainF\ord{r,s}\uv,
 \end{equation}
where $\plainF\ord0$ is the free field value.
We may write the conformal partial wave expansion for $\plainF$, as in \eqref{eq:CPW-FDl}, 
with $\Delta\goesto\D{I}$, $I$ labelling different operators with identical spin $\ell$ and scale dimension $\Delta\ord0$ at $\epsilon=0$,
and letting
 \begin{equation} \label{eq:eps-DaExpn}
  \D{I} = \Delta\ord0 + \epsilon \D{I,1} + \epsilon^2 \D{I,2} + \dotsb \,,
\qquad
  \adl{\D{I}}\ell = \adlI{\Delta\ord0}\ell{I} + \epsilon \adlI[b]{\Delta\ord0}\ell{I} + \dotsb \,.
 \end{equation}

\subsection{Loop Integrals and Perturbative Amplitudes} \label{se:Phi}

Results in the perturbative theory are given by conformal loop integrals $\loopint{L}$,
which in the manner of \cite{rf:0412335} we will express in terms of the variables 
\ifthenelse{\value{xprimecount}=0}%
{$x',\xb'$, where
 \begin{equation} \label{eq:xxDefn}
  u = \frac{x'\xb'}{(1-x')(1-\xb')} = x\xb,
\quad
  v = \frac1{(1-x')(1-\xb')} = (1-x)(1-\xb).
 \end{equation}
Clearly we have
 \begin{equation} \label{eq:x'Defn}
  x' = \frac{x}{x-1} \,.
 \end{equation} \stepcounter{xprimecount}}%
{$x,x'$ and their conjugates, as defined in \eqref{eq:xxDefn}, \eqref{eq:x'Defn}.}%
Then defining
 \begin{equation} \label{eq:Phi^}
  \loopint^L\xxb' = \loopint^L(\xb',x') = \loopint{L}\uv,
 \end{equation}
for $L=1,2$ we have
 \begin{equation} \label{eq:Phi12Values}
  \begin{aligned}
   v\loopint^1\xxb' &= -\lnxxb' \, \phlog1\xxb' +2\phlog2\xxb',
\\ v\loopint^2\xxb' &= \tfrac12 \lnxxb'[2] \, \phlog2\xxb' - 3\lnxxb' \, \phlog3\xxb' + 6\phlog4\xxb',
  \end{aligned}
 \end{equation}
where the $\phlog{n}$ are defined in terms of single variable polylogarithms,
 \begin{equation} \label{eq:plogDefn}
  \phlog{n}(x,\xb) = \frac{\Li{n}(x)-\Li{n}(\xb)}\xxb,
\qquad
  \Li{n}(x) = \sum_{r=1}^\infty \frac{x^r}{r^n},
\:
  \Li1(x) = -\ln(1-x).
 \end{equation}
From standard polylogarithm identities, it follows that the $\loopint{L}$ obey
 \begin{equation} \label{eq:PhiRefl}
  \begin{gathered}
   \loopint{L}\uv = \loopint{L}(v,u)
\iff
   \loopint^L\xxb' = \loopint^L(1/x',1/\xb'),
\\
   \fB\uv = \tfrac1v\,\fB(\tfrac{u}v,\tfrac1v) = \tfrac1u\,\fB(\tfrac1u,\tfrac{v}u).
  \end{gathered}
 \end{equation}

\subsection{$\Db{}{}{}{}$ Functions and Large $N$ Amplitudes} \label{se:Dbar}

Using the AdS/CFT correspondence at strong coupling, $n$-point correlation functions are given by integrals on $\AdS{d+1}$, 
as defined in \cite{rf:9903196}, of the form
 \begin{equation} \label{eq:D}
  \DInt{\D{1}\dots\D{n}}(x_1,\dots,x_n)
   = \frac1{\pi^{\frac12d}} \int_0^\infty \dd{z} \int \dd[d]x \frac1{z^{d+1}} \prod_{i=1}^n \bfrac{z}{z^2+(x-x_i)^2}^{\!\D{i}} 
 \end{equation}
We may express the $D{}$ functions in terms of the conformal invariants $u,v$ by defining, in the case $n=4$, $\OTTF[+]{\D{#1}} = 2\Sigma$,
 \begin{multline} \label{eq:Dbar}
  \DInt{\ottf[]\Delta}(\ottf{x}) \\
   = \frac{\Gamma(\Sigma-\frac12d)}{2\OTTF{\Gamma(\D{#1})}}
     \frac{\xx14^{\Sigma-\D1-\D4}\xx34^{\Sigma-\D3-\D4}}{\xx13^{\Sigma-\D4}\xx24^{\D2}} \DbD1234\uv\,.
 \end{multline}
In terms of the loop integrals of \secref{se:Phi}, $\Db1111\uv = \fB\uv$.
Letting
 \begin{equation}
  2s=\D1+\D2-\D3-\D4,
 \end{equation}
then for $s=0,1,\dotsc$ we can write, \cite{rf:0412335},
 \begin{equation} \label{eq:DDecomp}
  \DbD1234\uv = \ln u \, \DbD1234\uv\reg + \DbD1234\uv\sing,
 \end{equation} 
where $\Db{}{}{}{}{}\reg$ has a regular power expansions in $u, 1-v$,
and $\Db{}{}{}{}{}\sing$ has a similar expansion, but with the possibility of negative powers $u^{-r}$, 
$r\le s$, expressible as
 \begin{multline} \label{eq:Dlogfree}
  \DbD1234\uv\sing
   = u^{-s} \frac{\Gamma(\D1-s)\Gamma(\D2-s)\Gamma(\D3)\Gamma(\D4)}{\Gamma(\D3+\D4)}
\\ \times \sum_{m=0}^{s-1} (-1)^m (s-m+1)! \frac{\OTT{\pochhammer{\D{#1}-s}m}2\pochhammer{\D3}m\pochhammer{\D4}m}{m!\pochhammer{\D3+\D4}{2m}}
\\ \times u^m \hyperg{\D2-s+m}{\D3+m}{\D3+\D4+2m}{1-v}.
 \end{multline}
Cases where $s<0$ may be dealt with by means of the symmetry relations for $\Db{}{}{}{}$ functions
.
From the permutation symmetries of the definition \eqref{eq:D} we have several such relations, 
which we quote in the form given in \cite{rf:0212116}, namely
 \begin{equation} \label{eq:DSym}
  \begin{aligned}
   \DbD1234\uv
    &= v^{\D1+\D4-\Sigma}\,\DbD2143\uv
     = u^{\D3+\D4-\Sigma}\,\DbD4321\uv
\\  &= v^{-\D2}\DbD1243\uvmod*
     = v^{\D4-\Sigma}\,\DbD2134\uvmod*
\\  &= \DbD3214(v,u) \,,
  \end{aligned}
 \end{equation}
and
 \begin{equation} \label{eq:DRefl}
  \DbD1234\uv = \DbS{\Sigma-\D3}{\Sigma-\D4}{\Sigma-\D1}{\Sigma-\D2}\uv \,.
 \end{equation}
Additionally, for $\Db{}{}{}{}$ functions with different values of $\Sigma$, there are the relations
 \begin{equation} \label{eq:DUpDown}
  \begin{aligned}
   (\D2+\D4-\Sigma)\DbD1234\uv &= \DbS{\D1}{\D2+1}{\D3}{\D4+1}\uv - \DbS{\D1+1}{\D2}{\D3+1}{\D4}\uv \,,
\\ (\D1+\D4-\Sigma)\DbD1234\uv &= \DbS{\D1+1}{\D2}{\D3}{\D4+1}\uv - v\DbS{\D1}{\D2+1}{\D3+1}{\D4}\uv\,,
\\ (\D3+\D4-\Sigma)\DbD1234\uv &= \DbS{\D1}{\D2}{\D3+1}{\D4+1}\uv - u\DbS{\D1+1}{\D2+1}{\D3}{\D4}\uv\,,
  \end{aligned}
 \end{equation}
and we have the sum
 \begin{multline} \label{eq:DSum}
  \D4\DbD1234\uv = \DbS{\D1}{\D2}{\D3+1}{\D4+1}\uv + \DbS{\D1}{\D2+1}{\D3}{\D4+1}\uv \\ + \DbS{\D1+1}{\D2}{\D3}{\D4+1}\uv \,.
 \end{multline}
Taking this in the limit where one of the $\D{i}\tendsto0$ leads to
 \begin{multline} \label{eq:DLimit}
  \bigl(\DbS{\D1+1}{\D2}{\D3+1}{\D4}\uv + u\DbS{\D1+1}{\D2+1}{\D3}{\D4}\uv
\\ + \evalat{\D1+\D2+\D3=\D4}{v\DbS{\D1}{\D2+1}{\D3+1}{\D4}\uv\bigr)}
   = \OTT{\Gamma(\D{#1})}3 \,.
 \end{multline}

 \section{Thesis Outline}
 \label{se:Outline}

The structure of the thesis, then, is as follows.
In \secref{ch:4pt}, we begin by defining chiral primary \BPS2 operators in $\N=4$ SYM, and introduce their four point functions.
These we express in a form which manifestly satisfies the restrictions arising from the superconformal Ward identities \cite{rf:0407060},
and subsequently expand in terms of conformal partial waves.
We particularly consider correlation functions of two pairs of identical chiral primaries, 
one pair having the lowest possible scale dimension, $\Delta=2$.
Making use of crossing symmetries of these four point functions we determine their value in the free field limit,
up to three numeric constants, two of which we may fix exactly.
In this case we analyse the partial wave expansion in terms of contributions from long, short, and semi-short multiplets
belonging to different representations of the superconformal algebra.
We determine such contributions from the free field, which we write such as to remove the apparent presence of non-unitary multiplets.

In \secref{ch:Interacting}, we compare this analysis with established results at strong and weak coupling,
obtained from supergravity calculations via the AdS/CFT correspondence \cite{rf:linda} and perturbation theory \cite{rf:0504061} respectively.
In the large $N$, strong coupling limit, we find non-trivial cancellations between our free field values and the supergravity results, 
providing a persuasive consistency check.
In the perturbative case, the comparison enables us to obtain values for the anomalous dimensions of the lowest twist operators present,
up to second order, and also for the first order correction to the coupling.
These results require the expansion of certain hypergeometric and logarithmic functions in terms of conformal partial waves, 
which is performed in \secref{se:CPW-Coeff}.

In \secref{ch:Semishort}, we make use of results found in \cite{rf:0609179} to count shortened $\N=4$ SYM operators.
%
We begin by constructing, from fundamental fields, the most general operators belonging to certain $SU(4)_R$ representations at low twists. 
We then use a combinatoric approach to count the number of independent solutions to the necessary conditions imposed on such operators.
In \secref{se:Z}, we use an alternative method, starting with the partition functions for shortened operators, again given in \cite{rf:0609179}.
From these we derive generating functions $\generatingfunctionsymbol(\sigma)$ for the number of operators with spin $\ell=0,1,2,\dotsc$.
Explicit values are obtained, again, for specific $R$-symmetry representations at low twist, in various sectors of the theory.
In the more general case, we find leading order approximations for $\generatingfunctionsymbol(\sigma)$ as $\sigma\tendsto1$,
and consider their asymptotic behaviour at large twist.

Finally, in \secref{ch:OPE}, we consider the conformal field theory operator product expansion \eqref{eq:OPE}, which satisfies \eqref{eq:OPE-CCond}
and is crucial in determining expressions for the conformal partial waves.
Taking an alternative approach to \cite{rf:0011040},~\cite{rf:0309180}, we attempt to find solutions in terms of series expansions,
first in restricted cases (e.g. for scalar operators only, in two dimensions),
and then for the more general problem, which is analysed in \secref{se:OPE-Spin}.

\chapter{$\N=4$ Superconformal Four Point Functions}
 \label{ch:4pt}

 \section{Introduction}
 \label{se:4pt-introduction}

In $\N=4$ superconformal theory, chiral primary \BPS2 operators belonging to an $SU(4)_R$ representation with Dynkin labels $\dynk0p0$, 
which are spinless and have scale dimension $\Delta = p$, may be written as symmetric traceless $SO(6)$ tensor fields $\chiral\rpee(x)$, 
with $r_i=1,\dots,6$.
It is convenient, as in \cite{rf:0407060},\cite{rf:0412335}, to avoid the proliferation of invariant tensors as $p$ increases by
considering $\chiral\dimn{p}(x,t) = \chiral\rpee(x)\trp$,
homogeneous of degree $p$ in $t$, where $t_r$ is an arbitrary six-dimensional complex vector satisfying $t^2 = 0$.
Clearly $\chiral\rpee(x)$ may be recovered from $\chiral\dimn{p}$.
The four point function of chiral primary operators then becomes a homogeneous polynomial in $\ottf{t}$ of degree $\ottf{p}$ respectively.
Since the $t_i$ are null vectors, the conformally covariant four point function reduces to an invariant function $\Ggenp\uvst$, with $\sigma,\tau$
the two independent conformal invariants that may be formed from the $t_i$, given by
 \begin{equation} \label{eq:stDefn}
  \sigma = \frac{\dpt13\dpt24}{\dpt12\dpt34},
 \qquad
  \tau   = \frac{\dpt14\dpt23}{\dpt12\dpt34},
 \end{equation}
analogous to the spatial conformal invariants $u,v$, defined in \eqref{eq:uvDefn}.
%
We define
 \begin{multline} \label{eq:4ptG}
  \fourpt\hspt\chiral{p_1}\chiral{p_2}\chiral{p_3}\chiral{p_4}
\\ = \bfrac{\dpt14}{\xx14{}}^{\!p_1-E}\! \bfrac{\dpt24}{\xx24{}}^{\!p_2-E}\! \bfrac{\dpt12}{\xx12{}}^{\!E}\! \bfrac{\dpt34}{\xx34{}}^{\!p_3} \Ggenp\uvst,
 \end{multline}
where the parameter $E$ is the extremality of the four point function, defined by
 \begin{equation} \label{eq:4pt-EDefn}
  2E = p_1 + p_2 + p_3 - p_4,
\qquad
  p_1,p_2,p_3 \le p_4,
 \end{equation}
where without loss of generality we take $p_4$ to be maximal.
\pagebreak[3]%
We note that the right-hand side of \eqref{eq:4ptG} satisfies conformal covariance, and 
for $p_1,p_2\ge E$, the function $\Ggenp\uvst$ is a polynomial of degree $\min(E,p_3)$ in $\sigma,\tau$,
i.e. may be expanded in terms of the form $\sigma^r \tau^s, r+s\le E,p_3$.
\vspace{-1ex}
\paragraph{} \label{se:Ward}
It will often be convenient to consider, as an alternative to $u,v$,
\ifthenelse{\value{xprimecount}=0}%
{new variables $x,\xb$ defined by
 \begin{equation} \label{eq:xxDefn}
  u = x\xb,
 \quad
  v = (1-x)(1-\xb),
 \end{equation}}%
{the variables $x,\xb$ defined in \eqref{eq:xxDefn},}
and analogously in place of $\sigma,\tau$ using
 \begin{equation} \label{eq:aaDefn}
  \sigma = \alpha\alphb,
 \quad
  \tau = (1-\alpha)(1-\alphb).
 \end{equation}
In terms of these variables, $\Ggenp\uvst$ becomes a symmetric function of $x,\xb$ and $\alpha,\alphb$.
Furthermore, analysis of superconformal Ward identities, as given in \cite[(4.44)]{rf:0407060}, requires
 \begin{equation} \label{eq:Ward}
  \evalat{\alphb=\frac1\xb}{\Ggenp\uvst} = \f\xa = \fk + \amx\fhat\xa.
 \end{equation}
We may solve \eqref{eq:Ward} by writing
 \begin{multline} \label{eq:GDecomp}
   \Ggenp\uvst = \fk + \Gx\fhat\uvst
\\  + \ax\abx\axb\abxb\GH\uvst,
 \end{multline}
where the contribution from $\fhat\xa$ is given explicitly by
 \begin{multline} \label{eq:GfDefn}
  \Gx\fhat\uvst + 2\fk
\\ = \frac{\abx\axb\bigl(\f\xa+\f\xbab\bigr) - \ax\abxb\bigl(\f\xab+\f\xba\bigr)}{(\xxb)(\aab)}.
 \end{multline}
If we substitute $\fhat$ for $\f$, as in \eqref{eq:Ward}, the contributions from $\fk$ can be seen to cancel.
From the definitions \eqref{eq:xxDefn}, \eqref{eq:aaDefn}, we see that
 \begin{multline} \label{eq:4pt-sDefn}
  s\uvst = \ax\abx\axb\abxb
\\ = v + \sigma^2uv + \tau^2u + \sigma v(v-1-u) + \tau(1-u-v) + \sigma\tau u(u-1-v),
 \end{multline}
and we conclude that $\GH\uvst$ is polynomial in $\sigma,\tau$ of degree $E-2$ (or $p_3-2$).
In the extremal and next-to-extremal cases $E=0,1$, we must have $\GH$ identically zero.

Although not essential as far as the superconformal symmetry is concerned, dynamical considerations ensure that the value of $\fhat$ 
is always identical to the result obtained for free fields.
Non-trivial dynamic contributions to the four point function are contained entirely within $\GH$.
We may thus write \eqref{eq:GDecomp} as
 \begin{align} \label{eq:GFree+Dyn}
  \Ggenp\uvst &= \Ggenp\ord0\uvst + s\uvst\GH\dyn\uvst,
\\\intertext{where} \label{eq:G0Decomp}
  \Ggenp\ord0\uvst &= \fk + \Gx\fhat\uvst + s\uvst\GH\ord0\uvst.
\\\intertext{$\Ggenp\ord0, \GH\ord0$ denote the free field values, and $\GH\dyn$ is the dynamic part of $\GH$. Clearly, we have} \label{eq:HFree+Dyn}
  \GH\uvst &= \GH\ord0\uvst 
     + \GH\dyn\uvst.
 \end{align}
Thus there can be no dynamic contribution for the extremal and next-to-extremal four point functions (when we have $\GH\equiv0$).

 \section{Conformal Partial Waves}
 \label{se:CPW}

In four dimensions, the form of the conformal partial waves defined in \eqref{eq:CPW-GCond} 
--- which are 
essentially harmonic functions for $SO(4,2)$ ---
is given explicitly in \cite{rf:0011040},~\cite{rf:0309180}.
We express it here after the manner of \cite
{rf:0407060}, 
writing
 \begin{equation} \label{eq:CPW-Defn}
  u^j\Gpw\ell{2a+2j+\ell}uv{2\bt}{-2\bt*} = -\frac1\xxb \left[ \gbb*a{j+\ell+1}x \, \gbb*aj\xb - \swap{x}\xb \right],
 \end{equation}
with $x,\xb$ as in \eqref{eq:xxDefn}, and
where for $j=0,1,2,\dots$
 \begin{equation} \label{eq:CPW-gDefn}
  \gbb*{a}jx = (-x)^j \hyperg{a+\bt+j-1}{a+\bt*+j-1}{2a+2j-2}x.
 \end{equation}
It follows directly from the definition \eqref{eq:CPW-gDefn} that
 \begin{equation} \label{eq:CPW-gRefl}
  \gbb*{a}{j+n}x = (-x)^n \gbb*{a+n}jx,
 \end{equation}
and using the standard properties \cite{rf:tables} of hypergeometric functions,\footnote{%
Specifically, the Pfaff Transformation $(1-x)^{-a}\hyperg{a}bc{\frac{x}{x-1}} = \hyperg{a}{c-b}cx$}
we may write the additional relation
 \begin{equation} \label{eq:CPW-gx'}
  \begin{split}
   \gbb*{a}j{x'} &= (-1)^j(1-x)^{a+\bt-1}\gpw\bt{-\bt*}ajx
\\  &= (-1)^j(1-x)^{a+\bt*-1}\gpw{-\bt}{\bt*}ajx,
  \end{split}
 \end{equation}
where $x'$ is
\ifthenelse{\value{xprimecount}=0}%
{defined as
 \begin{equation} \label{eq:x'Defn}
  x' = \frac{x}{x-1} \,.
 \end{equation} \stepcounter{xprimecount}}%
{as defined by \eqref{eq:x'Defn}.}%

With reference to \eqref{eq:4ptG}, we define, for a four point function of scalar operators, dimensions $\ottf\Delta$
with $2\Sigma = \OTTF[+]{\D{#1}}$,
 \begin{equation} \label{eq:4pt-G}
  \corel{\OTTF[\,]{\sscpt{#1}}}
   = \frac{\xx14^{\Sigma-\D1-\D4}\xx24^{\Sigma-\D2-\D4}}{\xx12^{\Sigma-\D4}\xx34^{\D3}} \plainG\uv.
 \end{equation}
Clearly, by comparison with \eqref{eq:4pt},
 \begin{equation} \label{eq:4pt-GF}
  u^{\frac12\DD43} \plainG\uv = \plainF\uv.
 \end{equation}
Thus, setting $\Delta = 2t+\DD43+\ell$ in \eqref{eq:CPW-FDl}, we conclude that $\plainG\uv$ has conformal partial wave expansion
 \begin{equation} \label{eq:CPW-G}
  \plainG\uv = \sum_{t,\ell\ge0} a_{t\ell} \, u^t \Gpw\ell{2t+\DD43+\ell}uv{\DD21}{\DD43}.
 \end{equation}

For the $\N=4$ superconformal chiral four point function \eqref{eq:4ptG}, $\D{i}=p_i$, 
and by comparison with \eqref{eq:4pt-G}, we may extend \eqref{eq:CPW-G} to include $t_i$ dependence, obtaining
 \begin{equation} \label{eq:CPW-GExpn}
  \Ggenp\uvst = \sum_{t,\ell\ge0} a_{t\ell}\st* \, u^t \Gpw\ell{2t-2\bt*+\ell}uv{2\bt}{-2\bt*},
 \end{equation}
where we have defined
 \begin{equation} \label{eq:CPW-bgDefn}
  \bt = \tfrac12(p_2-p_1),
\quad
  \bt*= \tfrac12(p_3-p_4).
 \end{equation}


\subsection{Partial Wave Expansions} \label{se:CPWCPW}

We now let $F\uv$ be an arbitrary function, with a power series expansion in \mbox{$u,1-v$}, 
which we write as an expansion in conformal partial waves with some lowest twist $2a$,
 \begin{equation} \label{eq:CPW-F}
  F\uv = \sum_{j,\ell\ge0} a_{j,\ell} \, u^j \Gpw\ell{2a+2j+\ell}uv{2\bt}{-2\bt*}.
 \end{equation}
%
%
\eqref{eq:CPW-F} may be written as
 \begin{equation} \label{eq:CPW-xxF}
  (\xxb)F\uv = \sum_{k\ge0} F\p{k}(x) \, \xb^k = \sum_{j\ge0} F\g{j}(x) \, \gbb*{a}j\xb,
 \end{equation}
where
 \begin{equation} \label{eq:CPW-FjExpn-midpoint}
  F\g{j}(x) = \sum_{\ell=0}^{j-1} a_{j-\ell-1,\ell} \, \gbb*{a}{j-\ell-1}x - \sum_{\ell=0}^\infty a_{j,\ell} \, \gbb*{a}{j+\ell+1}x.
 \end{equation}
If we extend the definition of $a_{j\ell}$ to $\ell<0$ by letting
 \begin{equation} \label{eq:CPW-aRefl}
  a_{j,\ell}=-a_{j+\ell+1,-\ell-2},
 \qquad
  a_{j,-1}=0,
 \end{equation}
we may write \eqref{eq:CPW-FjExpn-midpoint} in the form
 \begin{equation} \label{eq:CPW-FjExpn}
  F\g{j}(x) = -\sum_{\subbox{\ell=-j-1}}^\infty a_{j,\ell} \, \gbb*{a}{j+\ell+1}x.
 \end{equation}
Using the power series expansion of the hypergeometric function, we may match powers of $\xb$ in \eqref{eq:CPW-xxF} to obtain
 \begin{gather} \label{eq:CPW-alphaExpn}
  F\p{k}(x) = \sum_{j=0}^k \alpha_{k,j} \, F\g{j}(x),
\displaybreak[0]\\ \label{eq:CPW-alpha}
  \alpha_{k,j} = (-1)^j \, \frac{\pochhammer{a+\bt+j-1}{k-j}\pochhammer{a+\bt*+j-1}{k-j}}{(k-j)!\,\pochhammer{2a+2j-2}{k-j}}.
 \end{gather}
The relation \eqref{eq:CPW-alphaExpn} may be inverted (see
below%
) giving
 \begin{gather} \label{eq:CPW-betaExpn}
  F\g{j}(x) = \sum_{k=0}^j \beta_{j,k} \, F\p{k}(x),
\\ \label{eq:CPW-beta}
  \beta_{j,k} = (-1)^k \, \frac{\pochhammer{a+\bt+k-1}{j-k}\pochhammer{a+\bt*+k-1}{j-k}}{(j-k)!\,\pochhammer{2a+j+k-3}{j-k}}.
 \end{gather}
In particular, by considering $F\p{k} = \delta_{kn}$, we have
 \begin{equation} \label{eq:CPW-xnExpn}
  x^n =
   \sum_{j=n}^\infty \beta_{j,n}\gbb*{a}jx
  .
 \end{equation}
Considering \eqref{eq:CPW-gx'}, we thus have
 \begin{equation}
  \begin{aligned}[b]
   x'^n &= \sum_{j=n}^\infty \beta_{j,n}\gbb*{a}j{x'}
 = \sum_{j=n}^\infty \beta_{j,n}(-1)^j(1-x)^{a+\bt*-1}\gpw{-\bt}{\bt*}{a}jx
\\ &= \sum_{j=n}^\infty (-1)^j\beta_{j,n}\gpw\bt{-\bt*}{1-\bt}jx = \sum_{j=n}^\infty (-1)^j\beta_{j,n}\gpw{-\bt}{\bt*}{1-\bt*}jx.
  \end{aligned}
 \end{equation}
and if we define
 \begin{equation} \label{eq:CPW-beta'}
  \beta'_{j,k} = \left. \beta_{j,k} \right|_{\bt\goesto-\bt}
  = (-1)^k \, \frac{\pochhammer{a-\bt+k-1}{j-k}\pochhammer{a+\bt*+k-1}{j-k}}{(j-k)!\,\pochhammer{2a+j+k-3}{j-k}},
 \end{equation}
then
 \begin{equation} \label{eq:CPW-x'nExpn}
  \begin{aligned}[b]
   x'^n &
  = \sum_{j=n}^\infty \beta'_{j,n}\gpw{-\bt}{\bt*}{a}j{x'}
= \sum_{j=n}^\infty (-1)^j\beta'_{j,n}\gbb*{1-\bt*}jx.
  \end{aligned}
 \end{equation}

\subsubsection{Proof of inversion formula} \label{se:CPW-InversionProof}

For \eqref{eq:CPW-alphaExpn}, \eqref{eq:CPW-betaExpn} to be consistent, we require $\alpha_{k,j}$, $\beta_{j,k}$ to satisfy
 \begin{equation} \label{eq:CPW-alphabetaCond}
  \sum_{k=\ell}^j \beta_{j,k}\,\alpha_{k,\ell} = \delta_{j\ell}, \qquad \ell\le j.
 \end{equation}
This is trivially true in the case that $\ell=j$, when $\beta_{k,k}\alpha_{k,k} = (-1)^k\,(-1)^k = 1$.
For the $\ell<j$ case, the $k$-dependent part of $\beta_{j,k}\,\alpha_{k,\ell}$ is
 \begin{equation}
  \frac{(-1)^k}{(j-k)!\,(k-\ell)!}\,\frac{\Gamma(2a+k+j-3)}{\Gamma(2a+k+\ell-2)}.
 \end{equation}
Substituting $N=j-\ell\ge1$, $n=k-\ell$, the condition \eqref{eq:CPW-alphabetaCond} becomes
 \begin{equation} \label{eq:CPW-Inversion-nNSum}
  \sum_{n=0}^N \frac{(-1)^n}{(N-n)!\,n!}\,\frac{\Gamma(2a+j+\ell+n-3)}{\Gamma(2a+2\ell+n-2)} = 0.
 \end{equation}
If we define $B=2a+2\ell-2$, the ratio of $\Gamma$-functions in \eqref{eq:CPW-Inversion-nNSum} becomes
 \begin{equation}
  \frac{\Gamma(B+N-1+n)}{\Gamma(B+n)} = \pochhammer{B+n}{N-1} \,,
 \end{equation}
a polynomial in $n$ of degree $(N-1)$; hence the sum \eqref{eq:CPW-Inversion-nNSum} vanishes, 
which we may see by considering, for $0\le r\le N-1$,
 \begin{equation*}
  \sum_{n=0}^N \binom{N}n (-1)^n n^r
   = \left(x\frac{d}{dx}\right)^r \sum_{n=0}^N \binom{N}n (-x)^n \biggr|_{x=1} \!\!\!
   = \left(x\frac{d}{dx}\right)^r (1-x)^N \biggr|_{x=1} \!\!\! = 0.
 \end{equation*}

\subsection{Jacobi Polynomials} \label{se:Jacobi}

We may further decompose the conformal expansion \eqref{eq:CPW-GExpn} into contributions from different $SU(4)_R$ representations%
 by writing,
for $p_1,p_2\ge E$,
 \begin{equation} \label{eq:CPWJac-GExpn}
  \Ggenp\uvst
  = \sum_{\subbox[0]{\substack{\lemn{E}\\t,\ell}}} a_{nm,t\ell} \, \JJacobi{p_1-E}{p_2-E}{n}m\st*  \, u^t \Gpw\ell{2t-2\bt*+\ell}uv{2\bt}{-2\bt*},
 \end{equation}
with $E$ as defined in \eqref{eq:4pt-EDefn}, and
where the $\JJab{n}m\st*$ are defined in terms of Jacobi polynomials by
 \begin{equation} \label{eq:Jac-2varDefn}
  \JJab{n}m\st* = \frac{\Jab{n+1}(y)\Jab{m}(\yb) - \Jab{m}(y)\Jab{n+1}(\yb)}\yyb,
 \end{equation}
where
 \begin{equation} \label{eq:yyDefn}
  y = 2\alpha-1, \quad \yb = 2\alphb-1.
 \end{equation}
From the above definition, the 2-variable Jacobi polynomials are manifestly symmetric in $y,\yb$, and possess the additional symmetry
 \begin{equation} \label{eq:Jac-Refl}
  \JJab{n}m\st* = -\JJab{m-1}{n+1}\st*,
\qquad
  \JJab{n}{n+1}\st* = 0.
 \end{equation}
Thus we may take $m\le n$.
The polynomial $\JJacobi{p_1-E}{p_2-E}{n}m\st*$ gives the contribution of operators belonging to $SU(4)_R$ representation
$\dynk{n-m}{p_1+p_2-2E+2m}{n-m}$;
the limits of the sum ensure there is one such contribution for each representation common to the tensor products
$\dynk0{p_1}0\otimes\dynk0{p_2}0, \dynk0{p_3}0\otimes\dynk0{p_4}0$,
as given by \eqref{eq:SU4-0p10x0p20-nm}.
We have already defined $2\bt=p_2-p_1, -2\bt*=p_4-p_3$, which with \eqref{eq:4pt-EDefn} gives
 \begin{equation} \label{eq:Jac-abParam}
  p_1-E = -\bt-\bt*,
\qquad
  p_2-E = \bt-\bt*.
 \end{equation}

Single-variable Jacobi polynomials satisfy a recurrence relationship \cite{rf:tables}, which may be conveniently expressed as
 \begin{equation} \label{eq:Jac-Recur}
  \tfrac12y\Jab{n}(y) = \gpl{n}\Jab{n+1}(y) + \gze{n}\Jab{n}(y) + \gmi{n}\Jab{n-1}(y),
 \end{equation}
where, with $a=-\bt-\bt*, b=\bt-\bt*$ as in \eqref{eq:Jac-abParam}, we have
 \begin{equation} \label{eq:Jac-gamma}
  \begin{gathered}
   \gpl{n} = \frac{(n+1)(n-2\bt*+1)}{2(n-\bt*+1)(2n-2\bt*+1)}; \qquad
   \gmi{n} = \frac{(n-\bt-\bt*)(n+\bt-\bt*)}{2(n-\bt*)(2n-2\bt*+1)}; \\
   \gze{n} = \frac{-\bt\,\bt*}{2(n-\bt*)(n-\bt*+1)}.
  \end{gathered}
 \end{equation}
It follows from \eqref{eq:Jac-2varDefn}, \eqref{eq:Jac-Recur} that
 \begin{equation} \label{eq:Jac-2varRecur}
  \begin{gathered}
   (\sigma-\tau)\Jab{nm}\st* = \sum\Jlim \gam{n+1}r\Jab{n+r,m}\st* + \gam{m}r\Jab{n,m+r}\st*, \\
   \tfrac12(\sigma+\tau-\tfrac12)\Jab{nm}\st* = \sum\Jlim[r,s] \gam{n+1}r\gam{m}s\Jab{n+r,m+s}\st*,
  \end{gathered}
 \end{equation}
where by \eqref{eq:yyDefn},
 \begin{equation}
  2(\sigma-\tau) = y+\yb,
\qquad
  2(\sigma+\tau-\tfrac12) = y\yb.
 \end{equation}
Note also that
 \begin{equation} \label{eq:Jac00}
  \JJbg00\st* = 1-\bt* = \frac1{2\gpl0}.
 \end{equation}

\subsection{Conformal Wave Recurrence Relations} \label{se:CPW-Recur}

We may write a similar recurrence relation for the $\gbb*{a}jx$, as given in \eqref{eq:CPW-gDefn},
by matching terms in the power expansion of the relevant hypergeometric functions.
We find
 \begin{equation} \label{eq:CPW-gRecur}
  -\tfrac1x\gbb*{a}{j+1}x = \gbb*{a}jx - (\tfrac12+\gze{a+j+\bt*-1})\gbb*{a}{j+1}x + c_{a+j+\bt*}\gbb*{a}{j+2}x,
 \end{equation}
with $c_j = \gpl{j-1}\gmi{j}$ and the $\gam{n}r$ as in \eqref{eq:Jac-gamma}.
This simplifies further if we set $a=1-\bt*$.
Thus
 \begin{equation} \label{eq:CPW-axaxgbg}
  \amx\abmx\gbb*{1-\bt*}{j+1}x = \sum_{t=j}^{j+4} c_{t,j}\gbb*{1-\bt*}{t-1}x,
 \end{equation}
where
 \begin{align} \label{eq:CPW-ctjValues}
 &\qquad\quad\:\begin{gathered}
   c_{j,j}   = 1; \qquad
   c_{j+1,j} = \tfrac12(y+\yb)-\gze{j}-\gze{j-1}; \\
   c_{j+2,j} = (\tfrac12y-\gze{j})(\tfrac12\yb-\gze{j})+c_j+c_{j+1}; \\
   c_{j+3,j} = c_{j+1}c_{j+2,j+1}; \qquad
   c_{j+4,j} = c_{j+1}c_{j+2}.
  \end{gathered}
\\\intertext{It follows that} \label{eq:CPW-cj0Values}
 &\begin{gathered}
   c_{2,0} = 2\gpl0^2\gpl1\JJbg11, \qquad
   c_{3,0} = 2\gpl0\gpl1c_1\JJbg10, \\
   c_{4,0} = 2\gpl0c_1c_2\JJbg00.
  \end{gathered}
 \end{align}


 \section{Correlation Functions for $\chiral\dimn2$, $\chiral\dimn{p}$}
 \label{se:2p}

Correlation functions for four identical chiral primaries, $p_1=\dots=p_4=p$, have been extensively researched.
Here, we examine four point functions for two pairs of operators, of scale dimension $2,p$ respectively.
The former correspond to the superconformal primary of the lowest $\N=4$, \BPS2 short multiplet $\short11$, containing the energy momentum tensor.

Unlike the case of identical operators, when we come to expand the correlator in terms of conformal partial waves, there are two distinct pairings to 
which we may apply the operator product expansion.
These we will refer to as the \ch22pp and \ch2p2p channels.

\subsubsection{The \ch22pp Channel} \label{se:22pp}

The \ch22pp channel corresponds to the case $p_1=p_2=2, p_3=p_4=p$.
Thus \eqref{eq:4pt-EDefn} gives $E=2$.
Following \eqref{eq:4ptG}, we write the correlation function as
 \begin{multline} \label{eq:22pp-4ptG}
  \fourpt\hspt\chiral2\chiral2\chiral{p}\chiral{p}
  = \frac{(\dpt12)^2(\dpt34)^p}{\xx12^2\xx34^p}\G*\uvst,
 \end{multline}
with $\G*\uvst\equiv\Ggen22pp\uvst$ a polynomial in $\sigma, \tau$ of degree 2.
Quantities associated with the \ch22pp channel will be denoted with a tilde;
thus the requirements of the Ward identity \eqref{eq:Ward} become
 \begin{equation} \label{eq:22pp-Ward}
  \evalat{\alphb=\frac1\xb}{\G*\uvst} = \f*\xa = \fk* + \amx\fhat*\xa
 \end{equation}
with
 \begin{equation} \label{eq:22pp-GDecomp}
  \G*\uvst = \fk* + \Gx*{\fhat*}\uvst + s\uvst\GH*\uv.
 \end{equation}
We see that it is necessary for $\f*\xa$ be at most quadratic in $\alpha$ --- thus $\fhat*\xa$ must be at most linear ---
and $\GH*$ must be independent of $\sigma,\tau$, and hence a function of $u,v$ only.
As mentioned previously, dynamic contributions to $\G*\uvst$ are contained exclusively in $\GH*\uv$;
$\f*\xa$ is always equal to the free field value, which we will find explicitly below.

As we are dealing with pairs of identical operators, we have additional restrictions from crossing symmetry,
the four point function in \eqref{eq:22pp-4ptG} being invariant under \switch12.
From the definitions \eqref{eq:uvDefn}, \eqref{eq:stDefn}, \eqref{eq:xxDefn}, \eqref{eq:aaDefn} we find that under $\swap12$,
 \begin{equation} \label{eq:swap12}
  \begin{aligned}
   u,v &\mapsto \tfrac{u}v,\tfrac1v,
\\ x,\xb &\mapsto x',\xb',
  \end{aligned}
   \qquad
  \begin{gathered}
   \swap\sigma\tau,
\\ \alpha,\alphb \mapsto (1-\alpha),(1-\alphb),
  \end{gathered}
 \end{equation}
where $x'$ is 
\ifthenelse{\value{xprimecount}=0}%
{defined by
 \begin{equation} \label{eq:x'Defn}
  x' = \frac{x}{x-1},
 \end{equation} \stepcounter{xprimecount}}%
{as defined by \eqref{eq:x'Defn},}
and similarly for $\xb'$.
Thus we require
 \begin{align} \label{eq:22pp-GCrossSym}
  \G*(\tfrac{u}v,\tfrac1v;\tau,\sigma) &= \G*\uvst.
\\\intertext{It follows from \eqref{eq:swap12} that $\amx \mapsto (1-\alpha)-\frac1{x'} = -\amx$;
thus as a consequence of \eqref{eq:22pp-Ward}, \eqref{eq:22pp-GCrossSym}} \label{eq:22pp-fhatCrossSym}
  \fhat*(x',1-\alpha) &= -\fhat*\xa,
\\\intertext{and $s\uvst\GH*\uv$ must be invariant, with $s\uvst$ as defined in \eqref{eq:4pt-sDefn}.
Since $s(\tfrac{u}v,\tfrac1v;\tau,\sigma) = \tfrac1{v^2}s\uvst$, crossing symmetry requires} \label{eq:22pp-HCrossSym}
  \GH*(\tfrac{u}v,\tfrac1v) &= v^2\,\GH*\uv.
 \end{align}

\subsubsection{The \ch2p2p Channel} \label{se:2p2p}

We represent the alternative \ch2p2p channel by the case that $p_1=p_3=2, p_2=p_4=p$.
As in the \ch22pp channel, \eqref{eq:4pt-EDefn} gives $E=2$.
\eqref{eq:4ptG} now becomes
 \begin{multline} \label{eq:2p2p-4ptG}
  \fourpt\hspt\chiral2\chiral{p}\chiral2\chiral{p} \\
  = \frac{(\dpt12\dpt34)^2(\dpt24)^{p-2}}{(\xx12\xx34)^2\xx24^{p-2}} \G\uvst,
 \end{multline}
where $\G\uvst\equiv\Ggen2p2p\uvst$ is a polynomial of degree 2 in $\sigma, \tau$, satisfying the Ward identity
 \begin{equation} \label{eq:2p2p-Ward}
  \evalat{\alphb=\frac1\xb}{\G\uvst} = \f\xa = \fk + \amx\fhat\xa
 \end{equation}
with
 \begin{equation} \label{eq:2p2p-GDecomp}
  \G\uvst = \fk + \Gx\fhat\uvst + s\uvst\GH\uv.
 \end{equation}

We note the left-hand side of \eqref{eq:2p2p-4ptG} takes \switch23 with respect to \eqref{eq:22pp-4ptG}.
Under $\swap23$,
 \begin{alignat}2 \label{eq:swap23}
  u,v &\mapsto \tfrac1u,\tfrac{v}u; \qquad & \sigma,\tau &\mapsto \tfrac1\sigma,\tfrac\tau\sigma
 \end{alignat}
It follows, then, that
 \begin{align} \label{eq:22pp-2p2p-G}
  \G\uvst &= \sigma^2u^2 \G*\uvstmod.
\\\intertext{We expect \eqref{eq:22pp-2p2p-G} to be satisfied by both the free field and full dynamic values of $\G,\G*$;
thus, considering \eqref{eq:GFree+Dyn}, \eqref{eq:HFree+Dyn}, we obtain} \label{eq:22pp-2p2p-sH}
  s\uvst\GH\dyn\uv &= \sigma^2u^2 s\uvstmod\GH*\dyn\uvmod.
\\\intertext{From the definition \eqref{eq:4pt-sDefn}, it follows that
$s\uvst = \sigma^2u^2\,s(\tfrac1u,\tfrac{v}u;\tfrac1\sigma,\tfrac\tau\sigma)$; 
hence the dynamic contributions to $\GH, \GH*$ are related by} \label{eq:22pp-2p2p-H}
  \GH\dyn\uv &= \GH*\dyn\uvmod.
 \end{align}

Additionally, the four point function in \eqref{eq:2p2p-4ptG} is invariant when we take \switch13, under which
 \begin{gather}
  \begin{gathered} \label{eq:swap13}
   \swap{u}v
\\ x,\xb \mapsto (1-x),(1-\xb)
  \end{gathered}
   \qquad
  \begin{gathered}
   \sigma,\tau \mapsto \tfrac\sigma\tau,\tfrac1\tau
\\ \alpha,\alphb \mapsto \tfrac\alpha{\alpha-1},\tfrac\alphb{\alphb-1};
  \end{gathered}
\\ \label{eq:swap13-txFactor}
 \frac{(\dpt12\dpt34)^2(\dpt24)^{p-2}}{\xx12^2\xx34^2\xx24^{p-2}}
  \mapsto \left(\frac{u}v\,\tau\right)^{\!2}\! \bfrac{(\dpt12\dpt34)^2(\dpt24)^{p-2}}{\xx12^2\xx34^2\xx24^{p-2}}.
 \end{gather}
Thus for the right-hand side of \eqref{eq:2p2p-4ptG} to satisfy crossing symmetry, we require that
 \begin{equation} \label{eq:2p2p-GCrossSym}
  \G(v,u;\tfrac\sigma\tau,\tfrac1\tau) = \bfrac{v}{u\tau}^2 \, \G\uvst.
 \end{equation}
The factor $\bfrac{v}{u\tau}^2$ appearing in \eqref{eq:2p2p-GCrossSym} means $\fhat, \GH$ do not obey independent crossing symmetry relations
analogous to \eqref{eq:22pp-fhatCrossSym}, \eqref{eq:22pp-HCrossSym};
however, we may see that $\f$ satisfies
 \begin{equation} \label{eq:2p2p-fCrossSym}
  x'^2(1-\alpha)^2\f(1-x,\alpha') = \f\xa.
 \end{equation}
We may also, in a similar manner to \eqref{eq:22pp-2p2p-H}, find a crossing symmetry restriction on the dynamic part of $\GH$.
Requiring \eqref{eq:2p2p-GCrossSym} to hold for the free field and interacting values of $\G$, and noting that 
under \eqref{eq:swap13}, $s\uvst \mapsto s(v,u;\tfrac\sigma\tau,\tfrac1\tau) = \tfrac1{\tau^2}s\uvst$,
we deduce that $\GH\dyn$ satisfies
 \begin{align} \label{eq:2p2p-HCrossSym}
  \GH\dyn(v,u) &= \frac{v^2}{u^2} \, \GH\dyn\uv.
\\\intertext{If we define} \label{eq:2p2p-FDefn}
  \GH\dyn\uv &= \frac{u}v\,\HF\uv
\\\intertext{then \eqref{eq:2p2p-HCrossSym} may be written as} \label{eq:2p2p-FCrossSym}
  \HF(v,u) &= \HF\uv.
 \end{align}

  \subsection{Free Field Results}
  \label{se:FF}

Following \eqref{eq:GFree+Dyn}, \eqref{eq:2p2p-GDecomp} we consider tree-level contributions, which will give us the free field function
 \begin{equation} \label{eq:2p2p-G0Decomp}
  \G\ord0\uvst = \fk + \Gx\fhat\uvst + s\uvst\GH\ord0\uv,
 \end{equation}
and the corresponding result for $\G*\ord0\uvst$.

The non-connected component, in the \ch22pp channel, is given by
\begin{subequations} \label{eq:22ppFF-Diags}
 \begin{equation} \label{eq:22ppFF-2ptDiag}
  \begin{gathered}[c]
   \usebox{\FFDiagI}
  \end{gathered}
  \: \propto \: \frac{(\dpt12)^2(\dpt34)^p}{(\xx12)^2(\xx34)^p}.
 \end{equation}
We note that this is the same factor multiplying $\G*\uvst$ in \eqref{eq:22pp-4ptG}.
We normalise the contribution from this diagram to $\G*\ord0$ to $1$;
thus the five possible diagrams remaining contribute
 \begin{equation} \label{eq:22ppFF-ConnectedDiags}
  \begin{gathered}
   \begin{gathered}[t]
    \usebox{\FFDiaga} \FFspace \usebox{\FFDiagaX}
   \\[-4pt]
    \underbrace{\makebox[0.3\columnwidth]{}}
   \\
    a\left(\sigma u + \frac{\vphantom{u^2}\tau u}v\right),
   \end{gathered}
   \FFspace
   \begin{gathered}[t]
    \usebox{\FFDiagb} \FFspace \usebox{\FFDiagbX}
   \\[-4pt]
    \underbrace{\makebox[0.3\columnwidth]{}}
   \\
    b\left(\sigma^2u^2 + \frac{\tau^2u^2}{v^2}\right),
   \end{gathered}
   \FFspace
   \begin{gathered}[t]
    \usebox{\FFDiagc}
   \\[-4pt]
    \underbrace{\makebox[0.13\columnwidth]{}}
   \\
    c\left(\frac{\sigma\tau u^2}v\right),
   \end{gathered}
  \end{gathered}
 \end{equation}
\end{subequations}
where $a,b,c$ are determined by symmetry and colour factors, investigated more fully below.
Thus we have
 \begin{equation} \label{eq:22ppFF-G}
  \G*\ord0\uvst = 1
   + a\left(\sigma u + \tau\frac{\vphantom{u^2}u}v\right)
   + b\left(\sigma^2u^2 + \tau^2\frac{u^2}{v^2}\right)
   + c\,\sigma\tau\frac{u^2}v.
 \end{equation}
Using \eqref{eq:22pp-2p2p-G}, it follows directly that in the \ch2p2p channel,
 \begin{equation} \label{eq:2p2pFF-G}
   \G\ord0\uvst = \sigma^2u^2
  + a\left(\sigma u + \frac{\sigma\tau u^2}v\right)
  + b\left(1 + \frac{\tau^2u^2}{v^2}\right)
  + c\, \frac{\tau\vphantom{u^2}u}v.
 \end{equation}

\subsection{Colour factors} \label{se:Colour}

Here we will briefly examine the coefficients $a,b,c$ appearing in \eqref{eq:22ppFF-G}, \eqref{eq:2p2pFF-G},
which arise from the $SU(N)$ gauge structure of the chiral primary operators.
The generators of $SU(N)$ are $\set{T_a}$, $a=1,\dots,\Ns$, where the $T_a$ are $N\times N$ matrices satisfying
 \begin{align} \label{eq:SU(N)-Conv}
  \tr(T_aT_b) &= \tfrac12\delta_{ab} ,
& \comm{T_a}{T_b} &= i f_{abc}T_c ,
& f_{abc}f_{abd} &= N\delta_{cd},
 \end{align}
from which follows
 \begin{equation} \label{eq:SU(N)-TaTa}
  T_aT_a = \tfrac1{2N}\Nsb1\ident.
 \end{equation}
For the purposes of counting diagrams, we introduce an adjoint scalar field $\adscalar$
with basic two point function $\corel{\adscalar_a \adscalar_b} = \delta_{ab}$.
Chiral primary operators $\chiral\dimn{p}$, which are gauge-invariant, then correspond to $X\ap\muSERIES[a]p\adscalar{}{}$,
where $X\ap$ is a totally symmetric, rank $p$ colour tensor.
The $\Delta=2$ operators $\chiral\dimn2$ thus have colour structure $X_{ab}=\tr(T_aT_b)=\tfrac12\delta_{ab}$.
We encode the structures defining the two $\Delta=p$ operators 
in the colour tensors $\XL\ap, \XR\ap$.
%
The contributions from the relevant diagrams are then:
\vspace{-2ex}%
\begin{subequations} \label{eq:Col-contractions}
 \begin{align}
  \begin{gathered} \usebox{\FFDiagI} \end{gathered}
   &\:\propto\: \frac1{2!}\cdot\frac1{p!}\; \COLcontractX{p}{}{}cdcd = \frac\Ns{8\,p!}\XLXR,
 \label{eq:Col-contractions-1} \tag{\ref{eq:Col-contractions}$\diamond$}
 \displaybreak[0] \\[10pt]
  \begin{gathered} \usebox{\FFDiaga} \end{gathered}
   &\:\propto\: \frac1{(p-1)!}\;            \COLcontractX{p-1}bdbccd \COLwibbleX = \frac1{4\,(p-1)!}\XLXR,
 \label{eq:Col-contractions-a} 
 \displaybreak[0] \\[-4pt]
  \begin{gathered} \usebox{\FFDiagb} \end{gathered}
   &\:\propto\: \frac1{2!}\cdot\frac1{2!}\cdot\frac1{(p-2)!}\; \COLcontractX{p-2}{bc}{de}bcde \COLwibbleX = \frac1{16\,(p-2)!}\dpr\XLb\XRd,
 \label{eq:Col-contractions-b} 
 \displaybreak[0] \\[-4pt]
  \begin{gathered} \usebox{\FFDiagc} \end{gathered}
   &\:\propto\: \frac1{(p-2)!}\;            \COLcontractX{p-2}{bc}{de}bdce \COLwibbleX = \frac1{4\,(p-2)!}\XLXR.
 \label{eq:Col-contractions-c} 
 \end{align}
\end{subequations}
Here we use a bar ${}_\rvert$ to denote an incomplete set of indices, and a dot $\cdot$ for contraction over omitted indices.
Thus in \eqref{eq:Col-contractions}
 \begin{equation} \label{eq:Col-Explanation}
  \XLXR \equiv \XL_{a_1\dots a_p}\XR_{a_1\dots a_p},
\qquad
  \dpr\XLb\XRd \equiv \XL_{a_1\dots a_{p-2}bb} \, \XR_{a_1\dots a_{p-2}dd} \,.
 \end{equation}
The coefficient of the disconnected graph is set to one, as a consequence of normalisation choice for the two point functions of BPS operators.
Thus $a,b,c$ are determined by the ratios of \eqref{eq:Col-contractions-1} with (\ref{eq:Col-contractions-a}--\ref{eq:Col-contractions-c}),
which give
 \begin{align} \label{eq:Col-abcExplicit}
  a &= \frac{2p}\Ns\,,
& b &= \frac{p(p-1)}{2\Nsb1}\bfrac{\dpr\XLb\XRd}\XLXR\,,
& c &= \frac{2p(p-1)}\Ns\,.
 \end{align}
Note that we have
 \begin{equation} \label{eq:Col-acRelation}
  c = (p-1)a,
 \end{equation}
and that exact values for $a,c$ are, with our normalisation convention, completely and straightforwardly determined for any given $p$ and $N$;
however $b$, which is given by
 \begin{equation} \label{eq:Col-bRelations}
  b = \frac14 \bfrac{\dpr\XLb\XRd}\XLXR c = \frac{p-1}4 \bfrac{\dpr\XLb\XRd}\XLXR a,
 \end{equation}
requires a non-trivial tensor calculation, dependant on the underlying colour structures of our $\Delta=p$ operators.
More explicit calculations for $b$ appear in \secref{se:Col}, 
in the cases that $p=2,3$, 
and in the large $N$ limit for general $p$ with specific choices for $\XL,\XR$.

\subsection{Ward Identities}

In the calculations that follow, it will on occasions be more straightforward to consider new variables $z,\zb$ defined by
 \begin{equation} \label{eq:zzDefn}
   z = \frac2x-1 ,\quad \zb = \frac2\xb-1,
 \end{equation}
and $y,\yb$ as given in \eqref{eq:yyDefn}.

\subsubsection{The \ch22pp Channel} \label{se:22ppFF}

In terms of $z,\zb,y,\yb$, \eqref{eq:22pp-Ward} may be written
 \begin{equation} \label{eq:22pp-fhat}
  \f*\xa = \fk* + \tfrac12\yz\fhat*\xa.
 \end{equation}
We have already concluded that $\f*\xa$ be at most quadratic in $\alpha$ (and hence also in $y$), and also that $\GH*$ be independent of $\sigma,\tau$.
Hence for some functions $\fg,\fh$
 \begin{equation} \label{eq:22pp-fgh}
  \f*\xa = \fk* + \yz\bigl(\fg*(z) + y\,\fh*(z)\bigr).
 \end{equation}
In terms of \eqref{eq:22pp-fgh}, \eqref{eq:22pp-GDecomp} becomes
 \begin{equation} \label{eq:22pp-GghDecomp}
  \begin{aligned}[b]
   \G*\uvst = \fk* &- \frac1\zzb
    \begin{aligned}[t]
     \biggl\{ \yz&\ybz\bigl(\fg*(z)+(y+\yb-\zb)\fh*(z)\bigr) \\ &- \yzb\ybzb\bigl(\fg*(\zb)+(y+\yb-z)\fh*(\zb)\bigr) \biggr\}
    \end{aligned}
\\ &+ \usqsxt\yz\ybz\yzb\ybzb\GH*\uv,
  \end{aligned}
 \end{equation}
and writing \eqref{eq:22ppFF-G} in terms of $z,y$, and their conjugates, we obtain
 \begin{multline} \label{eq:22ppFF-Gzy}
  \G*\ord0\uvst
   = 1 + a\left(\frac{(1+y)(1+\yb)}{(1+z)(1+\zb)}+\frac{(1-y)(1-\yb)}{(1-z)(1-\zb)}\right)
\\ + b\left(\frac{(1+y)^2(1+\yb)^2}{(1+z)^2(1+\zb)^2}+\frac{(1-y)^2(1-\yb)^2}{(1-z)^2(1-\zb)^2}\right)
   + c\frac{(1-y^2)(1-\yb^2)}{(1-z^2)(1-\zb^2)}.
 \end{multline}
Using \eqref{eq:22pp-Ward}, \eqref{eq:GFree+Dyn}, then \eqref{eq:22ppFF-G} implies
 \begin{multline} \label{eq:22pp-fValue}
  \begin{aligned}
   \f*\xa
    &= \evalat{\alphb=\frac1\xb}{\G*\uvst} = \evalat{\yb=\zb}{\G*\ord0\uvst} \\
    &= 1+a\left(\frac{1+y}{1+z}+\frac{1-y}{1-z}\right)+b\left(\frac{(1+y)^2}{(1+z)^2}+\frac{(1-y)^2}{(1-z)^2}\right)+c\,\frac{1-y^2}{1-z^2} \\
    &= \left[1+\frac{2a+c}{1-z^2}+\frac{2b(1+z^2)}{(1-z^2)^2}\right]-y\left[\frac{2az}{1-z^2}+\frac{8bz}{(1-z^2)^2}\right]
  \end{aligned} \\
     + y^2\left[\frac{2b(1+z^2)}{(1-z^2)^2}-\frac{c}{1-z^2}\right];
 \end{multline}
matching terms with \eqref{eq:22pp-fhat}, which expands to
 \begin{equation} \label{eq:22pp-fhatExpn}
  \f*\xa = \bigl[\fk*-z\fg*(z)\bigr] + y\bigl[\fg*(z)-z\fh*(z)\bigr] + y^2\,\fh*(z)
 \end{equation}
gives us
%
 \begin{multline} \label{eq:22pp-hgValues} 
  \fh*(z)
   = \frac1{(1-z^2)^2} \left((2b-c)+(2b+c)z^2\right) 
   = \tfrac14\left(b(x^2+x'^2)-c(x+x')\right),
\\\shoveleft{%
  \fg*(z)
   = \frac{z}{(1-z^2)^2} \left((2a+2b+c)z^2-(2a+6b+c)\right) }\\
   = \tfrac14\left(b(x^2-x'^2)+(2a+2b+c)(x-x')\right),
 \end{multline}
and
 \begin{equation} \label{eq:22pp-kValue}
  \fk* = 1+2a+2b+c,
 \end{equation}
%
where $x'$ is defined as in \eqref{eq:x'Defn}, satisfying
 \begin{equation} \label{eq:x'Satisfies}
  x+x' = xx';
\qquad
  z = \frac2x - 1 = 1 - \frac2{x'} \,.
 \end{equation}
From 
\eachlabelcase{{eq:22pp-hgkValues}{\eqref{eq:22pp-hgkValues}}{\eqref{eq:22pp-hgValues}}}
we obtain an explicit result for $\fhat*\xa = 2\bigl(\fg*(z)+y\,\fh*(z)\bigr)$,
 \begin{equation} \label{eq:22pp-fhatValue}
  \fhat*\xa = \bigl(b(x^2+x'^2) - c(x+x')\bigr)\alpha - \bigl(bx'^2 - (a+b)(x-x') - cx\bigr).
 \end{equation}
From \eqref{eq:22pp-GghDecomp}, \eqref{eq:22ppFF-Gzy}, we see that for $-y=\yb=1$ ($\implies \sigma=\tau=0$),
 \begin{multline} \label{eq:22ppFF-Gzz11}
  \evalat{-y=\yb=1}{\G*\ord0\uvst} = \G*\ord0\uvoo \\ =
   \fk* + \frac1\zzb\Bigl\{ (1-z^2)\bigl[\fg*(z)-\zb\,\fh*(z)\bigr] - (1-\zb^2)\bigl[\fg*(\zb)-z\,\fh*(\zb)\bigr] \Bigr\} \\
       + \usqsxt(1-z^2)(1-\zb^2)\GH*\ord0\uv,
 \end{multline}
whilst using \eqref{eq:22ppFF-G},
 \begin{equation} \label{eq:22ppFF-Guv00}
  \G*\ord0\uvoo = 1;
 \end{equation}
thus from the results 
\eachlabelcase{{eq:22pp-hgkValues}{\eqref{eq:22pp-hgkValues}}{\eqref{eq:22pp-hgValues}, \eqref{eq:22pp-kValue}}}%
, we find
 \begin{equation} \label{eq:22ppFF-u2H/16}
  \usqsxt\GH*\ord0\uv = \frac{(2b+c)(1+z\zb)^2 + (2b-c)(z+\zb)^2}{(1-z^2)^2(1-\zb^2)^2},
 \end{equation}
and hence
 \begin{equation} \label{eq:22ppFF-HValue}
  \GH*\ord0\uv = b\Bigl(1+\frac1{v^2}\Bigr)+c\,\frac1v \,.
 \end{equation}

\paragraph{}
We may check that the results we have obtained so far are compatible with the crossing symmetry requirements. 
In addition to \eqref{eq:swap12}, we have
 \begin{equation} \label{eq:swap12yz}
   z,\zb \mapsto -z,-\zb
\qquad
   y,\yb, \mapsto -y,-\yb.
 \end{equation}
Thus the condition \eqref{eq:22pp-fhatCrossSym} requires $\fg*(z)$ to be odd and $\fh*(z)$ even, i.e.
 \begin{equation} \label{eq:22pp-ghCrossSym}
  \fg*(-z) = -\fg*(z),
\qquad
  \fh*(-z) = \fh*(z),
 \end{equation}
which are indeed satisfied by the $\fg*(z), \fh*(z)$ given in 
\eachlabelcase{{eq:22pp-hgkValues}{\eqref{eq:22pp-hgkValues}}{\eqref{eq:22pp-hgValues}}}%
.
It is readily apparent that $\GH*\ord0\uv$ as given in \eqref{eq:22ppFF-HValue} satisfies \eqref{eq:22pp-HCrossSym}.
Thus our free field results are compatible with crossing symmetry.

\subsubsection{The \ch2p2p Channel} \label{se:2p2pFF}

We proceed from \eqref{eq:2p2pFF-G} as above;
seeing that for $z,y$ as in \eqref{eq:zzDefn}, \eqref{eq:yyDefn}, $\f\xa$ must be of the form
 \begin{equation} \label{eq:2p2p-fExpn}
  \f\xa = \fk + \tfrac12\yz\fhat\xa  = \fk + \yz\bigl(\fg(z) + y\,\fh(z)\bigr),
 \end{equation}
then with $x'$ as defined in \eqref{eq:x'Defn},
\begin{subequations} \label{eq:2p2p-hgkValues}
 \begin{gather}
  \fh(z) = \tfrac14\left(x^2+bx'^2-a(x+x')\right), \label{eq:2p2p-hValue} \\
  \fg(z) = \tfrac14\left(x^2-bx'^2+(2+3a)x-(a+2b+2c)x'\right), \label{eq:2p2p-gValue}\\
  \fk = 1 + 2a + 2b + c = \fk*; \label{eq:2p2p-kValue}
 \end{gather}
\end{subequations}
hence
 \begin{equation} \label{eq:2p2p-fhatValue}
  \fhat\xa = \left(x^2+bx'^2-a(x+x')\right)\alpha - \left(bx'^2-(1+2a)x+(b+c)x'\right).
 \end{equation}
Finally, we obtain
 \begin{equation} \label{eq:2p2pFF-HValue}
  \GH\ord0(u,v) = 1 + a\,\frac1v + b\,\frac1{v^2}.
 \end{equation}

\vspace{0.5ex}
We note that each term in \eqref{eq:2p2pFF-G} satisfies \eqref{eq:2p2p-GCrossSym};
thus the free field $\G\ord0\uvst$ possesses the required crossing symmetry.
We also observe that, as expected (and unlike in the \ch22pp channel),
the free field value $\GH\ord0$ as given by \eqref{eq:2p2pFF-HValue}
does \emph{not} satisfy the relation \eqref{eq:2p2p-HCrossSym} if we take $\GH\dyn\goesto\GH\ord0$.
\eqref{eq:2p2p-HCrossSym} is only true for the dynamic contribution $\GH\dyn$.

 \section{Superconformal Expansions}
 \label{se:Expansions}

\subsection{Wave Expansion in the $\ch2p2p$ Channel} \label{se:2p2pCPW}

We consider the expansions \eqref{eq:CPW-GExpn}, \eqref{eq:CPWJac-GExpn} in the case that $p_1=p_3=2, p_2=p_4=p$.
From \eqref{eq:CPW-bgDefn} it follows that $2\bt=-2\bt*=p-2$.
Thus we may expand $\G\uvst$, as defined in \eqref{eq:2p2p-4ptG}, in terms of conformal partial waves, as
 \begin{equation} \label{eq:2p2pCPW-GExpn}
   \G\uvst =
   \sum_{nm,t\ell} a_{nm,t\ell} \, \JJbb{n}m\st* \, u^t\Gpw\ell{2\bt+2t+\ell}uv{2\bt}{2\bt}.
 \end{equation}
As described in \secref{se:Jacobi}, the $\JJbb{n}m\st*$ are 2-variable Jacobi polynomials, defined in \eqref{eq:Jac-2varDefn},
corresponding to the six $SU(4)_R$ representations $\dynk{n-m}{2m+p-2}{n-m}$ appearing in the tensor product $\dynk020\otimes\dynk0p0$.
The twist of the contributing conformal waves is parametrised by $t$,
where $\Delta = 2\bt+2t+\ell$ and we have lowest twist contribution $\Twist \ge p-2 = 2\bt$.

Using \eqref{eq:CPW-xxF}, \eqref{eq:CPW-FjExpn}, we may write
 \begin{equation} \label{eq:2p2pCPW-GtDefn}
  (\xxb)\G\uvst
   = x \sum_t \Gt\xst \gbb\bt{t}\xb
   = -x\xb \sum_t \Gt\xst \gbb{1+\bt}{t-1}\xb
 \end{equation}
which with \eqref{eq:2p2pCPW-GExpn} implies that
 \begin{equation} \label{eq:2p2pCPW-GtExpn}
  \Gt\xst = \sum_{\ell=-t-1}^\infty \sum_{nm} a_{nm,t\ell} \, \JJbb{n}m\st* \, \gbb{1+\bt}{t+\ell}x.
 \end{equation}
Similarly, we let
 \begin{gather} \label{eq:2p2pCPW-HExpn}
  \GH\uv = \sum_{nm,j\ell} A_{nm,j\ell} \, \JJbb{n}m\st* \, u^j\Gpw\ell{4+2\bt+2j+\ell}uv{2\bt}{2\bt} ,
\\ \label{eq:2p2pCPW-HjDefn}
  (\xxb)\GH\uv = \frac1x\sum_{j=0}^\infty \GH\g{j}(x)\gbb{2+\bt}j\xb \,,
 \end{gather}
where from the results of \secref{se:Ward} we have
 \begin{equation} \label{eq:2p2pCPW-GDecomp}
   \G\uvst = \fk + \Gx\fhat\uvst + \underbrace{s\uvst\GH\uv}_{\textstyle=\Gx\GH\uvst}.
 \end{equation}
We note that $\GH$ is a function of $u,v$ only (\secref{se:2p2p}), independent of $\sigma,\tau$;
necessarily, then, $n=m=0$ identically in \eqref{eq:2p2pCPW-HExpn}.
Letting $A_{j\ell}\equiv A_{00,j\ell}$, we obtain
 \begin{equation} \label{eq:2p2pCPW-HjExpn}
  \GH\g{j}(x) = \sum_{\subbox{\ell=-j-1}}^\infty A_{j\ell} \, \JJbb00 \gbb{1+\bt}{j+\ell+2}x ,
 \end{equation}
with $\JJbb00$ a constant, as given in \eqref{eq:Jac00}.

\subsection{Contributions from Different Multiplets} \label{se:2p2pCPW-M}

Starting from \eqref{eq:2p2pCPW-GDecomp}, 
we will find, in the manner of \cite{rf:0412335}, \cite{rf:0601148},
the contribution to the $\corel{2p2p}$ correlation function from each supermultiplet, given by
 \begin{equation} \label{eq:2p2pCPW-GtM}
  \Gt\xst[\M;] = \sum_{nm,\ell} a_{nm,tl}(\M) \, \JJbb{n}m\st* \, \gbb{1+\bt}{t+\ell}x
 \end{equation}
for any supermultiplet $\M$.
\pagebreak[2]%
We make use of the relations detailed in \secref{se:Jacobi} and \secref{se:CPW-Recur}, where now we have
$2\bt=-2\bt*=p-2$.
Thus \eqref{eq:Jac-gamma} becomes
 \begin{equation} \label{eq:2p2pJac-gamma}
  \begin{gathered}
   \gpl{n} = \frac{(n+1)(n+2\bt+1)}{2(n+\bt+1)(2n+2\bt+1)} \,; \qquad
   \gmi{n} = \frac{n(n+2\bt)}{2(n+\bt)(2n+2\bt+1)} \,; \\
   \gze{n} = \frac{\bt^2}{2(n+\bt)(n+\bt+1)} \,.
  \end{gathered}
 \end{equation}
%
%
We first split $\Gt\xst$, defined in \eqref{eq:2p2pCPW-GtDefn}, contributions from $\fk$, $\fhat$, and $\GH$,
 \begin{equation} \label{eq:2p2pCPW-GtDecomp}
  \Gt\xst = \fk\,\Gt(\pp\short00;x) + \Gxt\fhat\xst + \Gxt\GH\xst.
 \end{equation}
To obtain $\Gxt\GH$ we use \eqref{eq:CPW-axaxgbg},
 \begin{equation} \label{eq:2p2pCPW-GHt}
  \Gxt\GH\xst = \amx\abmx\sum_{j\ge0} c_{t,j} \GH\g{j}(x) \,.
 \end{equation}
with the $c_{t,j}$ as in \eqref{eq:CPW-ctjValues}.
Applying \eqref{eq:CPW-axaxgbg} a second time, we may thus relate the expansions given in \eqref{eq:2p2pCPW-GtDefn}, \eqref{eq:2p2pCPW-HjDefn}.
The contribution of a long multiplet $\pp\Long{n}mk{2j+2\bt+k}$ is given by taking $\GHj\xst\goesto\gbb{1+\bt}{j+\ell+2}x\JJbb{n}m\st*$.
Then we obtain from \eqref{eq:2p2pCPW-GHt}, \eqref{eq:2p2pCPW-GtM}
 \begin{equation} \label{eq:2p2pCPW-GtA}
  \sum_{r,s}a_{rs,t\ell}(\pp\Long{n}mk{2j+2\bt+k})\JJbb{r}s = c_{t+\ell+1,j+k+1}c_{t,j}\JJbb{n}m,
 \end{equation}
where we require $j\le t\le j+4, j+k\le t+\ell\le j+k+4$.
Detailed expressions for $a_{rs,t\ell}$ are easily obtained by use of \eqref{eq:CPW-ctjValues}, \eqref{eq:Jac-2varRecur}.
As noted above, $\GH$ is independent of $\sigma,\tau$, so necessarily we have $n=m=0$ in \eqref{eq:2p2pCPW-GtA}.

The component $\Gt(\pp\short00)$ appearing in \eqref{eq:2p2pCPW-GtDecomp} gives the contribution of a constant term at twist $t$, determined by
 \begin{equation} \label{eq:2p2pCPW-GtB00Defn}
  \xxb = x\sum_t \Gt(\pp\short00;x) \gbb\bt{t}\xb,
 \end{equation}
which gives
 \begin{equation} \label{eq:2p2pCPW-GtB00Values}
  \G\g0(\pp\short00;x) = 1 = \gbb{1+\bt}0x,
\quad 
  \G\g1(\pp\short00;x) 
   = \tfrac12\left(\tfrac1x-\tfrac1{x'}\right) - \gze{-1} = -\gbb{1+\bt}{-1}x;
 \end{equation}
and consequently non-vanishing terms
 \begin{equation} \label{eq:2p2pCPW-aB00}
  a_{00,00}(\pp\short00) = a_{00,1\,-2}(\pp\short00) = 1.
 \end{equation}
This agrees with the contribution of the \BPS2 short multiplet $\pp\short00$ (see \secref{se:2p2pCPW-Short}).

We analyse the contributions from $\fhat\xa$ by first writing the expansion
 \begin{equation} \label{eq:2p2pCPW-fExpn}
  \fhat\xa
   = \sum_{\ell=0}^\infty \fhat_\ell(\alpha) \, \gbb{1+\bt}{\ell+1}x
   = \sum_{n,\ell} b_{n,\ell} \, \Jbb{n}(y) \gbb{1+\bt}{\ell+1}x .
 \end{equation}
We have
 \begin{multline} \label{eq:2p2p-GfDefn}
  \Gx\fhat\uvst + 2\fk \\ 
   = \frac{\abx\axb\bigl(\f\xa+\f\xbab\bigr) - \ax\abxb\bigl(\f\xab+\f\xba\bigr)}{(\xxb)(\aab)};
 \end{multline}
with $\f$ expressed in terms of $\fhat$ as in \eqref{eq:2p2p-Ward}, the contributions involving $\fk$ cancel.
This gives us
 \begin{equation} \label{eq:2p2pCPW-Gft}
  \begin{aligned}[b]
   \Gxt\fhat\xst &= -\amx\abmx\biggl(\frac{\fhat\xa-\fhat\xab}\aab \,\delta_{t\,0} \\
& \mspace{54mu} + \frac{(\frac12y-\gze{-1})\fhat\xa-(\frac12\yb-\gze{-1})\fhat\xab}\aab \,\delta_{t\,1} \biggr) \\
& \mspace{54mu} + \sum_{\ell\ge0} c_{t,\ell} \frac{(\alpha-\frac1x)\fhat\g\ell(\alpha)-(\alphb-\frac1x)\fhat\g\ell(\alphb)}\aab.
  \end{aligned}
 \end{equation}
From \eqref{eq:2p2pCPW-Gft}, the contribution from the $b_{0,0}$ in \eqref{eq:2p2pCPW-fExpn} is given by
 \begin{equation} \label{eq:2p2pCPW-Gtb00}
  \Gxt{b_{0,0}}\xst = -\biggl(\delta_{t1} \sum_{j=0}^4 c_{j,0} \, \gbb{1+\bt}{j-1}x \biggr) + c_{t,0}.
 \end{equation}
Writing
 \begin{equation} \label{eq:2p2pCPW-f0Defn}
  \fhat\xa = \fhat_0\xa + b_{0,0}\gbb{1+\bt}1x,
 \end{equation}
we can express \eqref{eq:2p2pCPW-Gft} as
\newcommand{\boo}{11}
 \begin{equation} \label{eq:2p2pCPW-Gtf0Defn}
  \Gxt\fhat\xst = b_{0,0}\Gt(\short00;x) - \Gxt\boo\xst + \Gxt{\fhat_0}\xst
 \end{equation}
where
 \begin{equation} \label{eq:2p2pCPW-Gt11Defn}
  \Gxt\boo\xst = 
  \begin{cases}
   \sum_{\ell=2}^4 c_{\ell,0}\,\gbb{1+\bt}{\ell-1}x\,, & t=1\,,
\\ -c_{t,0}\,, & t=2,3,4\,.
  \end{cases}
 \end{equation}
Using \eqref{eq:CPW-cj0Values}, it is evident that  the non-zero contributions from $\G\g\boo$
correspond to the results for $a_{nm,t\ell}(\pp\short11)$ given in \eqref{eq:2p2pCPW-aB11} for the \BPS2 short multiplet $\pp\short11$,
and we have
 \begin{equation} \label{eq:2p2pCPW-G11=GB11}
  \G\g\boo = 2\gpl0^2\gpl1\G(\pp\short11).
 \end{equation}

To consider the contribution of semi-short multiplets we identify for a function
$\something\xst=\sum_n\something\g{n}\st*\,\gbb{1+\bt}{n+1}x$
an expression for its twist $j$ contributions to $\GH$, namely
 \begin{equation} \label{eq:2p2pCPW-HFkjDefn}
  \GHj[\something,k]\twist{j} = \delta_{kj}\,\something\xst - \something\g{k}\st*\gbb{1+\bt}{j+1}x,
 \end{equation}
and using \eqref{eq:2p2pCPW-GHt}, the corresponding contributions to $\Gxt\GH$,
 \begin{equation} \label{eq:2p2pCPW-GFtjDefn}
  \Gxtj\something{j}\xst = \delta_{tj}\,\amx\abmx\something\xst - \sum_{k=0}^t c_{t,k}\,\something\g{k}\st*\gbb{1+\bt}{j-1}x.
 \end{equation}
The right-hand sides of \eqref{eq:2p2pCPW-HFkjDefn}, \eqref{eq:2p2pCPW-GFtjDefn} are such that they are compatible with \eqref{eq:CPW-aRefl}.
For the function $\fhat\xa$ we define
 \begin{equation}
  \begin{aligned} \label{eq:CPW-QjDefn}
   \fQ0j\xst &= - \frac{\Jbb{j}(\yb)\fhat\xa-\Jbb{j}(y)\fhat\xab}\aab,
\\ \fQ1j\xst &= - \frac{\frac12y\Jbb{j}(\yb)\fhat\xa-\frac12\yb\Jbb{j}(y)\fhat\xab}\aab,
  \end{aligned}
 \end{equation}
and, using the notation of \eqref{eq:2p2pCPW-GFtjDefn},
 \begin{multline} \label{eq:2p2pCPW-GftjDefn}
  \Gxtj*\fhat{j}\xst
   = \Gxtj{\fQ0j}j
   + \Gxtj{\fQ1j}{j+1} - \gze{j-1}\,\Gxtj{\fQ0j}{j+1} + \gmi{j}\,\Gxtj{\fQ0{j-1}}{j+1}
\\ + \gmi{j}\left(\Gxtj{\fQ1{j-1}}{j+2} - \gze{j}\,\Gxtj{\fQ0{j-1}}{j+2}\right) + c_j\,\Gxtj{\fQ0j}{j+2}
   + \gmi{j}c_{j+1}\,\Gxtj{\fQ0{j-1}}{j+3} \,.
 \end{multline}
By comparison with \eqref{eq:2p2pCPW-Gft}, we recognise that
 \begin{equation} \label{eq:2p2pCPW-Gft0}
  \begin{aligned}
   \Gxtj*\fhat0\xst
   &= \Gxtj{\fQ00}0\xst + \Gxtj{\fQ10}1\xst - \gze{-1}\,\Gxtj{\fQ00}1\xst
\\ &= \Gxt\fhat\xst.
  \end{aligned}
 \end{equation}

We denote by $\semi{n}m\ell$ a semi-short multiplet descended from a superconformal primary state belonging to $SU(4)$ representation
$\dynk{n-m}{2m}{n-m}$ and with spin $\ell$;
the shortening conditions then require $\Twist = 2n+2$.
The contribution to the partial wave expansion of a semi-short multiplet is given by
 \begin{equation} \label{eq:2p2pCPW-CjikContrib}
  \Gt\xst[\pp\semi{j-1}ik;] =
  \evalat{\textstyle\fhat\xa = \frac12\gbb{1+\bt}{j+k+2}x\Jbb{i}(y)}{\Gxtj*{\fhat}j\xst}
 \end{equation}
for $j\ge i+1, k\ge0$. With $\fhat\xa$ as in \eqref{eq:2p2pCPW-CjikContrib}, \eqref{eq:CPW-QjDefn} gives
 \begin{equation} \label{eq:2p2pCPW-QjSemishort}
  \begin{aligned}
   \fQ0j\xst &= \gbb{1+\bt}{j+k+2}x\,\JJbb{j-1}i\st*,
\\ \fQ1j\xst &= \gbb{1+\bt}{j+k+2}x\bigl(\textstyle\sum_{r=-1}^1 \gam{i}r \JJbb{j-1}{i+r}\st*\bigr),
  \end{aligned}
 \end{equation}
so it is clear that the lowest dimension contribution from the multiplet is given by
$a_{j-1\,i,jk}(\pp\semi{j-1}ik) = 1$.
Remaining contributions may be calculated using \eqref{eq:Jac-2varRecur}.
We may extend \eqref{eq:2p2pCPW-CjikContrib} to the case $k=-1$, using
 \begin{equation} \label{eq:CPW-Cji1Decomp}
  \Gt(\pp\semi{j-1\,}i{-1}) = 
  \begin{cases}
   \gpl{j}\Gt(\pp\short{j}i), & j>i,
\\ \gpl{j}\bigl(\Gt(\pp\short{j}j)-\gpl{j}\gpl{j+1}\Gt(\pp\short{j+1}{j+1})\bigr), & j=i,
  \end{cases}
 \end{equation}
where the first case follows from \eqref{eq:N=4-Cnm1}.
For $n>m$, $\short{n}m$ is a $\frac14$-BPS short multiplet, whereas $\short{n}n$ is a $\frac12$-BPS multiplet.
The contributions of short multiplets are given explicitly in \secref{se:2p2pCPW-Short}.

We may combine two contributions of the form \eqref{eq:2p2pCPW-GftjDefn} to form those of a long multiplet.
First, we note that
 \begin{multline} \label{eq:2p2pCPW-FDefn}
  \F\g{t}\twist{j} =
     \delta_{tj}\,\fQ0j
   + \delta_{t\,j+1} \left(\fQ1j-\gze{j-1}\fQ0j+\gmi{j}\fQ0{j-1}\right)
\\ + \delta_{t\,j+2} \left(\gmi{j}(\fQ1{j-1}-\gze{j}\fQ0{j-1})+c_j\fQ0j\right)
   + \delta_{t\,j+3}\, \gmi{j}c_{j+1}\fQ0{j-1}
 \end{multline}
satisfies
 \begin{equation} \label{eq:2p2pCPW-FCond}
  \F\g{t}\twist{j} + \gpl{j}\F\g{t}\twist{j+1} = c_{t,j}\fQ0j.
 \end{equation}
%
%
By comparison of the above with \eqref{eq:2p2pCPW-GftjDefn}, \eqref{eq:2p2pCPW-GHt}, we see that
 \begin{equation} \label{eq:2p2pCPW-CombineLong}
  \begin{split}
   \Gxtj*\fhat{j}\xst + \gpl{j}\Gxtj*\fhat{j+1}\xst &= \amx\abmx\sum_kc_{t,k}\GHj[\fQ0j,k]\twist{j}
\\ &= \evalat{\textstyle\GHj[k]=\GHj[\fQ0j,k]\twist{j}}{\Gxt\GH\xst} 
  \end{split}
 \end{equation}
reflecting the decomposition \eqref{eq:N=4-ADecomp}.
Using \eqref{eq:2p2pCPW-Gft0}, \eqref{eq:2p2pCPW-CombineLong}, we may rewrite \eqref{eq:2p2pCPW-GtDecomp}
by replacing $\Gxt\fhat$ with $\Gxtj*\fhat{J}, J=0,1,\dots$, if we compensate by progressively modifying $\GH$.
This reflects the ambiguity in the free theory, where contributions from pairs of semi-short operators are indistinguishable from those of
long operators at the unitarity bound.
Specifically,
 \begin{equation} \label{eq:2p2pCPW-GtDecompj}
  \Gt\xst = \fk\,\Gt(\pp\short00;x) + \gprod{J-1}\Gxtj*\fhat{J}\xst + \Gxt\GHhat\xst,
 \end{equation}
where
 \begin{equation} \label{eq:2p2pCPW-H^Defn}
  \GHhat\g{k}\xst = \GHj[k]\xst + \sum_{j=0}^{J-1}(-1)^j\gprod{j-1}\bigl(\delta_{kj}\fQ0j\xst-\fQ0{j,k}\st*\gbb.{1+\bt}{j+1}\bigr)
 \end{equation}
and we note
 \begin{equation} \label{eq:2p2pCPW-gammaprod}
  (-1)^j\gprod{j-1} = \gprod*.
 \end{equation}
For the $\corel{2p2p}$ correlation function, setting $J=2$ removes the contributions of any semi-short multiplets with non-unitary twist.

For the full decomposition into supermultiplet contributions, we need to separate those of short multiplets.
Extending \eqref{eq:2p2pCPW-fExpn}, we may write
 \begin{equation} \label{eq:2p2pCPW-fjDefn}
  \fhat\xa
  = \fhat_j\xa + \sum_{\ell=0}^j\fhat_\ell(\alpha)\gbb{1+\bt}{\ell+1}x
  = \fhat_j\xa + \sum_{\subbox{\substack{\ell=0,\dots,j\\n=0,\dots,\ell}}}b_{n,\ell}\Jbb{n}(y)\gbb{1+\bt}{\ell+1}x,
 \end{equation}
and using \eqref{eq:2p2pCPW-CjikContrib}, \eqref{eq:CPW-Cji1Decomp},
 \begin{align}
  \Gxtj*{\fhat_{j-1}}j - \Gxtj*{\fhat_j}j
  &= \evalat{\textstyle\fhat=\fhat_{j-1}-\fhat_j=\sum_{n=0}^jb_{n,j}\Jbb{n}\gbb.{1+\bt}{\ell+1}}{\Gxtj*{\fhat}j}
\notag\\
  &= 2\sum_{n=0}^j b_{n,j} \Gt(\pp\semi{j-1\,}n{-1})
\\\notag
  &= 2\gpl{j}\biggl(b_{j,j}\Bigl(\Gt(\short{j}j)-\gpl{j}\gpl{j+1}\Gt(\short{j+1}{j+1})\Bigr)+\sum_{i=0}^{j-1}b_{i,j}\Gt(\short{j}i)\biggr).
 \end{align}
We thus modify \eqref{eq:2p2pCPW-CombineLong},
 \begin{multline}
  \Gxtj*{\fhat_j}j + \gpl{j}\Gxtj*{\fhat_{j+1}}{j+1}
   = \Gxtj*{\fhat_j}j - \gpl{j}\left(\Gxtj*{\fhat_j}{j+1} - \bigl(\Gxtj*{\fhat_j}{j+1}-\Gxtj{\fhat_{j+1}}{j+1}\bigr)\right)
\\[2pt] = -2\gpl{j}\gpl{j+1}\biggl(b_{j+1,j+1}\bigl(\Gt(\short{j+1\,}{j+1})-\gpl{j+1}\gpl{j+2}\Gt(\short{j+2\,}{j+2})\bigr)
\\[-8pt] + \sum_{i=0}^jb_{i,j+1}\Gt(\short{j+1\,}i)\biggr) + \evalat{\textstyle\GHj[k] = \GHj[\fQ^0j,k]\twist{j}}{\Gxt{\GH}}
 \end{multline}
where with $\fhat\goesto\fhat_j$ we correspondingly take $\fQ0j\goesto\fQ^0j=\fQ0j-\sum_{k=0}^j\fQ0{j,k}$
(thus $\fQ^0{j,k}=0$ for $k\le j$).
Hence we may identify contributions from short, semi-short and long multiplets, writing
 \vspace{-2ex}%
 \begin{multline} \label{eq:2p2pCPW-GtMultDecomp}
  \Gt = (k+b_{0,0})\Gt(\short00)
   + 2\sum_{j=1}^{J-1}(-1)^j\gprod{j}(b_{j,j}+\gpl{j-1}b_{j-1,j-1})\Gt(\short{j}j)
\\ + 2(-1)^J\gprod{J}\gpl{J-1}b_{J-1,J-1}\Gt(\short{J}J)
\\ + 2\sum_{j=1}^J(-1)^j\gprod{j}\sum_{i=0}^{j-1}b_{i,j}\Gt(\short{j}i)
\\ + (-1)^J\gprod{J-1}\Gxtj*{\fhat_J}J
   + \Gxt{\GHhat}.
 \end{multline}
The contribution from semi-short multiplets is contained in
 \begin{equation}
  \Gxtj*{\fhat_J}J = 2\sum_{n,\ell\ge0}b_{n,J+\ell+1}\Gt(\semi{J-1\,}n\ell)
 \end{equation}
and that of long multiplets is given by $\Gxt{\GHhat}$, where $\GHhat$ is modified from \eqref{eq:2p2pCPW-H^Defn} by $\fQ0j\goesto\fQ^0j$
.
To account for contributions from short and semi-short multiplets in the \ch2p2p (or \ch22pp) channel, we identify $J=2$, giving
 \vspace{-1ex}%
 \begin{multline} \label{eq:2p2pCPW-J=2Gt}
  \Gt = (k+b_{0,0})\Gt(\short00)
   - 2\gpl0\gpl1(b_{1,1}+\gpl0b_{0,0})\Gt(\short11) + 2\gpl0\gpl1^2\gpl2b_{1,1}\Gt(\short22)
\\ - 2\gpl0\gpl1b_{0,1}\Gt(\short10) + 2\gpl0\gpl1\gpl2\bigl(b_{0,2}\Gt(\short20) + b_{1,2}\Gt(\short20)\bigr)
\\ + 2\sum_{\ell\ge0}\Bigl(b_{0,\ell+3}\Gt(\semi10\ell) + b_{1,\ell+3}\Gt(\semi11\ell)\Bigr)
   + \Gxt{\GHhat},
 \end{multline}
 \begin{equation} \label{eq:2p2pCPW-J=2H}
  \GHhat\g{j} = \GHj + \bigl(\delta_{j0}\fQ^00-\fQ^0{0,j}\gbb.{1+\bt}1\bigr)
  - \gpl0\bigl(\delta_{j1}\fQ^01-\fQ^0{1,j}\gbb.{1+\bt}2\bigr).
 \end{equation}
From the definition \eqref{eq:CPW-QjDefn}, writing $\fhat\xa$ as in \eqref{eq:2p2pCPW-fExpn}, we have
 \begin{equation} \label{eq:2p2pCPW-Q0jk}
  \fQ0{j,k}\st* = -2\sum_n b_{n,k}\JJbb{n-1}j\st*,
 \end{equation}
 \vspace{-1ex}%
and using \eqref{eq:Jac-Refl}, \eqref{eq:Jac00}, and the definition of $\fQ^0j$, find
 \begin{equation} \label{eq:2p2pCPW-Q0Values}
  \begin{aligned}
   \fQ^0{0,\ell} &= 
    \begin{cases}
     0, & \ell=0,
\\   -b_{1,\ell}/\gpl0, & \ell\ge1;
    \end{cases}
& \qquad
   \fQ^0{1,\ell} &=
    \begin{cases}
     0, & \ell=0,1,
\\   b_{0,\ell}/\gpl0, & \ell\ge2.
    \end{cases}
  \end{aligned}
 \end{equation}
The final contribution of long multiplets may thus be read from \eqref{eq:2p2pCPW-J=2H} via the expansion
 \begin{equation} \label{eq:2p2pCPW-HjlDefn}
  \GH\g{j}(x) = \sum_{\ell=0}^\infty \GH\g{j,\ell} \, \gbb{1+\bt}{\ell+1}x ,
 \end{equation}
where by comparison with \eqref{eq:2p2pCPW-HjExpn}, we see that $\GHj[j,\ell] = A_{j,\ell-j-1}\JJbb00$,
with $\JJbb00$ a constant, given in \eqref{eq:Jac00}.
Hence we make the modification $A_{j\ell}\goesto\hat{A}_{j\ell}=2\gpl0\GHhat_{j,\ell+j+1}$; the modified terms are then
 \begin{equation} \label{eq:2p2pCPW-Ahat}
  \begin{aligned}
   A_{0\ell} &\goesto \hat{A}_{0\ell} = A_{0\ell} - 2b_{1,\ell+1} = 2(\gpl0\GHj[0,\ell+1]-b_{1,\ell+1}),
\\ A_{1\ell} &\goesto \hat{A}_{1\ell} = A_{1\ell} - 2\gpl0b_{0,\ell+2} = 2\gpl0(\GHj[1,\ell+2]-b_{0,\ell+2}).
  \end{aligned}
 \end{equation}

\subsection{Results for Short and Semi-Short Multiplets} \label{se:2p2pCPW-Short}

We here work out more fully the contributions of a short or semi-short multiplet.
Taking $\fhat\xa = \frac12\gbb{1+\bt}{j+k+2}x\Jbb{i}(y)$, it follows that $\fQ0j, \fQ1j$ are as given in \eqref{eq:2p2pCPW-QjSemishort};
then \eqref{eq:2p2pCPW-GFtjDefn} gives
\WithSuffix\newcommand\Bob/{\notag\displaybreak[0]\\}
\WithSuffix\newcommand\Bob~{\notag\\}
\WithSuffix\newcommand\Bob0[4]{%
   \Gxtj{\fQ0{#1}}{#2} &= \bigl(\delta_{t\,#2}\textstyle\sum_n c_{n,j+k+1}\gbb.{1+\bt}{n-1} - c_{t,j+k+1}\gbb.{1+\bt}{#3}\bigr)\JJbb{#4}i%
}
\WithSuffix\newcommand\Bob1[4]{%
   \Gxtj{\fQ1{#1}}{#2} &= \bigl(\delta_{t\,#2}\textstyle\sum_n c_{n,j+k+1}\gbb.{1+\bt}{n-1} - c_{t,j+k+1}\gbb.{1+\bt}{#3}\bigr)%
   \bigl(\sum_r \gam{i}r\JJbb{#4}{i+r}\bigr)
}
 \begin{align} \label{eq:2p2pCPW-GQjtSemishort}
  \Bob0jj{j-1}{j-1}, \Bob/
  \Bob1j{j+1}j{j-1}, \Bob/
  \Bob0j{j+1}j{j-1}, \Bob/
  \Bob0{j-1}{j+1}j{j-2}, \Bob/
  \Bob1{j-1}{j+2}{j+1}{j-2}, \Bob/
  \Bob0{j-1}{j+2}{j+1}{j-2}, \Bob/
  \Bob0j{j+2}{j+1}{j-1}, \Bob/
  \Bob0{j-1}{j+3}{j+2}{j-2}.
 \end{align}
Hence \eqref{eq:2p2pCPW-GftjDefn}, \eqref{eq:2p2pCPW-CjikContrib}, lead to
 \begin{multline} \label{eq:2p2pCPW-GtCjik}
  \textstyle%
  \G\g{j+r}(\pp\semi{j-1}ik) = \sum_{s=0}^3 a_{r,s} \, \gbb.{1+\bt}{j+k+s},
\\\textstyle%
  \G\g{j+k+r+1}(\pp\semi{j-1}ik) = -\sum_{s=0}^3 a_{s,r} \, \gbb.{1+\bt}{j+s-1}
 \end{multline}
for $r=0,1,2,3$ with
 \begin{align} \label{eq:2p2pCPW-ars}
  a_{0,s} &= c_{j+k+s+1,j+k+1}\JJbb{j-1}i, \Bob/
  a_{1,s} &= c_{j+k+s+1,j+k+1}\bigl(\gpl{i}\JJbb{j-1}{i+1} + (\gze{i}-\gze{j-1})\JJbb{j-1}i + \gmi{i}\JJbb{j-1}{i-1} + \gmi{j}\JJbb{j-2}i\bigr), \Bob/
  a_{2,s} &= c_{j+k+s+1,j+k+1}\bigl(\gpl{j}(\gpl{i}\JJbb{j-2}{i+1} + (\gze{i}-\gze{j})\JJbb{j-2}i + \gmi{i}\JJbb{j-2}{i-1}) + c_j\JJbb{j-1}i\bigr), \Bob/
  a_{3,s} &= c_{j+k+s+1,j+k+1}\gmi{j}c_{j+1}\JJbb{j-2}i.
 \end{align}
The contribution of a semi-short multiplet to the conformal partial wave expansion is then given by
 \begin{equation} \label{eq:2p2pCPW-aCjik}
  \textstyle
  \sum_{n,m} a_{nm,j+r\,k+s+r}(\pp\semi{j-1}ik)\JJbb{n}m = a_{r,s}.
 \end{equation}

To acquire the conformal partial wave contribution of short multiplets, we use \eqref{eq:CPW-Cji1Decomp}, \eqref{eq:2p2pCPW-GtCjik}.
For $j>i$ and setting $k=-1$, we have
 \begin{equation} \label{eq:2p2pCPW-GtBji}
  \G\g{j+r}(\pp\short{j}i) = \sum_{s=0}^4 b_{r,s} \, \gbb.{1+\bt}{j+s-1},
 \qquad
  \gpl{j}b_{r,s} = a_{r,s} \! - a_{s,r},
 \end{equation}
which gives coefficients
\WithSuffix\newcommand\Bob2[2]{%
 \gpl{i}\JJbb{#1}{i+1} + (\gze{i}-\gze{#2})\JJbb{#1}i + \gmi{i}\JJbb{#1}{i-1}}
\WithSuffix\newcommand\Bob4[5][\quad]{%
 \gpl{i}\bigl(\gpl{i+1}\JJbb{#2}{i+2} + (\gze{i}+\gze{i+1}-\gze{#3}-\gze{#4})\JJbb{#2}{i+1}\bigr)
\Bob~ & #1 + \bigl((\gze{i}-\gze{#3})(\gze{i}-\gze{#4}) + c_{i} + c_{i+1} {#5} \bigr)\JJbb{#2}i
\Bob~ & #1 + \gmi{i}\bigl((\gze{i-1}+\gze{i}-\gze{#3}-\gze{#4})\JJbb{#2}{i-1} + \gmi{i-1}\JJbb{#2}{i-2}\bigr)}
 \begin{align} \label{eq:2p2pCPW-brs}
  b_{0,1} &= \JJbb{j}i, \Bob/
  b_{0,2} &= \gmi{j+1}\JJbb{j-1}i + \Bob2jj, \Bob/
  b_{0,3} &= \gmi{j+1}\bigl(\Bob2{j-1}{j+1}\bigr) + c_{j+1}\JJbb{j}i, \Bob/
  b_{0,4} &= c_{j+2}\gmi{j+1}\JJbb{j-1}i, \Bob/
  b_{1,2} &= \gmi{j+1}\bigl(\gmi{j}\JJbb{j-2}i + \Bob2{j-1}{j-1}\bigr) \Bob~
   & \quad + \Bob4j{j-1}j{- c_j}, \Bob/
  b_{1,3} &= \gmi{j+1}\Bigl(\gmi{j}\bigl(\Bob2{j-2}{j+1}\bigr) \Bob~
   & \quad + \Bob4{j-1}{j-1}{j+1}{} \Bob~
   & \quad + \gpl{j}\bigl(\Bob2{j}{j-1}\bigr)\Bigr), \Bob/
  b_{1,4} &= c_{j+2}\gmi{j+1}\bigl(\gmi{j}\JJbb{j-2}i + \Bob2{j-1}{j-1}\bigr), \Bob/
  b_{2,3} &= \gmi{j}\gmi{j+1}\Bigl(\Bob4{j-2}j{j+1}{- c_{j+1}}\Bigr) \Bob~
   & \quad + c_j\gmi{j+1}\bigl(\Bob2{j-1}j\bigr) + c_jc_{j+1}\JJbb{j-1}i, \Bob/
  b_{2,4} &= c_{j+2}\gmi{j+1}\Bigl(\gmi{j}\bigl(\Bob2{j-2}j\bigr) + c_j\JJbb{j-1}i\Bigr), \Bob/
  b_{3,4} &= c_{j+1}c_{j+2}\gmi{j}\gmi{j+1}\JJbb{j-2}i.
 \end{align}
We may then read off the contribution to the conformal partial wave expansion, given by
 \begin{equation} \label{eq:2p2pCPW-aBji}
  \textstyle
  \sum_{n,m} a_{nm,j+r\,s-r-1}(\pp\short{j}i)\JJbb{n}m = b_{r,s}.
 \end{equation}

For $j=i$, \eqref{eq:2p2pCPW-GtCjik} decomposes into contributions from two $\frac12$-BPS short multiplets, as given by \eqref{eq:CPW-Cji1Decomp}.
Making use of \eqref{eq:Jac-Refl} for $\JJbb{n}m, n<m$, we obtain
 \begin{equation} \label{eq:2p2pCPW-GtBjj}
  \G\g{j+r}(\pp\short{j}j) = \sum_{s=r+1}^3 \hat{b}_{r,s} \, \gbb.{1+\bt}{j+s-1} - \sum_{s=0}^{r-1} \hat{b}_{s,r} \, \gbb.{1+\bt}{j+s-1},
 \end{equation}
where
 \begin{equation} \label{eq:2p2pCPW-b^rs}
  \begin{aligned}
   \hat{b}_{0,1} &= \JJbb{j}j, \quad \hat{b}_{0,2} = \gmi{j}\JJbb{j}{j-1}, \quad \hat{b}_{0,3} = \gmi{j}\gmi{j+1}\JJbb{j-1}{j-1},
\\ \hat{b}_{1,2} &= \gmi{j-1}\gmi{j}\JJbb{j}{j-2}, \quad \hat{b}_{1,3} = \gmi{j-1}\gmi{j}\gmi{j+1}\JJbb{j-1}{j-2},
\\ \hat{b}_{2,3} &= \gmi{j-1}\gmi{j}^2\gmi{j+1}\JJbb{j-2}{j-2}.
  \end{aligned}
 \end{equation}
In the case $j=i=1$, we have
$\hat{b}_{0,1} = \JJbb11, \hat{b}_{0,2} = \gmi1\JJbb10, \hat{b}_{0,3} = \gmi1\gmi2\JJbb00$, 
$\hat{b}_{1,2} = \hat{b}_{1,3} = \hat{b}_{2,3} = 0$,
giving conformal partial wave contributions
 \begin{equation} \label{eq:2p2pCPW-aB11}
  \begin{aligned}
   a_{11,10}(\pp\short11) &= 1,
&  a_{10,11}(\pp\short11) &= \gmi1,
&  a_{00,12}(\pp\short11) &= \gmi1\gmi2,
  \end{aligned}
 \end{equation}
plus those given by symmetry \eqref{eq:CPW-aRefl}, which agree with \eqref{eq:2p2pCPW-Gt11Defn}, \eqref{eq:2p2pCPW-G11=GB11};
and at $j=i=0$,
 \begin{equation} \label{eq:2p2pCPW-aB00'}
  a_{00,00}(\pp\short00) = -a_{00,1\,-2}(\pp\short00) = 1,
 \end{equation}
agreeing with \eqref{eq:2p2pCPW-aB00}.

\subsection{Expansion in the \ch22pp Channel} \label{se:22ppCPW}

Performing a similar analysis for the $\corel{22pp}$ correlator, 
with $p_1=p_2=2, p_3=p_4=p$ we obtain from \eqref{eq:CPW-GExpn}, \eqref{eq:CPWJac-GExpn} 
a conformal partial wave expansion for $\G*\uvst$,
 \begin{equation} \label{eq:22ppCPW-GExpn}
   \G*\uvst =
   \sum_{\substack{\subbox{\lemn2}\\t,\ell}} a_{nm,t\ell} \, \JJacobi00{n}m\st* \, u^t\Gpw\ell{2t+\ell}{u}v00.
 \end{equation}
We note that \eqref{eq:22ppCPW-GExpn} is identical to \eqref{eq:2p2pCPW-GExpn} taken in the limit $\bt\goesto0$.
\pagebreak[2]
Thus all the results of \secref{se:2p2pCPW} hold, if we substitute $\bt=0$.
In particular, the contributing conformal waves have lowest twist zero, 
we have
 \begin{equation} \label{eq:22ppCPW-g}
  \gbb{a+\bt}jx \goesto \gpw00ajx \equiv \gpw*ajx,
\quad
  \gpw*1j{x'} = (-1)^j\gpw*1jx,
 \end{equation}
the identity for $\gpw*1j{x'}$ following from \eqref{eq:CPW-gx'}, and
 \begin{equation} \label{eq:22ppJac}
  \Jbb{n}(y), \JJbb{n}m\st* \goesto \Jacobi00{n}(y), \JJacobi00{n}m\st* = \Legendre{n}(y), \LL{n}m\st*,
 \end{equation}
where $\Legendre{n}$ are the Legendre polynomials, and $\LL{n}m$ are 2-variable versions, defined analogously to \eqref{eq:Jac-2varDefn}, 
giving the contribution of the six $SU(4)_R$ representation $\dynk{n-m}{2m}{n-m}$ present in the tensor product $\dynk020\otimes\dynk020$.
Notably, $\LL00=1$.
Correspondingly, \eqref{eq:2p2pJac-gamma} becomes
 \begin{equation} \label{eq:22ppJac-gamma}
  \begin{gathered}
   \gpl{n} = \frac{n+1}{2(2n+1)}; \qquad
   \gmi{n} = \frac{n}{2(2n+1)}; \qquad
   \gze{n} = 0.
  \end{gathered}
 \end{equation}
and \eqref{eq:2p2pCPW-gammaprod} becomes
 \begin{equation} \label{eq:22ppCPW-gammaprod}
  (-1)^j\gpl0\gpl1\dots\gpl{j-1} = \gprod* \xrightarrow{\bt=0} (-1)^j\frac{j!^2}{(2j)!}.
 \end{equation}
In terms of superconformal multiplets, $\pp\short00\goesto\idmult$, the unit operator, and
$\pp\short11\goesto\short11$, the lowest \BPS2 multiplet containing the energy momentum tensor.

 \section{Conformal Expansion in the Free Field}
 \label{se:FFCPW}

\newcommand{\FFCPWtitle}[4]{Free Field Expansion in the \ch{#1}{#2}{#3}{#4} Channel}

\subsection{\FFCPWtitle2p2p} \label{se:2p2pFF-CPW}

We now apply the conformal partial wave expansion (\secref{se:2p2pCPW}) to the free field result for the \ch2p2p channel.
We first consider the free field contribution $\GH\ord0\uv$.
We may write \eqref{eq:2p2pCPW-HjDefn} as
 \begin{equation} \label{eq:2p2pCPW-HjAlt}
  x\xb(\xxb)\GH\uv = -\sum_{j=0}^\infty \GH\g{j}(x)\gbb{1+\bt}{j+1}\xb ;
 \end{equation}
from \eqref{eq:2p2pFF-HValue} we have
 \begin{equation} \label{eq:2p2pFF-HAlt}
  x\xb(\xxb)\GH\ord0\uv = (x^2\xb-x\xb^2) - a(x'\xb-x\xb') - b(x'^2\xb'-x'\xb'^2).
 \end{equation}
Thus by \eqref{eq:CPW-xnExpn}, \eqref{eq:CPW-x'nExpn}, we have
 \begin{equation*} 
  \GH\free\g{j}(x) = \beta_{j+1,1}(ax'-x^2) + \beta_{j+1,2}\,x + (-1)^j\bigl[\beta'_{j+1,1}(ax-bx'^2) + \beta'_{j+1,2}\,bx'\bigr],
 \end{equation*}
with the $\beta_{j,k}, \beta'_{j,k}$ as in \eqref{eq:CPW-beta}, \eqref{eq:CPW-beta'}.
%
For $a = 1+\bt = 1-\bt* = \frac12p$ we have
 \begin{equation} \label{eq:2p2pCPW-bbValues}
  \begin{aligned}
   \beta_{j+1,k+1} &= (-1)^{k+1} \binom{j}k \frac{\pochhammer{2\bt+k+1}{j-k}}{\pochhammer{2\bt+j+k+1}{j-k}},
\\ \beta'_{j+1,k+1} & = (-1)^{k+1} \binom{j}k \frac{\pochhammer{k+1}{j-k}}{\pochhammer{2\bt+j+k+1}{j-k}}.
  \end{aligned}
 \end{equation}
Thus we find
 \begin{multline} \label{eq:2p2pFF-CPW-Hj}
  \GH\free\g{j}(x) = \frac{j!}{\pochhammer{2\bt+j+1}j}\Bigl[\tbinom{2\bt+j}j(x^2-ax') + \tbinom{2\bt+j+1}jj\,x
\\ + (-1)^j(bx'^2-ax) + (-1)^j(2\bt+j+1)j\,b\,x'\Bigr] .
 \end{multline}
Applying \eqref{eq:CPW-beta}, \eqref{eq:CPW-beta'} again we may make the further expansion \eqref{eq:2p2pCPW-HjlDefn} to obtain
 \begin{multline} \label{eq:2p2pFF-CPW-Hjl}
  \GH\free\g{j,\ell}
   = \frac{j!}{\pochhammer{2\bt+j+1}j} \frac{\ell!}{\pochhammer{2\bt+\ell+1}\ell} \biggl[
     \Bigl( \tbinom{2\bt+j}j\tbinom{2\bt+\ell+1}\ell\ell - \tbinom{2\bt+\ell}\ell\tbinom{2\bt+j+1}jj \Bigr)
\\ \begin{aligned}[b]
   &+ \Bigl( (-1)^j\tbinom{2\bt+\ell}\ell - (-1)^\ell\tbinom{2\bt+j}j \Bigr) a
\\ &+ (-1)^{j+\ell} \Bigl( (2\bt+j+1)j - (2\bt+\ell+1)\ell \Bigr) b \biggr]
   \end{aligned} ,
 \end{multline}
where 
$\GH\g{j,\ell}=A_{j\,\ell-j-1}\JJbb00=-\GH\g{\ell,j}$.

Similarly, to find the contribution of $\fhat\xa$, we may write \eqref{eq:2p2p-fhatValue} as
 \begin{multline} \label{eq:2p2pFF-CPW-fhat}
  \fhat\xa 
   = \frac1{2\bt+1}\Bigl(x^2-ax+bx'^2-ax')\Big)\Jbb1(y)
\\ + \frac1{2\bt+2}\Bigl(\tbt1x^2+\bigl(\tbt3a+\tbt2)\bigr)x
\\ - bx'^2-\bigl(\tbt1\tbt3a+\tbt2b\bigr)x'\Bigr)\Jbb0(y),
 \end{multline}
where we have made use of the results of \secref{se:Colour}, namely \eqref{eq:Col-acRelation}, which gives us $c=\tbt1a$.
Comparing with \eqref{eq:2p2pCPW-fExpn}, and again using \eqref{eq:CPW-xnExpn}, \eqref{eq:CPW-x'nExpn}, we find
 \begin{align} \label{eq:2p2pCPW-bnlValues}
  b_{0,\ell} &= \frac1{2\bt+2}\frac{\ell!}{\pochhammer{2\bt+\ell+1}\ell}
   \Bigl[\tbinom{2\bt+\ell}\ell\bigl((2\bt+\ell+2)(\ell-1)-(2\bt+3)a\bigr)
\notag\\
  b_{1,\ell} &= \!
  \begin{aligned}[b]
   - (-1)^\ell\bigl(\tbt1\tbt3a-(2\bt+\ell+2)(\ell-1)b\bigr)\Bigr]&,
\\ \frac1{2\bt+2}\frac{\ell!}{\pochhammer{2\bt+\ell+1}\ell}
   \Bigl[\tbinom{2\bt+\ell+1}\ell \ell + \tbinom{2\bt+\ell}\ell a - (-1)^\ell\bigl(a+(2\bt+\ell+1)\ell\,b\bigr)\Bigr]&.
  \end{aligned}
 \end{align}
These satisfy $b_{0,1} = b_{1,0} = 0$.
Thus from \eqref{eq:2p2pCPW-J=2Gt}, the lowest twist contribution from shortened multiplets is that of the 
\BPS2 multiplet $\pp\short00$, twist $2\bt$, given by
 \begin{equation} \label{eq:2p2pCPW-B00Coeff}
  k + b_{0,0} = b,
 \end{equation}
and at twist $2\bt+2$ we have the contribution of the \BPS2 multiplet $\pp\short11$,
 \begin{equation} \label{eq:2p2pCPW-B11Coeff}
  -2\gpl0\gpl1(b_{1,1}+\gpl0b_{0,0}) = \frac{4a}{\tbt3\tbt4}.
 \end{equation}
Other shortened multiplets contribute to the partial wave expansion with twist $2\bt+4$, namely 
the \BPS2 short multiplet $\pp\short22$, which appears with coefficient
 \begin{equation} \label{eq:2p2pCPW-B22Coeff}
  2\gpl0\gpl1^2\gpl2b_{1,1} = \frac{24(a+b+1)}{\tbt3\tbt4^2\tbt5\tbt6};
 \end{equation}
\BPS4 short multiplets $\pp\short20, \pp\short21$, coefficients
 \begin{gather} \label{eq:2p2pCPW-B20Coeff}
  2\gpl0\gpl1\gpl2b_{0,2} = \frac{12\blg(\tbt1\tbt2-\tbt1\tbt3a+2b\brg)}{\tbt2\tbt3\tbt4\tbt5\tbt6},
\\ \label{eq:2p2pCPW-B21Coeff}
  2\gpl0\gpl1\gpl2b_{1,2} = \frac{24\blg(\tbt2+\bt a-2b\brg)}{\tbt2\tbt4^2\tbt5\tbt6},
 \end{gather}
respectively;
and the semi-short multiplets $\pp\semi10\ell, \pp\semi11\ell$, respectively \BPS4, \BPS2 with coefficients
$2b_{0,\ell+3}, 2b_{1,\ell+3}$.

Contributions of long multiplets are in general given by $\hat{A}_{j\ell} = 2\gpl0\GHj[j,\ell+j+1]$, with $\GHj[j,\ell]$ as in \eqref{eq:2p2pFF-CPW-Hjl}.
As given by \eqref{eq:2p2pCPW-Ahat}, modifications occur at $j=0$, where we have
 \begin{equation} \label{eq:2p2pFF-CPW-A0}
  \hat{A}_{0\ell} = 2(\gpl0\GHj[0,\ell+1]-b_{1,\ell+1}) = 0,
 \end{equation}
thus removing any contribution from non-unitary long multiplets $\pp\Long00\ell\Delta$ with twist $\Twist\le2\bt$,
and for $j=1$, when
 \begin{equation} \label{eq:2p2pFF-CPW-A1}
  \hat{A}_{1\ell} = 2\gpl0(\GHj[1,\ell+2]-b_{0,\ell+2})
   = 2\gpl0 \tfrac{(\ell+2)!}{\pochhammer{2\bt+\ell+3}{\ell+2}}\left[\tbinom{2\bt+\ell+2}{\ell+2}+(-1)^\ell\tbt1)\right] a.
 \end{equation}

\subsection{\FFCPWtitle22pp}

From the results of \secref{se:22ppCPW}, \secref{se:22ppFF}, 
we analyse the partial wave contributions from the free field to the $\corel{22pp}$ correlation function.
First considering $\GH*\ord0\uv$, given by \eqref{eq:22ppFF-HValue}, 
from which we have
 \begin{equation} \label{eq:22ppFF-HAlt}
  x\xb(\xxb)\GH*\ord0\uv = b\bigl((x^2\xb-x\xb^2) - (x'^2\xb'-x'\xb'^2)\bigr) - (p-1)a(x'\xb-x\xb'),
 \end{equation}
as before using \eqref{eq:Col-acRelation}, this time in the form $c=(p-1)a$.
In this channel, \eqref{eq:2p2pCPW-HjAlt} becomes
 \begin{equation} \label{eq:22ppCPW-HjAlt}
  x\xb(\xxb)\GH*\uv = -\sum_{j=0}^\infty \GH*\g{j}(x)\gpw*1{j+1}\xb ;
 \end{equation}
now with
 \begin{equation}
  \beta_{j+1,k+1} = \beta'_{j+1,k+1} = (-1)^{k+1}\binom{j}k\frac{\pochhammer{k+1}j}{\pochhammer{j+1}j},
 \end{equation}
we proceed as in the previous section, first obtaining
 \begin{equation} \label{eq:22ppFF-CPW-Hj}
  \GH*\free\g{j} = \frac{j!^2}{(2j)!} \Bigl[b\bigl(x^2+(-1)^jx'^2\bigr) + \bigl(j(j+1)b-(-1)^j(p-1)a\bigr)\bigl(x+(-1)^jx'\bigr)\Bigr]
 \end{equation}
and then
 \begin{equation} \label{eq:22ppFF-CPW-Hjl}
  \GH*\free\g{j,\ell} = \frac{j!^2\ell!^2}{(2j)!(2\ell)!}\bigl(1-(-1)^{j+\ell}\bigr)\Bigl[b\bigl(\ell(\ell+1)-j(j+1)\bigr)+(-1)^j(p-1)a\Bigr].
 \end{equation}
Since $\LL00=1$, we have in this channel $\GH*\g{j,\ell} = A_{j\,\ell-j-1}$, and thus for $\ell$ even,
 \begin{equation} \label{eq:22ppFF-CPW-Atl}
  A_{t\ell} = 2\frac{t!^2(\ell+t+1)!^2}{(2t)!(2\ell+2t+2)!}\bigl[(\ell+1)(\ell+2t+2)b+(-1)^t(p-1)a\bigr],
 \end{equation}
with $A_{t\ell} = 0$ for $\ell$ odd.

For the $\fhat*\xa$ contribution, we write \eqref{eq:22pp-fhatValue} as
 \begin{multline} \label{eq:22ppFF-CPW-fhat}
   \fhat*\xa = \tfrac12\bigl(b(x^2+x'^2)-(p-1)a(x+x')\bigr)\Legendre1(y)
\\ + \tfrac12\bigl(b(x^2-x'^2)+((p+1)a+2b)(x-x')\bigr)\Legendre0(y),
 \end{multline}
which with
 \begin{equation}
  \fhat*\xa = \sum_{\subbox{\substack{\ell=0,1,\dots\\n=0,1}}} b_{n,\ell} \, \Legendre{n}(y)\gpw*1{\ell+1}x
 \end{equation}
gives
 \begin{equation}
  \begin{aligned}
   b_{0,\ell} &= \frac{\ell!^2}{(2\ell)!}\bigl[(\ell(\ell+1)-2)b - (p+1)a\bigr], & \ell &= 0,2,\dots \,,
\\ b_{1,\ell} &= \frac{\ell!^2}{(2\ell)!}\bigl[\ell(\ell+1)b + (p-1)a\bigr], & \ell &= 1,3,\dots \,.
  \end{aligned}
 \end{equation}

Thus with the results \eqref{eq:22ppJac-gamma}, \eqref{eq:22ppCPW-gammaprod}, we find that at zero twist,
the contribution of the identity operator $\idmult$ is given by
 \begin{equation} \label{eq:22ppCPW-ICoeff}
  \fk* + b_{0,0} = 1,
 \end{equation}
necessary for consistency with our normalisation.
At twist two, the contribution corresponding to the $\frac12\,$-BPS short multiplet $\short11$ containing the energy momentum tensor is
 \begin{equation} \label{eq:22ppCPW-B11Coeff}
  -2\frac{2!^2}{4!}\bigl(b_{1,1} + \gpl0b_{0,0}\bigr) = \frac13a.
 \end{equation}
At twist four, there is no contribution from $\short21$, since $b_{1,2}=0$;
we have contributions from \BPS2 $\short22$ and \BPS4 $\short20$, coefficients
 \begin{equation} \label{eq:22ppCPW-B2iCoeff}
  2\frac{3!^2}{6!}\gpl1b_{1,1} = \frac1{60}\bigl(2b+(p-1)a\bigr),
\quad
  2\frac{3!^2}{6!}b_{0,2} = \frac1{60}\bigl(4b-(p+1)a\bigr)
 \end{equation}
respectively.
Semi-short twist four multiplets $\semi10\ell, \semi11\ell$ contribute for $\ell$ odd, even respectively,
coefficients $2b_{0,\ell+3}, 2b_{1,\ell+3}$.

Contributions from long multiplets $\Long00\ell{2t+\ell}$ are given by $\hat{A}_{t\ell}$, 
where $A$ has been modified to cancel non-unitary contributions.
In particular, at zero twist
 \begin{equation} \label{eq:22ppFF-CPW-A0}
  \hat{A}_{0\ell} = A_{0\ell} - 2b_{1,\ell+1} = 0.
 \end{equation}
There is also a modification at twist two,
 \begin{equation} \label{eq:22ppFF-CPW-A1}
  \hat{A}_{1\ell} = A_{1\ell} - b_{0,\ell+2} = 2\frac{(\ell+2)!^2}{(2\ell+4)!}\,a = \frac{(\ell+1)!(\ell+2)!}{(2\ell+3)!}\,a.
 \end{equation}

 \section{Colour Contractions}
 \label{se:Col}

Here we attempt to calculate values, in some specific cases, for the constant $b$ described in \secref{se:Colour}.
Explicit values for the constants $a$ and $c$, which are trivially determined by \eqref{eq:Col-abcExplicit}, 
are also stated in these cases, for the sake of completeness.
As previously, $\set{T_a}$ are the generators of $SU(N)$, obeying \eqref{eq:SU(N)-Conv}, \eqref{eq:SU(N)-TaTa},
and also satisfying the fusion and fission rules,
\begin{subequations} \label{eq:SU(N)-FusionFission}
 \begin{align}
  \tr(T_aA)\tr(T_aB) &= \COLfusion{A}B, \label{eq:SU(N)-Fusion}
  \\
  \tr(T_aAT_aB) &= \COLfission{A}B. \label{eq:SU(N)-Fission}
 \end{align}
\end{subequations}

\subsection{The $p=2$ Case} \label{se:Col-p=2}

In the case that $p=2$, we are evaluating the correlation function of four \BPS2 operators $\chiral\dimn2$,
all with colour structure $X_{ab}=\tr(T_aT_b)=\tfrac12\delta_{ab}$; thus
 \begin{align} \label{eq:p=2Col-XLXR}
  \XLXR &= \tfrac14\delta_{ab}\delta_{ab} = \tfrac14\Nsb1, &
  \XL_{aa}\XR_{bb} &= \tfrac14\delta_{aa}\delta_{bb} = \tfrac14\Nsb1^2.
 \end{align}
Hence by \eqref{eq:Col-abcExplicit}
 \begin{align} \label{eq:p=2Col-abcValues}
  a &= c = \frac4\Ns\,, & b &= 1.
 \end{align}
This gives us for the \ch22pp channel, from \eqref{eq:22pp-kValue}, \eqref{eq:22pp-fhatValue}, \eqref{eq:22ppFF-HValue}
\begin{subequations} \label{eq:2222FF-kfhatHValues}
 \begin{gather}
  \fk* = 3(1+a), \\
  \fhat*\xa = \left(x^2+x'^2-a(x+x')\right)\alpha - \left(x'^2-(1+2a)x+(1+a)x'\right), \\
  \GH*\ord0\uv = 1+\frac1{v^2}+a\frac1v,
 \end{gather}
\end{subequations}
with $a$ as in \eqref{eq:p=2Col-abcValues}.
For the \ch2p2p channel, from \eqref{eq:2p2p-kValue}, \eqref{eq:2p2p-fhatValue}, \eqref{eq:2p2pFF-HValue} we find
 \begin{align} \label{eq:2222FF-Values}
  \fhat\xa &= \fhat*\xa,
& \GH\ord0\uv &= \GH*\ord0\uv,
 \end{align}
reflecting the fact that we have now a correlation function of four identical operators, with a correspondingly enhanced degree of crossing symmetry.
The results match those in \cite[(3.7),(3.8)]{rf:0412335} for the correlator of four $\Delta=2$ operators $\fourpt\spt\chiral2\chiral2\chiral2\chiral2$.

\subsection{The $p=3$ Case} \label{se:Col-p=3}

In the  $p=3$ case, the two $\Delta=3$ operators necessarily have a single-trace structure,
 \begin{equation} \label{eq:p=3Col-XL,XR}
  \XL_{abc} = \XR_{abc} = \COLtrtrace{a}bc = \tfrac12\tr(T_a\acomm{T_b}{T_c}).
 \end{equation}
It follows directly from \eqref{eq:Col-abcExplicit} that
 \begin{align} \label{eq:p=3Col-acValues}
  a &= \frac6\Ns\,,
& c &= 2a = \frac{12}\Ns\,.
 \end{align}
F%
rom \eqref{eq:Col-abcExplicit}, $b$ is proportional to
 \begin{equation}
  \XL_{aac}\,\XR_{bbc} = \COLtriple{a}acbbc,
 \end{equation}
and using the cyclic property of the trace, along with \eqref{eq:SU(N)-TaTa} and that the $SU(N)$ generators $T_a$ are themselves traceless, we see that
 \begin{equation}
  \COLtrtrace{a}ac = \tr(T_aT_aT_c) = \tr\left(\tfrac\Ns{2N}T_c\right) = 0.
 \end{equation}
Hence in this case
 \begin{equation} \label{eq:p=3Col-bValue}
  b=0.
 \end{equation}
Inserting these results into \eqref{eq:22pp-kValue}, \eqref{eq:22pp-fhatValue}, \eqref{eq:22ppFF-HValue} gives for the \ch22pp channel
\begin{subequations} \label{eq:2233FF-kfhatHValue}
 \begin{gather}
  \fk* = 1+4a, 
\\\fhat*\xa = a\bigl(3x-x'-2(x+x')\alpha\bigr), 
\\\GH*\ord0\uv = 2a\,\frac1v 
 \end{gather}
\end{subequations}
with $a=\frac6\Ns$ as in \eqref{eq:p=3Col-acValues};
and in the \ch2p2p channel, from \eqref{eq:2p2p-kValue}, \eqref{eq:2p2p-fhatValue}, \eqref{eq:2p2pFF-HValue},
\vspace{-1ex}%
\begin{subequations} \label{eq:2323FF-kfhatHValue}
 \begin{gather}
  \fk = \fk* = 1 + 4a, \\
  \fhat\xa = \left(x^2-a(x+x')\right)\alpha + \left(\vphantom{x^2}x+2a(x-x')\right), \\
  \GH\ord0(u,v) = 1 + a\, \frac1v\,.
 \end{gather}
\end{subequations}

Although not necessary for the calculation of $b$, we will also obtain an explicit value for $\XLXR$;
this will come in useful later when comparing results with those subject to different normalisation conventions.
We use the cyclic property of the trace and the fusion and fission rules, \eqref{eq:SU(N)-FusionFission}, to find firstly
 \begin{align} \label{eq:p=3Col-PreliminaryCalculation}
  \tr(\acomm{T_a}{T_b}\acomm{T_a}{T_b})
   &= 2\tr(T_aT_bT_aT_b) + 2\tr(T_aT_aT_bT_b) \Bob/
   &= 2\cdot\COLfission{T_b}{T_b} + 2\tr\left(\left[\tfrac1{2N}\Nsb1\right]^2\ident\right) \Bob/
   &= -\frac\Ns{2N} + \frac{\Nsb1^2}{2N} = \frac{\Nsb1\Nsb2}{2N}\,.
 \end{align}
Hence,
 \begin{align} \label{eq:p=3Col-XLXR}
  \XL_{abc}\,\XR_{abc}
   &= \tfrac14\tr(T_a\acomm{T_b}{T_c})\tr(T_a\acomm{T_b}{T_c}) \Bob/
   &= \tfrac14\cdot\COLfusion{\acomm{T_b}{T_c}}{\acomm{T_b}{T_c}} \Bob/
   &= \frac18\left(\frac{\Nsb1\Nsb2}{2N} - \frac1N\delta_{bc}\delta_{bc}\right) = \frac{\Nsb1\Nsb4}{16N}\,.
 \end{align}

\subsection{The Large $N$ Limit, $p\ge4$} \label{se:Col-LargeN}

For the general $p>3$ case, involved trace computations are required to exactly find the coefficient $b$.
Colour structures other than the single-trace operators which appear in the $p=2,3$ cases are also possible.
However, if we consider only single-trace operators, writing $\XL=\XR=\Xp$ where
 \begin{equation} \label{eq:Col-X(p)Defn}
  \Xp\ap = \Tr{\Tap()},
 \end{equation}
then we may successfully calculate the leading order behaviour for $b$ in the large $N$ limit.
As previously noted, the constants $a,c$ are independent of the colour structure, 
and their large $N$ values may be read directly from \eqref{eq:Col-abcExplicit}, giving
 \begin{align} \label{eq:LargeNCol-acValues}
  a &\Neq \frac{2p}{N^2}\,,
& c &= (p-1)a \Neq \frac{2p(p-1)}{N^2}\,,
 \end{align}
where $\Neq$ denotes that sub-dominant terms for large $N$ have been dropped.
Using the identities for $T_a$ (\ref{eq:SU(N)-Conv}--\ref{eq:SU(N)-TaTa}), 
and fusion and fission rules \eqref{eq:SU(N)-FusionFission},
we consider the two colour contractions appearing in the expression for $b$ in \eqref{eq:Col-abcExplicit}.
For all $p\ge2$,
 \begin{equation} \label{eq:LargeNCol-X(p)X(p)}
  \begin{aligned}[b]
   \dpr\Xp\Xp &= \Tr{\Tap()}\Tr{\Tap()} \\
    &\Neq \frac1{(p-1)!} \Tr{T_{a_1}\dots T_{a_p}}\Tr{T_{a_p}\dots T_{a_1}} \\
    &\Neq \frac12 \frac1{(p-1)!} \Tr{T_{a_1}\dots T_{a_{p-1}}T_{a_{p-1}}\dots T_{a_1}}
    = \frac{N^p}{2^p(p-1)!}\,,
  \end{aligned}
 \end{equation}
which we note gives the correct leading-order behaviour in \eqref{eq:p=2Col-XLXR}, \eqref{eq:p=3Col-XLXR} for $p=2,3$.
If $p\ge4$, we may write
 \begin{equation} \label{eq:LargeNCol-X(p)bbX(p)dd}
  \begin{aligned} [b]
   \dpr{\Xbb\Xp{b}}{\Xbb\Xp{d}} &= \Tr{\Tap[p-2]({}T_bT_{b)}}\Tr{\Tap[p-2]({}T_dT_{d)}} \\
    &\Neq \bfrac2{p-1}^2 \Tr{T_bT_b\Tap[p-2]()}\Tr{T_dT_d\Tap[p-2]()} \\
    &= \bfrac{N}{p-1}^2 \dpr{\Xp[p-2]}{\Xp[p-2]}
    \Neq \frac{4N^p}{2^p(p-1)!} \bfrac{p-2}{p-1}.
  \end{aligned}
 \end{equation}
From \eqref{eq:LargeNCol-X(p)X(p)}, \eqref{eq:LargeNCol-X(p)bbX(p)dd}, we deduce
 \begin{equation} \label{eq:LargeNCol-XbbXdd/XX}
  \frac{\dpr{\Xbb\Xp{b}}{\Xbb\Xp{d}}}{\dpr\Xp\Xp} \Neq 4\bfrac{p-2}{p-1},
 \end{equation}
which substituted into \eqref{eq:Col-abcExplicit} gives the large $N$ behaviour of the coefficient $b$, 
for single-trace operators $\Xp$, $p\ge4$, namely
 \begin{equation} \label{eq:LargeNCol-bValue}
  b \Neq \frac{2p(p-2)}{N^2} \Neq (p-2)a.
 \end{equation}
%
%
%
We may check the results \eqref{eq:LargeNCol-X(p)X(p)}, \eqref{eq:LargeNCol-X(p)bbX(p)dd} against explicit calculations for low $p$,
e.g. for $p=4,5,6$ we have
 \begin{align*}
  \dpr{\Xp[4]}{\Xp[4]} &= \tfrac{\Nsb1(N^4-6N^2+18)}{96N^2} \,,
& \dpr{\Xbb{\Xp[4]}b}{\Xbb{\Xp[4]}d} &= \tfrac{\Nsb1(2N^2-3)^2}{144N^2} \,,
\\\dpr{\Xp[5]}{\Xp[5]} &= \tfrac{\Nsb1\Nsb4(N^4+24)}{768N^3} \,,
& \dpr{\Xbb{\Xp[5]}b}{\Xbb{\Xp[5]}d} &= \tfrac{\Nsb1\Nsb4(N^2-2)^2}{256N^3} \,,
\\\dpr{\Xp[6]}{\Xp[6]} &= \tfrac{\Nsb1(N^8+6N^6-60N^4+600)}{7680N^4} \,,
& \dpr{\Xbb{\Xp[6]}b}{\Xbb{\Xp[6]}d} &= \tfrac{\Nsb1(4N^8-26N^6+155N^4-450N^2+450)}{9600N^4} \,.
 \end{align*}

We may also consider alternative colour structures, for example writing $\XL,\XR$ as products of the maximum number of traces.
Since the $T_a$ are traceless, this implies $\XL=\XR=\Xmp$, where
 \begin{equation} \label{eq:Col-Xm(p)Defn}
  \Xmp\ap =
  \begin{cases}
   \Tr{T_{a_1}T_{a_2}}\Tr{T_{a_3}T_{a_4}}\dotsm\Tr{T_{a_{p-1}}T_{a_p}}, & \text{$p$ even,}
\\ \Tr{T_{a_1}T_{a_2}T_{a_3}}\Tr{T_{a_4}T_{a_5}}\dotsm\Tr{T_{a_{p-1}}T_{a_p}}, & \text{$p$ odd.}
  \end{cases}
 \end{equation}
Considering leading order terms in the large $N$ limit gives
 \begin{align} \label{eq:LargeNCol-Xm(p)Xm(p)}
  \dpr\Xmp\Xmp &\Neq \left\{
   \begin{gathered}
    \frac{\bfrac{p}2!N^p}{2^{\frac{p}2}p!}, \quad \text{$p$ even,} \qquad\qquad
    \frac{3\bfrac{p-3}2!N^p}{2^{\frac{p+3}2}p!}, \quad \text{$p$ odd,}
   \end{gathered} \right.
\\\intertext{and} \label{eq:LargeNCol-Xm(p)bbXm(p)dd}
  \dpr{\Xbb\Xmp{b}}{\Xbb\Xmp{d}} &\Neq \left\{
   \begin{alignedat}2
    &\bfrac{N^2}{2(p-1)}^2 \dpr{\Xmp[p-2]}{\Xmp[p-2]}, &\quad& \text{$p$ even,}
\\  &\bfrac{(p-3)N^2}{2p(p-1)}^2 \dpr{\Xmp[p-2]}{\Xmp[p-2]}, &\quad& \text{$p$ odd.}
   \end{alignedat} \right.
\\\intertext{Thus we obtain}
  \frac{\dpr{\Xbb\Xmp{b}}{\Xbb\Xmp{d}}}{\dpr\Xmp\Xmp} &\Neq \left\{
   \begin{gathered}
    \frac{N^2}{p-1}, \quad \text{$p$ even,} \qquad\qquad
    \frac{(p-3)N^2}{p(p-1)}, \quad \text{$p$ odd,}
   \end{gathered} \right.
 \end{align}
giving $b$ leading order behaviour of $\orderof1$ for this structure of operator.
Again we may check \eqref{eq:LargeNCol-X(p)X(p)}, \eqref{eq:LargeNCol-X(p)bbX(p)dd} for explicit values of $p$;
we have
 \begin{align*}
  \dpr{\Xmp[4]}{\Xmp[4]} &= \tfrac{\Nsb1(N^2+1)}{48} \,,
& \dpr{\Xbb{\Xmp[4]}b}{\Xbb{\Xmp[4]}d} &= \tfrac{\Nsb1(N^2+1)^2}{144} \,,
\\\dpr{\Xmp[5]}{\Xmp[5]} &= \tfrac{\Nsb1\Nsb4(N^2+5)}{640N} \,,
& \dpr{\Xbb{\Xmp[5]}b}{\Xbb{\Xmp[5]}d} &= \tfrac{\Nsb1\Nsb4(N^2+5)^2}{6400N} \,,
\\\dpr{\Xmp[6]}{\Xmp[6]} &= \tfrac{\Nsb1(N^2+1)(N^2+3)}{960} \,,
& \dpr{\Xbb{\Xmp[6]}b}{\Xbb{\Xmp[6]}d} &= \tfrac{\Nsb1(N^2+1)(N^2+3)^2}{4800} \,,
\\\dpr{\Xmp[7]}{\Xmp[7]} &= \tfrac{\Nsb1\Nsb4(N^2+5)(N^2+7)}{26880N} \,,
& \dpr{\Xbb{\Xmp[7]}b}{\Xbb{\Xmp[7]}d} &= \tfrac{\Nsb1\Nsb4(N^2+5)(N^2+7)^2}{282240N} \,.
 \end{align*}

\chapter{Beyond the Free Theory}
 \label{ch:Interacting}

 \section{Large $N$, Strong Coupling Results}
 \label{se:SUGRA}

As previously noted, dynamical contributions to the four point function of chiral primary operators are contained entirely within the function $\GH$.
Results in the large $N$ limit may be obtained via the AdS/CFT correspondence, expressible in terms of conformal four point integrals
\eqref{eq:D},
from which we may define two-variable functions $\ottf[\Db]\Delta\uv$ of the conformal invariants for arbitrary $\D{i}$,
\eqref{eq:Dbar}.
The properties of these functions are well-known, and they satisfy various symmetries and other relations \cite{rf:0011040},\cite{rf:0212116},
which have been quoted in \secref{se:Dbar}.

\subsection{The \ch22pp Channel} \label{se:22ppSUGRA}

We write the four point function $\fourpt{\spt}\chiral2\chiral2\chiral{p}\chiral{p}$ in terms of $\G*\uvst$, as in \eqref{eq:22pp-4ptG}.
Simplification of the results \cite{rf:linda} from the supergravity approximation in this channel,
as detailed in \secref{se:DbarCalc},
gives
 \begin{multline} \label{eq:22ppSUGRA-G}
  \G*\uvst = 1 + \nN
   \biggl(\sigma u + \tau\frac{u}v + (p-1)\sigma\tau\frac{u^2}v \\- \frac1{(p-2)!}s\uvst u^p \Db{p}{p+2}22\uv \biggr),
 \end{multline}
where the disconnected contribution has been normalised to one.
As before, 
we may write $\G*\uvst$ in
the form \eqref{eq:22pp-GDecomp}, namely
 \begin{equation} \label{eq:22ppCPW-GDecomp}
  \G*\uvst = \fk* + \Gx*{\fhat*}\uvst + s\uvst\GH*\uv,
 \end{equation}
where 
$\fk*$, $\fhat*\xa$ receive contributions only from the free field in the conformal theory.
From \eqref{eq:22ppSUGRA-G} we obtain values
 \begin{equation} \label{eq:22ppSUGRA-kfhat}
  \fk* = 1 + \nN(p+1),
\quad
  \fhat*\xa = \nN\bigl((px-x')-(p-1)(x+x')\alpha\bigr).
 \end{equation}
We write
 \begin{equation} \label{eq:22ppSUGRA-H}
  \GH*\uv = \GH*\ord0\uv + \GH*\dyn\uv, 
 \end{equation}
where $\GH*\ord0$ is the free field component
, which 
\eqref{eq:22ppSUGRA-G} gives as
 \begin{equation} \label{eq:22ppSUGRA-H0}
  \GH*\ord0\uv
  = \nN(p-1)\frac1v
  = \sum_{\subbox{\substack{t=0,1,\dots\\\ell=0,2,\dots}}} A_{t\ell} \, u^t\Gpw*\ell{4+2t+\ell}uv.
 \end{equation}
A more detailed analysis of the free field contribution appears in \secref{se:DbarCalc}.
We observe that \eqref{eq:22ppSUGRA-kfhat}, \eqref{eq:22ppSUGRA-H0} match the 
values for the \ch22pp channel
obtained in \secref{se:FF},
if we take $a=\nN, b=0$ in \eqref{eq:22ppFF-G}.
This agrees with the large $N$ limit for $a$ given in \eqref{eq:LargeNCol-acValues},
although satisfactorily according the values for $b$ is more problematic.
The partial wave coefficients $A_{t\ell}$ are then given by \eqref{eq:22ppFF-CPW-Atl}, with the appropriate values for $a,b$.
The dynamic part of \eqref{eq:22ppSUGRA-G} is contained in $\GH*\dyn$, given by
 \begin{equation} \label{eq:22ppSUGRA-HI}
  \GH*\dyn\uv = -\nN \frac1{(p-2)!} u^p \Db{p}{p+2}22\uv = \AB*\uv + \orderof{u^p,u^p\ln u},
 \end{equation}
which \eqref{eq:DDecomp} enables us to decompose into $\AB*\uv$, free from $\ln u$ terms, plus higher order contributions.
The log terms correspond to the contributions of long multiplets that acquire an anomalous dimension in the interacting conformal theory, 
which thus do not appear at twists less than $2p$.
Using \eqref{eq:Dlogfree} we may write $\AB*$ explicitly as
 \begin{equation} \label{eq:22ppSUGRA-B}
  \begin{aligned}
   \AB*\uv
   & = -\nN \frac1{(p-2)!} u^p \Db{p}{p+2}22\uv\sing \\
   & =  \nN\sum_{m=1}^{p-1} (-u)^m \frac{(p-m-1)!(m!)^2}{(p-2)!\pochhammer{m+2}m} \hyperg{m+2}{m+1}{2m+2}{1-v},
  \end{aligned}
 \end{equation}
which we may then expand in terms of conformal partial waves, 
writing $\AB*\uv = \sum_{t,\ell} \AB*_{t\ell} \, u^t \Gpw*\ell{4+2t+\ell}uv$,
and expressing \eqref{eq:22ppSUGRA-B} as
 \begin{equation}
  \nN\sum_{\subbox{\substack{t=1,2,\dots\\\ell=0,2,\dots}}}
     \left\{\sum_{m=1}^{\subbox[0.86]{\min(t,p-1)}} (-1)^m \frac{m!}{{\pochhammer{p-m}{m-1}}\binom{2m+1}m} c_{m+2\,m+1;t-m,\ell} \right\}
     u^t \Gpw*\ell{4+2t+\ell}uv.
 \end{equation} 
The $c_{ab;j,\ell}$ here are coefficients in the conformal partial wave expansion of
 \begin{equation}
  u^{a-2}\hyperg{a}b{2b}{1-v} = \sum_{\subbox{\substack{j=0,1,\dots\\\ell=0,2,\dots}}} c_{ab;j,\ell} \, u^{j+a-2}\Gpw*\ell{2a+2j+\ell}uv,
 \end{equation}
where the first few cases, given in \cite{rf:0412335}, are
 \begin{align}
  c_{ab;0,\ell} &= r_{ab,\ell},
 \qquad
  c_{ab;1,\ell} = \frac{ab(a-2b)}{2(4b^2-1)}r_{a+1\,b+1,\ell} - r_{a-1\,b-1,\ell+2},
\displaybreak[0]\\\notag
  c_{ab;2,\ell} &= a(a+1)(b+1)\frac{(a-2b-1)(a-2b)}{16(2b+1)^2(2b+3)}r_{a+2\,b+2,\ell} - a\frac{a(a-3b-1)+b}{4(2a-1)(2b+1)}r_{ab,\ell+2},
 \end{align}
with
 \begin{equation}
  (-x)^a\hyperg{a}b{2b}x = \sum_{\subbox{\ell=0,2,\dots}} r_{ab,\ell} \, \gpw*1{\ell+a}x
\iff
  r_{ab,\ell}
 = \frac1{4^\ell(\frac\ell2)!}\frac{\pochhammer{a}\ell\pochhammer{a-b}{\ell/2}}{\pochhammer{a+\frac\ell2-\frac12}{\ell/2}\pochhammer{b+\frac12}{\ell/2}}.
 \end{equation}
Thus the lowest twist contribution from $\AB*\uv$ is at twist 2, when for $\ell$ even,
 \begin{equation} \label{eq:22ppSUGRA-B1}
  \AB*_{1\ell} = -\nN \, \frac13c_{32;0,\ell} = -\nN \frac{(\ell+1)!(\ell+2)!}{(2\ell+3)!}.
 \end{equation}
We find the total twist two contribution from long multiplets by adding the free field component $\hat{A}_{1\ell}$, 
modified to remove contributions from short and semi-short multiplets, which is given by \eqref{eq:22ppFF-CPW-A1}.
We observe that this exactly cancels $\AB*_{1\ell}$ as given by \eqref{eq:22ppSUGRA-B1} for $a$ as in \eqref{eq:LargeNCol-acValues},
in accordance with the expectation that long operators not acquiring anomalous dimensions should decouple from the theory.
%

The contribution at next-highest twist is
 \begin{equation} \label{eq:22ppSUGRA-B2}
  \AB*_{2\ell} = \nN\left(-\frac13c_{32;1,\ell} + \frac1{p-2}\frac15c_{43;0,l}\right) = \nN \bfrac{p}{p-2} \frac{(\ell+2)!(\ell+3)!}{(2\ell+5)!},
 \end{equation}
which added to the free field gives
 \begin{equation}
  A_{2\ell} + \AB*_{2\ell} = \nN \frac{(p+1)(p+2)}{3(p-2)} \frac{(\ell+3)!^2}{(\ell+6)!}.
 \end{equation}

\subsection{The \ch2p2p Channel} \label{se:2p2pSUGRA}

We use \eqref{eq:22pp-2p2p-H} and the symmetries of $\Db{}{}{}{}$ functions \eqref{eq:DSym} to obtain from \eqref{eq:22ppSUGRA-HI}
 \begin{equation} \label{eq:2p2pSUGRA-HI}
  \GH\dyn\uv = -\nN \frac1{(p-2)!} u^2 \Db2{p+2}2p\uv.
 \end{equation}
Again by application of \eqref{eq:DDecomp} we can write \eqref{eq:2p2pSUGRA-HI} as
 \vspace{-0.5ex}%
 \begin{equation} \label{eq:2p2pSUGRA-HAOln}
  \GH\dyn\uv = \AB\uv + \orderof{u^2,u^2\ln u}.
 \end{equation}
with logarithmic terms contributing only at twists of at least ${p+2}$.
Hence the lowest twist long multiplets, which have $\Twist=p$, do not acquire an anomalous dimension,
and would be expected to decouple from the theory.
$\AB\uv$ is proportional to $\Db2{p+2}2p\uv\sing$, which \eqref{eq:Dlogfree} gives in terms of a single hypergeometric function,
 \begin{equation} \label{eq:2p2pSUGRA-BValue}
  \begin{aligned}
   \AB\uv
    &= -\nN \bfrac{p-1}{p+1} u\hyperg2{p+1}{p+2}{1-v},
  \end{aligned}
 \end{equation}
with partial wave expansion
 \begin{equation} \label{eq:2p2pSUGRA-CPW-B}
  \AB\uv = \sum_{t,\ell} \AB_{t\ell} \, u^t \Gpw\ell{4+2\bt+2t+\ell}uv{2\bt}{2\bt},
\quad
  \bt = \tfrac12p-1.
 \end{equation}
To find $\AB_{t\ell}$ we need to expand hypergeometric functions of the form
 \begin{equation} \label{eq:2p2pCPW-HyperExpn}
  (-x)^{\alpha+2}\hyperg{\alpha+1}{n+\alpha}{n+2\alpha}x = \sum_{\ell=0}^\infty \Fr\alpha{n}\ell \, \gbb{1+\bt}{\ell+\alpha+2}x,
\quad
  \bt = \tfrac12n-1,
 \end{equation}
which, as shown in \secref{se:CPW-Hyper}, has solution
 \begin{equation} \label{eq:2p2pCPW-rValue}
  \Fr\alpha{n}\ell = \frac{\pochhammer{n+\alpha}\alpha}{\pochhammer{n+\ell+\alpha}{\ell+\alpha+1}}
  \Bigl[\pochhammer{n+\alpha-1}{\ell+1}+(-1)^\ell\pochhammer{\alpha+1}{\ell+1}\Bigr].
 \end{equation}
At lowest twist, this leads to the result
 \begin{equation} \label{eq:2p2pSUGRA-B1Value}
  \begin{aligned}
   \AB_{1\ell}
   &= -\nN \bfrac{p-1}{p+1} \Fr1{p}\ell
\\ &= -\nN \frac{p-1}{\pochhammer{p+\ell+1}{\ell+2}}\Bigl[\pochhammer{p}{\ell+1}+(-1)^\ell(\ell+2)!\Bigr]
\\ &= -\nN \frac{(\ell+2)!}{\pochhammer{2\bt+\ell+3}{\ell+2}}\left[\binom{2\bt+\ell+2}{\ell+2}+(-1)^\ell(2\bt+1)\right]
  \end{aligned}
 \end{equation}
which 
exactly cancels $\hat{A}_{1,\ell}$, the free field contribution from twist $p$ long multiplets \eqref{eq:2p2pFF-CPW-A1}.
This cancellation is non-trivial, providing a strong consistency check.

 \section{Perturbation Theory Results, \ch2p2p Channel}
 \label{se:Perturb}

 \label{se:PerturbResults}

It is convenient in studying the perturbation expansion for the $\corel{2p2p}$ correlation function to express results in terms of $\HF\uv$, 
as defined in \eqref{eq:2p2p-FDefn}, namely
 \begin{equation} \label{eq:2p2p-FSummary}
  \GH\dyn\uv = \frac{u}v \HF\uv ,
\quad
  \HF(v,u) = \HF\uv.
 \end{equation}
Explicit results for $\HF\uv$ may be computed via the $\N=1$ formulation.
Detailed results appear in \secref{se:N=1}, given to second order in perturbation theory.
With the definition above, \eqref{eq:2323-H1,H2} and \eqref{eq:LargeN-H1,H2} give,\footnote{%
The following definitions hold in the large $N$ limit for general $p$, and exactly for $p=3$ if we take $N^2\goesto \Ns$.}
for $p\ge3$, $\lambda = g^2N/\fps$
 \begin{gather} \label{eq:2p2p-F1Value}
  \HF\ord1\uv \Neq \nN \, \bigl(-\tfrac12\lambda \fB\uv\bigr)
\\ \label{eq:2p2p-F2Value}
  \HF\ord2\uv \Neq \nN \, \frac{\lambda^2\!}4 \bigl(\I\uv+\J\uv\bigr),
 \end{gather}
where we define
 \begin{equation} \label{eq:2p2p-IJValues}
  \begin{aligned}
   \I\uv &= \tfrac{v}4\fB\uv^2 + \tfrac1v\fP\left(\tfrac1v,\tfrac{u}v\right) + \tfrac12\fP\left(u,v\right),
\\ \J\uv &= \tfrac{u}4\fB\uv^2 + \tfrac1u\fP\left(\tfrac1u,\tfrac{v}u\right).
  \end{aligned}
 \end{equation}
The functions $\loopint{L}$ appearing in \eqref{eq:2p2p-F1Value}, \eqref{eq:2p2p-IJValues} are conformal loop integrals, 
properties of which are given in \secref{se:Phi}.

%
We now consider a perturbative analysis of the partial wave expansion.
It has been established that long multiplets contributing to the $\corel{2p2p}$ correlator 
are necessarily descended from superconformal primaries which are $SU(4)_R$ singlets, 
and that their contribution is given by $A\uv = 2\gpl0\GH\uv$.
Initially we examine contributions from lowest twist long multiplets, whose primaries have $\Twist[0]=p=2\bt+2$ in the free theory,
for which it suffices to consider only leading terms as $u\tendsto0$. 
We write
 \begin{equation} \label{eq:Prtb-A1Defn}
  2\gpl0 \frac{u}v \HF\uv = A\uv\pert \equiv A\ord1\uv + \orderof{u^2};
 \end{equation}
it is shown in \cite
{rf:0412335} that there are no terms in $\tfrac1u\fP\left(\tfrac1u,\tfrac{v}u\right)$ proportional to $\ln u$ for small $u$,
so contributions from $\J$ to $A\ord1$ may be neglected, 
and from \eqref{eq:2p2p-F1Value}, \eqref{eq:2p2p-F2Value} we have
 \begin{align} \label{eq:Prtb-A1Value}
  A\ord1\uv &= 2\gpl0 \nN \, \frac{u}v \left( -\frac\lambda2\fB\uv + \frac{\lambda^2}4\I\uv + \orderof{\lambda^3} \right),
\\\intertext{which we write in the form of \eqref{eq:eps-FExpn},} \label{eq:Prtb-A-frs}
  A\ord1\uv &\goeslike 2\gpl0 \nN \, u \sum_{r=1}^\infty \lambda^r \sum_{s=0}^r \ln^sx\xb \, f\ord{rs}(x)
\quad\text{as}\quad
  \xb\tendsto0.
 \end{align}
From \eqref{eq:Prtb-A1Value}, \eqref{eq:2p2p-IJValues}, and with the results \eqref{eq:Phi12Values} for $\fB,\fP$ we obtain
 \begin{align} \label{eq:Prtb-fValues}
  f\ord{11}(x) &= -\frac12\frac1{x(1-x)}\ln(1-x), \qquad f\ord{10}(x) = -\frac1{x(1-x)}\Li2(x),
\Bob/
  f\ord{22}(x) &= \frac1{16}\frac1{x(1-x)}\Bigl[\left(\tfrac1x-\tfrac12\right)\ln^2(x) + 3\Li2(x)\Bigr]
,
\Bob/
  f\ord{21}(x) &= \frac1{16}\frac1{x(1-x)}\Bigl[
   \begin{aligned}[t]-&\ln^3(1-x) + 4\left(\tfrac1x-\tfrac32\right)\ln(1-x)\Li2(x)
 - 6\left(\vphantom{\tfrac11}2\Li3(x)-\Li3(x')\right)\Bigr].\end{aligned}
 \end{align}

If we assume there is one operator with free field twist $2\bt+2$ contributing at each spin $\ell$, then from \eqref{eq:2p2pCPW-HExpn}, 
using $\Gpw\ell{\ell+2t}0v{2\bt}{2\bt}=\gbb{1+t}\ell{1-v}$, which follows from the definition \eqref{eq:CPW-Defn},
we must have as $\xb\tendsto0$
 \begin{equation} \label{eq:Prtb-A-eta}
  \hat{A}\uv+A\uv\pert \goeslike 2\gpl0 \nN \, u\sum_\ell a\ord\ell \, u^{\frac12\ad} \gbb{4+\bt+\frac12\ad}\ell{x} ,
 \end{equation}
where 
$\hat{A}\uv$ is the free field theory result, modified to remove short and semi-short contributions, 
and
$\ad$ is the anomalous dimension at each $\ell$, 
$\Delta\goesto2\bt+2+\ell+\ad$.
We write expansions
 \begin{equation} \label{eq:Prtb-eta,aExpn}
  \ad = \lambda\ado1+\lambda^2\ado2+\dotsb \,,
\qquad
  a\ord\ell = a\ord{\ell,0}\bigl(1+\lambda b\ord{\ell,1}+\dotsb\bigr) ,
 \end{equation}
where by considering the free field limit, we require
 \begin{equation} \label{eq:Prtb-aMatch}
  2\gpl0 \nN \, a\ord{\ell,0} = \hat{A}_{1\ell}, 
 \end{equation}
with $\hat{A}_{1\ell}$ as given in \eqref{eq:2p2pFF-CPW-A1}.
With the appropriate large $N$ value of $a$ from \eqref{eq:LargeNCol-acValues} we may obtain
 \begin{equation} \label{eq:Prtb-aValue}
  a\ord{\ell,0} \Neq \tfrac{(\ell+2)!}{\pochhammer{2\bt+\ell+3}{\ell+2}}\left[\tbinom{2\bt+\ell+2}{\ell+2}+(-1)^\ell(2\bt+1)\right] .
 \end{equation}
Expanding the summand in \eqref{eq:Prtb-A-eta} we obtain
 \begin{multline} \label{eq:Prtb-augExpn}
  a\ord\ell \cdot u^{\frac12\ad} \gbb{t+\frac12\ad}\ell{x}
   \begin{aligned}[t]
    ={}& a\ord{\ell,0} \biggl[ 1 + \lambda\Bigl(b\ord{\ell,1} + \tfrac12\ado1\tddt + \tfrac12\ado1\ln u\Bigr)
\\  &+ \lambda^2\Bigl(\dotsb + \bigl(\tfrac12b\ord{\ell,1}\ado1 + \tfrac14\ado1^2\tddt + \tfrac12\ado2\bigr)\ln u + \tfrac18\ado1^2\ln^2u\Bigr)
   \end{aligned}
\\ + \orderof{\lambda^3} \biggr]
   \gbb{t}\ell{x}.
 \end{multline}
Determining values for the anomalous dimensions thus requires matching orders of $\lambda^r\ln^su$ in \eqref{eq:Prtb-A-frs}, \eqref{eq:Prtb-augExpn};
hence we find, using \eqref{eq:CPW-gRefl},
 \begin{align}
  \sum_{\ell\ge2}a\ord{\ell-2,0} \, \tfrac12\ado[\ell-2]1 \, \gbb{1+\bt}{\ell+1}x &= -x^3 f\ord{11}(x),	\label{eq:Prtb-x3f11}
\\\sum_{\ell\ge2}a\ord{\ell-2,0} \, \tfrac18\ado[\ell-2]1^2 \, \gbb{1+\bt}{\ell+1}x &= -x^3 f\ord{22}(x).	\label{eq:Prtb-x3f22}
 \end{align}
%
%
The problem then reduces to finding the single variable partial wave expansions for the expressions on the right of
\eqref{eq:Prtb-x3f11}, \eqref{eq:Prtb-x3f22}, which we write as
 \begin{align}
  & \label{eq:CPW-f11} \begin{aligned}[b]
   x^3 f\ord{11}
    &= \tfrac12 \bigl[ x\,\ln(1-x) - x'\,\ln(1-x') \bigr]
\\  &= -\tfrac12\sum_{\ell=0}^\infty \tfrac{\ell!}{\pochhammer{2\bt+\ell+1}\ell} \left[\tbinom{2\bt+\ell}\ell\lna1-\lnap1\right] \gbb{1+\bt}{\ell+1}x ,
   \end{aligned}
\\
  & \label{eq:CPW-f22} \begin{aligned}[b]
   x^3 f\ord{22}
    &= -\tfrac1{16} \bigl[ 3(x\Li2(x)-x'\Li2(x')) - (\tfrac12x+x')\ln^2(1-x) \bigr]
\\  &= -\tfrac1{16}\sum_{\ell=0}^\infty \tfrac{\ell!}{\pochhammer{2\bt+\ell+1}\ell}
        \left[\tbinom{2\bt+\ell}\ell(3\lib1+\lnsa1)-(3\libp1-2\lnsap1)\right] \gbb{1+\bt}{\ell+1}x .
   \end{aligned}
 \end{align}
In terms of expansion coefficients in \eqref{eq:CPW-f11}, \eqref{eq:CPW-f22}, we thus have
 \begin{gather} \label{eq:Prtb-eta-l1}
  \ado[\ell-2]1 = \frac{\tbinom{2\bt+\ell}\ell\lna1-\lnap1}{\tbinom{2\bt+\ell}\ell+(-1)^\ell(2\bt+1)} \,,
\\ \label{eq:Prtb-eta-l1^2}
  \ado[\ell-2]1^2 = \frac{\tbinom{2\bt+\ell}\ell(\tfrac32\lib1+\tfrac12\lnsa1)-(\tfrac32\libp1-\lnsap1)}{\tbinom{2\bt+\ell}\ell+(-1)^\ell(2\bt+1)} \,.
 \end{gather}
Results for $\lnaq,\lnsaq,\libq$ and their primed counterparts appear in \secref{se:CPW-ln,Li2}.
It is notable that, as a consequence of \eqref{eq:CPW-gx'}, there is no trivial correspondence between terms in the expansions of $f(x)$ and $f(x')$,
in contrast to the case of the $\corel{pppp}$ four point function, which corresponds to the case $\bt=\bt*=0$.
Thus two sets of coefficients need to be independently calculated.

It appears that it is hard to determine a simple form for $\lnapq$, $\libpq$, etc., at least in the case of general $\bt$.
However, even without readily calculable expressions, it seems unlikely that \eqref{eq:Prtb-eta-l1}, \eqref{eq:Prtb-eta-l1^2} 
should in general give compatible values for $\ado1$.
This indicates the presence of more than one operator at spin $\ell$, leading to mixing effects.
In this case, we have rather results for $\expect{\ado1}, \expect{\ado1^2}$.

It is possible to evaluate the required coefficients at given (fixed) values of $\ell$,
e.g. for $\ell=2,3,4,5$ \eqref{eq:Prtb-eta-l1} gives
 \begin{equation}
  \begin{aligned}
   \expect{\ado[0]1} &= 1+\frac2{2\bt+1} \,, \qquad
   \expect{\ado[1]1}  = \frac32 + \frac92\frac1{2\bt+1} \,,
\\ \expect{\ado[2]1} &= \frac{11}6 + \frac{10}3\frac1{2\bt+1} + \frac{2\bt-4}{2\bt^2+3\bt+4} \,,
\\ \expect{\ado[3]1} &= \frac{25}{12} + \frac{125}{24}\frac1{2\bt+1} + \frac{25}{48}\frac{2\bt-10}{2\bt^2+7\bt+11} \,.
  \end{aligned}
 \end{equation}
Similar results may be found for \eqref{eq:Prtb-eta-l1^2}.
They appear to be consistent with the presence of a single operator, 
i.e. satisfy $\expect{\ado1}^2=\expect{\ado1^2}$, only in the following cases:
 \begin{itemize}
  \item $\bt=\frac12$ ($p=3$), with $\ell=0,1,2,3$ {or} $5$;
  \item $\bt=1$ ($p=4$), for $\ell=1$ only%
.
 \end{itemize}
Values for $\ado1$ in these cases are given in \tabref{tb:eta1Values}.
\begin{table}
 \newlength{\etaequalsfrac} \settowidth{\etaequalsfrac}{$\ado[1]1 = \frac{15}4$}%
 \centering
 \caption{Values of $\ado1$ for single operators} \label{tb:eta1Values}
 \begin{tabular}{|c|cccccc|}
  \hline
  & $\ell = 0$ \OTT{ & $\ell = #1$}{5}
\\\hline
  $\bt = \frac12$
  & $\ado[0]1 = 2$ & $\ado[1]1 = \frac{15}4$ & $\ado[2]1 = 3$ & $\ado[3]1 = \frac{35}8$ & & $\ado[5]1 = \frac{581}{120}$
  \bigstrut
\\$\bt = 1$
  & & \makebox[\etaequalsfrac][l]{$\ado[1]1 = 3$} & & & &
  \bigstrut[b]
\\\hline
 \end{tabular}
\end{table}

In the case that $\bt=\frac12$ --- i.e. the $\corel{2323}$ correlator ---
we are able to find explicit solutions to \eqref{eq:Prtb-eta-l1}, \eqref{eq:Prtb-eta-l1^2} for general $\ell$, giving
%
 \begin{equation} \label{eq:2323-eta-l1}
  \expect{\ado[\ell-2]1} =
  \begin{cases}
   \frac1{\ell-1} \left[2\ell\,\h(\ell+1)-(\ell+2)\right], & \text{$\ell$ odd,}
\\ \frac1{\ell+3} \left[2(\ell+2)\h(\ell)-\ell\right], & \text{$\ell$ even,}
  \end{cases}
 \end{equation}
 \begin{equation} \label{eq:2323-eta-l1^2}
  \expect{\ado[\ell-2]1^2} =
  \begin{cases}
   \frac1{\ell-1} \left[(4\ell+3)\h(\ell)^2 - (4\ell+\frac3{\ell+1})\h(\ell) + (\ell+3)\hb*2(\ell) + 2\ell + \frac1{\ell+1}\right], & \text{$\ell$ odd,}
\\ \frac1{\ell+3} \left[(4\ell+5)\h(\ell)^2 - (4\ell+\frac1{\ell+1})\h(\ell) + (\ell-1)\hb*2(\ell) + 2\ell\right], & \text{$\ell$ even,}
  \end{cases}
 \end{equation}
where the harmonic number $\h(n)$ and its generalisations $\h*r(n), \hb*r(n)$ are given by
 \begin{equation} \label{eq:hDefn}
  \h(n) = \sum_{k=1}^n \frac1k \,,
\qquad
  \h*r(n) = \sum_{k=1}^n \frac1{k^r} \,,
\qquad
  \hb*r(n) = \sum_{k=1}^n \frac{(-1)^k}{k^r} \,.
 \end{equation}
From \eqref{eq:2323-eta-l1}, \eqref{eq:2323-eta-l1^2} it is easy to derive an expression for $\expect{\ado1^2} - \expect{\ado1}^2$ 
in the $\bt=\frac12$ case.
For both odd and even $\ell$, we find that this tends to $1-\pi^2/12$ as $\ell\tendsto\infty$.

It is also possible to find \eqref{eq:Prtb-eta-l1}, \eqref{eq:Prtb-eta-l1^2} in the limit $\bt\tendsto\infty$.
Using the results of \secref{se:Prtb-Largeb}, we obtain
 \begin{align} \label{eq:Largeb-eta-l1}
  \expect{\ado1} &= \h(\ell+1),
\\\expect{\ado1^2} &= \h(\ell+1)\h(\ell+2) + \tfrac12\h*2(\ell+1). \label{eq:Largeb-eta-l1^2}
 \end{align}
In this case we have $\expect{\ado1^2} - \expect{\ado1}^2 = \frac1{\ell+2}\h(\ell+1)+\frac12\h*2(\ell+1)$,
which tends to ${\pi^2}/{12}$ in the large $\ell$ limit.

Returning to the expansions \eqref{eq:Prtb-A-frs}, \eqref{eq:Prtb-augExpn}, we may also write
 \begin{align}
  -x^3 f\ord{10}(x) &=	\label{eq:Prtb-x3f10} 
   \sum_{\ell\ge2} a\ord{\ell-2,0} \left(b\ord{\ell-2,1} \, \gbb{1+\bt}{\ell+1}x + \tfrac12\ado[\ell-2]1 \, \gbb'{1+\bt}{\ell+1}x\right),
\\-x^3 f\ord{21}(x) &=	\label{eq:Prtb-x3f21}
   \sum_{\ell\ge2} \tfrac12 a\ord{\ell-2,0}
    \left(\bigl(b\ord{\ell-2,1}\ado[\ell-2]1 + \ado[\ell-2]2\bigr)\gbb{1+\bt}{\ell+1}x + \tfrac12\ado[\ell-2]1^2 \, \gbb'{1+\bt}{\ell+1}x\right).
 \end{align}
We now have contributions from derivatives of the partial wave,
 \begin{equation} \label{eq:CPW-g'Defn}
  \gpw[']{\bt}{\bt*}t\ell{x} = \frac\partial{\partial t}\,\gbb*{t}\ell{x}.
 \end{equation}
Extracting values from these expansions is messy, but an attempt is made in \secref{se:Prtb-f10f21}.
We find that we may obtain from \eqref{eq:Prtb-x3f10}
 \begin{equation} \label{eq:Prtb-b0121Values}
  \begin{aligned}
   \expect{b\ord{0,1}} &= -\frac{2\bt+3}{2\bt+1}, \qquad
   \expect{b\ord{1,1}}  = -\frac54 - \frac35\frac1{2\bt+6} - \frac{93}{20}\frac1{2\bt+1},
\\ \expect{b\ord{2,1}} &= -\frac1{18}\frac{(2\bt+5)(196\bt^4+1932\bt^3+5641\bt^2+5232\bt+3280)}{(2\bt+1)(2\bt+7)(2\bt+8)(2\bt^2+3\bt+4)}.
  \end{aligned}
 \end{equation}
We note that in the limit $\bt\tendsto0$ we recover the results of \cite{rf:0412335} 
for the corrections to the coupling, $b\ord{0,1}=-3, b\ord{2,1}=-\frac{1025}{252}$.
Similarly, from \eqref{eq:Prtb-x3f21}, we obtain
 \begin{align} \label{eq:Prtb-eta-0122Values}
  \expect{\ado[0]2}
   &= -\frac74 - \frac3{2(2\bt+1)} + \frac4{(2\bt+1)^2}, \Bob/
  \expect{\ado[1]2}
   &= -\frac{69}{32} - \frac{39}{8\bt} + \frac{5247}{800(2\bt+1)} + \frac{837}{40(2\bt+1)^2} - \frac{21}{25(2\bt+6)},
\Bob/
  \expect{\ado[2]2}
   &= -\frac{2077}{864} - \frac{75991}{10584(2\bt+1)} + \frac{2050}{189(2\bt+1)^2} - \frac8{9(2\bt+7)} - \frac{405}{392(2\bt+8)}
\Bob~ &\hphantom{\displaystyle{} = -\frac{2077}{864}} + \frac{1396\bt+3049}{432(2\bt^2+3\bt+4)} - \frac{27\bt}{(2\bt^2+3\bt+4)^2}.
 \end{align}

\subsection{Restrictions from the \ch22pp Channel} \label{se:2p2p-22pp}

We observe that in the \ch22pp conformal expansion \eqref{eq:22ppCPW-GExpn}, 
the only $p$-dependence is through the coefficients $a_{nm,t\ell}$;
the parameters of the Legendre polynomials and conformal partial wave functions are independent of $p$.

Hence, the spectrum of operators able to contribute in this channel is not dependent on $p$, 
and any coefficients in $\G*\uvst$ related to measurable properties of these operators (such as their scale dimension)
must be similarly independent.
In particular, they must match the corresponding terms in the $\corel{2222}$ correlator,
which as the simplest four point function of identical chiral primary operators has been well studied,
and for which many results are known.
Using the prescription \eqref{eq:swap23}, we may translate such restrictions to the \ch22pp correlation function;
thus certain constraints are placed upon results in this channel even without further calculation.

We proceed by defining two alternative spatial conformal invariants, $s$ and $t$,
 \begin{equation} \label{eq:STDefn}
  s = \frac1u,
\qquad
  t = \frac{v}u.
 \end{equation}
With this definition, \eqref{eq:22pp-2p2p-H} may be written in the weak coupling limit as
 \begin{equation} \label{eq:22pp-2p2p-Hst}
  \GH*\ord{n}\st = \GH\ord{n}\left(u\st,v\st\right) = \GH\ord{n}(\tfrac1s,\tfrac{t}s),
\quad
  n > 0,
 \end{equation}
where $\GH$ has perturbation expansion
 \begin{equation} \label{eq:2p2p-HPerturb}
  \GH\uv = \GH\ord0\uv + g^2 \GH\ord1\uv + g^4 \GH\ord2\uv + \dotsb 
 \end{equation}
in the coupling $g^2$, and similarly for $\GH*$.
It follows that
 \begin{equation} \label{eq:22ppH-2p2pF}
  \GH*\dyn\st = \frac1t\F(\tfrac1s,\tfrac{t}s),
 \end{equation}
where $\F$ is as defined in \eqref{eq:2p2p-FDefn}.
With this definition, we construct a generalised form for $\F\uv=\F(v,u)$ satisfying the requirements of crossing symmetry \eqref{eq:2p2p-FCrossSym}.
We do this in terms of loop integrals $\loopint{L}$, which obey the identities \eqref{eq:PhiRefl}%
\eachlabelcase{{eq:Phi1Refl}{, \eqref{eq:Phi1Refl}}{}}.
Letting
 \begin{equation}
  \F\ord2\uv = \nN \, \frac{\lambda^2\!}4 \bigl(\I\uv+\J\uv\bigr),
 \end{equation}
we write
 \begin{equation} \label{eq:2p2p-IJ}
  \begin{aligned}
   \I\uv &= \tfrac14(A+Bv)\fB\uv^2 + D\tfrac1v\fP\left(\tfrac1v,\tfrac{u}v\right) + C\fP\left(u,v\right),
\\ \J\uv &= B\tfrac{u}4\fB\uv^2 + D\tfrac1u\fP\left(\tfrac1u,\tfrac{v}u\right),
  \end{aligned}
 \end{equation}
and from \eqref{eq:22ppH-2p2pF}, \eqref{eq:PhiRefl},\eachlabelcase{{eq:Phi1Refl}{ \eqref{eq:Phi1Refl},}{}} obtain
 \begin{equation}
  \GH*\ord2\st = \nN \, \frac{\lambda^2\!}4 \bigl(\widetilde\I\st+\widetilde\J\st\bigr)
 \end{equation}
where
 \begin{equation} \label{eq:22pp-IJ}
  \begin{aligned}
   \widetilde\I\st &= B\tfrac{1+t}4\fB\st^2 + D\left(\tfrac1t\fP\left(\tfrac1t,\tfrac{s}t\right) + \fP\st\right),
\\ \widetilde\J\st &= A\tfrac{s}4\fB\st^2 + C\tfrac1s\fP\left(\tfrac1s,\tfrac{t}s\right).
  \end{aligned}
 \end{equation}

We previously observed that the terms in $\I,\widetilde\I$ are those that contribute to the expansion which determines anomalous dimensions.
By the arguments above, then, the coefficients in $\widetilde\I$ must be chosen such that they agree with the (known) $\corel{2222}$ correlator.
Thus (up to normalisation) we fix $B=D=1$ --- although $A,C$ remain undetermined by such considerations.

\subsection{
 $\HF\uv$ from results in $\N=1$ formulation} \label{se:N=1}

\newcommand{\OL}{\Op_L^{11;3}}
\newcommand{\OR}{{\Op_R^\dagger}^{22;3}}
\newcommand{\CL}{\CC_{11}^\dagger}
\newcommand{\CR}{\CC_{22}}

\newcommand{\nb}{\bar{n}}

In \cite{rf:0504061}, D'Alessandro and Genovese express the four-point function
 \begin{equation} \label{eq:N=1-4ptGp}
  \corel{\OL(x_1)\,\CL(x_2)\,\OR(x_3)\,\CR(x_4)}
=
  G\dimn{p}\xxxx
 \end{equation}
as a perturbation expansion,
 \vspace{-0.8ex}%
 \begin{multline} \label{eq:N=1-GpPerturb}
  G\dimn{p}\xxxx = G\dimn{p}\ord0\xxxx + g^2\,G\dimn{p}\ord1\xxxx \\ + g^4\,G\dimn{p}\ord2\xxxx + \dots \,.
 \end{multline}
The operators appearing in \eqref{eq:N=1-4ptGp} are particular choices of flavour representatives (in the $\N=1$ formulation)
for \BPS2 chiral primaries; two belonging to the $\dynk020$ representation of $SU(4)_R$,
 \begin{align} \label{eq:N=1-C11C22Defn}
  \CL &= \tr(\cscal_1^\dagger \cscal_1^\dagger),
 &
  \CR &= \tr(\cscal_2 \cscal_2),
 \end{align}
and two in the $\dynk0p0$ representation,
 \vspace{-0.5ex}%
 \begin{align} \label{eq:N=1-O113O223Defn}
  \OL &= \XL_{abc_1\dots c_{p-2}}\cscal_1^a\cscal_1^b\cscal_3^{c_1}\dots\cscal_3^{c_{p-2}},
 \\
  \OR &= \XR_{abc_1\dots c_{p-2}}{\cscal_2^\dagger}^a{\cscal_2^\dagger}^b{\cscal_3^\dagger}^{c_1}\dots{\cscal_3^\dagger}^{c_{p-2}}.
 \end{align}
The six real scalars $\chiral_i$ used previously have been combined into three complex fields,
$\cscal_I = \sq(\chiral_I + i\chiral_{I+3})$ and $\cscal_I^\dagger = \sq(\chiral_I - i\chiral_{I+3})$, for $I=1,2,3$.
The tensors $\XL, \XR$, as before, describe the colour structure of the operators.

We will compare the results obtained for $G\dimn{p}\xxxx$ with our calculations of the correlator from \secref{se:2p}.
From \eqref{eq:2p2p-4ptG}, we have\footnote{%
Note that the correlator in \eqref{eq:N=1-4ptGp} has $\swap12, \swap34$ with respect to \eqref{eq:2p2p-4ptG};
this leaves $\uvst*$ unchanged.}
 \vspace{-1.2ex}%
 \begin{multline} \label{eq:p2p2-4ptG}
  \fourpt\hspt\chiral{p}\chiral2\chiral{p}\chiral2 \\
   = \frac{(\dpt12\dpt34)^2(\dpt13)^{p-2}}{(\xx12\xx34)^2(\xx13)^{p-2}} \G\uvst.
 \end{multline}
Making a choice of flavour representatives corresponds to fixing particular values for the $t_i$, and hence for $\sigma,\tau$.
If we take a basis for the $t_i$, $\{n_I,\nb_I\}$, given by
 \vspace{-0.4ex}%
 \begin{equation} \label{eq:N=1-n123Defn}
  \begin{aligned} n_1 &= \sq(1,0,0,i,0,0), \\ n_2 &= \sq(0,1,0,0,i,0), \\ n_3 &= \sq(0,0,1,0,0,i), \end{aligned}
 \qquad
  \begin{aligned} \nb_1 &= \sq(1,0,0,-i,0,0), \\ \nb_2 &= \sq(0,1,0,0,-i,0), \\ \nb_3 &= \sq(0,0,1,0,0,-i), \end{aligned}
 \end{equation}
satisfying
 \begin{equation}
  \dpr{n_I}{\nb_J} = \delta_{IJ},
  \qquad
  \dpr{n_I}{n_J} = \dpr{\nb_I}{\nb_J} = 0,
 \end{equation}
then our flavour choices \eqref{eq:N=1-C11C22Defn}, \eqref{eq:N=1-O113O223Defn}, are equivalent to taking
 \begin{equation} \label{eq:N=1-ttttDefn}
  \begin{alignedat}{2}
   t_1 &= n_1+n_3, \qquad \qquad
&  t_2 &= \nb_1,
\\ t_3 &= \nb_2+\nb_3,
&  t_4 &= n_2.
  \end{alignedat}
 \end{equation}
Thus
 \begin{equation} \label{eq:N=1-st}
  \begin{gathered}
   \dpt12 = \dpt13 = \dpt34 = 1, \quad
   \dpt14 = \dpt23 = \dpt24 = 0,
  \\
   \implies\: \sigma = \tau = 0 \;
   \implies\: s\uvst = v,
  \end{gathered}
 \end{equation}
and \eqref{eq:p2p2-4ptG} reduces to give, once normalisation (below) is taken into account,
 \begin{equation} \label{eq:N=1-matched}
  G\dimn{p}\xxxx
 =
  \frac1{(\xx12\xx34)^2(\xx13)^{p-2}} \G\uvoo.
 \end{equation}
Following \eqref{eq:GFree+Dyn}, \eqref{eq:HFree+Dyn}, we may write $\G\uvst$ as a perturbation expansion in $g^2$,
 \begin{align} \label{eq:2p2p-GPerturb}
  \G\uvst &= \fk + \Gx\fhat\uvst + s\uvst\GH\uv, \notag\\
  \GH\uv  &= \GH\ord0\uv + g^2 \GH\ord1\uv + g^4 \GH\ord2\uv + \dots \,,
 \end{align}
and we can match the left- and right-hand sides of \eqref{eq:N=1-matched} at each order, giving
 \begin{gather} \label{eq:N=1-match0}
  (\xx12\xx34)^2(\xx13)^{p-2} G\dimn{p}\ord0 = \G\ord0\uvoo ;
\\ \label{eq:N=1-matchn}
  (\xx12\xx34)^2(\xx13)^{p-2} G\dimn{p}\ord{n} = s\uvoo\GH\ord{n}\uv = v\,\GH\ord{n}\uv,
\quad n>0.
 \end{gather}
To match normalisation conventions, we will compare the free field results, for which we have obtained an exact value.
\eqref{eq:2p2pFF-G}, \eqref{eq:Col-abcExplicit} give us
 \begin{equation} \label{eq:2p2pFF-Guv00}
   \G\ord0\uvoo = b = \frac{p(p-1)}{2\Nsb1} \bfrac{\dpr\XLb\XRd}\XLXR ;
 \end{equation}
whilst \cite[(24)]{rf:0504061} gives
 \begin{equation} \label{eq:N=1-Gp0}
  G\dimn{p}\ord0\xxxx = \frac{(p-2)!}{\fps^{p+2}} \frac{Y_{ab|cd}\,\delta^{ab}\delta^{cd}}{(\xx13)^{p-2}(\xx12\xx34)^2},
 \end{equation}
where
 \begin{gather} \label{eq:N=1-YDefn}
  Y_{ab|cd} = \XL_{abc_1\dots c_{p-2}}\,\XR_{cdc_1\dots c_{p-2}}
\\ \implies
  Y_{ab|cd}\,\delta^{ab}\delta^{cd} = \dpr\XLb\XRd \,.
 \end{gather}
Thus in order for \eqref{eq:N=1-match0} to hold, we are required to divide the results of \cite{rf:0504061} by a normalisation factor of
 \begin{equation} \label{eq:N=1-normfactor}
  \frac{2\Nsb1}{p(p-1)}\,\frac{(p-2)!}{\fps^{p+2}}\,\XLXR.
 \end{equation}

\subsubsection{The $p=3$ case} \label{se:N=1p=3}

For $p=3$, D'Alessandro and Genovese make the choice of representatives
 \begin{equation}
  \begin{aligned}
   \OL &= \tfrac14d_{abc}\cscal_1^a\cscal_1^b\cscal_3^c, \quad
  &
   \OR &= \tfrac14d_{abc}{\cscal_2^\dagger}^a{\cscal_2^\dagger}^b{\cscal_3^\dagger}^c,
  \end{aligned}
 \end{equation}
where $d_{abc}$ is the totally symmetric tensor arising from the anticommutators of $SU(N)
$ generators,
 \begin{equation} \label{eq:SU(N)-Anticomm}
  \acomm{T_a}{T_b} = \frac1N\delta_{ab}\,\ident + d_{abc}T_c.
 \end{equation}
We see that this is equivalent to the definition \eqref{eq:p=3Col-XL,XR}, since
 \begin{equation*}
   \tfrac12\tr(T_a\acomm{T_b}{T_c})
 = \tfrac12\Tr{\tfrac1N\delta_{ab}T_a + d_{bcd}T_aT_d}
 = \tfrac12d_{bcd}\cdot\tfrac12\delta_{ad}
 = \tfrac14d_{abc}.
 \end{equation*}
From \cite[(47--50)]{rf:0504061}, taking into account the relevant normalisation factor \eqref{eq:N=1-normfactor}, we obtain
 \begin{equation} \label{eq:p=3N=1-G012}
  \begin{aligned}[b]
   G\dimn3\ord0\xxxx &= 0, \\
   G\dimn3\ord1\xxxx &= - \frac3\Ns \frac{N}\fps \frac{u}{\xx13(\xx12\xx34)^2} \fB\uv, \\
   G\dimn3\ord2\xxxx &= \begin{aligned}[t]
    &\frac3\Ns \frac{N^2}{2\fps^2} \frac1{\xx13(\xx12\xx34)^2}
     \biggl\{ \frac{u}4\fB\uv^2(u+v) \\
    &+ \Bigl[ \fP\left(\tfrac1u,\tfrac{v}u\right) + \frac{u}v\fP\left(\tfrac1v,\tfrac{u}v\right) + \frac{u}2\fP\left(u,v\right) \Bigr] \biggr\}.
   \end{aligned}
  \end{aligned}
 \end{equation}
\eqref{eq:N=1-match0} then implies
 \begin{equation}
  \G\ord0\uvoo = 0;
 \end{equation}
which matches what we would expect from \eqref{eq:2p2pFF-Guv00}, \eqref{eq:p=3Col-bValue}.
Now taking
 \begin{equation} \label{eq:lambdaDefn}
  \lambda = \frac{g^2N}{4\pi^2} \,,
 \end{equation}
we find that \eqref{eq:N=1-matchn}, \eqref{eq:p=3N=1-G012} gives
\begin{subequations} \label{eq:2323-H1,H2}
 \begin{align}
     g^2\,\GH\ord1\uv &= \frac{-3\lambda}\Ns \, \frac{u}v \, \fB\uv, \\
     g^4\,\GH\ord2\uv &= \begin{aligned}[t] & \frac{3\lambda^2}{2\Nsb1} \, \frac{u}v \, \biggl\{ \frac{u+v}4\fB\uv^2 \\
&\qquad + \Bigl[ \tfrac1u\fP\left(\tfrac1u,\tfrac{v}u\right) + \tfrac1v\fP\left(\tfrac1v,\tfrac{u}v\right) + \tfrac12\fP\left(u,v\right) \Bigr] \biggr\}.
   \end{aligned}
 \end{align}
\end{subequations}

\subsubsection{The Large $N$ Limit}

We may also use the results of \cite[(24--26)]{rf:0504061} and \secref{se:Col-LargeN}
to find expressions for the first and second order contributions to $\GH\dyn$ at large $N$.
We will require the large $N$ limit of the colour contractions \cite[(23)]{rf:0504061}, which are given by
 \begin{equation} \label{eq:LargeN-YH1,YH2,ZL}
  \begin{gathered}
  \dpr{Y}{H_1} \Neq \frac{N}{p-1}\,\dpr{X}X \,,
\qquad\qquad
  \dpr{Y}{H_2} \Neq \frac{N^2}{2(p-1)}\,\dpr{X}X \,,
\\
  \dpr{Z}{L} \Neq \frac{-N^2}{2(p-1)(p-2)}\,\dpr{X}X \,.
  \end{gathered}
 \end{equation}
Thus we find, at leading order in $\frac1N$,
\begin{subequations} \label{eq:LargeN-H1,H2}
 \begin{align}
    g^2\,\GH\ord1\uv &= -\frac{\lambda p}{N^2} \, \frac{u}v \, \fB\uv, \\
    g^4\,\GH\ord2\uv &= \begin{aligned}[t] & \frac{\lambda^2p}{2N^2} \, \frac{u}v \, \biggl\{ \frac{u+v}4\fB\uv^2 \\
&\qquad + \Bigl[ \tfrac1u\fP\left(\tfrac1u,\tfrac{v}u\right) + \tfrac1v\fP\left(\tfrac1v,\tfrac{u}v\right) + \tfrac12\fP\left(u,v\right) \Bigr] \biggr\}.
   \end{aligned}
 \end{align}
\end{subequations}

\subsubsection{Results for the \ch22pp Channel} \label{se:22ppN=1}

We may translate the results obtained so far to the \ch22pp channel, making use of the relation \eqref{eq:22pp-2p2p-Hst}.
We have previously found lower-order contributions to $\GH\dyn$ in the \ch2p2p channel for the case $p=3$, 
\eqref{eq:2323-H1,H2}, giving
 \begin{multline} \label{eq:2323-Hdyn}
  \GH\dyn\uv = \frac3\Ns \, \frac{u}v \Biggl( -\lambda\,\fB\uv + \tfrac12\lambda^2 \biggl\{ \frac{u+v}4\fB\uv^2 \\
   + \tfrac1u\fP\left(\tfrac1u,\tfrac{v}u\right) + \tfrac1v\fP\left(\tfrac1v,\tfrac{u}v\right) + \tfrac12\fP\left(u,v\right) \biggr\}
   + \orderof{\lambda^3} \Biggr).
 \end{multline}
Using \eqref{eq:PhiRefl},\eachlabelcase{{eq:Phi1Refl}{ \eqref{eq:Phi1Refl},}{}} we derive
 \begin{multline} \label{eq:2233-Hdyn}
  \GH*\dyn\st \begin{aligned}[t]&= \GH\dyn\bigl(u\st,v\st\bigr) = \GH\dyn(\tfrac1s,\tfrac{t}s) \\&
   = \frac3\Ns \, \frac{s}t \Biggl( -\lambda\,\fB\st 
   + \tfrac12\lambda^2 \biggl\{ \frac{1+t}4\fB\st^2 \end{aligned}\\
   + \fP\st + \tfrac1t\fP(\tfrac{s}t,\tfrac1t) + \tfrac1{2s}\fP(\tfrac1s,\tfrac{t}s) \biggr\}
   + \orderof{\lambda^3} \Biggr).
 \end{multline}
We obtain a similar result in the large $N$ limit, replacing $\frac3\Ns$ with $\frac{p}{N^2}$.

 \section{Results for $\Db{}{}{}{}$ Functions}
 \label{se:DbarCalc}

We here use the standard relations for the $\Db{}{}{}{}$ functions given in \secref{se:Dbar} to simplify the results of \cite{rf:linda} 
to the form quoted in \eqref{eq:22ppSUGRA-G}, which is manifestly compatible with \eqref{eq:22pp-GDecomp}.
\newcommand{\fo}{\tilde{o}}%
\newcommand{\fa}[1]{\tilde{a}_{#1}}%
\newcommand{\fb}[1]{\tilde{b}_{#1}}%
\newcommand{\fc}{\tilde{c}}%
\newcommand{\norm}{\mathfrak{P}}
We begin by introducing an alternative expression for $\G*$ in the interacting theory, writing
 \begin{multline} \label{eq:22pp-GAlt}
  \G*\uvst = \fo\uv
   + \biggl( \sigma u \, \fa1\uv  + \tau\frac{\vphantom{u^2}u}v \, \fa2\uv \biggr)
\\ + \biggl( \sigma^2u^2 \, \fb1\uv + \tau^2\frac{u^2}{v^2} \, \fb2\uv \biggr)
   + \sigma\tau\frac{u^2}v \, \fc\uv,
 \end{multline}
where crossing symmetry in the \ch22pp channel, as given by \eqref{eq:22pp-GCrossSym}, requires that
 \begin{equation} \label{eq:22pp-abcCrossSym}
  \fo\uv = \fo\uvmod*,
\quad
  \fa1\uv = \fa2\uvmod*,
\quad
  \fb1\uv = \fb2\uvmod*,
\quad
  \fc\uv = \fc\uvmod*.
 \end{equation}
In the free field limit $\G*\tendsto\G*\ord0$, so by comparison with \eqref{eq:22ppFF-G} we have
 \begin{equation} \label{eq:22ppFF-abc}
  \fo\uv\tendsto1,
\quad
  \fa{i}\uv\tendsto a,
\quad
  \fb{i}\uv\tendsto b,
\quad
  \fc\uv\tendsto c,
 \end{equation}
where $a,b,c$ are the constants defined in \eqref{eq:Col-abcExplicit}
,
and as previously, we have set the disconnected contribution to 1.
It follows from \eqref{eq:4pt-sDefn}, \eqref{eq:GFree+Dyn}, \eqref{eq:22ppFF-abc} that
 \begin{equation} \label{eq:22pp-abcH}
  \begin{aligned}
   \fo\uv &= 1 + v\GH*\dyn\uv,
&  \fa1\uv&= a + \tfrac{v}u(v-u-1)\GH*\dyn\uv,
\\ \fb1\uv&= b + \tfrac{v}u\GH*\dyn\uv,
&  \fc\uv &= c + \tfrac{v}u(u-v-1)\GH*\dyn\uv,
  \end{aligned}
 \end{equation}
with corresponding results for $\fa2\uv$ and $\fb2\uv$ determined by \eqref{eq:22pp-abcCrossSym} 
and the crossing symmetry relation for $\GH*\dyn\uv$, \eqref{eq:22pp-HCrossSym}.
Results from the effective supergravity action, initially written in a form compatible with known values in the $p=3$ case, give
 \begin{align} \label{eq:22ppSUGRA-abcValues}
  \fo'\uv &= \, \norm \, u\Bigl(\Db11pp - (p+1)u\Db22pp - (1-u+v)\Db22{p+1}{p+1}\Bigr),
\Bob/
  \fa1\uv &= \begin{aligned}[t] & \norm
   \Bigl(-2\Db11pp - 2(p-1)u\Db12{p-1}p - 2\bigl(\Db21p{p+1}-\Db21{p+1}p\bigr) + (p+2)u\Db22pp
\\ & \quad - u\bigl(\Db31pp - \Db12p{p+1} + (p-1)(\Db22{p-1}{p+1} - u\Db23{p-1}p)\bigr) + 4u\Db32p{p+1}\Bigr),\end{aligned}
\Bob/
  \fb1\uv &= \begin{aligned}[t] & \frac\norm{(p+1)(p+2)}
   \Bigl(2(p-1)^2(p+2)\Db12{p-1}p - p(p+1)(p+2)\Db22pp
\\ & \quad + 2p(p-1)u\Db33{p-1}{p+1} + 4\Db31{p+1}{p+1} + 4p(1-u-v)\Db23p{p+1}
\\ & \quad + (p-2)(p+1)\bigl(\Db31pp-\Db12p{p+1}+(p-1)(\Db22{p-1}{p+1}-u\Db23{p-1}p)\bigr)\Bigr),\end{aligned}
\Bob/
  \fc \uv &= \begin{aligned}[t] & \norm \, v
   \Bigl(-2(p-1)^2\bigl(\Db12{p-1}p+\Db12p{p-1}\bigr) + 6p\,\Db22pp - 4\bigl(\Db23p{p+1}+\Db32p{p+1}\bigr)
\\ & \quad + (p-1)\bigl(\Db22{p-1}{p+1}+\Db22{p+1}{p-1}\bigr) - \bigl(\Db12p{p+1}+\Db12{p+1}p\bigr)
\\ & \quad + \bigl(\Db31pp+\Db13pp\bigr) - (p-1)u\bigl(\Db23{p-1}p+\Db23p{p-1}\bigr)\Bigr).\end{aligned}
\notag\\[-22pt]
 \end{align}
Here $\norm$ is a normalisation constant, which for consistency with our convention above is given by
 \begin{equation} \label{eq:SUGRA-norm}
  \norm = \bfrac{-p}{\Gamma(p-1)} \, \frac1{N^2} \,,
 \end{equation}
and we write $\fo'$ to denote the value of $\fo$ when the contribution of the disconnected component has been removed;
thus from \eqref{eq:22pp-abcH}, $\fo'\uv = v\GH*\dyn\uv$.
It is easy to check that $\fo'\uv,\fc\uv$ satisfy \eqref{eq:22pp-abcCrossSym}, 
as do $\fa1\uv,\fb1\uv$ with respect to the corresponding values of $\fa2\uv,\fb2\uv$, which are given explicitly in \cite{rf:linda}.

We proceed to use the $\Db{}{}{}{}$ identities to demonstrate that \eqref{eq:22ppSUGRA-abcValues} is equivalent to \eqref{eq:22ppSUGRA-HI}.
Using \eqref{eq:DUpDown} to rewrite $u\Db22pp$, $u\Db11{p+1}{p+1}$, $v\Db11{p+1}{p+1}$, we obtain
 \vspace{-0.8ex}%
 \begin{multline} \label{eq:DFaff-1}
  \Db11pp - (p+1)u\Db22pp - (1-u+v)\Db22{p+1}{p+1}
\\  = p^2\Db11pp - (2p+1)\Db11{p+1}{p+1} + \Db11{p+2}{p+2}
\\  + \Db21p{p+1} - \Db22{p+1}{p+1} - \Db31p{p+2}.
 \end{multline}
From \eqref{eq:DSum} we have
 \begin{equation} \label{eq:D11pp,D21pp+1}
  \begin{aligned}[b]
    p\,\Db11pp &= \Db21p{p+1} + \Db12p{p+1} + \Db11{p+1}{p+1}
\\&\begin{aligned}
   \implies
   \Db11{p+2}{p+2} &= (p+1)\Db11{p+1}{p+1} - \Db21{p+1}{p+2} - \Db12{p+1}{p+2}, 
\\ (p+1)\Db21p{p+1} &= \Db31p{p+2} + \Db22p{p+2} + \Db21{p+1}{p+2},
\\ (p+1)\Db12p{p+1} &= \Db13p{p+2} + \Db22p{p+2} + \Db12{p+1}{p+2},
   \end{aligned}
  \end{aligned}
 \end{equation}
which substituting for $\Db11pp, \Db11{p+2}{p+2}, \Db31p{p+2}$ in \eqref{eq:DFaff-1} gives
 \begin{multline} \label{eq:DFaff-2}
  \Db11pp - (p+1)u\Db22pp - (1+u-v)\Db22{p+1}{p+1}
\\  = p\,\Db12p{p+1} - \Db22{p+1}{p+1} - \Db12{p+1}{p+2} + \Db22p{p+2}.
 \end{multline}
From \eqref{eq:DUpDown} we have $\Db12p{p+1} = \Db13p{p+2} - \Db22{p+1}{p+1}$, 
which together with the last identity of \eqref{eq:D11pp,D21pp+1} gives
 \begin{equation} \label{eq:D12pp+1}
  p\,\Db12p{p+1} = \Db22{p+1}{p+1} + \Db22p{p+2} + \Db12{p+1}{p+2}.
 \end{equation}
Substituting into \eqref{eq:DFaff-2} and cancelling terms, we then have
 \begin{equation} \label{eq:22ppSUGRA-o'Value}
  \fo'\uv = 2 \norm u \Db22p{p+2}\uv = -\nN \frac1{(p-2)!} \, u^pv\Db{p}{p+2}22\uv = v\GH*\dyn\uv,
 \end{equation}
where we have made use of the symmetries \eqref{eq:DSym} and $\norm$ is as in \eqref{eq:SUGRA-norm}.
Thus we recover the result \eqref{eq:22ppSUGRA-HI}.

We may similarly show that the other expressions in \eqref{eq:22ppSUGRA-abcValues} are compatible with 
\eqref{eq:22pp-abcH} and \eqref{eq:22ppSUGRA-o'Value},
where we will now need to use \eqref{eq:DLimit} to obtain the constant terms $a,b,c$.
E.g. for $\fa1\uv$, we first take $\D4=p+2$ in \eqref{eq:DLimit}, which gives for appropriate choices of $\OTT[,]{\D{#1}}{3}$,
 \vspace{-0.2ex}%
 \begin{equation} \label{eq:DGammas}
  \begin{gathered}
   \Db21{p+1}{p+2} + u\Db22p{p+2} + v\Db12{p+1}{p+2} = \Gamma(p),
\\ \Db31p{p+2} + u\Db32{p-1}{p+2} + v\Db22p{p+2} = \Gamma(p-1),
\\ \Db22p{p+2} + u\Db23{p-1}{p+2} + v\Db13p{p+2} = \Gamma(p-1).
  \end{gathered}
 \end{equation}
Thus we have that
 \begin{multline} \label{eq:vu1DFaff-1}
  (v-u-1)\Db22p{p+2} =
   -\Gamma(p) + \bigl(v\Db13p{p+2}-\Db31p{p+2}\bigr)
   + \bigl(v\Db12{p+1}{p+2}-u\Db32{p-1}{p+2}\bigr)
\\ + u\Db23{p-1}{p+2} + \Db21{p+1}{p+2};
 \end{multline}
applying \eqref{eq:DUpDown} to \eqref{eq:vu1DFaff-1} and cancelling terms, we find
 \begin{multline} \label{eq:vu1DFaff-2}
  (v-u-1)\Db22p{p+2} =
   -\Gamma(p) + p\bigl(\Db21{p-1}{p+2}-\Db12{p-1}{p+2}\bigr) \\ + 2\bigl((p-1)\Db21p{p+1} + u\Db32p{p+1}\bigr),
 \end{multline}
and hence, recalling $v\GH*\dyn = \fo'$, we may write
 \begin{multline} \label{eq:aoFaff-1}
  \fa1 - \bfrac{v-u-1}u\fo'
   = \norm \Bigl( (p+2)u\Db22pp - pu\Db22{p-1}{p+1} + pu\Db12p{p+1}
\\\begin{aligned}[b]
   &- 2(2p-1)\Db21p{p+1} + 2\Db21{p+1}p - p(p-1)u\Db12{p-1}p
\\ &- 2\Db11pp - 2p\bigl(\Db21{p-1}{p+2}-\Db12{p-1}{p+2}\bigr) + 2\Gamma(p)\Bigr).
  \end{aligned}
 \end{multline}
We now apply \eqref{eq:D12pp+1} to the $\Db12{p-1}p$ term; use
 \begin{equation}
  \Db11{p-1}{p+1} = \Db12{p-1}{p+2} - \Db21p{p+1} = \Db21{p-1}{p+2} - v\Db12p{p+1},
 \end{equation}
which follows from \eqref{eq:DUpDown},
together with
the first identity of \eqref{eq:DGammas} (taking $p\goesto p-1$), 
and finally the first identity given in \eqref{eq:D11pp,D21pp+1}, 
we obtain
 \begin{equation}
  a = \fa1\uv - \bfrac{v-u-1}u\fo'\uv = -2\norm \Gamma(p-1) = \nN,
 \end{equation}
consistent with the large $N$ value of $a$ given in \eqref{eq:LargeNCol-acValues}
(which is simply the limit of the (exact) value \eqref{eq:Col-abcExplicit} as $N\tendsto\infty$),
and in agreement with the form of $\G*$ given in \eqref{eq:22ppSUGRA-G}.
The corresponding term from $\fa2$ obviously follows by symmetry.

Similar arguments may be performed for the other terms in \eqref{eq:22ppSUGRA-abcValues}.
In the case of $\fc$, we likewise obtain
 \begin{equation}
  c = \fc\uv - \bfrac{u-v-1}u\fo'\uv = -2\norm \Gamma(p) = \nN(p-1),
 \end{equation}
again agreeing with \eqref{eq:22ppSUGRA-G} and with the values of $c$ in the large $N$ limit from 
\eqref{eq:Col-abcExplicit}, \eqref{eq:LargeNCol-acValues}.
For $\fb{i}$, we find
 \begin{equation}
  \fb1\uv = \tfrac1u\fo'\uv \implies b=0.
 \end{equation}
This is apparently at odds with the large $N$ value of $b$, at least for single-trace operators, $p\ge4$, given in \eqref{eq:LargeNCol-bValue}.
We note however that $b$ is dependent on the colour structure of the operators, and an exact value is in general hard to obtain.
In the specific case that $p=3$ we do have $b=0$.

 \section{Required Partial Wave Expansions}
 \label{se:CPW-Coeff}

\subsection{Hypergeometric functions} \label{se:CPW-Hyper}

We here derive the necessary coefficients \eqref{eq:2p2pCPW-rValue} for the expansion \eqref{eq:2p2pCPW-HyperExpn},
 \vspace{-0.3ex}%
 \begin{equation} \label{eq:2p2pCPW-HyperExpn-Repeat}
  (-x)^{\alpha+2}\hyperg{\alpha+1}{n+\alpha}{n+2\alpha}x = \sum_{\ell=0}^\infty \Fr\alpha{n}\ell \, \gbb{1+\bt}{\ell+\alpha+2}x,
 \end{equation}
with $\bt = \frac12n-1$,
and show that these enable us to find the coefficients $\AB_{t\ell}$ in \eqref{eq:2p2pSUGRA-CPW-B}.
We start by expanding the related function
 \vspace{-0.3ex}%
 \begin{equation} \label{eq:2p2pCPW-HyperExpn-Tilde}
  (-x)^{\alpha+1}\hyperg{\alpha+1}{n+\alpha}{n+2\alpha}x = \sum_{\ell=0}^\infty \Fr*\alpha{n}\ell \, \gbb{1+\bt}{\ell+\alpha+1}x.
 \end{equation}
Writing $\hyperg{\alpha+1}{n+\alpha}{n+2\alpha}x$ firstly as a power series in $x$, 
then in terms of conformal partial waves using \eqref{eq:CPW-xnExpn}, 
we obtain
 \begin{equation}
  \Fr*\alpha{n}\ell
   = \frac{\pochhammer{\alpha+1}\ell\pochhammer{n+\alpha}{\ell-1}}{\pochhammer{n+2\alpha}{2\ell-1}}
     \sum_{m=0}^\ell (-1)^m \frac{(n+m+\alpha-1)}{m!(\ell-m)!} \pochhammer{n+m+2\alpha}{\ell-1}.
 \end{equation}
The sum may readily be calculated, giving $(-1)^\ell$ unless $\ell=0$.
Thus
 \begin{equation}
  \Fr*\alpha{n}0 = 1,
\qquad
  \Fr*\alpha{n}\ell = (-1)^\ell \frac{\pochhammer{\alpha+1}\ell\pochhammer{n+\alpha}{\ell-1}}{\pochhammer{n+2\alpha}{2\ell-1}}.
 \end{equation}
The expansions \eqref{eq:2p2pCPW-HyperExpn-Repeat}, \eqref{eq:2p2pCPW-HyperExpn-Tilde} are related by a factor of $(-x)$,
and hence by means of the recurrence formula \eqref{eq:CPW-gRecur} we may derive
 \begin{equation} \label{eq:2p2pCPW-rRecur}
  \begin{aligned}
   \Fr\alpha{n}0 &= \Fr*\alpha{n}0, \qquad
   \Fr\alpha{n}1  = \Fr*\alpha{n}1 + (\gze{\alpha+1}+\tfrac12)\Fr*\alpha{n}0,
\\ \Fr\alpha{n}\ell &= \Fr*\alpha{n}\ell + (\gze{\alpha+\ell}+\tfrac12)\Fr\alpha{n}{\ell-1} - c_{\alpha+\ell}\Fr\alpha{n}{\ell-2},
  \end{aligned}
 \end{equation}
where $\gze{j}$ and $c_j=\gpl{j-1}\gmi{j}$ are as given by \eqref{eq:2p2pJac-gamma} for the appropriate value of $\bt$.
We then note that \eqref{eq:2p2pCPW-rValue} satisfies \eqref{eq:2p2pCPW-rRecur} as required.

Now we consider
 \vspace{-0.3ex}%
 \begin{equation} \label{eq:2p2pCPW-Hyper-Fc}
  u^\alpha \hyperg{\alpha+1}{n+\alpha}{n+2\alpha}{1-v} = \sum_{j,\ell} \Fc\alpha{n}j\ell u^{j+\alpha-1} \Gpw\ell{4+2\bt+2j+\ell}uv{2\bt}{2\bt},
 \end{equation}
and, using \eqref{eq:xxDefn} to write $1-v$ in terms of $x,\xb$, let
 \begin{equation} \label{eq:2p2pCPW-Hyper-Fk}
  (\xxb)\hyperg{\alpha+1}{n+\alpha}{n+2\alpha}{x+\xb(1-x)} = x \sum_{k=0}^\infty F\p{k}(x) \xb^k.
 \end{equation}
We may use the hypergeometric identities
 \begin{equation}
  (1-x)\bfrac{c}{ab}\diff{x}\hyperg{a}bcx = \hyperg{a}bcx - \frac{(c-a)(c-b)}{c(c+1)}x\,\hyperg{a+1}{b+1}{c+2}x,
 \end{equation}
 \begin{multline}
  \frac1x \hyperg{a}bcx
   = \frac1x \hyperg{a-1}{b-1}{c-2}x
   + \frac{c(a+b-1)-2ab}{c(c-2)} \hyperg{a}bcx \\
   + ab\frac{(c-a)(c-b)}{c^2(c^2-1)}x\,\hyperg{a+1}{b+1}{c+2}x,
 \end{multline}
to write $F\p{k}(x)$ in the form
 \begin{equation}
  F\p{k}(x) = \sum_{r=-1}^k F\p{k}_r x^r\hyperg{\alpha+r+1}{n+\alpha+r}{n+2\alpha+2r}x.
 \end{equation}
Using the results of \secref{se:CPWCPW}, it follows that
 \begin{equation}
  \Fc\alpha{n}j\ell = \sum_{k=0}^j \sum_{r=-1}^k \beta_{j+1,k+1} F\p{k}_r \Fr{\alpha+r}n{j+\ell-\alpha-r},
 \end{equation}
with $\beta_{j+1,k+1}$ as in \eqref{eq:2p2pCPW-bbValues}.
Hence we may find the $\AB_{t\ell}$ in \eqref{eq:2p2pSUGRA-CPW-B}.
In particular, to find the lowest twist contribution to \eqref{eq:2p2pCPW-Hyper-Fc},
we note the \eqref{eq:2p2pCPW-Hyper-Fk} gives
 \begin{equation}
  F\p0(x) = \hyperg{\alpha+1}{n+\alpha}{n+2\alpha}x.
 \end{equation}
Thus we have $\Fc\alpha{n}1\ell = \Fr\alpha{n}\ell$, and the result \eqref{eq:2p2pSUGRA-B1Value} follows.

\subsection{Logarithm and Polylogarithm expansions} \label{se:CPW-ln,Li2}

We here determine coefficients for the conformal partial wave expansions of logarithmic and polylogarithmic functions 
required in \secref{se:PerturbResults}.
\subsubsection{Logarithm and dilogarithm}
We begin by writing expansions the logarithm and dilogarithm,
 \begin{alignat}{2} \label{eq:CPW-ln}
  (-x)^q \ln(1-x)
  &= \sum_{\subbox[0.8]{n=q+1}}^\infty (-1)^{q+1} \mspace{10mu} \frac{x^n}{n-q}
 &&= \sum_{\ell=0}^\infty \tfrac{\pochhammer{2\bt+1}\ell}{\pochhammer{2\bt+\ell+1}\ell} \lnaq \gbb{1+\bt}{\ell+1}x \,,
\\ \label{eq:CPW-Li2}
  -(-x)^q \Li2(x)
  &= \sum_{\subbox[0.8]{n=q+1}}^\infty (-1)^{q+1} \frac{x^n}{(n-q)^2}
 &&= \sum_{\ell=0}^\infty \tfrac{\pochhammer{2\bt+1}\ell}{\pochhammer{2\bt+\ell+1}\ell} \libq \gbb{1+\bt}{\ell+1}x \,.
 \end{alignat}
Note that $\frac{\pochhammer{2\bt+1}\ell}{\pochhammer{2\bt+\ell+1}\ell} \tendsto \frac{\ell!^2}{(2\ell)!}$ as $\bt\tendsto0$,
in which limit we recover the expansions given in \cite{rf:0412335}.
%
%
From \eqref{eq:CPW-ln}, we obtain
 \begin{equation} \label{eq:CPW-ln-a}
  \begin{aligned}[b]
  \pochhammer{2\bt+1}\ell \, \lnaq
   &= \sum_{k=q}^\ell \frac{(-1)^{q+1}}{k-q+1} \pochhammer{2\bt+\ell+1}\ell \, \beta_{\ell+1,k+1}
\\ &= \sum_{k=q}^\ell \frac{(-1)^{k+q}}{k-q+1} \binom\ell{k} \pochhammer{2\bt+k+1}\ell \,, \qquad \ell\ge q.
  \end{aligned}
 \end{equation}
For $q=0$ this has solution
 \begin{equation} \label{eq:CPW-ln-a0}
  \lna0 = \frac{2\bt}{2\bt+\ell}\,\frac1{\ell+1} \,,
 \end{equation}
and for $q\ge1$,
 \begin{multline} \label{eq:CPW-ln-aq}
  \frac{(q-1)!\pochhammer{2\bt+1}{q-1}}{\pochhammer{l-q+2}{q-1}\pochhammer{2\bt+\ell+1}{q-1}} \, \lnaq =
   \Bigl(\h(2\bt+\ell)-\h(2\bt+q-1)\\[-8pt]+\h(\ell)-\h(q-1)\Bigr),
 \end{multline}
where $\h(n)$ denotes the harmonic numbers defined in \eqref{eq:hDefn}.
For small values of $q$, \eqref{eq:CPW-ln-aq} gives
 \begin{equation} \label{eq:CPW-ln-a123}
  \begin{aligned}
   \lna1 &= \h(2\bt+\ell)-\h(2\bt)+\h(\ell) \,,
\\ \lna2 &= \frac{(2\bt+\ell+1)\ell}{2\bt+1} \Bigl(\h(2\bt+\ell)-\h(2\bt+1)+\h(\ell)-1\Bigr) \,,
\\ \lna3 &= \frac{(2\bt+\ell+2)(2\bt+\ell+1)\ell(\ell-1)}{2(2\bt+1)(2\bt+2)} \Bigl(\h(2\bt+\ell)-\h(2\bt+2)+\h(\ell)-\tfrac32\Bigr) \,.
  \end{aligned}
 \end{equation}
Similarly to \eqref{eq:CPW-ln-a}, we have for the dilogarithm
 \begin{equation} \label{eq:CPW-Li2-b}
  \pochhammer{2\bt+1}\ell \, \libq
   = \sum_{k=q}^\ell \frac{(-1)^{k+q}}{(k-q+1)^2} \binom\ell{k} \pochhammer{2\bt+k+1}\ell \,, \qquad \ell\ge q.
 \end{equation}
We find
 \begin{equation}
  \lib0 = \frac{2\bt}{2\bt+\ell}\,\frac1{\ell+1} \Bigl(\h(2\bt+\ell-1)-\h(2\bt-1)+\h(\ell+1)\Bigr) \,,
 \end{equation}
and defining further generalisations of the harmonic numbers by
 \begin{equation} \label{eq:hbqDefn}
  \h*r_b(n) = \h*r(b+n)-\h*r(b),
\qquad
  \h*r_{b,q}(n) = \h*r_b(n)-\h*r_b(q),
 \end{equation}
we have
 \begin{equation}
  \begin{aligned}[b]
   \lib1 &= \tfrac12 \bigl[\h_{2\bt}(\ell)+\h(\ell)\bigr]^2 - \tfrac12 \bigl[\h*2_{2\bt}(\ell)-\h*2(\ell)\bigr] \,,
\\ \lib2 &= \frac{(2\bt+\ell+1)\ell}{2\bt+1} \biggl(\tfrac12 \bigl[\h_{2\bt,1}(\ell)+\h_{0,1}(\ell)\bigr]^2 - \tfrac12 \bigl[\h*2_{2\bt,1}(\ell)-\h*2_{0,1}(\ell)\bigr] + 1\biggr) \\&\mspace{240mu} - \Bigl[\h_{2\bt}(\ell)+\h(\ell)\Bigr] \,.
  \end{aligned}
 \end{equation}

\subsubsection{Logarithm and dilogarithm of $x'$}

As has been noted, when $\bt=0$ the relation \eqref{eq:CPW-gx'} reduces to a form that makes finding coefficients for 
partial wave expansions of functions of $x'$ trivial, in terms of the corresponding expansion for functions of $x$.
However in the more general case the coefficients must be independently determined. We define the $\lnapq$ by
 \begin{equation} \label{eq:CPW-ln-x'}
  (-x')^q \ln(1-x')
   = \sum_{\ell=0}^\infty \tfrac{\ell!}{\pochhammer{2\bt+\ell+1}\ell} \lnapq \gbb{1+\bt}{\ell+1}x \,,
 \end{equation}
and thus using \eqref{eq:CPW-beta'}, \eqref{eq:CPW-x'nExpn} obtain
 \begin{equation} \label{eq:CPW-ln-a'}
  (2\bt+\ell)! \, \lnapq
= \sum_{k=q}^\ell \frac{(-1)^{k-q+\ell+1}}{k-q+1} \binom\ell{k} \pochhammer{k+1}{2\bt+\ell} \,, \qquad \ell\ge q.
 \end{equation}
Using $\ln(1-x')=-\ln(1-x)$, the $\lnap0$ are easily found; for $q>0$, the first few results may be written in the form
 \begin{equation} \label{eq:CPW-ln-a'12}
  \begin{aligned}
   \lnap1 &= (-1)^{\ell+1} \bigl[\h_{2\bt}(\ell)+\h(\ell)\bigr] + \Ap1{2\bt}\ell \,,
\\ \lnap2 &= (-1)^{\ell+1} \,(2\bt+\ell+1)\ell\, \bigl[\h_{2\bt,1}(\ell)+\h_{0,1}(\ell)\bigr] - \Ap2{2\bt}\ell \,,
  \end{aligned}
 \end{equation}
where $\Ap1{2\bt}\ell, \Ap2{2\bt}\ell$ are given by a sum ranging over $1,\dots,2\bt$,
and thus may be expressed as explicit functions of $\ell$ for a given value of $\bt$,
 \begin{equation} \label{eq:CPW-ln-A12}
  \begin{aligned}
   \Ap1{2\bt}\ell &= \sum_{a=1}^{2\bt} \left(\frac{\pochhammer{2\bt-a+1}\ell}{\pochhammer{a+1}\ell}-(-1)^\ell\right)\frac1a \,,\\
   \Ap2{2\bt}\ell &= \sum_{a=1}^{2\bt} \left(\frac{\pochhammer{2\bt-a+1}\ell}{\pochhammer{a+1}\ell}+(-1)^\ell\left\{\frac{(2\bt+\ell+1)\ell}{a+1}-1\right\}\right) \,.
  \end{aligned}
 \end{equation}
For the $p=3$ ($\implies2\bt=1$) case, \eqref{eq:CPW-ln-A12} becomes
 \begin{equation}
  \begin{alignedat}{2}
   \Ap11\ell &= \begin{cases}
    \hphantom-\frac{\ell+2}{\ell+1}, & \text{$\ell$ odd,} \\
    -\frac\ell{\ell+1}, & \text{$\ell$ even;} \end{cases}
& \qquad
   \Ap21\ell &= \begin{cases}
    -\frac12 \, \frac{\ell-1}{\ell+1} \, (\ell+2)^2, & \text{$\ell$ odd,} \\
    \hphantom-\frac12 \, \frac{\ell+3}{\ell+1} \, \ell^2, & \text{$\ell$ even.} \end{cases}
  \end{alignedat}
 \end{equation}
For the dilogarithm, with a definition for $\libpq$ analogous to \eqref{eq:CPW-ln-x'}, we find that
 \begin{equation} \label{eq:CPW-Li2-x'}
  \libp0 = \frac{(-1)^{\ell+1} - \binom{2\bt+\ell-1}{\ell+1}}{(\ell+1)(2\bt+\ell)} \,.
 \end{equation}
Again, it is possible to determine explicit expressions for higher $q$ if we first fix the value of $\bt$.
Then we are able to write the required sum
 \begin{equation} \label{eq:CPW-thingybob}
  \sum_{k=q}^\ell \frac{(-1)^{k-q}}{(k-q+1)^2} \binom\ell{k}
   \frac{\pochhammer{k+1}{2\bt+\ell}}{(2\bt+\ell)!}
 \end{equation}
as
 \begin{equation}
  \sum_{k=q}^\ell \frac{(-1)^{k-q}}{(k-q+1)^2} \binom\ell{k}
   \left( \frac{\pochhammer{2\bt+k+1}\ell}{\pochhammer{2\bt+1}\ell} \times \frac{\pochhammer{k+1}{2\bt}}{(2\bt)!} \right).
 \end{equation}
The term $\frac1{(2\bt)!}{\pochhammer{k+1}{2\bt}}$ becomes, for fixed values of $2\bt$ and $q$, a polynomial in $(k-q+1)$;
hence our series for $\libpq$ reduces to a combination of those in
\eqref{eq:CPW-ln-a}, \eqref{eq:CPW-Li2-b}, given by $\lnaq, \libq$,
plus, potentially, some involving non-negative powers of $k$ in the summand, 
the contributions of which may also be readily calculated\footnote{%
For example, by using
$
  \sum_{k=q}^\ell (-1)^{k-q+1} (k-q+1)^n \binom\ell{k} \pochhammer{2\bt+k+1}\ell
\\ =
  \left(\diff{x}\right)^\ell
  \left[ x^{2\bt+\ell+q-1} \cdot \left(x\diff{x}\right)^n \left((1-x)^\ell-\Simgaq(x)\right) \right]_{x=1}
$
where $\Simgaq(x) \equiv \sum_{r=0}^{q-1}{\textstyle\binom\ell{r}}(-x)^r$.%
}.
In the $2\bt=1$ case, this gives us
 \begin{equation}
  (-1)^{\ell+1}\,\libpq = \lnaq + q\,\libq \,.
 \end{equation}

\subsubsection{Logarithm squared}

For the expansion of $f\ord{22}$, \eqref{eq:CPW-f22}, we need coefficients for conformal wave expansions of $x'\Li2(x)$, $x\Li2(x')$.
We make use of the identity
 \begin{equation} \label{eq:ln^2Li2}
  -\ln^2(1-x) = 2\Li2(x) + 2\Li2(x'),
 \end{equation}
and define $\lnsaq$ by
 \begin{equation}
  \tfrac12(-x)^q\ln^2(1-x)
 = (-x)^q \sum_{n=2}^\infty \h(n-1)\,\frac{x^n}n
 = \sum_{\ell=0}^\infty \tfrac{\pochhammer{2\bt+1}\ell}{\pochhammer{2\bt+\ell+1}\ell} \lnsaq \gbb{1+\bt}{\ell+1}x \,.
 \end{equation}
Thus
 \begin{equation}
  \pochhammer{2\bt+1}\ell \lnsaq = \sum_{k=q+1}^\ell \frac{(-1)^{k-q+1}}{k-q+1} \binom{\ell}k \pochhammer{2\bt+k+1}\ell \, \h(k-q) \,.
 \end{equation}
From our knowledge of $\lib0, \libp0$ we deduce that
 \begin{equation}
  \lnsa0 = \frac{2\bt}{(2\bt+\ell)(\ell+1)}\left(\h_{2\bt}(\ell)+\h(\ell)-\tfrac{(-1)^\ell}{2\bt\binom{2\bt+\ell}\ell}\right).
 \end{equation}
If $\hb(n)$ denotes the alternating harmonic number, as given in \eqref{eq:hDefn}, we find for $\bt=0$
 \begin{equation}
  \evalat{\bt=0}{\lnsa1} = 2\left(\h(\ell)^2+\hb*2(\ell)\right).
 \end{equation}
For general $\bt$, with the generalisation $\hb*r_{b,q}(n)$ defined as for the harmonic number in \eqref{eq:hbqDefn}, we may write
 \begin{multline}
  \lnsa1
   = \lib1 + \bigl(\hb*2_{2\bt}(\ell)+\hb*2(\ell)\bigr) + \bigl(\h*2_{2\bt}(\ell)-\h*2(\ell)\bigr)
 + \h(2\bt-1)\bigl(\hb_{2\bt}(\ell)+\hb(\ell)\bigr) 
\\ + \sum_{a=1}^{2\bt-1} \binom{2\bt}a \left(\frac1{2\bt-a}-\frac1a\right) \hb_a(\ell) ,
 \end{multline}
with an explicit solution determinable for a given value of $\bt$.
%
%
Proceeding as for the logarithm and dilogarithm, we then define
 \begin{gather}
  \tfrac12(-x')^q\ln^2(1-x')
 = \sum_{\ell=0}^\infty \tfrac{\ell!}{\pochhammer{2\bt+\ell+1}\ell} \lnsapq \gbb{1+\bt}{\ell+1}x
\displaybreak[2]\\ \implies
  (2\bt+\ell)! \, \lnsapq = \sum_{k=q+1}^\ell \frac{(-1)^{k-q+\ell}}{k-q+1} \binom{\ell}k \pochhammer{k+1}{2\bt+\ell} \, \h(k-q) .
 \end{gather}
We find that for $2\bt=1$ there is a solution in terms of $\lnsa1$, namely
 \begin{equation}
  \evalat{\bt=\half}%
  {(-1)^{\ell+1}\lnsap1} = \lnsa1 +
  \begin{cases}
   2\bigl(1-\tfrac1{\ell+1}\bigr)\h(\ell+1) - \bigl(1+\tfrac1{\ell+1}\bigr), & \text{$\ell$ odd,} \\
   2\bigl(1+\tfrac1{\ell+1}\bigr)\h(\ell) - \bigl(1-\tfrac1{\ell+1}\bigr), & \text{$\ell$ even.}
  \end{cases}
 \end{equation}

\subsection{Results for large $\bt$} \label{se:Prtb-Largeb}

To obtain the results \eqref{eq:Largeb-eta-l1}, \eqref{eq:Largeb-eta-l1^2}, we first observe that
 \begin{equation} \label{eq:Largeb-Limits}
  \lim_{\bt\tendsto\infty} \frac{\pochhammer{2\bt+\ell+1}k}{\pochhammer{2\bt+\ell}\ell} = 0,
\quad
  k < \ell;
\qquad
  \lim_{\bt\tendsto\infty} \frac{\pochhammer{2\bt+k+1}\ell}{\pochhammer{2\bt+\ell}\ell} = 1,
\quad
  \forall k.
 \end{equation}
Substituting \eqref{eq:CPW-ln-a}, \eqref{eq:CPW-ln-a'} directly into \eqref{eq:Prtb-eta-l1} gives a sum for $\expect{\ado[\ell-2]1}$;
using \eqref{eq:Largeb-Limits}, we eliminate the terms that vanish as $\bt\tendsto\infty$, and are left with
 \begin{equation}
  \expect{\ado[\ell-2]1} = \left[\sum_{k=1}^\ell (-1)^{k+1} \binom\ell{k} \frac1k\right] - \frac1\ell.
 \end{equation}
Similarly from \eqref{eq:Prtb-eta-l1^2} we obtain
 \begin{equation}
  \expect{\ado[\ell-2]1^2}
  = \left[\sum_{k=1}^\ell (-1)^{k+1} \binom\ell{k} \left(\frac3{2k^2} - \frac{\h(k-1)}{2k}\right)\right]
    - \left(\frac3{2\ell^2} + \frac{\h(\ell-1)}\ell\right).
 \end{equation}
These sums may be calculated using the results
 \begin{equation}
  \begin{gathered}
   \sum_{k=1}^\ell (-1)^{k+1} \binom\ell{k} \frac1k = \h(\ell),
\quad
   \sum_{k=1}^\ell (-1)^{k+1} \binom\ell{k} \frac1{k^2} = \tfrac12\bigl(\h*2(\ell) + \h(\ell)^2\bigr),
\\
   \sum_{k=1}^\ell (-1)^{k+1} \binom\ell{k} \frac{\h(k-1)}k = \tfrac12\bigl(\h*2(\ell) - \h(\ell)^2\bigr),
  \end{gathered}
 \end{equation}
yielding \eqref{eq:Largeb-eta-l1}, \eqref{eq:Largeb-eta-l1^2}.

\subsection{Results from $f\ord{10}$, $f\ord{21}$} \label{se:Prtb-f10f21}

To obtain information from \eqref{eq:Prtb-x3f10}, \eqref{eq:Prtb-x3f21}, 
we require an expression for the derivative of partial waves, \eqref{eq:CPW-g'Defn}.
Using the definition \eqref{eq:CPW-gDefn} of $\gbb*t\ell{x}$ in terms of a hypergeometric function,
we may write $\gbb't\ell{x}$ as a power series expansion,
 \begin{multline} \label{eq:CPW-g'Expn}
  \gbb'{t}\ell{x}
   = (-x)^\ell \sum_{n=0}^\infty \bigl(\h_{t+\ell+\bt-2}(n)+\h_{t+\ell-\bt-2}(n)-2\h_{2t+2\ell-3}(n)\bigr)
\\[-10pt]  \times \frac{\pochhammer{t+\ell+\bt-1}n\pochhammer{t+\ell-\bt-1}n}{n! \, \pochhammer{2t+2\ell-2}n} \, x^n,
 \end{multline}
where the $\h_b(n)$ are as defined in \eqref{eq:hbqDefn}.
From the value of $f\ord{10}$ given in \eqref{eq:Prtb-fValues}, we may write, making use of \eqref{eq:ln^2Li2},
 \vspace{-0.5ex}%
 \begin{equation} \label{eq:CPW-f10}
  \begin{aligned}[b]
  x^3 f\ord{10}
    &= (x+x')\Li2(x) = x\Li2(x)-x'\Li2(x')-\tfrac12x'\ln^2(1-x')
\\  &= \sum_{\ell=0}^\infty \tfrac{\ell!}{\pochhammer{2\bt+\ell+1}\ell} \left[\tbinom{2\bt+\ell}\ell\lib1-\libp1+\lnsap1\right] \gbb{1+\bt}{\ell+1}x,
  \end{aligned}
 \end{equation}
where $\lib1,\libp1,\lnsa1$ are the partial wave expansion coefficients defined previously.
Taking the value of $a\ord{\ell-2,0}\,\ado[\ell-2]1$ from \eqref{eq:Prtb-eta-l1}, 
and using \eqref{eq:CPW-g'Expn}, 
we may expand the partial waves 
in \eqref{eq:Prtb-x3f10}, \eqref{eq:CPW-f10}
as power series;
matching powers of $x$, we are left with the condition
 \begin{multline} \label{eq:Prtb-bl1Cond}
  \sum_{\ell=2}^m \tfrac{(-1)^\ell(2\bt+2\ell+1)\ell!}{\pochhammer{2\bt+m+1}{\ell+1}} \tbinom{m}\ell
    \Bigl\{
     \bigl[\tbinom{2\bt+\ell}\ell+(-1)^\ell(2\bt+1)\bigr]b\ord{\ell-2,1}
   + \bigl[\tbinom{2\bt+\ell}\ell\lib1-\libp1+\lnsap1\bigr]
\\[-6pt]
   + \tfrac12\bigl[\tbinom{2\bt+\ell}\ell\lna1-\lnap1\bigr]
     \bigl(\h_{2\bt+\ell}(m-\ell) + \h_{2\bt}(m-\ell) - 2\h_{2\bt+2\ell+1}(m-\ell)\bigr)
    \Bigr\} = 0.
 \end{multline}
Evaluating \eqref{eq:Prtb-bl1Cond} for $m=2,\dots,L$ gives a system of $L-1$ equations which may be solved to find
$\expect{b\ord{0,1}},\dots,\expect{b\ord{L-2,1}}$.
(At finite $\ell$, the partial wave coefficients may be found from their summation form if no more concise formula is known.)
Explicit results for $b\ord{0,1},b\ord{1,1},b\ord{2,1}$ are given in \eqref{eq:Prtb-b0121Values}.

To obtain similar results from \eqref{eq:Prtb-x3f21} we need an expansion for $f\ord{21}$, which starting from \eqref{eq:Prtb-fValues}, 
and using \eqref{eq:ln^2Li2} to express $x'\ln^3(1-x)$ in terms of $x'\ln(1-x)\Li2(x)$ and $x'\ln(1-x)\Li2(x')$, we write as
 \begin{equation} \label{eq:CPW-f21}
  \begin{aligned}[b]
   x^3 f\ord{21}  
    &= \tfrac18\Bigl[\tfrac12x\ln^3(1-x) + 3\,x\ln(1-x)\Li2(x) + x'\ln(1-x')\Li2(x')
\\[-4pt]&\mspace{80mu} + 3(x+x')\left(2\Li3(x)-\Li3(x')\right)\Bigr]
\\  &= -\frac18 \sum_{\ell=0}^\infty \tfrac{\ell!}{\pochhammer{2\bt+\ell+1}\ell}
        \Bigl[ 3\tbinom{2\bt+\ell}\ell \Bigl(\tfrac12\lnca1+\lnlie1-2(\lic1-\licb1)\Bigr)
\\[-6pt]&\mspace{188mu} + \Bigl(\lnliep1+3(\licp1-\licbp1)\Bigr) \Bigr] \gbb{1+\bt}{\ell+1}x,
   \end{aligned}
 \end{equation}
where we have new expansion coefficients defined by
 \vspace{-0.8ex}%
 \begin{align}
  \tfrac13(-x)^q\ln^3(1-x) = \sum_{\ell=0}^\infty \tfrac{\pochhammer{2\bt+1}\ell}{\pochhammer{2\bt+\ell+1}\ell} \lncaq \gbb{1+\bt}{\ell+1}x;&
\displaybreak[1]\\
   (-x)^q\ln(1-x)\Li2(x) = \sum_{\ell=0}^\infty \tfrac{\pochhammer{2\bt+1}\ell}{\pochhammer{2\bt+\ell+1}\ell} \lnlieq \gbb{1+\bt}{\ell+1}x;&
\displaybreak[1]\\\begin{aligned}
   -(-x)^q\Li3(x) = \sum_{\ell=0}^\infty \tfrac{\pochhammer{2\bt+1}\ell}{\pochhammer{2\bt+\ell+1}\ell} \licq \gbb{1+\bt}{\ell+1}x&,
\\  (-x')^q\Li3(x) = \sum_{\ell=0}^\infty \tfrac{\pochhammer{2\bt+1}\ell}{\pochhammer{2\bt+\ell+1}\ell} \licbq \gbb{1+\bt}{\ell+1}x&.
  \end{aligned}&
 \end{align}
Primed versions $\lnliepq,\licpq,\licbpq$ are defined analogously as before.
Writing $\ln^3(1-x)$ as $\ln(1-x)\ln^2(1-x)$, we have the power expansion
 \begin{equation}
  -\ln^3(1-x)
   = \biggl(\sum_{r=1}^\infty \frac{x^r}r\biggr)\biggl(\sum_{s=2}^\infty 2\h(s-1)\frac{x^s}s\biggr)
   = \sum_{n=3}^\infty \frac3n\left(\h(n-1)^2-\h*2(n-1)\right) x^n,
 \end{equation}
which gives
 \begin{equation}
  \pochhammer{2\bt+1}\ell \lncaq = \sum_{k=q+2}^\ell \frac{(-1)^{k-q}}{k-q+1} \binom\ell{k}
   \pochhammer{2\bt+k+1}\ell
   \left(\h(k-q)^2-\h*2(k-q)\right),
 \end{equation}
and similarly we have the expansion for $\ln(1-x)\Li2(x)$
 \begin{equation}
  -\ln(1-x)\Li2(x)
   = \biggl(\sum_{r=1}^\infty \frac{x^r}r\biggr)\biggl(\sum_{s=1}^\infty \frac{x^s}{s^2}\biggr)
   = \sum_{n=2}^\infty \frac1n\left(\h*2(n-1)+\frac2n\h(n-1)\right) x^n,
 \end{equation}
giving
 \begin{equation}
  \pochhammer{2\bt+1}\ell\lnlieq = \sum_{k=q+1}^\ell \frac{(-1)^{k+q}}{k-q+1} \binom\ell{k} \pochhammer{2\bt+k+1}\ell 
  \left(\h*2(k-q)+\frac{2\h(k-q)}{k-q+1}\right).
 \end{equation}
%
Finally, from the power expansion of $\Li3(x)$ \eqref{eq:plogDefn} we find
 \begin{equation}
  \pochhammer{2\bt+1}\ell\licq = \sum_{k=q}^\ell \frac{(-1)^{k+q}}{(k-q+1)^3} \binom\ell{k} \pochhammer{2\bt+k+1}\ell,
 \end{equation}
and writing $-x' = x+x^2+x^3+\dotsb$, we have
 \begin{equation}
  -x'\Li3(x)
   = \biggl(\sum_{r=1}^\infty x^r\biggr)\biggl(\sum_{s=1}^\infty \frac{x^s}{s^3}\biggr)
   = \sum_{n=2}^\infty \h*3(n-1)\,x^n,
 \end{equation}
and thus
 \begin{equation}
  \pochhammer{2\bt+1}\ell\licb1 = \sum_{k=1}^\ell (-1)^{k+1} \binom\ell{k} \pochhammer{2\bt+k+1}\ell \, \h*3(k).
 \end{equation}

Following a similar procedure to that for $f\ord{10}$, and using \eqref{eq:Prtb-eta-l1}, \eqref{eq:Prtb-eta-l1^2}, \eqref{eq:Prtb-bl1Cond}
to obtain values for $\ado[\ell-2]1, \ado[\ell-2]1^2, b\ord{\ell-2,1}$, we match powers of $x$ in \eqref{eq:Prtb-x3f21}, \eqref{eq:CPW-f21}.
This gives an expression similar to \eqref{eq:Prtb-bl1Cond}, which yields a system of equations that may be solved for $\ado[\ell-2]2$ at finite $\ell$.
Results for $\ado[0]2,\ado[1]2,\ado[2]2$ are given in \eqref{eq:Prtb-eta-0122Values}.

\chapter{Short and Semi-Short Operators}
 \label{ch:Semishort}

 \section{Construction of semi-short operators}
 \label{se:SS-Construct}

The basic operators in $\N=4$ supersymmetric Yang Mills are gauge singlets formed from traces of products of the fundamental fields, 
given in \secref{se:N=4-SYM}, 
and their derivatives.
These make up the content of supermultiplets, each of which is descended from a unique highest weight operator as detailed in \secref{se:N=4-Multiplets}.
For the various multiplets which obey shortening conditions, we may restrict to the traces of those fields annihilated by the appropriate supercharges.
We follow the procedures of \cite{rf:0609179}, Section~8, to construct \shorten88 semi-short operators, the field content of which consists of
 \begin{equation} \label{eq:SS-Fields}
  \partial^n \bigl(Z,Y,\lm,\lm*\bigr),
\quad
  n = 0,1,2,\dotsc
 \end{equation}
where, letting the fundamental scalar fields $\adscalar_r \goesto \adgscalar_{ij} = -\adgscalar_{ji}$ by use of the $SU(4)$ gamma matrices,
and $\partial_{\aad} = (\sigma^\mu)_{\alpha\alphd}\partial_\mu$,
 \begin{equation} \label{eq:SS-FieldDefn}
  Z = \adgscalar_{34},
\quad
  Y = \adgscalar_{42},
\quad
  \lm = \gaugino[1]4,
\quad
  \lm* = \gaugino*[2]1,
\quad
  \partial = \partial_{12}.
 \end{equation}
The supercharges $\Q[2]1,\q[2]1,\Q*[1]4,\q*[1]4$ act trivially on $Z,Y,\lm,\lm*$.
The weights of the fields with respect to the dilation, spin, and $SU(4)_R$ Cartan generators are given in \tabref{tb:ZYlld},
from which we see that the operator corresponding to $\partial^n\bigl(Z^uY^v\lm^{s+1}\lm*^{t+1}\bigr)$
belongs to representation $\rep{t+v}{u-v}{s+v}{\frac12(s+n)}{\frac12(t+n)}$ with dimension $\Delta = \frac32(s+t)+u+v+n+2$.
Hence the twist is $s+t+u+v+2$, compatible with \eqref{eq:N=4-Semishort}.
\begin{table}
 \centering
 \caption{Associated weights of the \shorten88 fields} \label{tb:ZYlld}
 \begin{tabular}{|c|c|}
  \hline
  Field & $(\Delta; J_3,\Jb_3; H_1,H_2,H_3)$
\\\hline
  $Z$ & $(1;0,0;0,1,0)$
\\
  $Y$ & $(1;0,0;1,-1,1)$
\\
  $\lm$ & $(\frac32;\frac12,0;0,0,1)$ \bigstrut[b]
\\
  $\lm*$ & $(\frac32;0,\frac12;1,0,0)$ \bigstrut[b]
\\
  $\partial$ & $(1;\frac12,\frac12;0,0,0)$ \bigstrut[b]
\\
  \hline
 \end{tabular}
\end{table}

We may determine the number of operators belonging to $\nrep[\ell]{q}pq$
by writing the most general combination of fields \eqref{eq:SS-Fields} with the given 
$SU(4)_R$ representation, and imposing appropriate commutation conditions, 
namely that the operator be annihilated by the remaining non-trivial $\So[i],\St*[j]$ conformal supercharges,
and be an $SU(2)$ highest weight state, where $(Z,Y)$ form an $SU(2)$ doublet.
In addition to \eqref{eq:SS-Fields} we will use the notation
 \begin{equation}
  \Zi = (Z,Y),
\quad
  \So[i] = (\q[1]2,\q[1]3),
\quad
  \St*[i] = (\q*[2]2,\q*[2]3).
 \end{equation}
It is sufficient to check that an $SU(2)$ top state commutes with $\So,\St*$, 
where from the full superconformal algebra we obtain the necessary relations
 \begin{alignat}{2} \label{eq:SS-S1Algebra}
 &\hphantom{\St*}
  \begin{aligned}
   \comm\So{\dZ[n]} &= 2in\dl*[n-1],
\\ \acomm\So{\dl[n]} &= -4(n+1)\dY,
  \end{aligned} \quad
&&\hphantom{\St*}
  \begin{aligned}
   \comm\So{\dY[n]} &= 0,
\\ \acomm\So{\dl*[n]} &= 0,
  \end{aligned}
\\\intertext{and} \label{eq:SS-S2Algebra}
 &\hphantom{\So}
  \begin{aligned}
   \comm{\St*}{\dZ[n]} &= 2in\dl[n-1],
\\ \acomm{\St*}{\dl[n]} &= 0,
  \end{aligned}
&&\hphantom{\So}
  \begin{aligned}
   \comm{\St*}{\dY[n]} &= 0,
\\ \acomm{\St*}{\dl*[n]} &= 4(n+1)\dY[n].
  \end{aligned} \:
 \end{alignat}
The results will be compared with those given in Appendix~C of \cite{rf:0609179}.

\subsection{Twist 2, 3}

At twists 2 and 3, we have only single-trace operators belonging to $\nrep000$, $\nrep010$ respectively.
For the former, a general operator may be written
 \begin{equation} \label{eq:SS-t2-Op}
  \Op = \sum_{r+s=n} a_{rs} \Tr{\dl[r]\dl*[s]} - \sum_{\subbox{r+s=n+1}} i \, b_{rs} \Tr{\dZ[r]\dY}.
 \end{equation}
For this to be an $SU(2)$ highest weight state requires $b_{rs} + b_{sr} = 0$.
From \eqref{eq:SS-S1Algebra}, \eqref{eq:SS-S2Algebra} we obtain
 \begin{align}
  \comm\So\Op &= \sum_{r,s} 2\left((r+1)b_{r+1\,s} - 2(s+1)a_{sr}\right) \Tr{\dl*[r]\dY},
\\\comm{\St*}\Op &= \sum_{r,s} 2\left((r+1)b_{r+1\,s} - 2(s+1)a_{rs}\right) \Tr{\dl[r]\dY}.
 \end{align}
Hence we require
 \begin{equation} \label{eq:SS-t2-Cond}
  \begin{gathered}[b]
   (r+1)b_{r+1\,s} = 2(s+1)a_{rs} = 2(s+1)a_{sr}
\\\implies
   a_{rs} = a_{sr},
\quad
   (s+1)^2b_{s+1\,r} = (r+1)^2b_{r+1\,s}.
  \end{gathered}
 \end{equation}
thus $a_{rs}$ is completely determined by $b_{rs}$, and writing
 \begin{equation} \label{eq:SS-t2-BDefn}
  b_{rs} = \binom{n+1}s^2 B_{rs},
\quad
  r+s = n+1,
 \end{equation}
we see that \eqref{eq:SS-t2-Cond} implies $B_{rs} = -B_{r-1\,s+1} = \dotsb$. 
This is compatible with $B_{rs} = -B_{sr}$ only when $n$ is even, when we have a unique solution (up to normalisation).

Similarly, a highest weight twist 3 operator may be written
 \begin{equation}
  \Op = \sum_{r,s,t} a_{rst} \Tr{\dZ[r]\dl\dl*} - \at_{rst} \Tr{\dZ[r]\dl*\dl} - i \, b_{rst} \Tr{\dZ[r]\dZ[s]\dY[t]},
 \end{equation}
where $b_{rst} + b_{str} + b_{trs} = 0$.
Commutation with $\So,\St*$ imposes the conditions
 \begin{equation} \label{eq:SS-t3-Cond}
  \begin{gathered}
   (r+1)a_{r+1\,st} - (t+1)\at_{t+1\,sr} = (r+1)a_{r+1\,st} - (s+1)\at_{s+1\,rt} = 0,
\\ \begin{aligned}
    (t+1)b_{t+1\,r\,s} &= 2(s+1)a_{rst},
&   (s+1)b_{r\,s+1\,t} &= 2(t+1)a_{rst,}
\\  (t+1)b_{r\,t+1\,s} &= 2(s+1)\at_{rst},
&   (s+1)b_{s+1\,r\,t} &= 2(s+1)\at_{rst}.
   \end{aligned}
  \end{gathered}
 \end{equation}
If we let
 \begin{equation} \label{eq:SS-t3-BDefn}
  b_{rst} = \bfrac{(n+1)!}{r!\,s!\,t!}^2 B_{rst},
\quad
  r+s+t = n+1,
 \end{equation}
it suffices to find the number of independent $B_{rst}$ satisfying $B_{rst}+B_{str}+B_{trs}=0$, $B_{r+1\,st} = B_{s\,t+1\,r}$.
The number of operators at twists two and three is given by
 \begin{align} \label{eq:SS-t2t3-number}
  \Num{\nrep000} &=
   \begin{cases}
    1 & \text{$n$ even,}
\\  0 & \text{$n$ odd;}
   \end{cases}
\quad&
  \Num{\nrep010} &=
   \begin{cases}
    \tfrac13(n+3) & \text{for $n=3m$,}
\\  \tfrac13(n+5) & \text{for $n=3m+1$,}
\\  \tfrac13(n+1) & \text{for $n=3m+2$,}
   \end{cases}
 \end{align}
with generating functions
 \begin{align}
  \gen000(t) &= \frac1{1-t^2}\,,
&
  \gen010(t) &= \frac{1+t-t^2}{(1-t)(1-t^3)}\,.
 \end{align}
This gives the number of operators for the first few values of $n$ to be
 \begin{equation}
  \begin{aligned}
   \text{\emph{Twist 2:} } & 1, 0, 1, 0, 1, 0, 1, \dots
\\ \text{\emph{Twist 3:} } & 1, 2, 1, 2, 3, 2, 3, \dots
  \end{aligned}
 \end{equation}
As only single-trace operators are present, these numbers should --- and do --- agree with both the figures in
 Tables 10 \& 11 (listing single-trace operators)
and
 Tables 6--9 (giving all operators)
in Appendix~C of \cite{rf:0609179}.

\subsection{Twist 4}

At twist 4, operators may belong to $\nrep020$ or $\nrep101$, and additionally may have single- or double-trace structure.

\subsubsection{Operators in the $\dynk020$ representation}

We construct single-trace, twist-4 \shorten88 semi-short operators belonging to $\nrep020$,
beginning with the most general form for such an operator
 \begin{multline}
  \Op = \sum_{q,r,s,t} a_{qrst}\Tr{\dZ\dZ[r]\dl\dl*} + \at_{qrst}\Tr{\dZ\dZ[r]\dl*\dl}
\\[-8pt] + \ah_{qrst}\Tr{\dZ\dl\dZ[r]\dl*} - ib_{qrst}\Tr{\dZ\dZ[r]\dZ[s]\dY[t]}.
 \end{multline}
By considering the action of the $SU(2)$ raising operator on $\Op$, we have the condition
 \begin{equation} \label{eq:SS-020t4-1bcyc}
  b_{qrst} + b_{rstq} + b_{stqr} + b_{tqrs} = 0.
 \end{equation}
From \eqref{eq:SS-S1Algebra}, \eqref{eq:SS-S2Algebra}, we obtain the commutators of $\Op$ with $\So,\St*$,
 \begin{align}
 &\begin{aligned}[b]
   \comm\So\Op &= \sum_{q,r,s,t}
       2i\bigl((r+1)a_{q\,r+1\,st}+(t+1)\at_{t+1\,qsr}\bigr)\Tr{\dZ\dl*[r]\dl\dl*}
\\[-12pt]&\qquad\qquad\begin{aligned}[b]
     + 2i\bigl((r+1)\at_{q\,r+1\,st}+(t+1)\ah_{t+1\,qsr}\bigr)&\Tr{\dZ\dl*[r]\dl*\dl}
\\   - 2i\bigl((r+1)\ah_{q\,r+1\,st}-(t+1)a_{t+1\,qsr}\bigr)&\Tr{\dZ\dl\dl*[r]\dl*}
\\   + 2\left((t+1)b_{t+1\,qrs}-2(s+1)a_{qrst}\right)&\Tr{\dZ\dZ[r]\dY\dl*}
\\   + 2\left((t+1)b_{qr\,t+1\,s}+2(s+1)\at_{qrst}\right)&\Tr{\dZ\dZ[r]\dl*\dY}
\\   + 2\left((t+1)b_{r\,t+1\,qs}-2(s+1)\ah_{qrst}\right)&\Tr{\dZ\dY\dZ[r]\dl*},
   \end{aligned}
  \end{aligned}
\\
 &\begin{aligned}[b]
   \comm{\St*}\Op &= \sum_{q,r,s,t}
       2i\bigl((s+1)a_{s+1\,qrt}+(r+1)\at_{q\,r+1\,st}\bigr)\Tr{\dZ\dl[r]\dl*\dl}
\\[-12pt]&\qquad\qquad\begin{aligned}[b]
     + 2i\bigl((s+1)\at_{s+1\,qrt}+(r+1)\ah_{r+1\,qst}\bigr)&\Tr{\dZ\dl*\dl[r]\dl}
\\   - 2i\bigl((s+1)\ah_{q\,s+1\,rt}-(r+1)a_{q\,r+1\,st}\bigr)&\Tr{\dZ\dl[r]\dl\dl*}
\\   + 2\left((s+1)b_{qr\,s+1\,t}-2(t+1)a_{qrst}\right)&\Tr{\dZ\dZ[r]\dl\dY[t]}
\\   + 2\left((s+1)b_{s+1\,qrt}+2(t+1)\at_{qrst}\right)&\Tr{\dZ\dZ[r]\dY[t]\dl}
\\   + 2\left((s+1)b_{q\,s+1\,rt}-2(t+1)\ah_{qrst}\right)&\Tr{\dZ\dl\dZ[r]\dY[t]},
   \end{aligned}
  \end{aligned}
 \end{align}
and thus require
 \begin{gather}
  \begin{aligned} \label{eq:SS-020t4-1set1}
   (r+1)a_{q\,r+1\,st}+(t+1)\at_{t+1\,qsr} &= (s+1)a_{s+1\,qrt}+(r+1)\at_{q\,r+1\,st} = 0,
\\ (r+1)\at_{q\,r+1\,st}+(t+1)\ah_{t+1\,qsr} &= (s+1)\at_{s+1\,qrt}+(r+1)\ah_{q\,r+1\,st} = 0,
\\ (r+1)\ah_{q\,r+1\,st}+(t+1)a_{t+1\,qsr} &=  (s+1)\ah_{q\,s+1\,rt}-(r+1)a_{q\,r+1\,st} = 0,
  \end{aligned}
\\\begin{aligned} \label{eq:SS-020t4-1set2}
   2(s+1)(t+1)a_{qrst} &= (t+1)^2b_{t+1\,qrs} = (s+1)^2b_{qr\,s+1\,t},
\\-2(s+1)(t+1)\at_{qrst} &= (t+1)^2b_{qr\,t+1\,s} = (s+1)^2b_{s+1\,qrt},
\\ 2(s+1)(t+1)\ah_{qrst} &= (t+1)^2b_{r\,t+1\,qs} = (s+1)^2b_{q\,s+1\,rt}.
  \end{aligned}
 \end{gather}
To solve these, we set
 \begin{equation} \label{eq:SS-020t4-1Bdef}
  b_{qrst} = \bfrac{(n+1)!}{q!\,r!\,s!\,t!}^2B_{qrst},
\quad
  q+r+s+t=n+1,
 \end{equation}
then \eqref{eq:SS-020t4-1set1}, \eqref{eq:SS-020t4-1set2} are satisfied if
 \begin{equation} \label{eq:SS-020t4-1Beqn}
  B_{qrst} = B_{rs\,t+1\,q-1} = B_{s\,t+1\,q\,r-1} = B_{t+1\,qr\,s-1}.
 \end{equation}
The cyclic identity \eqref{eq:SS-020t4-1bcyc} becomes
 \begin{equation} \label{eq:SS-020t4-1Bcyc}
  B_{qrst} + B_{rstq} + B_{stqr} + B_{tqrs} = 0.
 \end{equation}
We may now find the number of semi-short operators with spin $n$ by counting the degrees of freedom in choosing
$B_{qrst}$, subject to \eqref{eq:SS-020t4-1Bdef}, \eqref{eq:SS-020t4-1Beqn}, \eqref{eq:SS-020t4-1Bcyc}.
Without imposing any restrictions, the initial number of $qrst$ satisfying \eqref{eq:SS-020t4-1Bdef} is
 \begin{equation} \label{eq:SS-020t4-1initDeg}
  \binom{n+4}3 = \frac16(n+2)(n+3)(n+4).
 \end{equation}
Next we consider the number of distinct cycles within the set of all such $qrst$, and hence the number of restrictions coming from \eqref{eq:SS-020t4-1Bcyc}.
This turns out to be given by
 \begin{equation} \label{eq:SS-020t4-1cycRestric}
  \begin{cases*}
   \tfrac1{24}(n+5)(n^2+4n+9) & \text{when $n+1=4m$,}
\\ \tfrac1{24}(n+3)(n^2+6n+11)& \text{when $n+1=4m+2$,}
\\ \tfrac1{24}(n+2)(n+3)(n+4) & \text{for $n+1$ odd.}
  \end{cases*}
 \end{equation}
Finally, we divide the set of $B_{qrst}$ into equivalence classes of those set equal by application of \eqref{eq:SS-020t4-1Beqn}.
By considering the number and size of the resulting classes, we find the number of restrictions arising from \eqref{eq:SS-020t4-1Beqn} to be:
 \begin{equation} \label{eq:SS-020t4-1eqnRestric}
  \begin{cases*}
   \tfrac18(n+1)(n+3)(n+4) & \text{for $n+1$ even,}
\\ \tfrac18(n+4)(n^2+4n+2) & \text{when $n+1=4m+1$,}
\\ \tfrac18(n^3+8n^2+18n+4)& \text{when $n+1=4m+3$.}
  \end{cases*}
 \end{equation}
Thus, subtracting \eqref{eq:SS-020t4-1cycRestric}, \eqref{eq:SS-020t4-1eqnRestric} for \eqref{eq:SS-020t4-1initDeg}, we obtain
 \begin{equation} \label{eq:SS-020t4-1number}
  \Num[\ntr]{\nrep020} = 
  \begin{cases}
   \tfrac18(n+4)^2 & \text{for $n=4m$,}
\\ \tfrac18(n+3)^2 & \text{for $n=4m+1$,}
\\ \tfrac18(n+4)^2+\tfrac12 & \text{for $n=4m+2$,}
\\ \tfrac18(n+3)^2-\tfrac12 & \text{for $n=4m+3$.}
  \end{cases}
 \end{equation}

Similarly, we construct the most general double-trace $\nrep020$ operator,
 \begin{multline}
  \Op = \sum_{q,r,s,t} a_{qrst}\Tr{\dZ\dZ[r]}\Tr{\dl\dl*} + \ah_{qrst}\Tr{\dZ\dl}\Tr{\dZ[r]\dl*}
\\[-8pt] - ib_{qrst}\Tr{\dZ\dZ[r]}\Tr{\dZ[s]\dY[t]}.
 \end{multline}
Because of the cyclic nature of the trace, we need consider only $a_{(qr)st}, b_{(qr)st}$, symmetric in the first two indices.
By considering the action of the $SU(2)$ raising operator, we obtain the condition
 \begin{equation} \label{eq:SS-020t4-b2cyc}
  b_{(qr)st} + b_{(qr)ts} + b_{(st)qr} + b_{(st)rq} = 0.
 \end{equation}
Commutators of $\Op$ with $\So,\St*$ give conditions
 \begin{equation} \label{eq:SS-020t4-b2}
  \begin{gathered}
   (r+1)\ah_{q\,r+1\,st} = (t+1)\ah_{q\,t+1\,sr}, \qquad (r+1)\ah_{r+1\,qst} = (s+1)\ah_{s+1\,qrt},
\\ 2(r+1)a_{(q\,r+1)st} = (t+1)\ah_{t+1\,qsr} = (s+1)\ah_{q\,s+1\,rt},
\\ 2(s+1)(t+1)a_{(qr)st} = (t+1)^2b_{(qr)t+1\,s} = (s+1)^2b_{(qr)s+1\,t},
\\ (s+1)(t+1)\ah_{qrst} = (t+1)^2b_{(r\,t+1)qs} = (s+1)^2b_{(q\,s+1)rt}.
  \end{gathered}
 \end{equation}
We find that with $B$ defined as in \eqref{eq:SS-020t4-1Bdef}, all the \eqref{eq:SS-020t4-b2cyc}, \eqref{eq:SS-020t4-b2} 
are satisfied if
 \begin{gather} \label{eq:SS-020t4-B2cyc}
  B_{(qr)(st)} + B_{(st)(qr)} = 0,
\\ \label{eq:SS-020t4-B2eqn}
  B_{(qr)st} = B_{(qr)t+1\,s-1} = B_{(s\,t+1)q\,r-1} = B_{(s\,t+1)r\,q-1}.
 \end{gather}
Taking the number of independent $B_{(qr)st}$ for fixed $n$, and subtracting the number of restrictions arising from 
\eqref{eq:SS-020t4-B2cyc}, \eqref{eq:SS-020t4-B2eqn}
gives the number of double-trace operators with spin $n$, namely
 \begin{align} \label{eq:SS-020t4-2number}
  \Num[{\ntr[double]}]{\nrep020} &=
  \begin{cases}
   \tfrac1{16}(n+6)^2-\tfrac14 & \text{for $n=4m$,}
\\ \tfrac1{16}(n+3)^2 & \text{for $n=4m+1$,}
\\ \tfrac1{16}(n+6)^2 & \text{for $n=4m+2$,}
\\ \tfrac1{16}(n+3)^2-\tfrac14 & \text{for $n=4m+3$.}
  \end{cases}
\\\intertext{%
Combining \eqref{eq:SS-020t4-1number}, \eqref{eq:SS-020t4-2number}, 
we find the total number of twist-4 \shorten88 semi-short operators belonging to $\nrep020$,%
}
 \label{eq:SS-020t4-number}
  \Num{\nrep020} &= 
  \begin{cases}
   \tfrac1{16}(n+4)(3n+16) & \text{for $n=4m$,}
\\ \tfrac3{16}(n+3)^2 & \text{for $n=4m+1$,}
\\ \tfrac1{16}(n+4)(3n+16)+\tfrac34 & \text{for $n=4m+2$,}
\\ \tfrac3{16}(n+3)^2-\tfrac34 & \text{for $n=4m+3$.}
  \end{cases}
 \end{align}

\subsubsection{Operators in the \dynk101 representation}

A single trace operator in the $\nrep101$ representation takes the general form
 \begin{multline}
  \Op = \sum_{q,r,s,t}
   a_{qrst}\ep\Tr{\DZ[q]i\DZ[r]j\dl\dl*}
  + \at_{qrst}\ep\Tr{\DZ[q]i\DZ[r]j\dl*\dl}
\\[-12pt]\qquad
  + \ah_{qrst}\ep\Tr{\DZ[q]i\dl\DZ[r]j\dl*}
  + \tfrac12i\,b_{qrst}\ep\ep[kl]\Tr{\DZ[r]i\DZ[s]j\DZ[t]k\DZ[q]l}
\\+ i\,c_{qrst}\Tr{\dl[q]\dl\dl*[r]\dl*}
  + i\,\hat{c}_{qrst}\Tr{\dl[q]\dl*[r]\dl\dl*},
 \end{multline}
where we can only determine $b,\hat{c}$ up to $b_{qrst}+b_{stqr}$, $\hat{c}_{qrst}+\hat{c}_{stqr}$,
so may impose that they are symmetric with respect to these indices.
%
%
%
By construction, $\Op$ is an $SU(2)$ highest weight state.
Commutation with $\So,\St*$ requires
 \vspace{-1.06ex}%
 \begin{equation} \label{eq:SS-101t4-Cond}
 \begin{gathered}
  \begin{aligned}
   (t+1)a_{t+1\,qsr} + (r+1)\ah_{q\,r+1\,st} &= 2(q+1)c_{qrst},
\\-(t+1)\ah_{t+1\,qsr} + (r+1)\at_{q\,r+1\,st} &= 2(q+1)c_{srqt},
\\ (t+1)\at_{t+1\,qsr} - (r+1)a_{q\,r+1\,st} &= 2(q+1)\bigl(\hat{c}_{qrst} + \hat{c}_{stqr}\bigr),
\\ (s+1)\ah_{q\,s+1\,rt} - (r+1)a_{q\,r+1\,st} &= 2(q+1)c_{rtsq},
\\-(s+1)\at_{s+1\,qrt} - (r+1)\ah_{r+1\,qst} &= 2(q+1)c_{rqst},
\\ (s+1)a_{s+1\,qrt} - (r+1)\at_{q\,r+1\,st} &= 2(q+1)\bigl(\hat{c}_{sqrt} + \hat{c}_{rtsq}\bigr),
  \end{aligned}
\\\begin{gathered}
   (s+1)a_{qrst} + (r+1)\ah_{qsrt} = -(s+1)\at_{qrts} - (r+1)\ah_{sqtr} = \tfrac12(t+1)b_{t+1\,qrs},
\\ (r+1)\at_{sqrt} - (s+1)\ah_{rqst} = -(r+1)a_{sqtr} + (s+1)\ah_{qrts} = \tfrac12(t+1)b_{q\,t+1\,rs}.
  \end{gathered}
 \end{gathered}
 \end{equation}
%
The $\dynk101$ double trace operator may be written
 \vspace{-1.06ex}%
 \begin{multline}
  \Op = \sum_{q,r,s,t}
   a_{qrst} \ep\Tr{\DZ[q]i\dl}\Tr{\DZ[r]j\dl*}
  + \ah_{qrst} \Tr{\dZ\dY[r]}\Tr{\dl\dl*}
\\[-12pt]\quad
  - i \, b_{qrst} \Tr{\dZ\dZ[r]}\Tr{\dY\dY[t]}
  - i \, \bh_{qrst} \Tr{\dZ\dY}\Tr{\dZ[r]\dY[t]}
\\
  + i \, c_{qrst} \Tr{\dl[q]\dl}\Tr{\dl*[r]\dl*}
  + i \, \hat{c}_{qrst} \Tr{\dl[q]\dl*[r]}\Tr{\dl\dl*},
 \end{multline}
where we may impose $b_{qrst} = b_{(qr)(st)}, \bh_{qrst} = \bh_{rqts}, c_{qrst} = -c_{srqt} = -c_{qtsr} = c_{stqr}, \hat{c}_{qrst} = \hat{c}_{stqr}$.
With this convention, the $SU(2)$ condition gives $\ah_{qrst} = -\ah_{rqst}$ and $2b_{qrst} + \bh_{qsrt} + \bh_{rsqt} = 0$,
and the commutation relations require
 \vspace{-1.06ex}%
 \begin{gather} \label{eq:SS-101t4-2Cond}
  \begin{gathered}
  \begin{aligned}
   (r+1)\ah_{r+1\,qst} - (t+1)a_{t+1\,qsr} &= 2(q+1)\hat{c}_{qrst},
\\ (q+1)\ah_{q+1\,rst} + (s+1)a_{r\,s+1\,qt} &= 2(r+1)\hat{c}_{qrst},
\\ (r+1)a_{q\,r+1\,st} - (t+1)a_{q\,t+1\,sr} &= 2(q+1)c_{qrst},
\\ (s+1)a_{s+1\,rqt} - (q+1)a_{q+1\,rst} &= 2(r+1)c_{qrst},
  \end{aligned}
\\
  \begin{aligned}
  -(t+1)a_{sqrt} - (s+1)a_{tqsr} = (t+1)a_{qsrt} + (s+1)a_{qtrs} &= \tfrac12(r+1)b_{q\,r+1\,st},
\\ (s+1)a_{qtsr} + (t+1)\ah_{qstr} = (t+1)\ah_{qsrt} - (s+1)a_{tqrs} &= \tfrac12(r+1)\bh_{q\,r+1\,st}.
  \end{aligned}
  \end{gathered}
 \end{gather}
%
%
%
The number of independent solutions to \eqref{eq:SS-101t4-Cond}, \eqref{eq:SS-101t4-2Cond} are given in \tabref{tb:n4}
%
%
, along with the results 
\eqref{eq:SS-020t4-1number}, \eqref{eq:SS-020t4-2number}, \eqref{eq:SS-020t4-number}
previously obtained for twist 4 operators%
.
Corresponding generating functions are given in \tabref{tb:g4}.

\begin{table}[!b]
 \centering
 \caption{Number of twist-4 \shorten88 operators} \label{tb:n4}
 \begin{tabular}{|c||c|c||c|}
  \hline
   & $\nrep020$ & $\nrep101$ & Total
  \uberstrut
\\\hline\hline
   Single trace
&   \begin{tablecell}
     \frac18(n+4)^2
 \\  \frac18(n+3)^2
 \\  \frac18(n+4)^2+\frac12
 \\  \frac18(n+3)^2-\frac12
    \end{tablecell}
&   \begin{tablecell}
     \frac18(3n^2+12n+8)
 \\  \frac18(3n^2+14n+23)
 \\  \frac18(3n^2+12n+12)
 \\  \frac18(3n^2+14n+19)
    \end{tablecell}
&   \begin{tablecell}
     \frac12(n+2)(n+3)
 \\  \frac12(n+2)(n+3)+1
 \\  \frac12(n+2)(n+3)+1
 \\  \frac12(n+2)(n+3)
    \end{tablecell}
\\\hline
   Double trace
&   \begin{tablecell}
     \frac1{16}(n+6)^2-\frac14
 \\  \frac1{16}(n+3)^2
 \\  \frac1{16}(n+6)^2
 \\  \frac1{16}(n+3)^2-\frac14
    \end{tablecell}
&   \begin{tablecell}
     \frac1{16}(3n^2+8n)
 \\  \frac1{16}(3n^2+14n+31)
 \\  \frac1{16}(3n^2+8n+4)
 \\  \frac1{16}(3n^2+14n+27)
    \end{tablecell}
&   \begin{tablecell}
     \frac14(n+2)(n+3)+\frac12
 \\  \frac14(n+2)(n+3)+1
 \\  \frac14(n+2)(n+3)+1
 \\  \frac14(n+2)(n+3)+\frac12
    \end{tablecell}
\\\hline\hline
   Total
&   \begin{tablecell}
     \frac1{16}(3n^2+28n+64)
 \\  \frac1{16}(3n^2+18n+27)
 \\  \frac1{16}(3n^2+28n+76)
 \\  \frac1{16}(3n^2+18n+15)
    \end{tablecell}
&   \begin{tablecell}
     \frac1{16}(9n^2+32n+16)
 \\  \frac1{16}(9n^2+42n+77)
 \\  \frac1{16}(9n^2+32n+28)
 \\  \frac1{16}(9n^2+42n+65)
    \end{tablecell}
&   \begin{tablecell}
     \frac34(n+2)(n+3)+\frac12
 \\  \frac34(n+2)(n+3)+2
 \\  \frac34(n+2)(n+3)+2
 \\  \frac34(n+2)(n+3)+\frac12
    \end{tablecell}
\\\hline
 \end{tabular}
\\\footnotesize{The four entries in each cell indicate the results for $n\equiv0,1,2,3\pmod4$.}
\end{table}

\begin{table}[!b]
 \centering
 \caption{Generating functions for number of twist-4 \shorten88 operators} \label{tb:g4}
 \begin{tabular}{|c||c|c||c|}
  \hline
   & $\nrep020$ & $\nrep101$ & Total
  \uberstrut
\\\hline\hline
   Single trace
&   \begin{formcell}
     \frac{2+t^2-t^3-t^4+t^5}{(1-t)^3(1+t)^2(1+t^2)}
    \end{formcell}
&   \begin{formcell}
     \frac{1+4t+t^3-t^5+t^6}{(1-t)^3(1+t)^2(1+t^2)}
    \end{formcell}
&   \begin{formcell}
     \frac{3-2t+2t^2-2t^3+t^4}{(1-t)^3(1+t^2)}
    \end{formcell}
\\\hline
   Double trace
&   \begin{formcell}
     \frac{2-t+t^2-t^3-t^4+t^5}{(1-t)^3(1+t)^2(1+t^2)}
    \end{formcell}
&   \begin{formcell}
     \frac{3t-t^2+t^3-t^5+t^6}{(1-t)^3(1+t)^2(1+t^2)}
    \end{formcell}
&   \begin{formcell}
     \frac{2-2t+2t^2-2t^3+t^4}{(1-t)^3(1+t^2)}
    \end{formcell}
\\\hline\hline
   Total
&   \begin{formcell}
     \frac{4-t+2t^2-2t^3-2t^4+2t^5}{(1-t)^3(1+t)^2(1+t^2)}
    \end{formcell}
&   \begin{formcell}
     \frac{1+7t-t^2+2t^3-2t^5+2t^6}{(1-t)^3(1+t)^2(1+t^2)}
    \end{formcell}
&   \begin{formcell}
     \frac{5-4t+4t^2-4t^3+2t^4}{(1-t)^3(1+t^2)}
    \end{formcell}
\\\hline
 \end{tabular}
\end{table}

As at lower twists, we compare the figures given by these formul\ae{} for low $n$ with those in \cite{rf:0609179}, Appendix~C,
now considering separately the single-trace and overall operator counts.
%
The overall count agrees for both $SU(4)_R$ representations; 
however, in order to compare single-trace $\nrep101$ operators we must first take account of the fact that the tables list only primary operators,
whilst our count also includes descendant operators arising as a result of the decomposition of a long supermultiplet into semi-short contributions
at the unitarity bound, \eqref{eq:N=4-ADecomp}.
\begin{figure}
 \centering
 \caption{Descendant $\nrep101$ operators} \label{fg:desc101}
 \begin{tabular}{c@{\qquad\qquad}c}
  \xymatrix@!=14pt{
   & {\nrep[n+1]000} \ar[ld]\ar[rd]
\\  {\hrep100n{n+1}} \ar[rd]&& {\hrep001{n+1}n} \ar[ld]
\\ & *+[F]{\nrep101} }
 &
  \xymatrix@!=14pt{
   & {\hrep001{n+1}n} \ar[ld]\ar@{.}[rd(0.5)] && {\hrep100n{n+1}} \ar[rd]\ar@{.}[ld(0.5)]
\\  *+[F]{\nrep101} \ar@{.}[rd(0.5)] &&&& *+[F]{\nrep101} \ar@{.}[ld(0.5)]
\\ &&&}
 \end{tabular}
\end{figure}

\figref{fg:desc101} shows the ways in which $\nrep101$ operators may appear as descendants.
To check the number of primary operators, we thus need the number of $\nrep000$ operators, given in \eqref{eq:SS-t2t3-number},
and also the number of \mitx{\hrep001{n+1}n} operators and their conjugates, which may be computed as in the previous twist 3 case, 
giving
 \begin{equation} \label{eq:SS-100t3-number}
  \Num{\hrep001{n+1}n} = \Num{\hrep100n{n+1}} =
  \begin{cases}
   \frac13n & \text{for $n=3m$,}
\\ \frac13(n+2) & \text{for $n=3m+1$,}
\\ \frac13(n+4) & \text{for $n=3m+2$.}
  \end{cases}
 \end{equation}
Note that this figure itself includes both primary operators and those themselves descended from \mitx{\nrep[n+1]000}.
Next, we find the number of descendant operators in the \dynk101 representation,
 \begin{equation}
  \begin{aligned}
   \Num[\ntr\desc]{\nrep101}
   & = \Num[\prim]{\nrep[n+1]000} + \Num[\prim]{\hrep001{n+1}n} + \Num[\prim]{\hrep100n{n+1}}
\\ & = 2\cdot\Num{\hrep001{n+1}n} - \Num{\nrep[n+1]000}
\\[4pt] & =
   \begin{doublecases}
    \frac23n & n=6m,  & \frac23(n-\frac32) & n=6m+3,
\\  \frac23(n+\frac12) & n=6m+1, & \frac23(n+2) & n=6m+4,
\\  \frac23(n+4) & n=6m+2, & \frac23(n+\frac52) & n=6m+5.
   \end{doublecases}
  \end{aligned}
 \end{equation}
Thus by subtraction, the number of primaries may be calculated, dependent on \mbox{$n\bmod12$}:
 \begin{equation}
  \Num[\ntr\prim]{\nrep101} = \frac1{24}\times
  \begin{doublecases}[-2pt]
   \scriptstyle 9n^2+20n+24 & \scriptstyle n=12m,     & \scriptstyle 9n^2+20n+36 & \scriptstyle n=12m+6,
\\ \scriptstyle 9n^2+26n+61 & \scriptstyle n=12m+1,   & \scriptstyle 9n^2+26n+49 &  \quad\cdot
\\ \scriptstyle 9n^2+20n-28 &  \quad\cdot & \scriptstyle 9n^2+20n-40 &  \quad\cdot 
\\ \scriptstyle 9n^2+26n+81 &  \quad\cdot & \scriptstyle 9n^2+26n+93 &  \quad\cdot 
\\ \scriptstyle 9n^2+20n-8  &  \quad\cdot & \scriptstyle 9n^2+20n+4  &  \quad\cdot 
\\ \scriptstyle 9n^2+26n+29 &  \quad\cdot & \scriptstyle 9n^2+26n+17 & \scriptstyle n=12m+11.
  \end{doublecases}
 \end{equation}
This gives $1,4,2,10,9,\dots$, as required to agree with \cite{rf:0609179}.
The generating function for primary $\dynk101$ single-trace operators is then
 \begin{equation}
  \gen[s.t.prim]101(t) = \frac{1+4t+t^2+5t^3+2t^4+t^5+4t^6}{(1-t^2)(1-t^3)(1-t^4)}\,.
 \end{equation}

 \section{Analysis of Partition Functions}
 \label{se:Z}

Here we analyse partition functions for free $\N=4$ super Yang Mills with gauge group $SU(N)$, large $N$, in terms of characters, 
as in \cite{rf:0609179}, Section~7, 
in order to find generating functions for the number of \shorten88 semi-short multiplets in the theory which belong to different representations.
We start from the expansion
 \begin{equation} \label{eq:Z-Msum}
  \partition = \sum_\M N_\M \character*_\M,
 \end{equation}
where $Z$ is a partition function, and the $N_\M = \Num\M$ are integers counting the number of multiplets $\M$ in the relevant sector of the theory.
When $\M$ corresponds to a \shorten88 semi-short multiplet with representation \mitx{\rep{m+2\jb+2}n{m+2j+2}j\jb},
the character $\character*_\M$ is given by $\ch8{nm}\jj(z,y,-a,-b)$, where
 \begin{equation} \label{eq:Z-ch8Defn}
  \ch8{nm}\jj\zyab = \frac{(z+a)(z+b)(y+a)(y+b)}{1-\sigma} a^{2\jb+1}b^{2j+1} \ch4{n+m}m(z,y),
 \end{equation}
with 
$\sigma=\frac{ab}{zy}$ and
 \begin{equation} \label{eq:Z-ch4Defn}
  \ch4{n+m}m(z,y) = (zy)^m \frac{z^{n+1}-y^{n+1}}{z-y} = z^{m+n}y^m+\dots+z^my^{m+n}.
 \end{equation}
It follows from \eqref{eq:Z-ch8Defn}, \eqref{eq:Z-ch4Defn} that
 \begin{equation} \label{eq:Z-ch8Eqv}
  \ch8{nm}\jj\zyab 
   = \sigma^{2J}\,\ch8{n\,m+2J}{j-J\,\jb-J}\zyab, \quad J\le j,\jb\,;
 \end{equation}
thus the character corresponding to $\M\goesto\nrep[\ell]qpq$ is
 \begin{equation} \label{eq:Z-ch8qpq}
  \ch*qpq\ell\zyab = \ch8{p\,q-\ell-2}{\frac12\ell\frac12\ell}\zyab = \sigma^\ell\cdot\ch8{p\,q-2}{00}\zyab,
 \end{equation}
and for $\nrep[\ell]kpq$, $k\ge q$,
 \begin{equation} \label{eq:Z-ch8kpq}
   \ch*kpq{j,\jb}\zyab = \sigma^\ell\cdot\ch8{p\,q-2}{0\,\frac12\!(k-q)}\zyab
 \end{equation}
with $\ell=2j=2\jb-k+q$.
The corresponding result for $k\le q$ is obvious by symmetry.
Thus the contribution to 
the right-hand side of \eqref{eq:Z-Msum}
of all multiplets with $SU(4)$ representation $\dynk{k}pq$ is given by
 \begin{multline} \label{eq:Z-Rsum}
  \sum_\ell \Num{\hrep{k}pq\ell{\ell+k-q}} \cdot \sigma^\ell \cdot \ch8{p\,q-2}{0\,\frac12\!(k-q)}\zyab-
  \\[-6pt] = \ch8{p\,q-2}{0\,\frac12\!(k-q)}\zyab- \cdot \gen{k}pq(\sigma)
 \end{multline}
for $k\ge q$, and similarly for $q\ge k$ with ${}_{0\,\frac12\!(k-q)}\goesto{}_{\frac12\!(q-k)\,0}$.
By symmetry, $\gen{q}pk(\sigma) = \gen{k}pq(\sigma)$.
Summing \eqref{eq:Z-Rsum} over all $k,p,q$ recovers the full partition function,
 \begin{multline} \label{eq:Z-chsum}
  Z\zyab = 
  \sum_{p,q\ge0} \biggl[ \;
   \ch8{p\,q-2}{00}\zyab- \cdot \gen{q}pq(\sigma)
\\[-8pt]
 + \sum_{k>q}
    \Bigl(\ch8{p\,q-2}{0\,\frac12\!(k-q)}\zyab-\\[-12pt]+\ch8{p\,q-2}{\frac12\!(k-q)\,0}\zyab-\Bigr) \gen{k}pq(\sigma)
  \biggr].
 \end{multline}
We may use \eqref{eq:Z-ch8Defn} to write the terms in \eqref{eq:Z-chsum} as
 \begin{equation} \label{eq:Z-ch8zyab}
  \ch8{p\,q-2}{0\jb}\zyab = \frac\sigma{1-\sigma}\zeta\zyab \, a^{2\jb} \, \ch4{p+q}q(z,y),
 \end{equation}
and similarly for the conjugates, with $a^{2\jb}\goesto b^{2j}$,
where $\zeta\zyab$ is given by
 \begin{equation} \label{eq:Z-zetaDefn}
  \begin{aligned}
   \zeta\zyab &= \frac{(z+a)(z+b)(y+a)(y+b)}{zy} \\ &= (1+\sigma)^2zy+\sigma(z^2+y^2)+(1+\sigma)(a+b)(z+y)+(a^2+b^2).
  \end{aligned}
 \end{equation}
Thus \eqref{eq:Z-chsum} becomes
 \begin{multline} \label{eq:Z-ch4sum}
  Z\zyab = \frac\sigma{1-\sigma}\zeta\zyab-
\\ \times \sum_{p,q\ge0} \biggl[ \gen{q}pq(\sigma) + \sum_{r>0} (-1)^r\bigl(a^r+b^r\bigr) \gen{q+r}pq(\sigma) \biggr] \ch4{p+q}q(z,y).
 \end{multline}
The function $\zeta\zyab$ is quadratic in $\zyab*$; 
thus terms of order $\tau$ in $\zyab*$ correspond to the contributions from twist $\tau=k+p+q+2$ multiplets.
For the lowest few twists, the terms in the expansion are given by
 \begin{multline} \label{eq:Z-Expansion}
  Z\zyab = \frac\sigma{1-\sigma}\zeta\zyab-
\times \biggl(
  \gen000(\sigma)
  + (z+y)\gen010(\sigma) \\ - (a+b)\gen100(\sigma)
  + (z^2+zy+y^2)\gen020(\sigma) - (a+b)(z+y)\gen110(\sigma) \\ + zy\,\gen101(\sigma) + (a^2+b^2)\gen200(\sigma)
  + \dotsb \biggr).
 \end{multline}
Thus the generating functions at a given twist $\tau$ may be extracted from a given partition function $Z\zyab$ 
by calculating all order $\tau$ terms present in $Z$ and matching coefficients on both sides.
To allow for the restriction placed on $\zyab*$ by the definition of $\sigma$, and taking account of symmetry, 
it is sufficient to match only the sum of coefficients of terms of the form
 \begin{equation}
  z^uy^va^{\tau-u-v} = \frac1\sigma z^{u-1}y^{v-1}a^{\tau-u-v+1}b = \dots = \frac1{\sigma^v}z^{u-v}a^{\tau-u}b^v
 \end{equation}
with $v\le\min(u,\tau-u)$.

\subsection{Finding generating functions}


For multi-trace operators, the partition function is given by
 \vspace{-0.5ex}%
 \begin{equation} \label{eq:Z-Z8mt}
  \Zmt88\zyab{} - \Zmt22\zyab,
 \end{equation}
where
 \vspace{-1ex}%
 \begin{align} \label{eq:Z-Z88mt}
  \Zmt88\zyab &=
      \prod_{m=1}^\infty \cfrac1{1-\frac{z^m-y^m+a^m+b^m}{1-\sigma^m}}
     \times
      \prod_{n=0}^\infty \frac{(1-z\,\sigma^n)(1-y\,\sigma^n)}{(1-a\,\sigma^n)(1-b\,\sigma^n)}
\,,
\\ \label{eq:Z-Z22mt}
  \Zmt22\zyab &= \frac{(z-a)(z-b)}{(1-\sigma)z(z-y)} \prod_{m=2}^\infty \frac1{1-z^m} \:+\: \swap{z}y \,.
 \end{align}
To match powers of $\zyab*$, the first term in \eqref{eq:Z-Z88mt} may be expanded using $\sum x^n = \frac1{1-x}$,
and the second term by the use of
 \vspace{-0.3ex}%
 \begin{equation} \label{eq:Z-Sproducts}
  \prod_{n=0}^\infty (1-z\,\sigma^n) = \sum_{r=0}^\infty (-z)^r \textbracket{\pip*[i]r} ,
\qquad
  \prod_{m=0}^\infty \frac1{1-a\,\sigma^m} = \sum_{s=0}^\infty a^s \textbracket{\pip[j]s} .
 \end{equation}
We may expand \eqref{eq:Z-Z22mt} as
 \begin{multline}
  \Zmt22\zyab
   = 1 + \frac1{1-\sigma} \sum_{n=2}^\infty \bigl(\parp(n)-\parp(n-1)\bigr)
   \Bigl(\ch4n0(z,y) - (a+b)\ch4{n-1}0(z,y) \\[-8pt] + \sigma\,\ch4{n-2}1(z,y)\Bigr),
 \end{multline}
where $\ch4{n+m}m(z,y)$ is as in \eqref{eq:Z-ch4Defn} and $\parp(n)$ counts the number of integer partitions of $n$.
%
%
%
Putting this in \eqref{eq:Z-Expansion} we obtain, for $SU(4)$ representations $\dynk0p0$ at twist $2,3,4$,
 \begin{equation} \label{eq:Z-gmtLow}
  \begin{gathered}
   \gen[m.t.]000(\sigma) = \frac1{1-\sigma^2},
\qquad
   \gen[m.t.]010(\sigma) = \frac{1+\sigma-\sigma^2}{(1-\sigma)(1-\sigma^3)},
\\
   \gen[m.t.]020(\sigma) = \frac{4-\sigma+2\sigma^2-2\sigma^3-2\sigma^4+2\sigma^5}{(1-\sigma)(1-\sigma^2)(1-\sigma^4)},
  \end{gathered}
 \end{equation}
which agree with the results of \secref{se:SS-Construct}.
For the representation $\nrep[\ell]101$ we obtain
 \begin{equation} \label{eq:Z-gmt101}
  \gen[m.t.]101(\sigma) = \frac{2-\sigma+5\sigma^2+\sigma^3+2\sigma^5}{\sigma(1-\sigma)(1-\sigma^2)(1-\sigma^4)}
   = \frac2\sigma + \frac{1+7\sigma-\sigma^2+2\sigma^3-2\sigma^5+2\sigma^6}{(1-\sigma)(1-\sigma^2)(1-\sigma^4)},
 \end{equation}
which agrees with the result in \tabref{tb:g4} after subtracting the singular term $2/\sigma$, 
which corresponds to the presence of $\Delta=4$ operators belonging to the \shorten44 short multiplet \mitx{\rep20200}.
Similar singularities 
count other \shorten44 and \shorten48 operators,
where following from \cite[(7.34,38)]{rf:0609179} we have the prescription, analogous to \eqref{eq:N=4-Cnm1},
 \begin{equation} \label{eq:Z-shorten}
  \begin{alignedat}{3}
   &\rep{k}pq{\text-\frac12}\jb &&\longrightarrow \text{\shorten48 semi-short operator $\rep{k+1}pq0\jb$,} &\quad\Delta&=\jb+k+p+q+2,
\\ &\rep{q}pq{\text-\frac12}{\text-\frac12} &&\longrightarrow \text{\shorten44 short operator $\rep{q+1}p{q+1}00$,} &\Delta&=p+2q+2.
  \end{alignedat}
 \end{equation}

In the supergravity sector of the $\AdS{}$ dual theory the partition function is given by
 \begin{equation} \label{eq:Z-Z8sugra}
  \Zsugra88\zyab{} - \Zmt22\zyab,
 \end{equation}
 \begin{equation} \label{eq:Z-Z88sugra}
  \Zsugra88\zyab = \prod_{n=0}^\infty \frac{%
   \prod_{k,l=0,k+l\ge1}^\infty (1-a\,z^k\,y^l\,\sigma^n)(1-b\,z^k\,y^l\,\sigma^n) }{%
   \prod_{k,l=0,k+l\ge2}^\infty (1-z^k\,y^l\,\sigma^n) \prod_{k,l=0}^\infty (1-ab\,z^k\,y^k\,\sigma^n)},
 \end{equation}
which may be expanded in powers of $\zyab*$ using \eqref{eq:Z-Sproducts}.
We obtain at lowest twist
 \vspace{-2ex}%
 \begin{equation} \label{eq:Z-gsugraLow}
  \begin{gathered}
   \gen[sugra]000(\sigma) = \gen[sugra]010(\sigma) = 0,
\qquad
   \gen[sugra]020(\sigma) = \frac1{1-\sigma^2},
\\
   \gen[sugra]101(\sigma) = \frac1{\sigma(1-\sigma^2)} = \frac1\sigma + \frac\sigma{1-\sigma^2}.
  \end{gathered}
 \end{equation}
For operators formed from products of symmetrised traces, we have partition functions for single and multi-trace operators
 \vspace{-0.8ex}%
 \begin{align} \label{eq:Z-Z88stsym}
  \Zsym[s.t.]88\zyab &= \prod_{n=0}^\infty \frac{(1-a\,\sigma^n)(1-b\,\sigma^n)}{(1-z\,\sigma^n)(1-y\,\sigma^n)}
   - \frac{z+y-a-b}{1-\sigma} - 1,
\displaybreak[1]\\\label{eq:Z-Z88mtsym}
  \Zsym88\zyab &= \exp \biggl(\sum_{m=1}^\infty \frac1m \Zsym[s.t.]88\zyab[m] \biggr).
 \end{align}
These may be expanded in powers of $\zyab*$ using \eqref{eq:Z-Sproducts}.
In this case we obtain generating functions
 \begin{equation} \label{eq:Z-gsymLow}
  \begin{gathered}
   \gen[m.t.,sym]000(\sigma) = \frac1{1-\sigma^2},
\qquad
   \gen[m.t.,sym]010(\sigma) = \frac{1+\sigma-\sigma^3}{(1-\sigma^2)(1-\sigma^3)},
\\
   \gen[m.t.,sym]020(\sigma) = \frac{3+2\sigma+3\sigma^2-2\sigma^3-2\sigma^4-2\sigma^5+2\sigma^7}{(1-\sigma^2)(1-\sigma^3)(1-\sigma^4)},
\\
   \gen[m.t.,sym]101(\sigma) = \frac1\sigma + \frac{4\sigma+2\sigma^2+4\sigma^3+\sigma^6+\sigma^8}{(1-\sigma^2)^2(1-\sigma^5)}.
  \end{gathered}
 \end{equation}

\subsubsection{Counting long multiplets}

As previously, it is a crucial aspect of the theory that compatible semi-short multiplets may combine into long multiplets.
These are free from the protection afforded by shortening conditions, and thus may acquire anomalous dimensions in the interacting theory.
Identifying which semi-short multiplets may combine is therefore vital if we are to know where these anomalous dimensions may arise.
It is established in \cite{rf:0609179} that all multiplets do combine except those coming from \mitx{\Zsugra88}, 
and we have the further relation
 \begin{equation} \label{eq:Z-Z8long}
  \Zmt88\zyab{}-\Zsugra88\zyab=\left(1-\frac{zy}a\right)\left(1-\frac{zy}b\right) \sum_\M N_{\text{long},\M} \character*_\M,
 \end{equation}
where the factor on the right-hand side may be written
 \begin{equation} \label{eq:Z-longRHS}
  \left(1-\frac{zy}a\right)\left(1-\frac{zy}b\right) = \frac1\sigma\bigl(\sigma-(a+b)+zy\bigr).
 \end{equation}
Thus in an expansion of \eqref{eq:Z-Z8long} in the manner of \eqref{eq:Z-Expansion}, 
generating functions of twist $\tau$ multiplets will contribute at orders $\tau$, $\tau+1$ and $\tau+2$ in $\zyab*$.
Hence calculating all terms up to order $\tau$ yields a system of equations which may be solved 
to determine values for the generating functions of multiplets with twist $\le\tau$.
For twist $2,3,4$ we obtain:
 \begin{equation} \label{eq:Z-glongLow}
  \begin{gathered}
   \gen[long]000(\sigma) = \gen[m.t.]000(\sigma),
\qquad
   \gen[long]010(\sigma) = \gen[m.t.]010(\sigma),
\\
   \gen[long]020(\sigma) = \frac{3+2\sigma^2-2\sigma^3-\sigma^4+\sigma^5}{(1-\sigma)(1-\sigma^2)(1-\sigma^4)},
\\
   \gen[long]101(\sigma) = \frac{1+6\sigma+3\sigma^2+8\sigma^3+3\sigma^4+2\sigma^5+4\sigma^6}{(1-\sigma^2)(1-\sigma^3)(1-\sigma^4)}.
  \end{gathered}
 \end{equation}

\subsection{Behaviour in the $\sigma\tendsto1$ limit}

We now consider the leading order behaviour of the generating functions in the limit as $\sigma\tendsto1$, i.e. for small values of $\epsilon=1-\sigma$.
In this limit $\sigma\goeslike1$, $1-\sigma^m\goeslike m\,\epsilon$; thus from \eqref{eq:Z-ch8zyab} 
we may write
 \begin{gather} \label{eq:Z-zetaLO}
   \zeta\zyab = (z+y+a+b)^2 + O(\epsilon),
 \end{gather}
and the quantities in \eqref{eq:Z-Sproducts} become
 \begin{equation} \label{eq:Z-SproductsLO}
  \pip*r \quad\goeslike\quad  \pip{r} \quad\goeslike\quad \frac1{r!\,\epsilon^r} \,.
 \end{equation}
We will also make use of the multinomial expansion,
 \begin{equation}
  \sum_{u+v+s+t=n} \frac{z^u\,y^v\,(-a)^s(-b)^t}{u!\,v!\,s!\,t!} = \frac{(z+y-a-b)^n}{n!} \,.
 \end{equation}
For the multi-trace partition function \eqref{eq:Z-Z88mt}, we then find that
 \begin{equation} \label{eq:Z-Z8mtLO}
  \evalat{\text{order $\tau$ term}}{\Zmt88\zyab}\goeslike\left(\sum_{n=0}^\tau\frac{(-1)^n}{n!}\right)\bfrac{z+y-a-b}\epsilon^\tau.
 \end{equation}
Using \eqref{eq:Z-zetaLO} and matching terms in \eqref{eq:Z-ch4sum}, \eqref{eq:Z-Z8mtLO}
we find that at twist $\tau=k+p+q+2$, the generating functions are to leading order given by
 \begin{equation} \label{eq:Z-gmtLO}
  \gen[m.t.]kpq(\sigma)
   \goeslike
  \frac{p+1}{\tau-1}\binom{\tau-1}k\binom{\tau-1}q\frac{\pexp\tau(-1)}{(1-\sigma)^{\tau-1}} \,,
  \qquad \sigma\tendsto1,
 \end{equation}
where $\pexp{n}(x)$ is the partial exponential, defined by
 \begin{equation} \label{eq:Z-pexpDefn}
  \pexp{n}(x) = \sum_{r=0}^n \frac{x^r}{r!} \,.
 \end{equation}
\begin{figure}
 \centering
 \caption{Multi-trace generating functions} \label{fg:Z-gmtPlot}
 \begin{tabular}{cc}
  \multicolumn2l{\small Twist 2}
\\\multicolumn2c{%
  \includegraphics[width=0.45\textwidth]{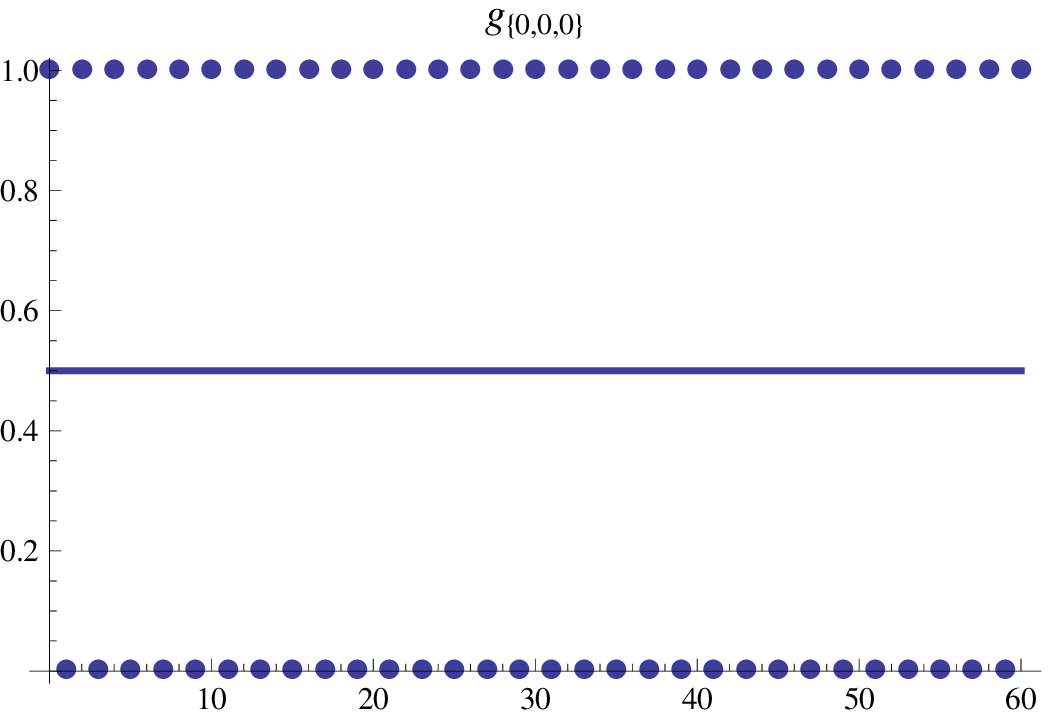}}
\\
  \multicolumn2l{\small Twist 3}
\\\includegraphics[width=0.45\textwidth]{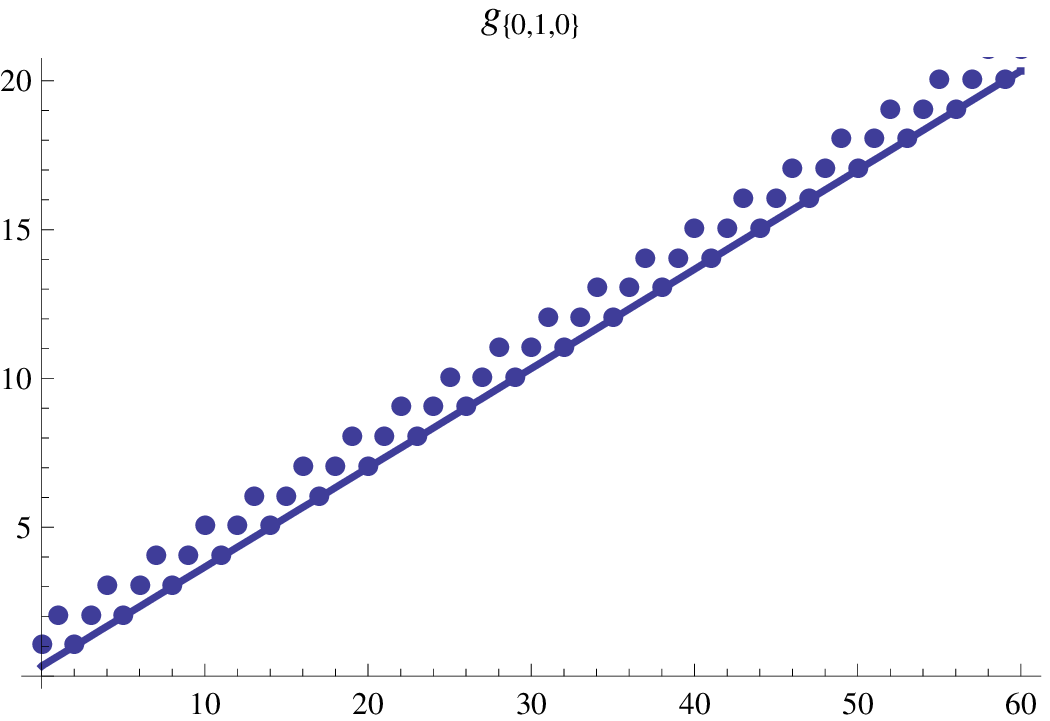}
& \includegraphics[width=0.45\textwidth]{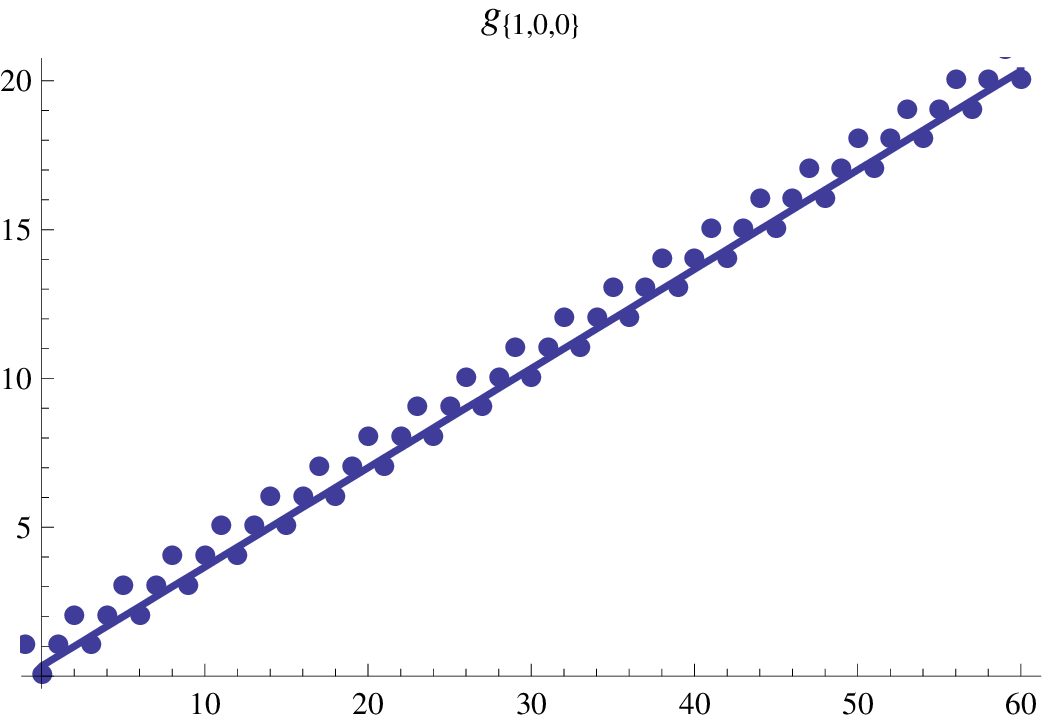}
\\
  \multicolumn2l{\small Twist 4}
\\\includegraphics[width=0.45\textwidth]{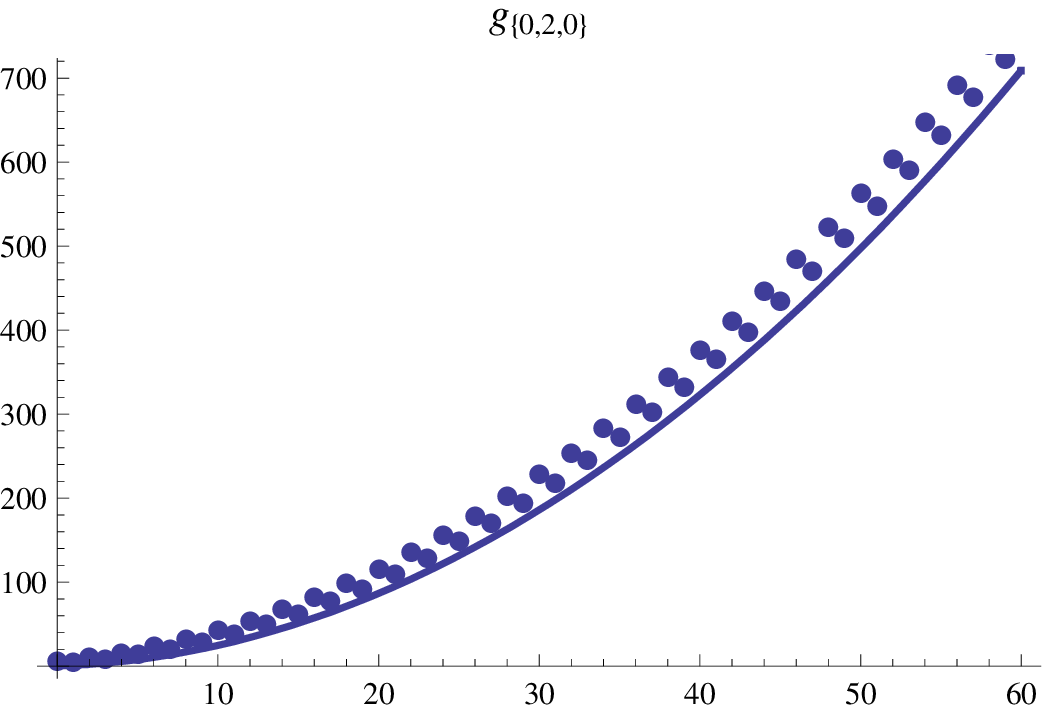}
& \includegraphics[width=0.45\textwidth]{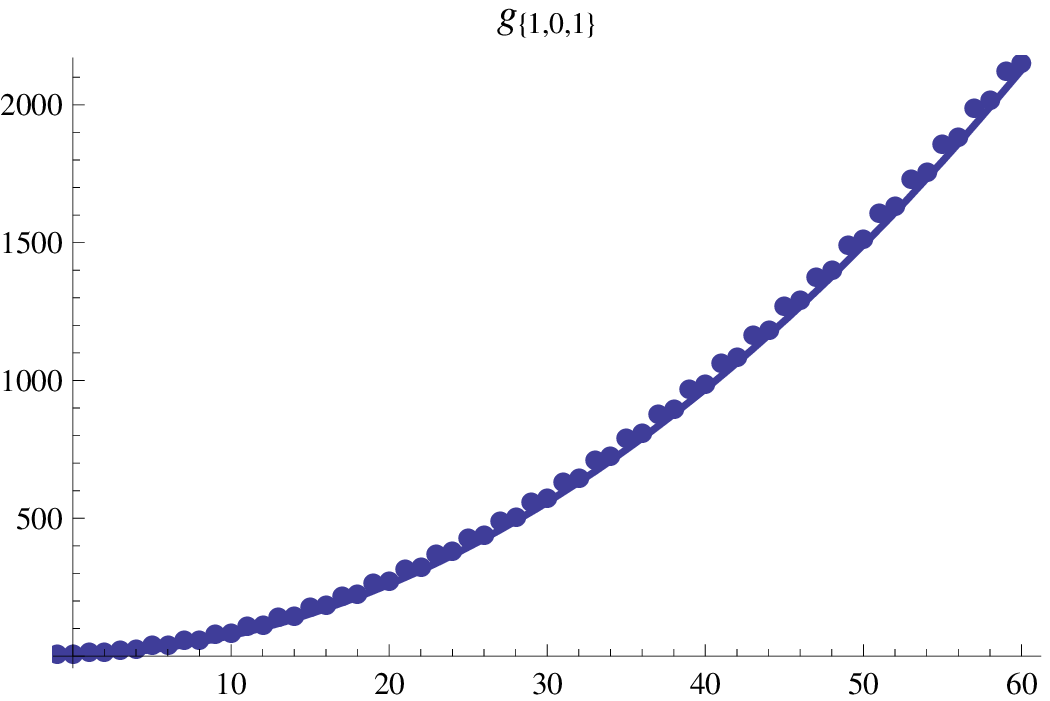}
\\\includegraphics[width=0.45\textwidth]{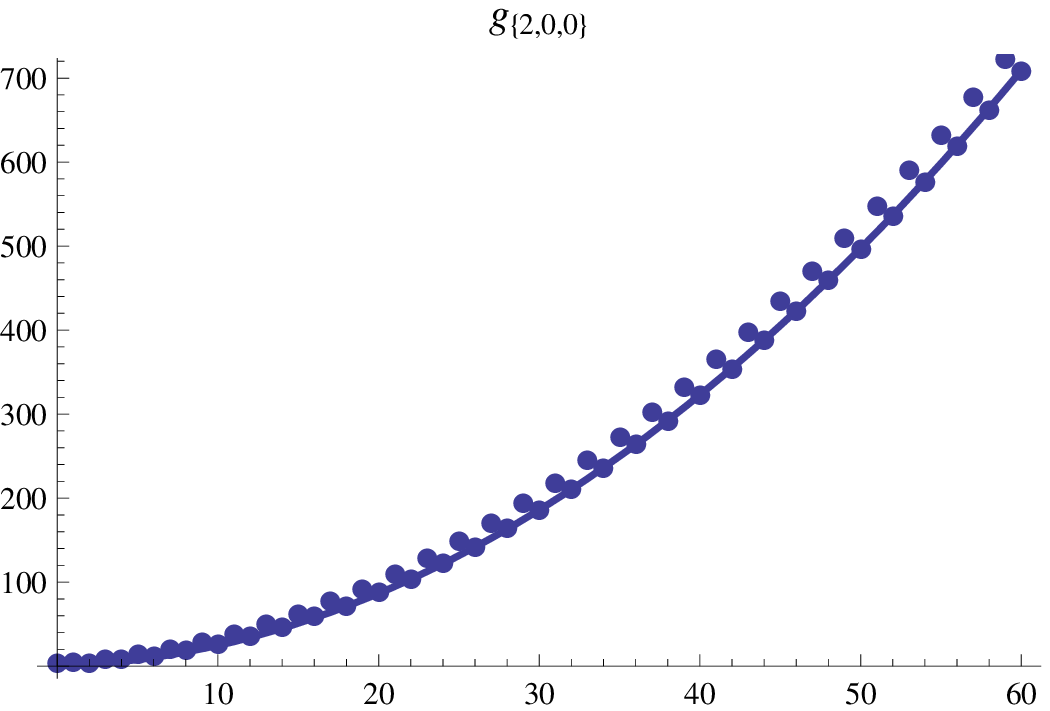}
& \includegraphics[width=0.45\textwidth]{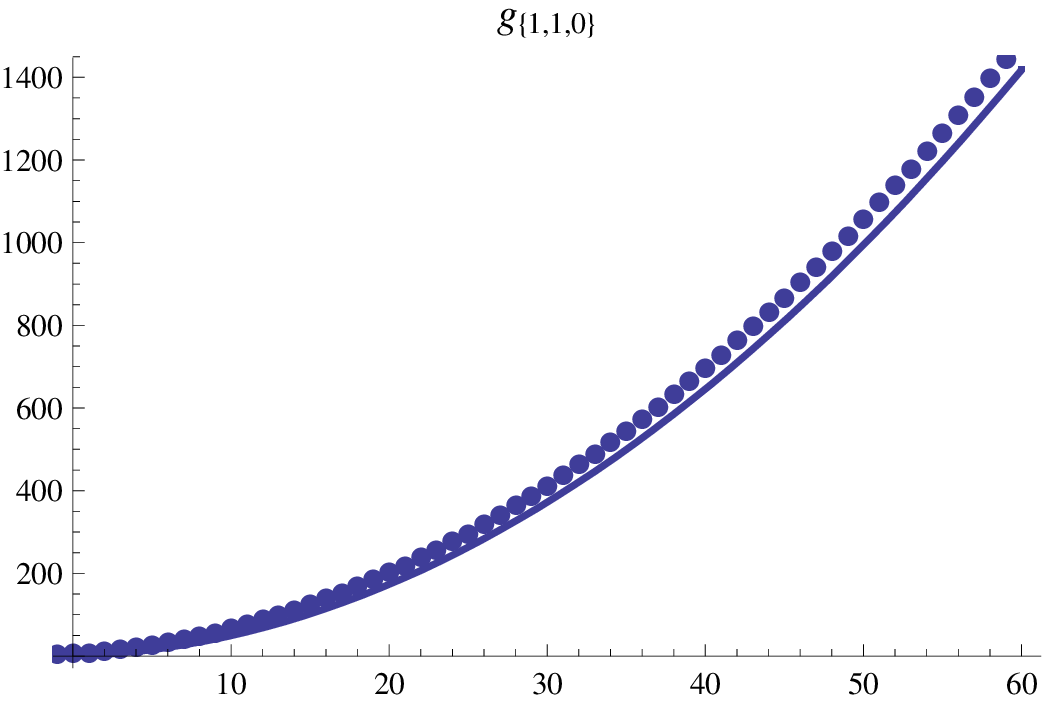}
 \end{tabular}
\end{figure}
For comparison in \figref{fg:Z-gmtPlot} we plot the first few coefficients (up to $\ell=60$) 
in the series expansions of the generating functions found in \eqref{eq:Z-gmtLow}
alongside the corresponding results from their lowest-order approximations \eqref{eq:Z-gmtLO}, 
the latter obtained using the power expansion of $(1-\sigma)^{-\kappa}$,
 \begin{equation} \label{eq:Z--kExpn}
  \frac1{(1-\sigma)^\kappa}
   = \sum_{\ell=0}^\infty \binom{\kappa+\ell-1}\ell \sigma^\ell
   = \sum_{\ell=0}^\infty \binom{\kappa+\ell-1}{\kappa-1} \sigma^\ell.
 \end{equation}

\begin{figure}[th!]
 \centering
 \caption{Supergravity sector generating functions} \label{fg:Z-gsugraPlot}
 \begin{tabular}{cc}
  \multicolumn2l{\small Twist 4}
\\\includegraphics[width=0.45\textwidth]{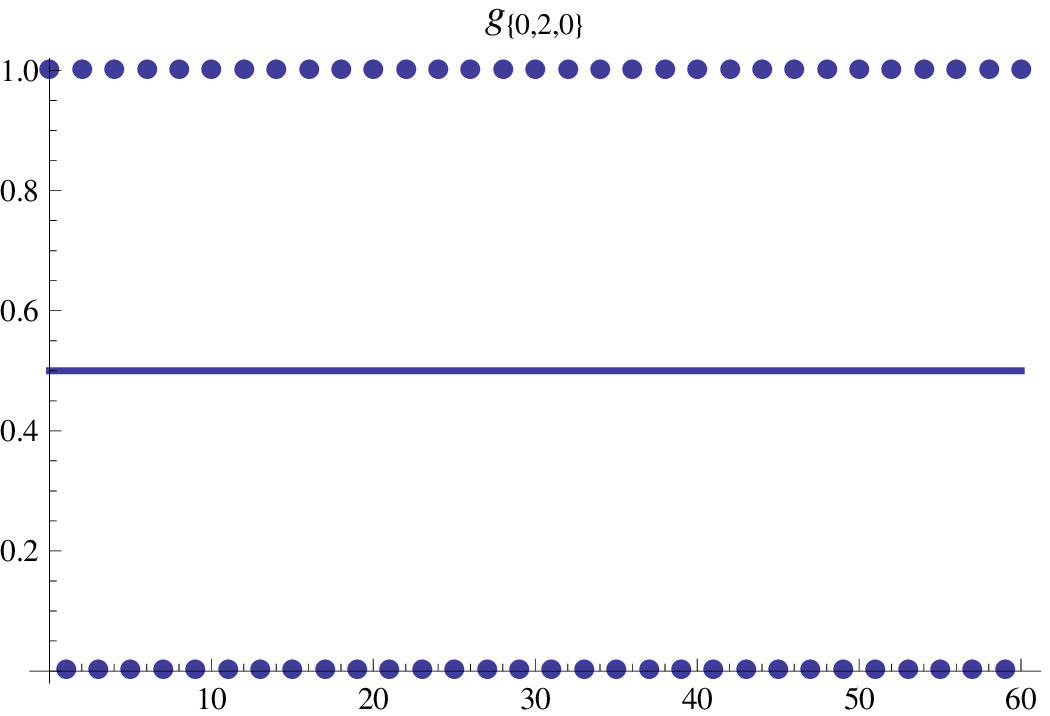}
& \includegraphics[width=0.45\textwidth]{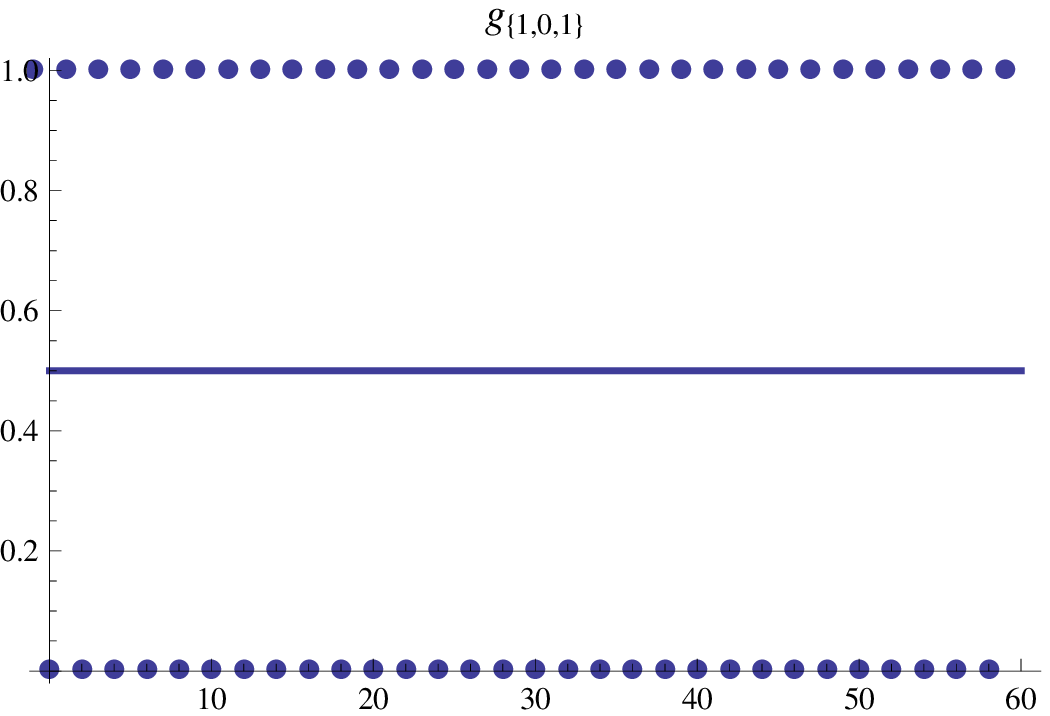}
\\
  \multicolumn2l{\small Twist 5}
\\\includegraphics[width=0.45\textwidth]{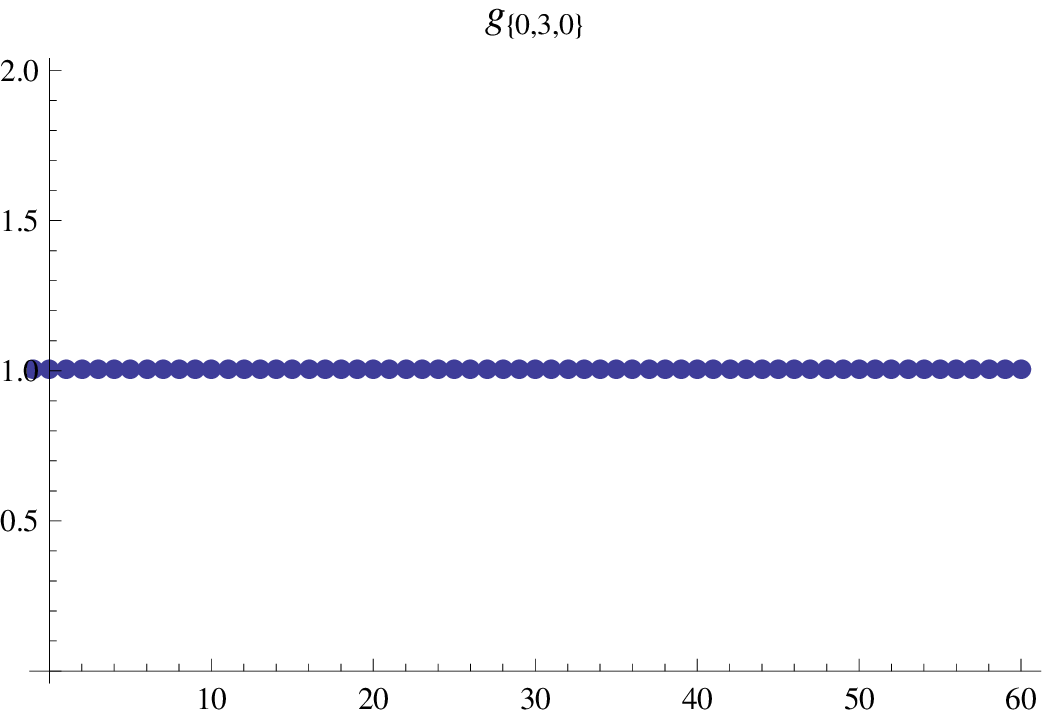}
& \includegraphics[width=0.45\textwidth]{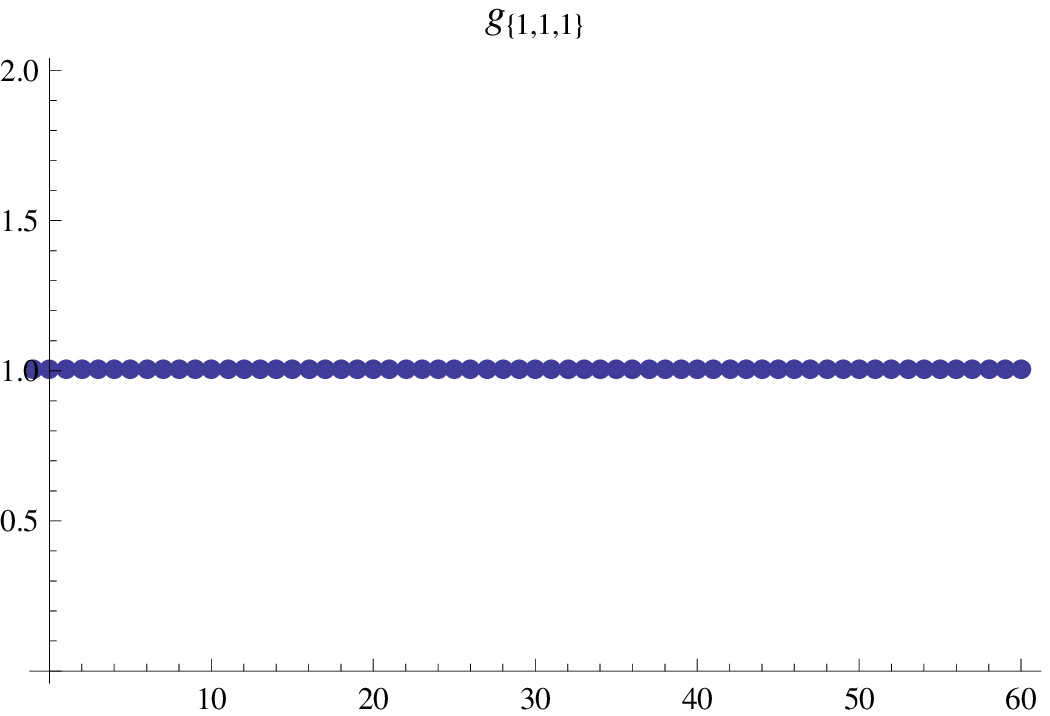}
 \end{tabular}
\end{figure}

In the supergravity sector in the $\sigma\tendsto1$ limit, \eqref{eq:Z-Z8sugra} gives
 \begin{multline} \label{eq:Z-Z8sugraLO}
  \evalat{\text{order $\tau$ term}}{\Zsugra88\zyab}\\\goeslike
  \begin{cases}
\displaystyle
   \frac{(z+y)^{\frac\tau2}}{\bfrac\tau2!}\bfrac{z+y-a-b}\epsilon^{\frac\tau2}
  & \text{$\tau$ even,}
\\[12pt]\displaystyle
   \frac{(z^2+zy+y^2)(z+y)^{\frac{\tau-3}2}}{\bfrac{\tau-3}2!}\bfrac{z+y-a-b}\epsilon^{\frac{\tau-1}2}
  & \text{$\tau$ odd.}
  \end{cases}
 \end{multline}
Thus, matching terms, for $k\ge q$ we have
\newcommand{\Feven}{F_{\text{even}}} \newcommand{\Fodd}{F_{\text{odd}}}
 \begin{equation} \label{eq:Z-gsugraLO}
  \gen[sugra]kpq(\sigma)
   \goeslike
  \begin{cases}
   0 & \tau\le3,
\\[4pt]\displaystyle
    \frac{\Feven\left(\frac\tau2-2,k-q,q\right)-\Feven\left(\frac\tau2-2,k-q,q-1\right)}{\bfrac\tau2!(1-\sigma)^{\frac\tau2-1}}
  & \text{$\tau\ge4$, even,}
\\[12pt]\displaystyle
    \frac{\Fodd\left(\frac{\tau-1}2-2,k-q,q\right)-\Fodd\left(\frac{\tau-1}2-2,k-q,q-1\right)}{\bfrac{\tau-3}2!(1-\sigma)^{\frac{\tau-1}2-1}}
  & \text{$\tau\ge5$, odd.}
  \end{cases}
 \end{equation}
where $\Feven, \Fodd$ are given by
 \begin{align} \label{eq:Z-FevenDefn}
  \Feven(n,r,u) &= \binom{n}r\binom{n+2}u\hyperF32{-n,-(n-r),-u}{r+1,n+3-u}{-1}
\\\label{eq:Z-FoddDefn}
  \Fodd(n,r,u) &= \binom{n}r
   \Biggl\{\binom{n+2}{u-1}\hyperF32{-n,-(n-r),-(u-1)}{r+1,n+4-u}{-1}
\notag\\ & \qquad \qquad \qquad + \binom{n+1}u\hyperF32{-n,-(n-r),-u}{r+1,n+2-u}{-1}\Biggr\}.
 \end{align}
For the general result in \eqref{eq:Z-gsugraLO}, we replace $k-q,q$ with $\abs{k-q},\min(k,q)$.
Note that $\Feven,\Fodd$ are proportional to the binomial coefficient $\binom{n}r$, meaning $\gen[sugra]kpq$ identically vanishes when
$\abs{k-q}>\frac\tau2-2$.
This holds for the exact result, as well as at leading order in the $\sigma\tendsto1$ limit.
Comparison plots between the supergravity sector generating functions \eqref{eq:Z-gsugraLow} and \eqref{eq:Z-gsugraLO} 
appear in \figref{fg:Z-gsugraPlot}.
\begin{figure}[th!]
 \centering
 \begin{tabular}{cc}
  \multicolumn2l{\small Twist 6}
\\\includegraphics[width=0.45\textwidth]{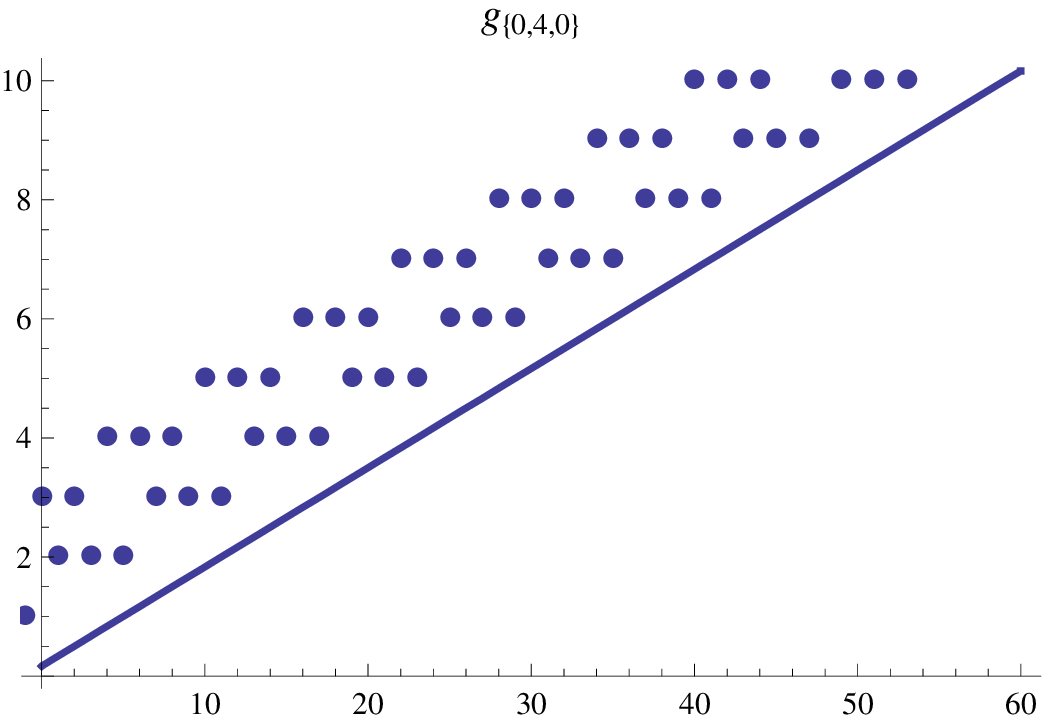}
& \includegraphics[width=0.45\textwidth]{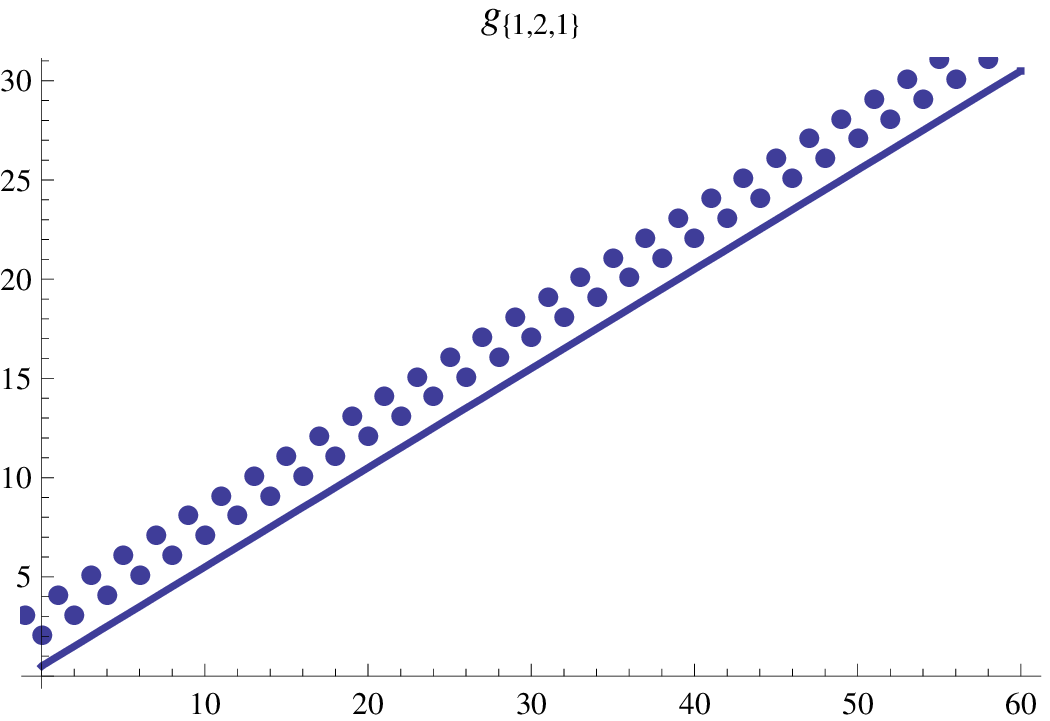}
\\\includegraphics[width=0.45\textwidth]{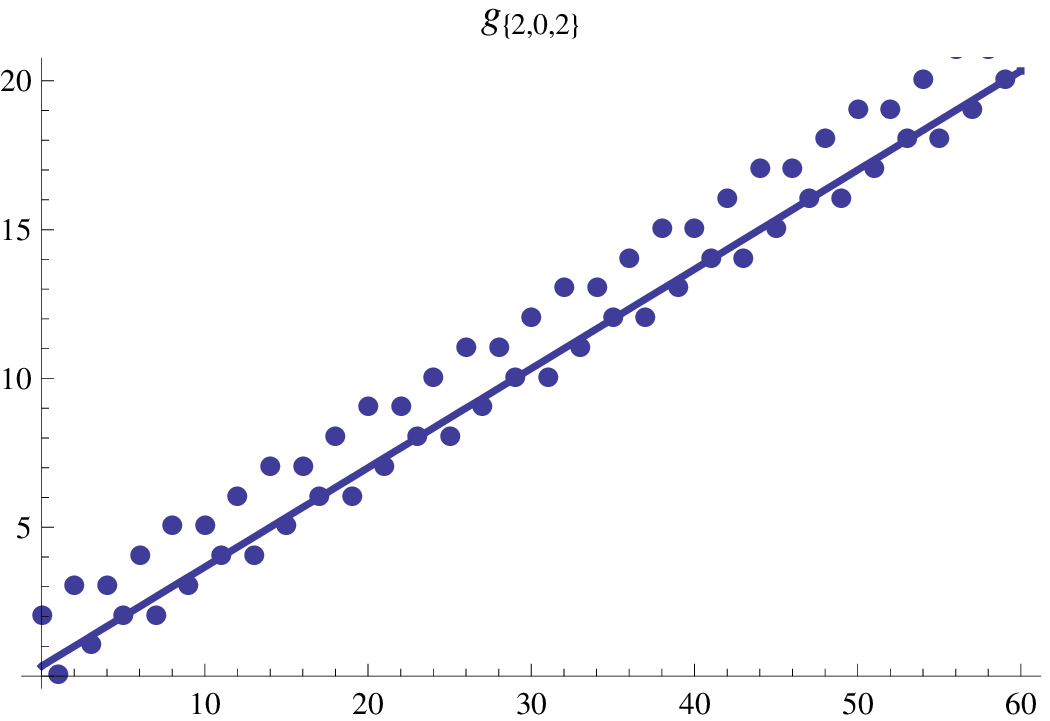}
& \includegraphics[width=0.45\textwidth]{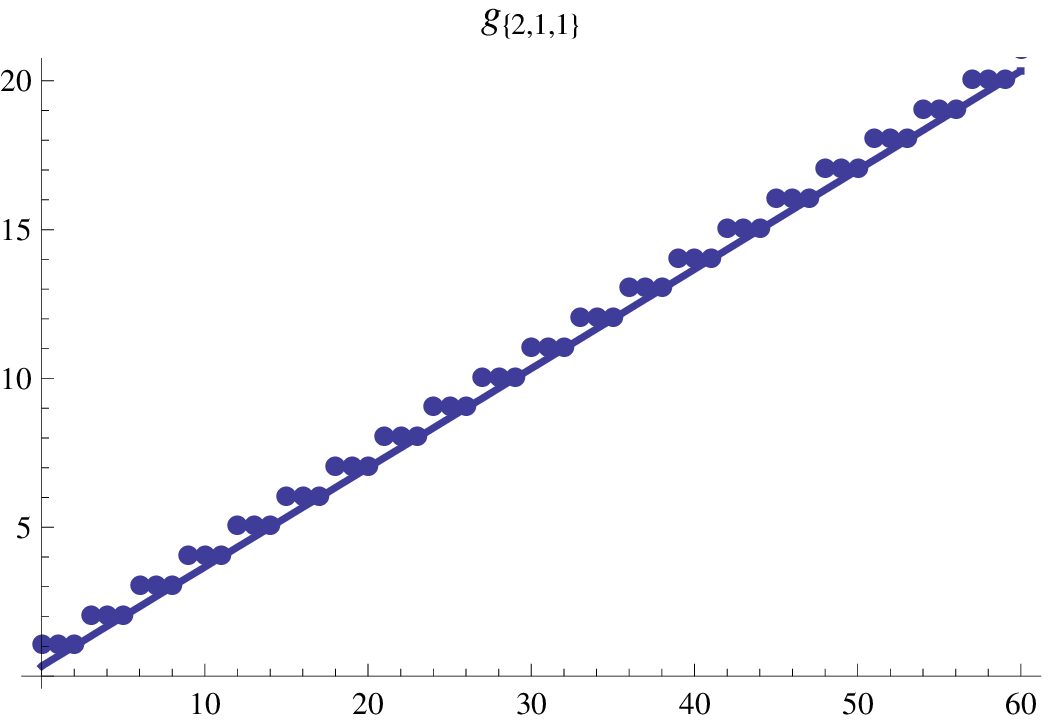}
\\\multicolumn2c{%
  \includegraphics[width=0.45\textwidth]{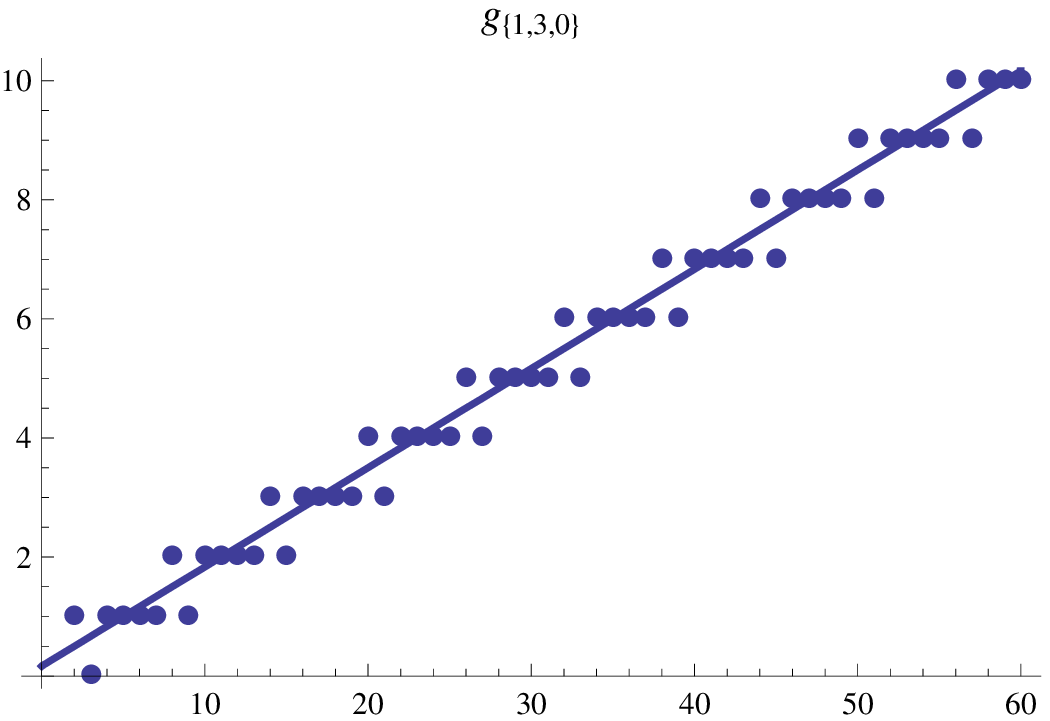}}
 \end{tabular}
\end{figure}

In the case of long multiplets arising from combined semi-short contributions, 
since $\frac\tau2<\tau$, we see from \eqref{eq:Z-Z8long} that comparing \eqref{eq:Z-Z8mtLO}, \eqref{eq:Z-Z8sugraLO},
the leading order contributions to $\gen[long]kpq$ will come from $\Zmt88$ as $\sigma\tendsto1$;
hence we deduce
 \begin{equation} \label{eq:Z-glongLO}
  \gen[long]kpq(\sigma)\goeslike\gen[m.t.]kpq(\sigma)\goeslike \orderof{(1-\sigma)^{1-\tau}}, \quad \sigma\tendsto1.
 \end{equation}

For operators formed from symmetrised traces, \eqref{eq:Z-Z88mtsym} in the $\sigma\tendsto1$ limit gives us
 \begin{equation} \label{eq:Z-Z8symLO}
  \evalat{\text{order $\tau$ term}}{\Zsym88\zyab}
   \goeslike
  \sum_{\substack{r_1+\dots+r_m=\tau,\\r_1,\dots,r_m\ge2}}\frac1{m!\prod_{i=1}^rr_i!}
   \times
  \bfrac{z+y-a-b}\epsilon^\tau,
 \end{equation}
proportional to the corresponding term in \eqref{eq:Z-Z8mtLO} for $\Zmt88$.
Thus in this limit $\gen[m.t.,sym]kpq(\sigma)$ is proportional to $\gen[m.t.]kpq(\sigma)$, with
\newcommand{\Fsym}{F_{\text{sym}}}
\newcommand{\Bell}[1]{B_{#1}}
 \begin{equation} \label{eq:Z-gsymLO}
  \gen[m.t.,sym]kpq(\sigma)
   \goeslike
  \frac{p+1}{\tau-1}\binom{\tau-1}k\binom{\tau-1}q\frac{\Fsym(\tau)}{(1-\sigma)^{\tau-1}} \,,
 \end{equation}
where $\Fsym(\tau)$ is the sum over $r_1,\dots,r_m$ appearing in \eqref{eq:Z-Z8symLO}, 
which may alternatively be expressed in terms of a sum of Bell numbers $\Bell{n}$ by
 \begin{equation} \label{eq:Z-FsymDefn}
  \Fsym(n)
   =
  \frac1{n!}\sum_{r=0}^n(-1)^{n-r}\binom{n}r\Bell{r}
   =
  \frac{(-1)^n}{n!}\left(1-\sum_{r=0}^{n-1}(-1)^r\Bell{r}\right),
 \end{equation}
where the $\Bell{n}$ count the number of ways a set of $n$ elements can be partitioned into nonempty subsets \cite{rf:MW-Bell}.

\subsection{Large $\tau$ behaviour}

We now examine the behaviour at large twists of the $\sigma\tendsto1$ limits for generating functions derived above,
which for convenience we will denote $\lgen{k}pq$, i.e.
 \begin{equation} \label{eq:Z-glimDefn}
  \gen{k}pq(\sigma)\goeslike\lgen{k}pq(\sigma), \quad \sigma\tendsto1.
 \end{equation}
Considering first the multi-trace generating function limit $\lgen[m.t.]kpq$ given in \eqref{eq:Z-gmtLO},
we note that for large $\tau$, $\pexp\tau(-1)\tendsto e^{-1}$.
The asymptotic behaviour of the binomial coefficients $\binom{\tau-1}k\binom{\tau-1}q$ may be derived for given $k,q$, 
using Stirling's approximation
 \begin{equation} \label{eq:Z-Stirling}
  N!\goeslike\sqrt{2\pi}\,N^{N+\frac12}\,e^{-N};
 \end{equation}
however, it may be more natural to consider the generating function for the total number of multiplets in a particular sector of the theory
at a given twist, summing over all representations present.
This we denote $\generatingfunctionform{}\tau$,
 \begin{equation} \label{eq:Z-gtauDefn}
  \sum_{\subbox{k+p+q=\tau-2}}\gen{k}pq(\sigma) = \generatingfunctionform{}\tau(\sigma) \goeslike \tgen(\sigma), \quad \sigma\tendsto1.
 \end{equation}
To find $\lgen[m.t.]kpq$, we require the sum
 \begin{equation} \label{eq:Z-mtBinomSum}
  \sum_{\subbox[0.9]{k+p+q=\tau-2}}\frac{p+1}{\tau-1}\binom{\tau-1}k\binom{\tau-1}q = \binom{2\tau-3}{\tau-1}
  \goeslike \frac{4^\tau}{8\sqrt{\pi\tau}},
 \end{equation}
where the large-$\tau$ behaviour of the right-hand side has been approximated using \eqref{eq:Z-Stirling}.
Application to \eqref{eq:Z-gmtLO} gives the result
 \begin{equation} \label{eq:Z-gmtLOLT}
  \tgen[m.t.](\sigma) \goeslike \frac{4^\tau}{8e\sqrt{\pi\tau}}(1-\sigma)^{1-\tau}, \quad \tau\tendsto\infty.
 \end{equation}
\eqref{eq:Z-mtBinomSum} may also be applied to \eqref{eq:Z-gsymLO} to find the number of operators formed from symmetric traces at large twists.
Here we also need the asymptotic behaviour of $\Fsym(\tau)$; we derive this using \eqref{eq:Z-FsymDefn}, 
and an approximation for the Bell numbers \cite{rf:MW-Bell}, 
$\Bell{n} \goeslike n^{-\frac12} (n/\plog(n))^{n+\frac12} e^{n/\plog(n)-n-1}$,
where the Lambert $W$-function or product-log $\plog(z)$ is the inverse of $\plog\goesto z=\plog e^\plog$.
We obtain
 \begin{equation} \label{eq:Z-gsymLOLT}
  \tgen[m.t.,sym](\sigma) \goeslike \frac{\left(4/\plog(\tau)\right)^{\tau-\frac12}}{4\sqrt2e\pi\tau^2}e^{\tau/\plog(\tau)}
  \, (1-\sigma)^{1-\tau},
 \end{equation}

\subsubsection{The supergravity sector}

In the supergravity sector, firstly for even $\tau$, we note from \eqref{eq:Z-gsugraLO} that the contribution in \eqref{eq:Z-gsugraLO} to 
$\tgen[sugra]$ of all multiplets with fixed $k-q=r>0$ is proportional to
 \begin{equation} \label{eq:Z-sugra-FevenSum}
  \sum_{q=0}^{\subbox{\frac\tau2-\floor{\frac{r}2}-1}}
   \Bigl(\Feven(\tfrac\tau2-2,r,q)-\Feven(\tfrac\tau2-2,r,q-1)\Bigr)
   =\Feven(\tfrac\tau2-2,r,\tfrac\tau2-\floor{\tfrac{r}2}-1),
 \end{equation}
where $\floor{x}$ denotes the integer part of $x$; 
we will drop this from here on, as its significance rapidly decreases at large $\tau$.
Letting $N=\frac\tau2-2$, we write $\Feven$ as a sum, expanding the hypergeometric function in \eqref{eq:Z-FevenDefn} to obtain
 \begin{equation} \label{eq:Z-sugra-FevenExpn}
  \Feven(N,r,N-\tfrac{r}2+1) = \sum_{i=0}^{N-r} \binom{N}i\binom{N}{r+i}\binom{N+2}{\frac{r}2+i+1}.
 \end{equation}
Thus $\tgen[sugra]$, which contains exactly one contribution from a representation $\dynk{k}pq$, \mbox{$k-q=r$}, for each $r=-N,\dots,N$,
is proportional to
 \begin{equation} \label{eq:Z-sugra-Ubersum}
  \sum_{r=-N}^N \Feven(N,\abs{r},N-\frac{r}2+1)
   = \sum_{\subbox{\substack{\abs{r}\le N,\\\abs{s}\le\frac12(N-\abs{r})}}} f(r,s)
 \end{equation}
where 
the summand
 \begin{align} \label{eq:Z-sugra-frs}
  f(r,s) &= \binom{N}{\frac{N-r}2+s}\binom{N}{\frac{N+r}2+s}\binom{N+2}{\frac{N}2+s+1}
\intertext{%
manifestly assumes its maximal value at $r=s=0$, which asymptotically obeys%
}
 \label{eq:Z-sugra-f00Asym}
  f(0,0) &= \binom{N}{\frac{N}2}^2\binom{N+2}{\frac{N}2+1} \goeslike 8^N\,\sqrt{2\bfrac4{\pi N}^3}.
 \end{align}
The second derivatives of $f$ at this point with respect to $r,s$ are respectively
 \begin{equation} \label{eq:Z-sugra-fr2s200}
  \begin{gathered}
   \pdiff[2]rf(r,s) = -f(0,0)\,\polygamma1(\tfrac{N}2+1) \goeslike -2f(0,0) \, \frac1N,
\\
   \pdiff[2]sf(r,s) = -2f(0,0)\left(3\polygamma1(\tfrac{N}2+1)-\frac4{(N+2)^2}\right) \goeslike -2f(0,0) \, \frac6N,
  \end{gathered}
 \end{equation}
where $\polygamma{i}$ is the polygamma function, which has leading order behaviour
 \begin{equation}
  \polygamma0(N+1)\goeslike\log N,
\qquad
  \polygamma1(N+1)\goeslike\frac1N.
 \end{equation}
We then may approximate $f(r,s)$ by a bell-shaped curve,
 \begin{equation} \label{eq:Z-sugra-fAsym}
  f(r,s) \goeslike 8^N\,\sqrt{2\btfrac4{\pi N}^3}\,e^{-(r^2+6s^2)/N}.
 \end{equation}
Near the limits of our sum, $r$ and $s$ are of $\order(N)$, and so the summand is exponentially suppressed.
Thus we may at large $N$ replace the sums with integrals, and take the limits to infinity.
The resulting double Gaussian integral gives the asymptotic behaviour of \eqref{eq:Z-sugra-Ubersum} as
 \begin{equation} \label{eq:Z-sugra-NevenResult}
  \sum_{r,s} f(r,s)
  \goeslike \iint_{\thereals^2} f(r,s) \, \dd{r}\dd{s}
  \goeslike 8^N\,\sqrt{\frac{128}{\pi^3N^3}}\,\frac{\pi N}{\sqrt6} = \frac{8^{N+1}}{\sqrt{3\pi N}} \,.
 \end{equation}
We substitute $N=\frac\tau2-2$ and divide by $\bfrac\tau2!(1-\sigma)^{\frac\tau2-1}$ to obtain $\tgen[sugra](\sigma)$ for $\tau$ even.
For the case of $\tau$ odd, we obtain a similar sum to \eqref{eq:Z-sugra-Ubersum};
however, from \eqref{eq:Z-gsugraLO} we have $N=\frac{\tau-1}2-2$, and from the definition of $\Fodd$ \eqref{eq:Z-FoddDefn}, the summand is now
 \begin{multline} \label{eq:Z-sugra-Odd}
  \binom{N}{\frac{N-r}2+s}\binom{N}{\frac{N+r}2+s}\left[\binom{N+2}{\frac{N}2+s+2}+\binom{N+1}{\frac{N}2+s}\right]
\\
  = f(r,s) \times \left[\frac{\frac{N}2-s+1}{\frac{N}2+s+2}+\frac{\frac{N}2+s+1}{N+2}\right]
  \goeslike \tfrac32 f(r,s),
 \end{multline}
where the approximation is made for $s\ll N$, the source of the dominant contribution to the sum/integral at large $N$.
Thus the result for $\tau$ odd is $\frac32$ times the even result \eqref{eq:Z-sugra-NevenResult}, 
with the substitution $N=\frac{\tau-1}2-2$;
to obtain $\tgen[sugra](\sigma)$, we then divide through by $\bfrac{\tau-3}2!(1-\sigma)^{\frac{\tau-3}2}$.
The full result is hence
 \begin{equation} \label{eq:Z-gsugraLOLT}
  \tgen[sugra](\sigma) \goeslike
  \begin{cases}
   \displaystyle
   \frac{(16e/\tau)^{\frac\tau2}}{(4\pi\sqrt6)\tau}\,(1-\sigma)^{1-\frac\tau2}
   & \text{$\tau\ge4$, even},
\\[12pt] \displaystyle
\frac{(16e/\tau)^{\frac\tau2}}{64\pi\sqrt{\nicefrac2{3\tau}}}\,(1-\sigma)^{\frac32-\frac\tau2}
   & \text{$\tau\ge5$, odd}.
  \end{cases}
 \end{equation}

\subsubsection{Contribution of symmetric $SU(4)_R$ representations}

\begin{figure}
 \centering
 \caption{Large $\tau$ comparison plots} \label{fg:Z-gLargeTPlot}
 \begin{tabular}{c}
  $\tgen[m.t.], \, \tgen[m.t.,sym]$
\\\includegraphics[width=0.90\textwidth]{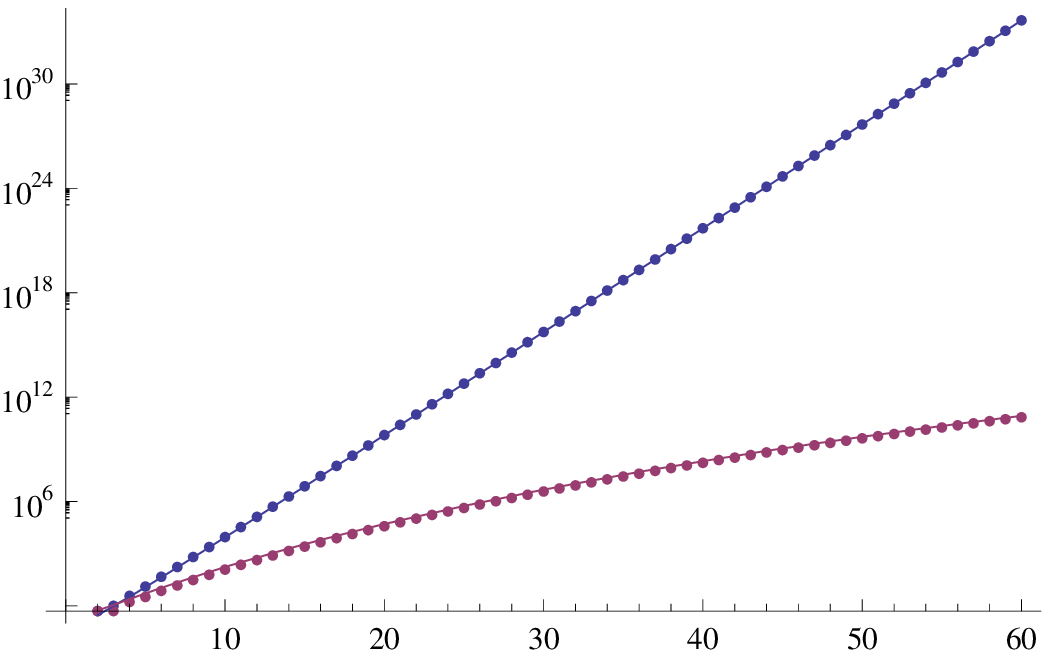}
\\ \\
  $\tgen[sugra]$ (even and odd $\tau$)
\\\includegraphics[width=0.90\textwidth]{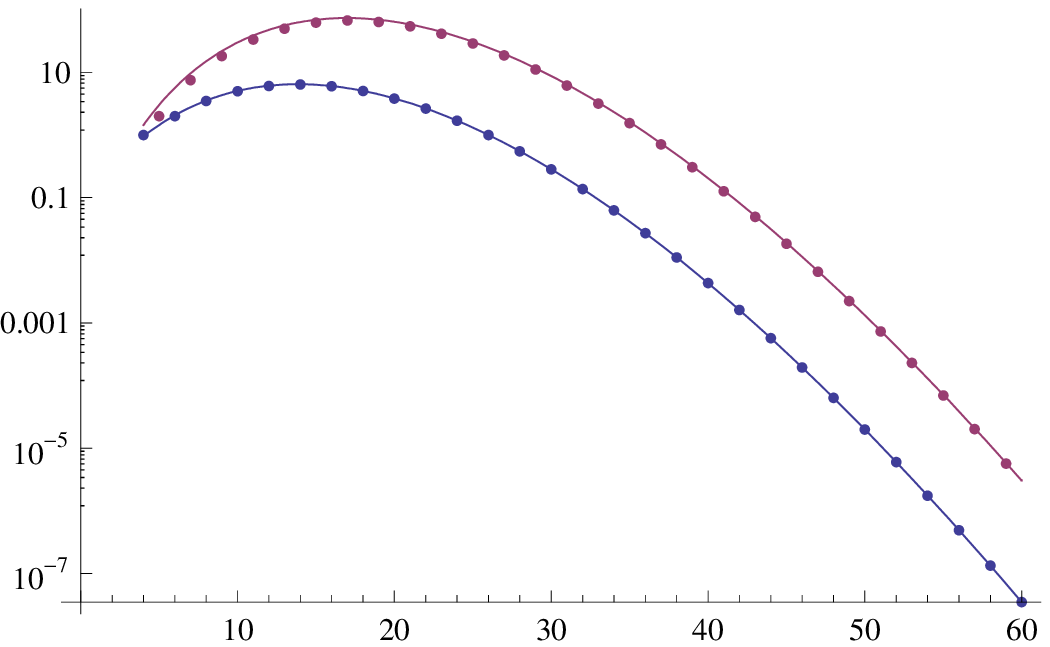}
 \end{tabular}
\end{figure}

We may be interested in counting shortened operators with symmetric $SU(4)_R$ representations, i.e. Dynkin labels $\dynk{q}pq$.
It is notably such representations, arising from the tensor product \eqref{eq:SU4-0p10x0p20},
that appear in the operator product expansion of the chiral primary operators studied in \secref{ch:4pt}.
We proceed by replacing \eqref{eq:Z-gtauDefn} with
 \begin{equation} \label{eq:Z-gqtauDefn}
  \sum_{\subbox{2q+p=\tau-2}}\gen{q}pq(\sigma) = \generatingfunctionform{}{\tau,\text{sym}}(\sigma) \goeslike \qgen(\sigma), \quad \sigma\tendsto1.
 \end{equation}
In the multi-trace case, \eqref{eq:Z-mtBinomSum} becomes
 \begin{equation} \label{eq:Z-mtqBinomSum}
  \sum_{q=0}^{\floor{\frac\tau2}-2}\frac{\tau-2q-1}{\tau-1}\binom{\tau-1}q^2 = \binom{\tau-2}{\floor{\frac\tau2}-1}^2
  \goeslike \frac{4^\tau}{8\pi\tau} \,;
 \end{equation}
hence we divide \eqref{eq:Z-gmtLOLT}, \eqref{eq:Z-gsymLOLT} by $\sqrt{\pi\tau}$ to obtain
 \begin{align} \label{eq:Z-gmtqLOLT}
  \qgen[m.t.](\sigma) &\goeslike \frac{4^\tau}{8e\pi\tau}(1-\sigma)^{1-\tau},
\\\qgen[m.t.,sym](\sigma) &\goeslike \frac{\left(4/\plog(\tau)\right)^{\tau-\frac12}}{4e\sqrt{2\pi^3\tau^5}}e^{\tau/\plog(\tau)}
  \, (1-\sigma)^{1-\tau}.
 \end{align}
To find the corresponding result for operators in the supergravity sector, we drop the summation over $r$ in \eqref{eq:Z-sugra-Ubersum}, 
and evaluate $\sum_s f(0,s)$ with $r=0$.
In our approximation \eqref{eq:Z-sugra-NevenResult} the double integral reduces to a single Gaussian; 
hence we divide \eqref{eq:Z-gsugraLOLT} by $\sqrt{\pi N}\goeslike\sqrt{\smash[b]{\pi\tau/2}}$
to obtain
 \begin{equation} \label{eq:Z-gsugraqLOLT}
  \tgen[sugra](\sigma) \goeslike
  \begin{cases}
   \displaystyle
   \frac{(16e/\tau)^{\frac\tau2}}{4\sqrt{3\pi^3\tau^3}}\,(1-\sigma)^{1-\frac\tau2}
   & \text{$\tau\ge4$, even},
\\[12pt] \displaystyle
\frac{(16e/\tau)^{\frac\tau2}}{64\sqrt{\pi^3/3}}\,(1-\sigma)^{\frac32-\frac\tau2}
   & \text{$\tau\ge5$, odd}.
  \end{cases}
 \end{equation}
Plots of the leading order coefficients in $\tgen(\sigma)$ alongside the approximations
\eqref{eq:Z-gmtLOLT}, \eqref{eq:Z-gsymLOLT}, \eqref{eq:Z-gsugraLOLT} for $\tau\le60$ are shown in \figref{fg:Z-gLargeTPlot}.

\chapter{Results for Operator Product Expansions}
 \label{ch:OPE}

 \section{Series Expansion of $C\xdx{}{}$, Spinless Case}
 \label{se:OPE-SeriesExpn}

We seek an alternative expression for the operator $C\xdx{}{}$ describing the operator product expansion \eqref{eq:OPE} 
in a $d$-dimensional conformal field theory,
which defines the conformal partial waves.
Results in two and four dimensions were presented in \cite{rf:0011040}, and for four and six dimensions in \cite{rf:0309180}.

We start from the standard conformal algebra \eqref{eq:ConformalAlgebra} 
and the operator product expansion for two scalar fields \eqref{eq:OPE}, 
which we write as
 \begin{equation} \label{eq:ope}
  \phi(x)\Op(y) = \CC(s,\partial_y)\Op'(y),
 \end{equation}
where $s=x-y$.
We deduce that for $\phi(x)$ acting on a state
$\kop \equiv \Op(0)\kvac$,
 \begin{equation} \label{eq:ope-state}
  \phi(x)\kop =
  \CC(s,\partial_y)\Op'(y)\ket{0}|_{y=0} =
  \CC(x,-iP)\kopp.
 \end{equation}
We consider the contribution in the case that the operator $\Op'$ is spinless,
assuming $\CC(x,-iP)$ is then also a scalar, commuting with the rotation generator $M_{\mu\nu}$.
First we consider the action of $\Dt$, the generator of dilations in the conformal algebra,\footnote{%
We define $\Dt=-iD$ with respect to \eqref{eq:ConformalAlgebra};
thus $\comm{\Dt}{f(P)}=\PdP f(P)$ and $\Dt\kop = \Dop\kop$.} %
on both sides of \eqref{eq:ope-state}, giving
 \begin{equation}
  (\xd + \Dphi + \Dop)\phi(x)\kop =
  \PdP \, \CC(x,-iP)\kopp + \Dopp\,\CC(x,-iP)\kopp,
 \end{equation}
where 
operators $\phi, \Op, \Op'$ have scale dimensions $\Dphi, \Dop\,, \Dopp$ respectively.
Inserting the OPE \eqref{eq:ope-state} on the left-hand side gives
 \begin{equation} \label{eq:C-action}
  (\xd + \Delta)\CC(x,-iP) = \PdP \, \CC(x,-iP),
 \end{equation}
where $\Delta\equiv\Dphi+\Dop-\Dopp$.
If as in \eqref{eq:OPE} we make the substitution
$\CC = C_{\phi\Op\Op'}(x^2)^{-\frac12\Delta}C$,
with $C$ a function of $x, P$ such that $\evalat{x=P=0}C=1$, 
then \eqref{eq:C-action} becomes
 \begin{equation}
  \Bigl(\xd* - \PdP*\Bigr)C = 0
  \implies
  \deg_x(C) = \deg_P(C).
 \end{equation}
Hence we deduce that $C$ is a function of $\ixp$ and $\xps$ only, which we posit may be expanded as a series
 \begin{equation} \label{eq:C-expn}
  \Cxp=\sum_{n,m}C_{nm}\ixp^n\xps^m.
 \end{equation}

Next, by considering the action of $K_\mu$ on \eqref{eq:ope-state}, and again substituting \eqref{eq:ope} for $\phi(x)\kop$, we obtain
 \begin{equation} \label{eq:K-action-C}
  \comm{K_\mu}{\CC}\kopp =
  i\K\mu{2\Dphi}\CC\kopp.
 \end{equation}
Remembering that
 \begin{equation}
  \partial_\mu\CC =
  \partial_\mu\Bigl(C_{\phi\Op\Op'}(x^2)^{-\frac\Delta2}C\Bigr) =
  \Cmult(x^2)^{-\frac\Delta2}\Bigl(\partial_\mu - \tfrac{\Delta}{x^2}x_\mu\Bigr)C,
 \end{equation}
\eqref{eq:K-action-C} becomes
 \begin{equation} \label{eq:K-action}
  \comm{K_\mu}{C}\kopp =
  i\K\mu{2a}C\kopp,
 \end{equation}
where we have defined $a \equiv \half\left(\Dphi - \Dop + \Dopp\right)$.
%
%
First we consider the right-hand side of \eqref{eq:K-action}, by looking at how a general term of the expansion \eqref{eq:C-expn}
acts under $\K\mu{2a}$, namely
 \begin{multline}
  \K\mu{2a} \left(\ixp^n\xps^m\right)\\ =
  -in \, x^2P_\mu\ixp^{n-1}\xps^m - 2x_\mu(m+n+a)\ixp^n\xps^m;
 \end{multline}
and thus obtaining
 \begin{multline} \label{eq:RHS}
  i\K\mu{2a}C\kopp \\=
  \sum_{n,m}\left((n+1)C_{n+1\,m} \, x^2P_\mu - 2i(n+m+a)C_{nm} \, x_\mu\right)\ixp^n\xps^m.
 \end{multline}
Now we consider the left-hand side of \eqref{eq:K-action}.
In general, 
recalling the standard commutators for the conformal generators and their actions on $\kop$, we can write
 \begin{multline} \label{eq:K-com-op}
  \comm{K_\mu}{P_{\nu_1} \dots P_{\nu_r}}\kopp =
  2 \bigl( (r-1+\Dopp) \met\mu{(\nu_1} P_{\nu_2} \dots P_{\nu_r)}
\\  - \met{(\nu_1}{\nu_2} P_{\nu_3} \dots P_{\nu_r)} P_\mu \bigr)\kopp.
 \end{multline}
The brackets ${}_{(\dots\!)}$ denote a summation over all non-equivalent permutations,
recalling that the $P$'s commute and that the metric $\met\mu\nu$ is symmetric and Euclidean, $\tr{\metric}=d$.
Thus there are $\binom{r}1=r$ contributions from first term, and $\binom{r}2=\frac12 r(r-1)$ from the second.
We may deduce the value of the commutator $\comm{K_\mu}{\ixp^n\xps^m}$ acting on $\kopp$ by contracting \eqref{eq:K-com-op} with
 \begin{equation}
  (-i)^n(x^2)^m\met{\nu_1}{\nu_2}\dots\met{\nu_{2m-1}}{\nu_{2m}}x_{\nu_{2m+1}}\dots x_{\nu_{2m+n}} \qquad \text{(with $r=n+2m$)},
 \end{equation}
to obtain
 \begin{equation}
  \begin{aligned}
  \comm{K_\mu}{\ixp^n\xps^m}\kopp &=
   4m(m+\Dopp-\tfrac{d}{2})x^2P_\mu\ixp^n\xps^{m-1}\kopp
\\   & \quad - 2in(n+2m-1+\Dopp)x_\mu\ixp^{n-1}\xps^m\kopp
\\   & \quad + n(n-1)x^2P_\mu\ixp^{n-2}\xps^m\kopp,
  \end{aligned}
 \end{equation}
and thus finding
 \begin{multline} \label{eq:LHS}
  \comm{K_\mu}{\Cxp}\kopp
\\ \begin{aligned}[b]
   = \sum_{n,m}
   & \Bigl( \bigl(4(m+1)(m+\thing)C_{n\,m+1} + (n+1)(n+2)C_{n+2\,m}\bigr) x^2P_\mu
\\[-10pt] & \quad - 2i(n+1)(n+2m+\Dopp)C_{n+1\,m} \, x_\mu\Bigr) \ixp^n \xps^m \kopp.
 \end{aligned}
\end{multline}
By comparing coefficients in the series expansions \eqref{eq:LHS} and \eqref{eq:RHS}, we find
 \begin{equation}
  \begin{gathered}
   (n+1)(n+2m+\Dopp\,)C_{n+1\,m} = (n+m+a)C_{nm} \,,\\
   4(m+1)(m+\thing)C_{n\,m+1} + (n+1)(n+2)C_{n+2\,m} = (n+1)C_{n+1\,m} \,;
  \end{gathered}
 \end{equation}
thus
 \begin{align} \label{eq:n-plus}
  C_{n+1\,m} &= \frac{m+n+a}{2m+n+\Dopp} \, \frac{C_{nm}}{n+1}
\\\intertext{and} \label{eq:m-plus}
  4(m+1)(m+\thing)C_{n\,m+1}
   &= \frac{m+n+a}{2m+n+\Dopp}\left(1 - \frac{m+n+1+a}{2m+n+1+\Dopp}\right)C_{nm}
\notag\\
   &= \frac{m+n+a}{2m+n+\Dopp}\frac{m+b}{2m+n+1+\Dopp}C_{nm},
 \end{align}
where we define $b \equiv \Dopp - a$.
These relations commute sensibly, i.e.
$C_{nm}\rightarrow C_{n+1\,m}\rightarrow C_{n+1\,m+1}$
and
$C_{nm}\rightarrow C_{n\,m+1}\rightarrow C_{n+1\,m+1}$
are equivalent. Thus we deduce
 \begin{equation}
  C_{nm}
  = \frac{\pochhammer{m+a}{n}}{\pochhammer{2m+\Dopp}{n}} \, \frac{C_{0m}}{n!}
  = \frac{\pochhammer{n+a}{m}\pochhammer{b}{m}}{\pochhammer{n+\Dopp}{2m}\pochhammer{\thing}{m}} \, \frac{C_{n0}}{4^mm!} \,,
 \end{equation}
and hence for general $n,m \in \set{0,1,2,\dotsc}$,
 \begin{equation}
  C_{nm} =
  \frac1{4^mn!m!}\frac{\pochhammer{a}{n+m}\pochhammer{b}{m}}{\pochhammer{\Dopp}{n+2m}\pochhammer{\thing}{m}}C_{00},
 \end{equation}
where since we have defined $C$ such that $C(0,0)=1$, we may substitute $C_{00}=1$.
%
Hence we obtain
 \begin{multline} \label{eq:full}
  \phi(x)\kop = \frac\Cmult{(x^2)^\frac\Delta2} \sum_{m,n \geq 0} \frac1{4^mn!m!} \left(
  \frac{\pochhammer{a}{n+m}\pochhammer{b}{m}}{\pochhammer{\Dopp}{n+2m}\pochhammer{\thing}{m}}
  \right)
  \\ \times
  \ixp^n\xps^m\kopp.
 \end{multline}

\newcommand{\intvar}{w}
As an exercise, we compare \eqref{eq:full} with the result found in \cite
{rf:AP76/161}
, which for the spinless case gives
 \begin{equation} \label{eq:ope-(5.6)}
  \begin{aligned}
   C &= \int_0^1\dd\intvar\, \intvar^{\frac12(\Dphi-\Dop+\Dopp)-1}{(1-\intvar)}^{\frac12(\Dop-\Dphi+\Dopp)-1}\,e^{\intvar\xd}
        \hyperf01{\Dopp-1;-\tfrac{x^2}{4}\intvar(1-\intvar)\partial^2}
\\   &= \int_0^1\dd\intvar\, \intvar^{a-1}{(1-\intvar)}^{b-1} \biggl(\sum_n \frac1{n!}\intvar^n\ixp^n\biggr)
          \biggl(\sum_m \frac{\intvar^m{(1-\intvar)}^m}{4^mm!\pochhammer{\Dopp-1}{m}}\xps^m\biggr).
  \end{aligned}
 \end{equation}
If we extract from the series expansion the coefficient of $\ixp^n\xps^m$, we find
 \begin{equation}
  \begin{aligned}[b]
   \evalat{\begin{subarray}{l} \ixp^n\\ \xps^m \end{subarray}}{C\vphantom{\big|}}
    & = \int_0^1 \dd\intvar\, \intvar^{a-1}{(1-\intvar)}^{b-1}
        \left(\frac1{n!}\intvar^n\right)\!\left(\frac{\intvar^m{(1-\intvar)}^m}{4^mm!\pochhammer{\Dopp-1}{m}}\right)\\
    & = \frac1{4^mn!m!}\frac1{\pochhammer{\Dopp-1}{m}} \int_0^1 \dd\intvar\, \intvar^{a+(n+m)-1}{(1-\intvar)}^{b+m-1}.
  \end{aligned}
 \end{equation}
The integral here is just the Beta function $\Beta{a+n+m}{b+m}$; from standard identities of the Beta and Gamma functions, we have
 \begin{gather}
  \Beta{a+n+m}{b+m}
   = \frac{\Gamma(a+m+n)\Gamma(b+m)}{\Gamma(a+b+n+2m)}
   = \frac{\pochhammer{a}{n+m}\pochhammer{b}{m}}{\pochhammer{\Dopp}{n+2m}}\Beta{a}{b}
\\
  \implies C_{nm} =
  \frac1{4^mn!m!}
  \frac{\pochhammer{a}{n+m}\pochhammer{b}{m}}
  {\pochhammer{\Dopp-1}{m}\pochhammer{\Dopp}{n+2m}},
 \end{gather}
from which we recover our result in \eqref{eq:full} in the case that $d=4$.
In the $x^2 \rightarrow 0$ limit, \eqref{eq:full} reduces to
 \begin{equation}
  \phi(x)\kop = \frac\Cmult{(x^2)^\frac\Delta2}\,\hyperf11{a;\Dopp\,;-ix \cdot P}\kopp,
 \end{equation}
which reproduces \cite
{rf:AP76/161}, again in the spinless case.

\subsection{Differential Equation} \label{se:OPE-DiffEqn}

As an alternative approach, we let 
 \begin{equation} \label{eq:def-ab}
  \alpha \equiv \ixp*,
\qquad \qquad
  \beta  \equiv \xps*, 
 \end{equation}
and look to write a differential equation satisfied by $C(\alpha,\beta)$
equivalent to the condition \eqref{eq:K-action}.
Equating \eqref{eq:RHS} and \eqref{eq:LHS}, and multiplying both sides by $x_\mu P^2/2i\beta$,
such that $x^2P_\mu\rightarrow\half\alpha$, $x_\mu \rightarrow -\tfrac{i}{2}$,
leads to the series equation
\pagebreak[2]
 \begin{multline} \label{eq:OPE-CmnabSeries}
  \sum_{n,m}\Bigl\{
  \bigl[2C_{n\,m+1}(m+1)(m+\thing) \\[-8pt]
  \begin{aligned}[b]
   &+ \half C_{n+2\,m}(n+1)(n+2) - \half C_{n+1\,m}(n+1)\bigr]\alpha\\
   &+ \bigl[C_{nm}(n+m+a) - C_{n+1\,m}(n+1)(n+2m+\Dopp)\bigr]
   \Bigr\}\alpha^n\beta^m = 0.
  \end{aligned}
 \end{multline}
Bearing in mind the action of differentiation on a power series,
\eqref{eq:OPE-CmnabSeries}
may be equivalently written
 \begin{multline} \label{eq:diff-ab}
  \Bigl\{
  \half \alpha\daa*
  + 2\beta\dab*
  - 2\alpha\beta\dbb*
  + \left(\Dopp-\half\alpha\right)\da* \\[-2pt]
  - \left(2\alpha(\thing)+\beta\right)\db*
  - a
  \Bigr\}C(\alpha,\beta) = 0.
 \end{multline}
We check that this is satisfied by \eqref{eq:ope-(5.6)}
, namely
 \begin{equation}
  C(\alpha,\beta) =
  \int_0^1 \dd\intvar\, \intvar^{a-1}(1-\intvar)^{b-1}e^{\intvar\alpha}\hyperf01{\thing;\quarter\intvar(1-\intvar)\beta};
 \end{equation}
substituting into \eqref{eq:diff-ab} gives
 \begin{align}
  & \int_0^1 \dd\intvar\, \intvar^{a-1}(1-\intvar)^{b-1}e^{\intvar\alpha} \biggl[\bigl(\half\alpha\intvar^2 + (\Dopp-\tfrac{\alpha}{2})\intvar-a\bigr)
  \hyperf01{\thing;\quarter\intvar(1-\intvar)\beta} \Bob~
  & \qquad + \bigl(2\beta\intvar - (2\alpha(\thing) + \beta)\bigr)\quarter\intvar(1-\intvar)\HYPERF01'(\thing;\quarter\intvar(1-\intvar)\beta) \Bob~
  & \qquad - 2\alpha\beta \tfrac1{16}\intvar^2(1-\intvar)^2\HYPERF01''(\thing;\quarter\intvar(1-\intvar)\beta) \biggr] \Bob/
  =
  & \int_0^1 \dd\intvar\, \intvar^{a-1}(1-\intvar)^{b-1}e^{\intvar\alpha}
    \biggl[\bigl(\Dopp\,\intvar-a-\alpha\intvar(1-\intvar)\bigr)\HYPERF01 + \quarter\intvar(1-\intvar)(2\intvar-1)\beta\HYPERF01' \Bob~
  & \qquad - \half\alpha\intvar(1-\intvar)\Bigl\{\quarter\intvar(1-\intvar)\beta\HYPERF01'' + (\thing)\HYPERF01' - \HYPERF01\Bigr\} \biggr].
\\\intertext{%
Here, $\HYPERF01'$ denotes $\tfrac{d}{dz}\hyperf01{a;z}$, etc; thus the term $\left\{\cdots\right\}$ vanishes, as this is the differential 
equation satisfied by the confluent hypergeometric limit function $\HYPERF01$.
Hence, expressing factors of $\alpha$ and $\beta$ in terms of derivatives with respect to $\intvar$, the above is equal to} \label{eq:diff-int}
  & \int_0^1 \dd\intvar\, \intvar^{a-1}(1-\intvar)^{b-1}
    \Bigl[\Dopp\,\intvar - a - \intvar(1-\intvar)\tfrac{\partial}{\partial\intvar}\Bigr] e^{\intvar\alpha}\HYPERF01 \,, \\
\intertext{which we may integrate by parts; $\intvar^a(1-\intvar)^be^{\intvar\alpha}\HYPERF01$ vanishes on the boundary, 
and thus \eqref{eq:diff-int} is equal to}
  & \int_0^1 \dd\intvar\, \intvar^{a-1}(1-\intvar)^{b-1}\bigl((a+b)\intvar-a\bigr)e^{\intvar\alpha}\HYPERF01 \notag\\
  & \qquad \qquad + \int_0^1 \dd\intvar\, \intvar^{a-1}(1-\intvar)^{b-1}\bigl(a(1-\intvar)-b\intvar\bigr)e^{\intvar\alpha}\HYPERF01
  \:\; = \:\; 0 \QED.
 \end{align}

 \section{The $d=2$ Case}
 \label{se:OPE-2d}

\subsection{The Conformal Algebra in Two Dimensions}

In two dimensions, we can define complex coordinates $\xpm \equiv x_1 \pm ix_2$.
We wish to choose momentum operators $P_+, P_-$ such that $\comm{\Ppm}{\Op} = i\dpm\Op$.
Observing that $\dpm = \half\left(\partial_1 \mp i \partial_2\right)$, we define $\Ppm \equiv \half\left(P_1 \mp i P_2\right)$,
and similarly, $\Kpm \equiv \half\left(K_1 \mp i K_2\right)$.
We then note, from our definitions of $\xpm$, $\Ppm$, that
 \begin{equation} \label{eq:2d-xP}
  x^2=x_+x_-,
  \qquad
  P^2=4P_+P_-,
  \qquad
  x \cdot P=x_+P_++x_-P_-.
 \end{equation}
The angular momentum generator $M_{\mu\nu}$ has only one independent component, which we shall call $J$,
defined such that its commutator with $\Ppm$ is given by $\comm{J}{P_{\pm}} = \pm P_{\pm}$.
The full set of non-vanishing commutators of conformal generators is:
 \begin{equation} \label{eq:2d-algebra}
  \begin{gathered}
   \begin{aligned}
    \comm{J}{\Ppm} &= \pm\Ppm,
    & \comm{\Dt}{\Ppm} &= \Ppm, \\
    \comm{J}{\Kpm} &= \pm\Kpm,
    & \comm{\Dt}{\Kpm} &= -\Kpm,
 \end{aligned} \\
\comm{\Kpm}{\Pmp} = \mp J + \Dt.
\end{gathered}
\end{equation}
Acting on a quasi-primary field $\Op(x_+,x_-)$ of scale dimension $\Dop$, which we will no longer take to be spinless but to have spin $\lop$,
 \begin{equation} \label{eq:2d-ops}
  \begin{gathered}
   \comm{\Ppm}{\Op} = i\dpm\Op, \\
   \begin{aligned}
    \comm{J}{\Op}    &= (x_+\dpl - x_-\dmi + \lop)\Op, \\
    \comm{\Dt}{\Op}  &= (x_+\dpl + x_-\dmi + \Dop)\Op,
 \end{aligned} \\
\comm{\Kpm}{\Op} = i\left(x_+x_-\dpm - \xmp(x_+\dpl + x_-\dmi + \Dop + \lop)\right)\Op.
\end{gathered}
\end{equation}
Thus the state $\kop$ satisfies
 \begin{equation} \label{eq:2d-states}
  J\kop = \lop\kop,
  \qquad
  \Dt\kop = \Dop\kop,
  \qquad
  \Kpm\kop = 0.
 \end{equation}
We define variables
 \begin{align} \label{eq:def-ggb}
  \gamma &\equiv -ix_+P_+,
  &
  \gammb &\equiv -ix_-P_-;\\[-4pt]
\intertext{\vspace{-1ex}then by comparison with \eqref{eq:2d-xP} we have that} \label{eq:ab-ggb}
  \alpha &= \ixp = \gpgb,
  &
  \beta &= \xps = -4\ggb
.\\
\intertext{Noting that $\alpha^2 + \beta = (\gmgb)^2$, we then have}
  \da* &= \frac1\gmgb\left(\gamma\dg - \gammb\dgb\right),
  &
  \db* &= \frac14\frac1\gmgb\left(\dg - \dgb\right),
 \end{align}
where $\dg\equiv\dg*$, $\dgb\equiv\dgb*$.

\subsection{The Spinless Case} \label{se:OPE-2d-Spin0}

We note that $\ggb*$ are well-defined by \eqref{eq:ab-ggb} in any number of dimensions $d$;
thus we may write $\Cgamma = C\bigl(\alpha\ofggb,\beta\ofggb\bigr)$,
and the differential equation \eqref{eq:diff-ab} for the OPE contribution of a spinless operator then becomes
 \begin{multline} \label{eq:diff-ggb-nospin}
  \biggl\{
  \gamma\dgg + \gammb\dgbgb
  + (\Dopp - \gamma)\dg + (\Dopp - \gammb)\dgb \\[-4pt]
  - (1-\tfrac{d}{2})\left(\frac\gpgb\gmgb\right)(\dg-\dgb)
  - 2a
  \biggr\}\Cgamma = 0.
 \end{multline}
In the case that $d=2$, the $1 - \tfrac{d}{2}$ term vanishes identically, and the $\gamma$, $\gammb$ terms in \eqref{eq:diff-ggb-nospin} separate.
If we factorise the solution, $\Cgamma=\chip(\gamma)\chim(\gammb)$, we find
 \begin{gather}
  \bigl\{ \gamma\chip'' + (\Dopp - \gamma)\chip' \bigr\} \chim
  + \chip \bigl\{ \gammb\chim'' + (\Dopp - \gammb)\chim' \bigr\}
  - 2a \chip \chim = 0\\ \label{eq:chip-chim-spin0}
  \implies
  \chip^{-1}\Bigl(\gamma\chip'' + (\Dopp - \gamma)\chip'\Bigr) - a =
  -\chim^{-1}\Bigl(\gammb\chim'' + (\Dopp + \gammb)\chim'\Bigr) + a.
 \end{gather}
The left-hand side of \eqref{eq:chip-chim-spin0} depends on $\gamma$ only, the right-hand side on $\gammb$.
Thus they must both equal some quantity independent of $\ggb*$, namely a constant, $\kappa$, giving
 \begin{equation}
  \begin{aligned}
   \gamma\chip'' + (\Dopp - \gamma)\chip' - (a + \kappa)\chip &= 0,\\
   \gammb\chim'' + (\Dopp - \gammb)\chim' - (a - \kappa)\chim &= 0.
 \end{aligned}
\end{equation}
We recognise the confluent hypergeometric equation;
we note also that $\Cgamma$ must be symmetric in $\gamma$ and $\gammb$, 
and hence necessarily $\kappa = 0$.
Thus we obtain, in the spinless $d=2$ case, the result
 \begin{displaystretch}{0.6}%
 \begin{equation}
  \Cgamma =
\hyperf11{a
; \Dopp; \gamma}
\hyperf11{a
; \Dopp; \gammb}.
 \end{equation}
 \end{displaystretch}%

\subsection{With Spin} \label{se:OPE-2d-Spin}

We repeat the analysis made in \secref{se:OPE-SeriesExpn} of the product of two scalar operators
$\phi$ and $\Op$,
where now we expand in terms of a spin-$\lopp$ operator $\Op'$, and work in $d=2$.
We begin with equation \eqref{eq:ope-state}, 
  $\phi(x)\kop = \CC(x,-iP)\kopp$. 
%
\subsubsection{Commutator with $\Dt$ and $J$}
%
Firstly, we consider the action 
\eqref{eq:2d-algebra}, \eqref{eq:2d-ops}, \eqref{eq:2d-states} 
of operators $\Dt$ and $J$
, giving
 \begin{displaystretch}{0.6}%
 \begin{equation} \label{eq:DJ-action-C-spin}
  \begin{aligned}
   \Dt &:
   & \left(x_+\dpl + x_-\dmi + \Delta\right)\CC\kopp
   &= \left(P_+\partial_{P_+} + P_-\partial_{P_-}\right)\CC\kopp,  \\
   J &:
   & \left(x_+\dpl - x_-\dmi - \lopp\right)\CC\kopp
   &= \left(P_+\partial_{P_+} - P_-\partial_{P_-}\right)\CC\kopp,
  \end{aligned}
 \end{equation}
 \end{displaystretch}%
where as previously, $\Delta \equiv \Dphi + \Dop - \Dopp$. 
We now write $\CC = \Cmult\xhxh\Chat$; then
 \begin{equation} \label{eq:chat-deriv}
  \dpm\CC = \dpm\left(\Cmult\xhxh\Chat\right) = \Cmult\xhxh\left(\tfrac{\hpm}{\xpm} + \dpm\right)\Chat.
 \end{equation}
Hence \eqref{eq:DJ-action-C-spin} implies
 \begin{displaystretch}{0.6}%
 \begin{equation} \label{eq:DJ-action-spin}
  \begin{aligned}
   \left(\hpl + \hmi + x_+\dpl + x_-\dmi + \Delta\right)\Chat\kopp
   &= \left(P_+\partial_{P_+} + P_-\partial_{P_-}\right)\Chat\kopp,  \\
   \left(\hpl - \hmi + x_+\dpl - x_-\dmi - \lopp\right)\Chat\kopp
   &= \left(P_+\partial_{P_+} - P_-\partial_{P_-}\right)\Chat\kopp.
  \end{aligned}
 \end{equation}
 \end{displaystretch}%
Thus if we set $\hpm = -\half(\Delta\mp\lopp) \implies \hpl+\hmi=-\Delta$, $\hpl-\hmi=\lopp$, we deduce
 \begin{gather}
  \begin{aligned}
   \left(x_+\dpl + x_-\dmi\right)\Chat\kopp
   &= \left(P_+\partial_{P_+} + P_-\partial_{P_-}\right)\Chat\kopp,  \\
   \left(x_+\dpl - x_-\dmi\right)\Chat\kopp
   &= \left(P_+\partial_{P_+} - P_-\partial_{P_-}\right)\Chat\kopp.
 \end{aligned}\\ \notag
\implies \quad
\deg_{\xpm}(\Chat) = \deg_{\Ppm}(\Chat)
\quad \implies \quad
\Chat = \Cgamma, \label{eq:chat-expn}
\end{gather}
where $\ggb*$ are defined as in \eqref{eq:def-ggb}.
Relations for $\xpm,\dpm$ with $\ggb*$ are given in \secref{se:2dCoords}.

\subsubsection{Commutator with $\Kpm$}

Considering the action of $\Kpm$ on \eqref{eq:ope-state}, we obtain
 \begin{align}
  \comm{\Kpm}{\CC}\kopp &= i\left(x_+x_-\dpm - \xmp(x_+\dpl + x_-\dmi + \Dphi)\right)\CC\kopp.
\\\intertext{Using \eqref{eq:chat-deriv}, this gives us} \label{eq:K-action-spin}
  \comm{\Kpm}{\Chat}\kopp &
  \begin{aligned}[t]
   &= i\Bigl(\hpm\xmp + x_+x_-\dpm 
    - \xmp(x_+\dpl + x_-\dmi + \hpl + \hmi + \Dphi)\Bigr)\Chat\kopp
  \end{aligned} \notag\\
 &\begin{aligned}[b]
   &= i\left(x_+x_-\dpm - \xmp(x_+\dpl + x_-\dmi + \hmp + \Dphi)\right)\Chat\kopp \\
   &= -i\xmp\left(\xmp\dmp + a \mp \half\lopp\right)\Chat\kopp \\
   &= -i\xmp\left(\xmp(-i\Pmp)\dgbg + a \mp \half\lopp)\right)\Chat\kopp \\
   &= -i\xmp\left(\gorgb\dgbg + a \mp \half\lopp\right)\Chat\kopp.
  \end{aligned}
 \end{align}
where we have used $\dpm=(\dpm\gamma)\dg+(\dpm\gammb)\dgb=(-i\Ppm)\dgbg$,
and $a$ is defined as in \eqref{eq:K-action}, namely $a\equiv\half(\Dopp+\Dphi-\Dop)$.
From the two-dimensional conformal algebra \eqref{eq:2d-algebra}, it is straightforward to show by induction that
 \begin{equation} \label{eq:2d-KPP}
  \comm{\Kpm}{\Ppm^{\qpm}\Pmp^{\qmp}} = \qmp\Ppm^{\qpm}\Pmp^{\qmp-1}\left(\qmp - 1 + \Dt \mp J\right).
 \end{equation}
Details are given in \secref{se:OPE-2d-KP}.
Multiplying \eqref{eq:2d-KPP} by $x_+^{\qnb}x_-^{\qba}$ and acting on $\kopp$ gives
 \vspace{-1ex}%
 \begin{equation}
  \begin{gathered}
  \comm{K_+}{\gamma^{\qnb}\gammb^{\qba}}\kopp =
   -ix_-\qba(\qba-1+\Dopp-\lopp)\gamma^{\qnb}\gammb^{(\qba-1)}\kopp,
\\\comm{K_-}{\gamma^{\qnb}\gammb^{\qba}}\kopp =
   -ix_+\qnb(\qnb-1+\Dopp+\lopp)\gamma^{(\qnb-1)}\gammb^{\qba}\kopp.
 \end{gathered}
\end{equation}
By considering the commutator of $\Kpm$ with a function expanded as a polynomial in $\ggb*$,
$f\ofggb = \sum_{\qnb,\qba}\Gqqb\gqgqb$,
we deduce that
 \begin{equation} \label{eq:LHS-spin}
  \comm{\Kpm}{f\ofggb} = -i\xmp\left(\gorgb\dgbg^2 + (\Dopp\mp\lopp)\dgbg\right)f\ofggb.
 \end{equation}
Substituting \eqref{eq:LHS-spin} into \eqref{eq:K-action-spin} (with $f=\Chat$) gives
 \begin{align}
  -i\xmp\Bigl(\gorgb\dgbg^2 + (\Dopp\mp\lopp)\dgbg\Bigr) = -i\xmp\Bigl(\gorgb\dgbg + a \mp \half\lopp\Bigr) &\\[4pt]
  \implies
  \begin{aligned}
   \Bigl\{ \gamma\dgg + (\Dopp+\lopp-\gamma)\dg - (a+\half\lopp) \Bigr\}\Cgamma &= 0,\\
   \Bigl\{ \gammb\dgbgb + (\Dopp-\lopp-\gammb)\dgb - (a-\half\lopp) \Bigr\}\Cgamma &= 0.
 \end{aligned}&
\end{align}
Substituting $\Cgamma=\chip(\gamma)\chim(\gammb)$ gives two confluent hypergeometric equations,
 \begin{equation}
  \begin{aligned}
   \gamma\chip'' + (\Dopp+\lopp-\gamma)\chip' - (a+\half\lopp)\chip &= 0,\\
   \gammb\chim'' + (\Dopp-\lopp-\gammb)\chim' - (a-\half\lopp)\chim &= 0.
 \end{aligned}
\end{equation}
Thus the full solution in two dimensions is given by
 \begin{align}
  \phi(x)\kop
   &= \Cmult x_+^{-\frac12(\Delta-\lopp)}x_-^{-\frac12(\Delta+\lopp)} \chip(\gamma)\chim(\gammb) \kopp\\\notag
   &= \frac{\Cmult x_+^{\lopp}}{(x^2)^{\frac12(\Delta+\lopp)}} \, \hyperf11{a+\half\lopp;\Dopp+\lopp;x_+\dpl}
   \, \hyperf11{a-\half\lopp;\Dopp-\lopp;x_-\dmi} \kopp.
 \end{align}

 \section{The General Case}
 \label{se:OPE-Spin}

In the most general case, working in $d$ dimensions, we expand $\phi(x)\kop$ in terms of a spin-$\ell$ operator $\Op'\muel$, writing
 \begin{equation} \label{eq:l-ope}
  \phi(x)\kop = \CC\muel(x,P)\kopp\muel.
 \end{equation}
The state $\kopp\muel$ obeys the $\SO{d}$ spin algebra,
 \begin{equation}
  M_{\mu\nu}\kopp\muel = i\ell\left(\met\mu{(\mu_1}\mopp{\moo2)\nu} - \met\nu{(\mu_1}\mopp{\moo2)\mu}\right).
 \end{equation}
By considering the action of conformal generators $M_{\mu\nu}$ and $K_\mu$ on \eqref{eq:l-ope}, 
we obtain two equations satisfied by $\CC\muel(x,P)$, namely
 \begin{equation} \label{eq:l-comm-1}
   \bigl(x_\mu\partial_\nu - x_\nu\partial_\mu\bigr)\CC\muel
    = \bigl(P_\nu\dP\mu - P_\mu\dP\nu\bigr)\CC\muel - \ell\bigl(\met\mu{(\mu_1}\CC_{\moo2)\nu} - \met\nu{(\mu_1}\CC_{\moo2)\mu}\bigr)
 \end{equation}
and
 \begin{multline} \label{eq:l-comm-2}
  i\K\mu{\Dphi}\CC\muel 
    = 2\bigl(\Dopp + \PdP\bigr)\dP\mu\CC\muel - P_\mu(\dP{})^2\CC\muel \\
    - 2\ell\,\dP\nu\bigl(\met\mu{(\mu_1}\CC_{\moo2)\nu} - \met\nu{(\mu_1}\CC_{\moo2)\mu}\bigr).
 \end{multline}
We may combine \eqref{eq:l-comm-1}, \eqref{eq:l-comm-2} to eliminate the terms involving symmetrisation over the indices of $\metric$, giving
 \begin{multline} \label{eq:l-MK}
  i\K\mu{\Dphi}\CC\muel \\
   = 2\Bigl((\Dopp-d+1)\dP\mu + \half P_\mu(\dP{})^2 + x_\mu(\dP{}\cdot\partial) - (x\cdot\dP{})\Bigr)\CC\muel.
 \end{multline}
We write $\CC\muel(x,P)$ in a form that is manifestly $M$- and $D$-covariant,
 \begin{equation} \label{eq:l-C-expn}
  \CC\muel(x,P) = 
  \frac1{(x^2)^{\frac12(\Delta+\ell)}}
  \left(\sum_{m=0}^\ell (ix^2)^m \stf{P^mx^{\ell-m}} C_{m+1}(\alpha,\beta)\right),
 \end{equation}
where we use $\stf{P^mx^{\ell-m}}$ to denote $P_{\{\mu_1}\dots P_{\mu_m}x_{\mu_{m+1}}\dots x_{\mu_\ell\}}$, 
with $\tensor_{\{\moo1\}}$ the symmetrised, trace-free part of a rank-$\ell$ $\SO{d}$ tensor $\tensor\muel$.
The definition \eqref{eq:def-ab} enables us to write the action of $x$ and $P$ derivatives on $C_{m+1}(\alpha,\beta)$ in terms of $\da*,\db*$;
thus substituting \eqref{eq:l-C-expn} for $\CC\muel$ in \eqref{eq:l-MK} gives on the left-hand side
 \begin{multline}
  -2i\stf{P^mx^{\ell-m}}x_\mu\left(\alpha\da* + \beta\db* + a\right) \\
  + x^2\stf{P^mx^{\ell-m}}P_\mu\da* 
  + i(\ell-m)x^2\stf{P^mx^{\ell-m-1}\metric}_\mu ,
 \end{multline}
and on the right
 \begin{multline}
  -2i\stf{P^mx^{\ell-m}}x_\mu\left(\alpha\daa* + 2\beta\dab* + (\Dopp-\ell)\da*\right) \\
  + x^2\stf{P^mx^{\ell-m}}P_\mu\left(\daa* + 4\beta\dbb* + 4(\Dopp-\tfrac{d}2+m+1)\db*\right) \\
  + 4(\ell-m)x^2\stf{P^{m+1}x^{\ell-m-1}}x_\mu\db* \qquad\qquad\qquad \\ \qquad\:
  + 2m(\Dopp-d+1-\ell+m)\stf{P^{m-1}x^{\ell-m}\metric}_\mu \\
  + 2i(\ell-m)x^2\stf{P^mx^{\ell-m-1}\metric}_\mu\left(\da* - 2\alpha\db*\right),
 \end{multline}
where $\stf{P^mx^{\ell-m-1}\metric}_\mu$ denotes $P_{\{\mu_1}\dots P_{\mu_m}x_{\mu_{m+1}}\dots x_{\mu_{\ell-1}}\metric_{\mu_\ell\}\mu}$, etc., 
and both expressions should be understood to be inserted into the sum
 \begin{equation}
  \frac1{(x^2)^{\frac12(\Delta+\ell)}} \sum_{m=0}^\ell (ix^2)^m \Bigl[\quad\dotsb\quad\Bigr] C_{m+1}(\alpha,\beta).
 \end{equation}
Combining the two, we obtain
 \begin{align}
  0 =
  \sum_{m=0}^\ell (ix^2)^m &\Bigl[
  x^2\stf{P^mx^{\ell-m}}P_\mu \, \RomI{m+1}
  -  2i\stf{P^mx^{\ell-m}}x_\mu\RomII\ell \Bob~[-4pt]&
  + 4(\ell-m)x^2\stf{P^{m+1}x^{\ell-m-1}}x_\mu\db*  \Bob~[2pt]&
  - 2i(\ell-m)x^2\stf{P^mx^{\ell-m-1}\metric}_\mu\plusO \Bob~[2pt]&
  + 2m(\Dopp-d-\ell+m+1)\stf{P^{m-1}x^{\ell-m}\metric}_\mu \Bigr]C_{m+1}(\alpha,\beta),
 \end{align}
where, with $a = \half(\Dphi-\Dop+\Dopp+\ell)$
--- note we have taken $a \goesto a+\half\ell$ with respect to the definition used in \secref{se:OPE-2d-Spin} ---
we define
 \begin{equation}
  \begin{aligned}
   \RomI{n} &= \hphantom{2\alpha}\daa* + 4\beta\dbb* - \da* + 4(\Dopp-\tfrac{d}2+n)\db*, \\
   \RomII\ell &= \hphantom{2}\alpha\daa* + 2\beta\dab* - \alpha\da* - \beta\db* + (\Dopp+\ell)\da* - a, \\
   \plusO &= 2\alpha\db* - \da* + \frac12 \,.
 \end{aligned}
\end{equation}
\begin{subequations} \label{eq:l-diff}%
We therefore require the $C_m$ to satisfy
 \begin{equation} \label{eq:l-diff-1}
  \RomI1C_1 = \RomII\ell C_1 = 0,
 \end{equation}
and for $n = 2,\dots,\ell+1$,
 \begin{alignat}{2}
  \label{eq:l-diff-nI}
   \RomI{n}C_n &= \hphantom-0; && \\
  \label{eq:l-diff-nII}
   \RomII\ell C_n &= -2k^\ell_n\db* C_{n-1}, &\qquad k^\ell_n &= \ell-n+2; \\
  \label{eq:l-diff-nO}
   \plusO C_n &= \hphantom-K^\ell_{n+1}C_{n+1}, & K^\ell_n &= \tfrac1{k^\ell_n}(n-1)(\Dopp-d+n-\ell).
 \end{alignat}
\end{subequations}%
If we take $C_1$ satisfying \eqref{eq:l-diff-1} and define $C_2,\dots,C_{\ell+1}$ by the action of $\plusO$, 
\eqref{eq:l-diff-nO},
we find that 
\eqref{eq:l-diff-nI}, \eqref{eq:l-diff-nII} hold automatically
(following from the commutators
$\comm{\RomI{n}}{\plusO} = -4\plusO\db*$,
$\comm{\smash{\RomII\ell}}{\plusO} = \RomI{n} - 2(\Dopp-d+2n-\ell)\db*$,
$\comm{\plusO}{\smash{\db*}} = 0$,
and the recurrence relation $\RomI{n+1} = \RomI{n} + 4\db*$).
In terms of $\ggb*$ (recalling $\alpha = \gpgb$, $\beta = -4\ggb$),
 \begin{gather}
  \begin{aligned}
   \RomI{n}   &= \frac1\gmgb\Bigl[\gamma\dgg - \gammb\dgbgb - \gamma\dg + \gammb\dgb + (\Dopp-\eps+n-1)\left(\dg-\dgb\right)\Bigl],
\\ \RomII\ell &= \frac1\gmgb\Bigl[\gamma^2\dgg-\gammb^2\dgbgb-\gamma^2\dg+\gammb^2\dgb+(\Dopp+\ell)\left(\gamma\dg-\gammb\dgb\right)-a(\gmgb)\Bigr],
  \end{aligned} \Bob/ \label{eq:l-diff-ggb}
  \plusO = -\frac12\left[\dg+\dgb-1\right],
\qquad
  -2\db* = -\frac12\frac1\gmgb\left[\dg-\dgb\right],
 \end{gather}
where we define $\eps \equiv \half d - 1$.
By considering $\RomII\ell - \gammb\,\RomI1$, $\RomII\ell - \gamma\,\RomI1$ on $C_1$, we obtain
 \begin{equation} \label{eq:I-II-C1}
  \begin{aligned}
   \biggl[\gamma\dgg - \gamma\dg + (\Dopp-\eps)\dg + \frac{\ell+\eps}\gmgb\left(\gamma\dg-\gammb\dgb\right)\biggr]C_1 &= aC_1, \\
   \biggl[\gammb\dgbgb - \gammb\dgb + (\Dopp-\eps)\dgb + \frac{\ell+\eps}\gmgb\left(\gamma\dg-\gammb\dgb\right)\biggr]C_1 &= aC_1.
  \end{aligned}
 \end{equation}

\subsection{Polynomial expansions}

We look to expand the $C_n$ as polynomials for general values of $\eps, \ell$.
We may solve \eqref{eq:l-diff-1} in terms of powers of $\alpha,\beta$, to find
 \begin{equation} \label{eq:l-expn-ab}
  C_1 = \sum_{r,s} \frac1{4^sr!s!}\frac{\pochhammer{a}{s+r}\pochhammer{b}{s}}{\pochhammer{\Dopp+\ell}{2s+r}\pochhammer{\Dopp-\eps}{s}} \alpha^r\beta^s,
 \end{equation}
with $b=\Dopp+\ell-a$.
A direct expansion in terms of powers of $\ggb*$ proves harder to obtain, but we are able to express $C_1$ using Jack polynomials
$\Jack{\lambda}mn\ofggb$, properties of which are given in \cite
{rf:0309180}. 
In particular, they obey recurrence relations
 \begin{gather}
  \bigl(\dg+\dgb)\Jack\lambda{m}n\ofggb
   = \frac{(m-\lambda)(m-n)}{m-n+\lambda}\Jack\lambda{m-1}n\ofggb + \frac{n(m-n+2\lambda)}{m-n+\lambda}\Jack\lambda{m}{n-1}\ofggb,
\Bob/[4pt] \label{eq:Jack-Recur}
  (\ggb)^r\Jack\lambda{m}n\ofggb = \Jack\lambda{m+r}{n+r}\ofggb,
 \end{gather}
and the eigenfunction equations
 \begin{equation} \label{eq:Jack-Eigen}
  \begin{gathered}
   \bigl(\gamma\dg+\gammb\dgb\bigr)\Jack\lambda{m}n\ofggb = (m+n)\Jack\lambda{m}n\ofggb,
\\[4pt]
   \JD\lambda\Jack\lambda{m}n\ofggb = \bigl(m(m+2\lambda-1)+n(n-1)\bigr)\Jack\lambda{m}n\ofggb,
  \end{gathered}
 \end{equation}
where the differential operator $\JD\lambda$ is given by
 \begin{equation} \label{eq:Jack-JDDefn}
  \JD\lambda = \bigl(\gamma^2\dgg + \gammb\dgbgb\bigr) + \frac{2\lambda}\gmgb\bigl(\gamma^2\dg - \gammb^2\dgb\bigr).
 \end{equation}
Making the general expansion
$C_1\ofggb = \sum_{m,n} a_{mn} \Jack{\ell+\eps}mn\ofggb$
and expressing $\RomII\ell,\RomI{n}$ in terms of the operators appearing in \eqref{eq:Jack-Recur} and \eqref{eq:Jack-Eigen},
we find that \eqref{eq:l-diff-1} has solution
 \begin{equation} \label{eq:l-expn-Jack}
  C_1 = \sum_{m\ge n\ge0}
  \frac{\ell+\eps+m-n}{(m-n)!n!}
  \frac{\pochhammer{2\ell+2\eps}{m-n}}{\pochhammer{\ell+\eps}{m+1}}
  \frac{\pochhammer{a}m\pochhammer{a-\ell-\eps}n}{\pochhammer{\Dopp+\ell}m\pochhammer{\Dopp-\eps}n} \Jack{\ell+\eps}mn\ofggb.
 \end{equation}
Explicit formul\ae{} for the $\Jack\lambda{m}n$ at low $\lambda$ 
may be used to write $C_1$ for certain $\ell, \eps$ as power expansions in $\ggb*$,
which we may subsequently identify as combinations of hypergeometric functions, 
e.g. $\Jack1mn, \Jack2mn$ respectively give
\begin{itemize}
 \item{The spinless, 4-dimensional case ($\ell=0, \eps=1$):}
 \begin{equation}
  C_1\ofggb = \frac{\Dopp-1}\gmgb\Bigl(\hyperf11{a;\Dopp-1;\gamma}\hyperf11{a-1;\Dopp-1;\gammb} - \gswapgb\Bigr);
 \end{equation}
 \item{The spin-one case in 4 dimensions ($\ell=\eps=1$):}
 \begin{align}
  C_1 &= \frac{\Dopp(\Dopp-1)}{(\gmgb)^3}\Bigl(\E02 + \E20 - \frac2{a-1}\E12 - 2\frac{a-2}{a-1}\E11 \Bigr), \\
  \E{p}q &= \gamma\hyperf11{a-p;\Dopp-1;\gamma}\hyperf11{a-q;\Dopp-1;\gammb} - \gswapgb.
 \end{align}
\end{itemize}

\subsection{Equivalence in $\eps=0$ case with results of \secref{se:OPE-2d-Spin}} \label{se:OPE-2d-Spinl}

We will now check that, working in two dimensions, \eqref{eq:l-diff} and \eqref{eq:l-diff-ggb} imply that 
\mbox{$C\ofggb \equiv C_1 - 2\gammb C_2 + 4\gammb^2C_3 - \dots + (-2\gammb)^\ell C_{\ell+1}$} 
may be identified with the operator $\Chat\ofggb$ appearing in \secref{se:OPE-2d}, i.e. that $C$ satisfies
 \begin{equation} \label{eq:l-ssb}
  \begin{aligned}
   \Bigl[\gamma\dgg - \gamma\dg + (\Dopp+\ell)\dg - a\Bigr]C
    &\equiv \starl C = 0, \\
   \Bigl[\gammb\dgbgb - \gammb\dgb + (\Dopp-\ell)\dgb - (a-\ell)\Bigr]C
    &\equiv \starbl C = 0.
 \end{aligned}
\end{equation}
To show this, we note firstly that
 \begin{equation}
  \starl  = \RomII\ell - \gammb\,\RomI{\ell+1},
\qquad
  \starbl = \RomII\ell - \gamma\,\RomI{\ell+1} - \ell\bigl(\dgb-1\bigr).
 \end{equation}
Also, we observe that from the definition \eqref{eq:l-ssb} we have
$\comm{\starl}{\gammb^m} = 0$,
and that for $C_n$ satisfying \eqref{eq:l-diff}, 
$\RomI{m}C_n = \frac{m-n}\gmgb\left(\dg-\dgb\right)C_n$.
Hence, with $a_n=(-2)^{n-1}$, we have
 \begin{align}
  \starl C
  &= \sum_{n=1}^{\ell+1} a_n\starl\left(\gammb^{n-1}C_n\right)
   = \sum_{n=1}^{\ell+1} a_n\gammb^{n-1}(\RomII\ell - \gammb\,\RomI{\ell+1})C_n 
\Bob~
  &= \sum_{n=2}^{\ell+1} a_n\gammb^{n-1}\frac{(-\frac12k_n^\ell)}\gmgb \bigl(\dg-\dgb\bigr)C_{n-1}
   - \sum_{n=1}^{\ell+1} a_n\gammb^n
\frac{\ell-n+1}\gmgb
\bigl(\dg-\dgb\bigr)C_n
\Bob~
  &= \sum_{n=1}^\ell (\half a_{n+1}+a_n)\gammb^n\frac{n-\ell-1}\gmgb\bigl(\dg-\dgb\bigr)C_n,
 \end{align}
where in the last step we have used $k_n^\ell = \ell-n+2$.
Thus $\starl C$ vanishes identically, since $a_{n+1} = -2a_n$. 
To show that the same is true for $\starbl C$, we first find
 \begin{equation}
  \comm{\starbl}{\gammb^m} = m\gammb^{m-1}\bigl(\gammb(2\dgb-1) + (\Dopp-\ell+m-1)\bigr).
 \end{equation}
Hence we find
 \begin{align}
  \starbl C
  &= \sum_{n=1}^{\ell+1} a_n\Bigl(\gammb^{n-1}\starbl + \comm\starbl{\gammb^{n-1}} \Bigr)C_n\Bob/
  &= \sum_{n=1}^{\ell+1} a_n\biggl(\gammb^{n-1}\left[\frac{(-\frac12k_n^\ell)}\gmgb\bigl(\dg-\dgb\bigr)C_{n-1}
  - \gamma\frac{\ell-n+1}\gmgb\bigl(\dg-\dgb\bigr)C_n
  - \ell\bigl(2\dgb-1\bigr)C_n\right]
  \Bob~[-4pt]&\mspace{84mu}
  + (n-1)\gammb^{n-2}\bigl[\gammb(2\dgb-1) + (\Dopp-\ell+n-2)\bigr]C_n\biggr) \Bob/[-4pt]
  &= -\sum_{n=2}^{\ell+1} \half a_n \gammb^{n-1} \frac{\ell-n+2}\gmgb\bigl(\dg-\dgb\bigr)C_{n-1}
  - \sum_{n=1}^\ell a_n\gamma\gammb^{n-1}\frac{\ell-n+1}\gmgb\bigl(\dg-\dgb\bigr)C_n
  \Bob~[-8pt]&\mspace{84mu}
  - \sum_{n=1}^\ell a_n\gammb^{n-1}(\ell-n+1)\bigl(2\dgb-1\bigr)C_n
  \Bob~[-8pt]&\mspace{168mu}
  + \sum_{n=2}^{\ell+1} a_n\gammb^{n-2}\underbrace{(n-1)(\Dopp-\ell+n-2)}_{=k_n^\ell K_n^\ell}C_n \Bob/[-8pt]
  &= \sum_{n=1}^\ell -(\overbrace{\half a_{n+1}\gammb^n+a_n\gamma\gammb^{n-1}}^{=a_n(\gmgb)\gammb^{n-1}}) \frac{\ell-n+1}\gmgb\bigl(\dg-\dgb\bigr)C_n
  \Bob~[-10pt]&\mspace{84mu}
  - \sum_{n=1}^\ell a_n\gammb^{n-1}(\ell-n+1)\bigl(2\dgb-1\bigr)C_n
  + \sum_{n=1}^\ell a_{n+1}\gammb^{n-1}k_{n+1}^\ell K_{n+1}^\ell C_{n+1}\Bob/[-6pt]
  &= \sum_{n=1}^\ell -a_n\gammb^{n-1}(\ell-n+1)\underbrace{\bigl(\dg+\dgb-1\bigr)C_n}_{= -2\plusO C_n}
  + a_{n+1}(\ell-n+1)\underbrace{K_{n+1}^\ell C_{n+1}}_{= \plusO C_n} \Bob/[-10pt]
  &= \sum_{n=1}^\ell (\overbrace{2a_n+a_{n+1}}^{=0})(\ell-n+1)\gammb^{n-1}\plusO C_n = 0.
 \end{align}
As shown in \secref{se:OPE-2d-Spin}, we can write an explicit solution to \eqref{eq:l-ssb} in terms of hypergeometric functions,
 \begin{equation}
  C\ofggb = C_1 - 2\gammb C_2 + \dots = \hyperf11{a;\Dopp+\ell;\gamma} \hyperf11{a-\ell;\Dopp-\ell;\gammb}.
 \end{equation}

\subsection{Generalising the $\eps=0$ case to higher dimensions}

Let us define two new variables,
 \begin{equation} \label{eq:l-defn-Dl}
  \Dl = \Dopp + \ell \qquad\text{and}\qquad \lambda = \ell+\eps.
 \end{equation}
We note that the expansions on the right-hand sides of \eqref{eq:l-expn-ab} and \eqref{eq:l-expn-Jack} 
may both be re-written in terms of $\Dl, \lambda$ only, becoming respectively
 \begin{equation}
  \sum_{r,s} \frac1{4^sr!s!}\frac{\pochhammer{a}{r+s}}{\pochhammer\Dl{2s+r}}\frac{\pochhammer{\Dl-a}s}{\pochhammer{\Dl-\lambda}s} \alpha^r\beta^s
 \end{equation}
and
 \begin{equation}
  \frac{(\lambda-1)!}{(2\lambda-1)!} \sum_{m\ge n\ge0}
  \frac{m-n+\lambda}{(\lambda+m)!n!}\frac{\pochhammer{a}m\pochhammer{a-\lambda}n}{\pochhammer\Dl{m}\pochhammer{\Dl-\lambda}n}
  \pochhammer{m-n+1}{2\lambda-1} \Jack\lambda{m}n\ofggb.
 \end{equation}
We deduce that the value $C_1$ depends only on $\lambda$, rather than independently on $\ell, \eps$,
if we also characterise the field $\Op'$ in terms of $\Dl$, equal to the sum of its spin and its scaling dimension.
Hence $C_1$ in the general $\ell, \eps$ case is identical to the $C_1$ obtained in the 2-dimensional, spin-$\lambda$ case, 
for the contribution of an operator $\Op'$ with the same value of $\Dl$.

Furthermore, the operator $\plusO$, which by its action on the $C_n$ we may think of as a raising operator, 
is independent of $\ell, \eps,$ and $\Dopp$.
Hence, if for fixed $\Dl$ we write $C^\elep_n$ to denote $C_n$ in the spin-$\ell$, $2(\eps+1)$-dimensional expansion, 
using \eqref{eq:l-diff-nO} and $C^\elep_1 = C^\lamz_1$ we deduce, for $0\le n\le\lambda$,
 \begin{equation}
  K^\elep_{n+1}\dotsm K^\elep_2 C^\elep_{n+1}
  = \bigl[\plusO\bigr]^n C^\elep_1
  = \bigl[\plusO\bigr]^n C^\lamz_1
  = K^\lamz_{n+1}\dotsm K^\lamz_2 C^\lamz_2
 \end{equation}
where
 \begin{equation}
  K^\elep_n = \frac{n-1}{\ell-n+2}\bigl(\Dl-2(\ell+\eps+1)+n\bigr)
 \end{equation}
and hence
 \begin{equation}
  C^\elep_{n+1}
  = \frac{K^\lamz_{n+1}\dotsm K^\lamz_2}{K^\elep_{n+1}\dotsm K^\elep_2} \, C^\lamz_{n+1}
  = \frac{(\ell-n+1)\dotsm(\ell-1)\ell}{(\lambda-n+1)\dotsm(\lambda-1)\lambda} \, C^\lamz_{n+1}
  = \frac{\pochhammer{-\ell}n}{\pochhammer{-\lambda}n} \, C^\lamz_{n+1}.
 \end{equation}
The $C^\lamz_n$ are the 2-dimensional, spin-$\lambda$ expansions, which we recall from \secref{se:OPE-2d-Spinl} satisfy
 \begin{align}
  \begin{aligned}[b]
   C\ofggb &= C^\lamz_1 - 2\gammb C^\lamz_2 + 4\gammb^2 C^\lamz_3 - \dots + (-2\gammb)^\lambda C^\lamz_{\lambda+1} \\
   &= \hyperf11{\smash{a;\Dl;\gamma}}\hyperf11{\smash{a-\lambda;\Dl-2\lambda;\gammb}}.
 \end{aligned}& \\ 
\intertext{Since the $C_n$ were originally functions of $\alpha, \beta$, we know them to be symmetric in $\ggb*$; hence we may also write}
 \begin{aligned}[b]
 \Ct\ofggb
   = C(\gammb,\gamma)
  &= C^\lamz_1 - 2\gamma C^\lamz_2 + 4\gamma^2 C^\lamz_3 - \dots + (-2\gamma)^\lambda C^\lamz_{\lambda+1} \\
  &= \hyperf11{\smash{a-\lambda;\Dl-2\lambda;\gamma}}\hyperf11{\smash{a;\Dl;\gammb}}.
 \end{aligned} &
\end{align}

\subsection{Calculation of some $C_n$}

The above tells us that once we have found the $C_n$ for the $d=2$, spin-$\lambda$ case, 
we can find $C_n$ for all higher-, even-dimensional cases where $\ell + \eps = \lambda$ and parameters $a$ and $\Dl\equiv\Dopp+\ell$ remain fixed.
A pictorial representation of the process is given in \figref{fg:Cn}.
\begin{figure}[tb!]
 \caption{Calculating $C_n$} \label{fg:Cn}
 \begin{equation*}
  \begin{array}{ccccccccccc}
   d=2 && d=4 && && \makebox[0.7\width]{$d=2(\eps+1)$} && && \makebox[0.8\width]{$d=2(\lambda+1)$} \\
   \text{spin-}\lambda && \makebox[0.8\width]{spin-$(\lambda-1)$} && && \text{spin-}\ell && && \text{spinless} \\[6pt]
   \raisebox{0pt}[0pt][0pt]{\framebox[1.4\width]{\rule[-45mm]{0pt}{0cm}$C^\lamz_1$}}
   \raisebox{5pt}[0pt][0pt]{\makebox[0pt]{\hspace{-80pt}$\xymatrix{{}\ar@/_0.8pc/[d]_{\plusO\!}\\{}}$}}
   &=& C^{\lambda-1,1}_1 &=& \dotsb &=& C^\elep_1 &=& \dotsb &=& C^{0,\lambda}_1 \\[8pt]
   C^\lamz_2 &\propto& C^{\lambda-1,1}_2 &\propto& \dotsb &\propto& C^\elep_2 &\propto& \dotsb \\[8pt]
   C^\lamz_3 &\propto& C^{\lambda-1,1}_3 &\propto& \dotsb \\[6pt]
   \vdots & \vdots & \vdots & \raisebox{1.12ex}{$\colon$} & \iddots 
   && \multicolumn{5}{c}{\boxed{C^\elep_n = \tfrac{\pochhammer{-\ell}{n-1}}{\pochhammer{-\lambda}{n-1}}C^\lamz_n}} \\[6pt]
   C^\lamz_\lambda &\propto& C^{\lambda-1,1}_\lambda \\[8pt]
   C^\lamz_{\lambda+1} & \leftrightsquigarrow &
   \multicolumn{9}{l}{\boxed{C^\lamz_1 - 2\gammb C^\lamz_2 + 4\gammb^2 C^\lamz_3 - \dots + (-2\gammb)^\lambda C^\lamz_{\lambda+1} = C\ofggb}}
 \end{array}
 \end{equation*}
\end{figure}
We find explicit results in the cases $\lambda=1,2$:

\paragraph{\boldmath $\lambda=1$ ($\text{Spinless}, d=4 \leftrightarrow \text{Spin-1}, d=2$):}

$C = C_1 - 2\gammb C_2$, and hence
 \begin{equation}
  (\gmgb)C_1 = \gamma C - \gammb\Ct
  = \gamma\hyperf11{\smash{a;\Dl;\gamma}}\hyperf11{\smash{a-1;\Dl-2;\gammb}} - \gswapgb.
 \end{equation}
Using standard identities for the confluent hypergeometric function $\HYPERF11$ (see \secref{se:OPE-F11}), 
we hence find, for $\eps=1, \ell=0$ (and thus $\Dl=\Dopp+\ell=\Dopp$),
 \begin{equation}
  C_1\ofggb = \frac{\Dopp-1}\gmgb\Bigl(\hyperf11{a;\Dopp-1;\gamma}\hyperf11{a-1;\Dopp-1;\gammb} - \gswapgb\Bigr)
 \end{equation}
and for $\eps=0, \ell=1 \implies \Dl=\Dopp+1$,
 \begin{equation}
  \begin{alignedat}{2}
   C_1\ofggb &=\, &\frac{\Dopp-1}\gmgb&\Bigl(\hyperf11{a;\Dopp;\gamma}\hyperf11{a-1;\Dopp;\gammb} - \gswapgb\Bigr), \\
   C_2\ofggb &= &\frac1\gmgb&\Bigl(\hyperf11{a;\Dopp+1;\gamma}\hyperf11{a-1;\Dopp-1;\gammb} - \gswapgb\Bigr).
 \end{alignedat}
\end{equation}

\paragraph{\boldmath $\lambda=2$ ($\text{Spin-1}, d=4 \leftrightarrow \text{Spin-2}, d=2$):}

For $\lambda=2$, we have $C=C_1-2\gammb C_2+4\gammb^2C_3=\hyperf11{\smash{a;\Dl;\gamma}}\hyperf11{\smash{a-2;\Dl-4;\gammb}}$, 
and similarly for $\Ct$.
Elimination gives
 \begin{equation}
  \begin{alignedat}{2}
   C&-\Ct &&= (\gmgb)\left(2C_2-4(\gpgb)C_3\right), \\
   \gamma C&-\gammb\Ct &&= (\gmgb)\left(C_1-4\ggb C_3\right), \\
   \gamma^2C&-\gammb^2\Ct &&= (\gmgb)\left((\gpgb)C_1-2\ggb C_2\right).
  \end{alignedat}
 \end{equation}
These three are not independent; however we may apply the raising operator $\plusO$ to the last of them, giving
 \begin{align}
  \plusO\left(\frac{\gamma^2 C-\gammb^2\Ct}\gmgb\right)
  &= \comm\plusO\gpgb C_1 + (\gpgb)\plusO C_1 - 2\comm\plusO\ggb C_2 - 2\ggb\plusO C_2 \Bob~[-8pt]
  &= -C_1 + \half(\gpgb)(\Dl-4)C_2 + (\gpgb)C_2 - 4\ggb(\Dl-3)C_3 \Bob~
  &= -C_1 + \half(\Dl-2)(\gpgb)C_2 - 4\ggb(\Dl-3)C_3
  \equiv Q.
 \end{align}
We may use $Q$ together with $C$, $\Ct$, to eliminate each of the components.
In particular, we find (noting $\comm\plusO\gmgb=0$)
 \begin{equation}
  \begin{gathered}
   (\Dl-2)(\gpgb)\frac{\gamma^2C-\gammb^2\Ct}\gmgb - 4\ggb\left((\Dl-3)\frac{\gamma C-\gammb\Ct}\gmgb - Q\right)
   = (\Dl-2)(\gmgb)^2C_1 \\
  \implies
   C_1
   =\frac1{(\gmgb)^3}\left(\Bigl(\gpgb+\frac{4\ggb}{\Dl-2}\plusO\Bigr)(\gamma^2C-\gammb^2\Ct)-4\ggb\frac{\Dl-3}{\Dl-2}(\gamma C-\gammb\Ct)\right).
  \end{gathered}
 \end{equation}
In terms of hypergeometric functions, if we denote 
$\Fg{a}\Dl\equiv\hyperf11{\smash{a;\Dl;\gamma}}$, $\Fgb{a}\Dl\equiv\hyperf11{\smash{a;\Dl;\gammb}}$, etc, 
then by substituting our explicit expression for $\plusO$ from \eqref{eq:l-diff-ggb} into the above, we may obtain
 \begin{equation}
  C_1
  = \frac1{(\gmgb)^3}\left(\gamma^2(\gamma-3\gammb)\Fg{a}\Dl\Fgb{a-2}{\Dl-4}
  - \tfrac{2\gamma^3\gammb}{\Dl-2}\Bigl(\tfrac{a-\Dl}\Dl\Fg{a}{\Dl+1}\Fgb{a-2}{\Dl-4}
  + \tfrac{a-2}{\Dl-4}\Fg{a}\Dl\Fgb{a-1}{\Dl-3}\Bigr) - \gswapgb \right).
 \end{equation}
We may subsequently apply $\plusO$ again, leading to
 \begin{multline}
  \plusO C_1 = \frac1{(\gmgb)^3}\biggl( 3\ggb\Fg{a}\Dl\Fgb{a-2}{\Dl-4} 
  - \tfrac{(\Dl-4)\gamma-3\Dl\gammb}{\Dl-2}\gamma^2
    \left(\tfrac{a-\Dl}\Dl\Fg{a}{\Dl+1}\Fgb{a-2}{\Dl-4}+\tfrac{a-2}{\Dl-4}\Fg{a}\Dl\Fgb{a-1}{\Dl-3}\right) \\
  + \tfrac\gammb{\Dl-2}\gamma^3\left(\tfrac{(a-\Dl)(a-\Dl-1)}{\Dl(\Dl+1)}\Fg{a}{\Dl+2}\Fgb{a-2}{\Dl-4}+%
    \tfrac{2(a-\Dl)(a-2)}{\Dl(\Dl-4)}\Fg{a}{\Dl+1}\Fgb{a-2}{\Dl-3}+\tfrac{(a-1)(a-2)}{(\Dl-3)(\Dl-4)}\Fg{a}\Dl\Fgb{a}{\Dl-2}\right) \\
  - \gswapgb\biggr).
 \end{multline}
In the spin-1, $d=4$ case, $C_2$ is given by $(\Dl-4)C_2=\plusO C_1$, with $\Dl=\Dopp+\ell=\Dopp+1$.

 \section{Useful Results}
 \label{se:OPE-Results}

\subsection{Complex co-ordinates in two dimensions} \label{se:2dCoords}

from the definition of $x_+, x_-$ in \secref{se:OPE-2d}, we have the results
 \begin{gather}
  \begin{alignedat}{2}
   x_+ &= z = x_1 + ix_2 ,
   &\qquad x_1 &= \half(x_+ + x_-) ,\\
   x_- &= \zb = x_1 - ix_2 , 
   & x_2 &= \tfrac1{2i}(x_+ - x_-) ;
 \end{alignedat} \\[6pt]
 \begin{alignedat}{2}
  \partial_+ &= \partial_z = \half(\partial_1 - i\partial_2) ,
  &\qquad \partial_1 &= \partial_+ + \partial_- ,\\
  \partial_- &= \partial_{\zb} = \half(\partial_1 + i\partial_2) ,
  & \partial_2 &= i(\partial_+ - \partial_-) ;
 \end{alignedat} \\[6pt]
 \begin{gathered}
  \xd = x_1\partial_1 + x_2\partial_2 = x_+\partial_+ + x_-\partial_- ,\\
  x_1\partial_2 - x_2\partial_1 = i(x_+\partial_+ - x_-\partial_-) ;
 \end{gathered} \\[6pt]
 \begin{alignedat}{2}
  \gamma &= x_+\partial_+ = -ix_+P_+ ,
  \quad
  \gammb &= x_-\partial_- = -ix_-P_- .
 \end{alignedat}
\end{gather}

\subsection{$\comm{K}{P^r}$ in two dimensions} \label{se:OPE-2d-KP}

We here derive the identity \eqref{eq:2d-KPP}, 
beginning by proving, by induction, $\comm{K_+}{P_-^r} = rP_-^{r-1}(r-1+\Dt-J)$.
The $r=1$ case follows directly from the two dimensional conformal algebra \eqref{eq:2d-algebra}, which gives
 \begin{equation}
  \comm{K_+}{P_-} = -J+\Dt = 1\cdot P_-^{1-1}(1-1+\Dt-J).
 \end{equation}
Then, assuming the hypothesis holds for $\comm{K_+}{P_-^r}$, we have
 \begin{equation}
  \begin{aligned}[b]
   \comm{K_+}{P_-^{r+1}}
   &= \comm{K_+}{P_-^r}P_- + P_-^r\comm{K_+}{P_-} \\
   &= rP_-^{r-1}(r-1+\Dt-J)P_- + P_-^r(-J+\Dt) \\
   &= r(r-1)P_-^r + (r+1)P_-^r(\Dt-J) + rP_-^{r-1}\underbrace{\comm{\Dt-J}{P_-}}_{=2P_-} \\[-2ex]
   &= (r+1)P_-^r(r+\Dt-J) \QED.
 \end{aligned}
\end{equation}
A similar result follows for $\comm{K_-}{P_+^r}$; thus
 \begin{equation}
  \comm{\Kpm}{\Pmp^r} = r\Pmp^{r-1}(r-1+\Dt\mp J).
 \end{equation}
The quoted result then follows, bearing in mind that $\comm{\Kpm}{\Ppm}\equiv 0$.

\subsection{Some useful results involving $\hyperf11{a;\Dopp;\gamma}$} \label{se:OPE-F11}
The confluent hypergeometric function $\hyperf11{a;\Dopp;\gamma}$ satisfies
 \begin{equation}
  \left[\gamma\dgg* + (\Dopp-\gamma)\dg* - a\right]\hyperf11{a;\Dopp;\gamma} = 0,
 \end{equation}
and has series expansion
 \begin{equation}
  \hyperf11{a;\Dopp;\gamma} = \sum_{k=0}^\infty\frac{\pochhammer{a}{k}}{\pochhammer\Dopp{k}}\frac{\gamma^k}{k!}.
 \end{equation}
We have recurrence relations
 \begin{align}
  \dg*\hyperf11{a;\Dopp;\gamma} &= \frac{a}\Dopp\hyperf11{a+1;\Dopp+1;\gamma}; \\
  \hyperf11{a;\Dopp-1;\gamma} &= \hyperf11{a;\Dopp;\gamma} + \gamma\frac{a}{\Dopp(\Dopp-1)}\hyperf11{a+1;\Dopp+1;\gamma}; \\
  \left(\textstyle\dg*-1\right)\hyperf11{a;\Dopp-1;\gamma}
  &= \frac{a}{\Dopp-1}\hyperf11{a+1;\Dopp;\gamma} - \hyperf11{a;\Dopp-1;\gamma} \notag\\
  &= \frac{a-\Dopp+1}{\Dopp-1}\hyperf11{a;\Dopp;\gamma}.
 \end{align}
Finally, writing $\Fg{a}\Delta \equiv \hyperf11{a;\Delta;\gamma}$, we have
 \begin{align}
  \gamma^n\Fg{a}\Delta
   &= \tfrac{\falling{(\Delta-1)}{n+1}}{\falling{(a-1)}n}\sum_{j=0}^n(-1)^{n-j}\textstyle\binom{n}j\falling{(\Delta-j-2)}{n-1}\Fg{a-n}{\Delta-n-j}
\notag\\
   &= \falling{(\Delta-1)}n\sum_{j=0}^n(-1)^j\textstyle\binom{n}j\Fg{a-j}{\Delta-n},
 \end{align}
where $\falling{x}n$ denotes the falling factorial of $x$, $\falling{x}n = x(x-1)\dots(x-n+1) = (-1)^n\pochhammer{-x}n$.

\backmatter
%

\setlength{\parskip}{0ex}
\begin{singlespacing}

\newcommand{\Bibentry}[5][.]{%
 \ifcomma{#2}%
 \ifcomma[\textit]{#3}%
 #4\ifthenelse{\equal{#4}{} \OR \equal{#5}{}}{}{, }{\small{\textsf{#5}}}#1}
\newcommand{\hep}[1]{\mbox{arXiv:hep-th/#1}}
\renewcommand{\bibname}{References}
\Bibitem{rf:AP76/161}{\Bibentry%
{S.~Ferrara, R.~Gatto and A.F.~Grillo}%
{Tensor Representations of Conformal Algebra}%
{Ann.~Phys.~76 (1973) 161}%
{}}
\Bibitem{rf:9108028}{\Bibentry%
{P.~Ginsparg}%
{Applied Conformal Field Theory}%
{}%
{\hep{9108028}}}
\Bibitem{rf:9711200}{\Bibentry%
{J.M.~Maldacena}%
{The Large $N$ Limit of Superconformal Field Theories and Supergravity}%
{Adv.~Theor.~Math.~Phys.~2 (1998) 231-252; Int.~J.~Theor.~Phys.~38 (1999) 1113-1133}%
{\hep{9711200}}}
\Bibitem{rf:9903196}{\Bibentry%
{E.~D'Hoker, D.Z.~Freedman, S.D.~Mathur, A.~Matsusis and L.~Rastelli}%
{Graviton exchange and complete four-point functions in the AdS/CFT correspondence}%
{Nucl.~Phys.~B562 (1999) 353-394}%
{\hep{9903196}}}
\Bibitem{rf:0006098}{\Bibentry%
{F.A.~Dolan and H.~Osborn}%
{Implications of N=1 Superconformal Symmetry for Chiral Fields}%
{Nucl.~Phys.~B593 (2001) 599-633}%
{\hep{0006098}}}
\Bibitem{rf:0011040}{\Bibentry%
{F.A.~Dolan and H.~Osborn}%
{Conformal Four Point Functions and the Operator Product Expansion}%
{Nucl.~Phys.~B599 (2001) 459-496}%
{\hep{0011040}}}
\Bibitem{rf:0209056}{\Bibentry%
{F.A.~Dolan and H.~Osborn}%
{On Short and Semi-Short Representations for Four Dimensional Superconformal Symmetry}%
{Annals Phys. 307 (2003) 41-89}%
{\hep{0209056}}}
\Bibitem{rf:0212116}{\Bibentry%
{G.~Arutyunov, F.A.~Dolan, H.~Osborn and E.~Sokatchev}%
{Correlation Functions and Massive Kaluza-Klein Modes in the AdS/CFT Correspondence}%
{Nucl.~Phys.~B665 (2003) 273-324}%
{\hep{0212116}}}
\Bibitem{rf:0309180}{\Bibentry%
{F.A.~Dolan and H.~Osborn}%
{Conformal Partial Waves and the Operator Product Expansion}%
{Nucl.~Phys.~B678 (2004) 491-507}%
{\hep{0309180}}}
\Bibitem{rf:0407060}{\Bibentry%
{M.~Nirschl and H.~Osborn}%
{Superconformal Ward Identities and their Solution}%
{Nucl.~Phys.~B711 (2005) 409-479}%
{\hep{0407060}}}
\Bibitem{rf:0412335}{\Bibentry%
{F.A.~Dolan and H.~Osborn}%
{Conformal Partial Wave Expansions for $\N=4$ Chiral Four Point Functions}%
{Annals Phys. 321 (2006) 581-626}%
{\hep{0412335}}}
\Bibitem{rf:0504061}{\Bibentry%
{M.~D'Alessandro and L.~Genovese}%
{A wide class of four point functions of BPS operators in $\N=4$ SYM at order $g^4$}%
{Nucl.~Phys.~B732 (2006) 64-88}%
{\hep{0504061}}}
\Bibitem{rf:0601148}{\Bibentry%
{F.A.~Dolan, M.~Nirschl and H.~Osborn}%
{Conjectures for Large $N$ $\N=4$ Superconformal Chiral Primary Four Point Functions}%
{Nucl.~Phys.~B749 (2006) 109-152}%
{\hep{0601148}}}
\Bibitem{rf:0609179}{\Bibentry%
{M.~Bianchi, F.A.~Dolan, P.J.~Heslop and H.~Osborn}%
{$\N=4$ Superconformal Characters and Partition Functions}%
{Nucl.~Phys.~B767 (2007) 163-226}%
{\hep{0609179}}}
\Bibitem{rf:MW-Bell}{\Bibentry[;\\]%
{L.~Lov\'asz}%
{Combinatorial Problems and Exercises}%
{(Amsterdam; Oxford: North-Holland, 1979)}%
{}%
\Bibentry%
{}%
{Bell Number}%
{Wolfram Mathworld}%
{http://mathworld.wolfram.com/BellNumber.html}}
\Bibitem{rf:tables}{\Bibentry%
{I.S.~Gradshteyn and I.M.~Ryzhik, ed.~A.~Jeffrey}%
{Table of Integrals, Series, and Products}%
{4th ed., (New York ; London : Academic Press, 1980)}%
{}}
\Bibitem{rf:linda}{\Bibentry%
{L.I.~Uruchurtu}%
{Correlators of different weight primaries in the supergravity approximation}%
{work in preparation}%
{}}

\TheBibliography

\end{singlespacing}

\end{document}